\documentclass[12pt]{iopart}
\usepackage{amssymb}
\usepackage{iopams}
\usepackage{makeidx}
\usepackage{graphicx}

\begin{document}

\title[How chemistry controls electron localization in 3d$^{1}$ perovskites]{%
How chemistry controls electron localization in 3d$^{1}$ perovskites: a
Wannier-function study.}
\author{E. Pavarini$^{1},$\ A. Yamasaki$^{2},$ J. Nuss$^{2},$ and O. K.
Andersen.$^{2}$}
\address{$^{1}$INFM and Dipartimento di Fisica ``A.Volta,'' 
         Universit\`{a} di Pavia, Via Bassi 6, I-27100 Pavia, Italy} 
\address{$^{2}$Max-Planck-Institut f\"{u}r Festk\"{o}rperforschung,
         Heisenbergstrasse 1, D-70569 Stuttgart, Germany} 
\eads{\mailto{eva.pavarini@pv.infn.it}, \mailto{A.Yamasaki@fkf.mpg.de},
      \mailto{oka@fkf.mpg.de}} 

\begin{abstract}
In the series of 3$d\left( t_{2g}\right) ^{1}$ perovskites, SrVO$_{3}$--CaVO$%
_{3}$--LaTiO$_{3}$--YTiO$_{3},$ the transition-metal $d$ electron becomes
increasingly localized and undergoes a Mott transition between CaVO$_{3}$
and LaTiO$_{3}.$ By defining a low-energy Hubbard Hamiltonian in the basis
of Wannier functions for the $t_{2g}$ LDA band and solving it in the
single-site dynamical mean-field (DMFT) approximation, it was recently shown
(Pavarini \textit{et al} (2004) \textit{Phys. Rev. Lett.} \textbf{92} 176403)
that simultaneously with the Mott transition there occurs a
strong suppression of orbital fluctuations due to splitting of the $t_{2g}$
levels. The present paper reviews and expands this work, in particular in
the direction of exposing the underlying chemical mechanisms by means of 
\textit{ab initio} LDA Wannier functions generated with the $N$th order
muffin-tin orbital (NMTO) method. The Wannier functions for the occupied
oxygen-$p$ band illustrate the importance of oxygen-$p\,$to\thinspace large
cation-$d$ covalency for the progressive GdFeO$_{3}$-type distortion along
the series. The oxygen-$p$ orbitals which $pd\sigma $-bond to the cations
are the same as those which $pd\pi $-bond to the transition-metal $t_{2g}$
orbitals. As a consequence, the Wannier functions for the $t_{2g}$ band
exhibit residual covalency between the transition-metal $t_{2g},$ the large
cation-$d,$ and the oxygen-$p$ states. This residual covalency, which
increases along the series, turns out to be responsible not only for the
splittings, $\Delta ,$ of the $t_{2g}$ levels, but also for non-cubic
perturbations of the hopping integrals, both of which are decisive for the
Mott transition. We find good agreement with the optical and photoemission
spectra for all four materials, with the crystal-field splittings and
orbital polarizations recently measured for the titanates, and with the
metallization volume (pressure) for LaTiO$_{3}.$ The metallization volume
for YTiO$_{3}$ is predicted and the role of the Jahn-Teller distortion is
discussed. For use in future many-body calculations, we tabulate the $t_{2g}$
on-site and hopping matrix elements for all four materials and give an
analytical expression for the orthorhombic Hamiltonian in the $\mathbf{k}+%
\mathbf{Q}$ representation. Using conventional super-exchange theory, our
on-site and hopping matrix elements reproduce the observed magnetic orders
in LaTiO$_{3}$ and YTiO$_{3},$ but the results are sensitive to detail, in
particular for YTiO$_{3}$ where, without the JT distortion, the magnetic
order would be antiferromagnetic C- or A-type, rather than ferromagnetic. It
is decisive that upon increasing the GdFeO$_{3}$-type distortion, the
nearest-neighbour hopping between the lowest and the upper-level Wannier
functions becomes stronger than the hopping between the lowest-level Wannier
functions. Finally, we show that the non-cubic perturbations responsible for
this behaviour make it possible to unfold the orthorhombic $t_{2g}$ LDA
bandstructure to a pseudo-cubic zone. In this zone, the lowest band is
separated from the two others by a direct gap and has a width, $W_{I},$
which is significantly smaller than that, $W,$ of the entire $t_{2g}$ band.
The progressive GdFeO$_{3}$-type distortion thus favours electron
localization by decreasing $W,$ by increasing $\Delta /W$, and by decreasing 
$W_{I}/W.$ 
Our conclusions concerning the roles of GdFeO$_{3}$-type and JT distortions
agree with those of Mochizuki and Imada 
(2003 \textit{Phys. Rev. Lett.} \textbf{91} 167203).
\end{abstract}

\submitto{\NJP} 
\tableofcontents
\title[How chemistry controls electron localization in 3d$^{1}$ perovskites]{}
\maketitle

\section{Introduction}

\begin{figure}[th]
\par
\begin{center}
\includegraphics[width=\textwidth]{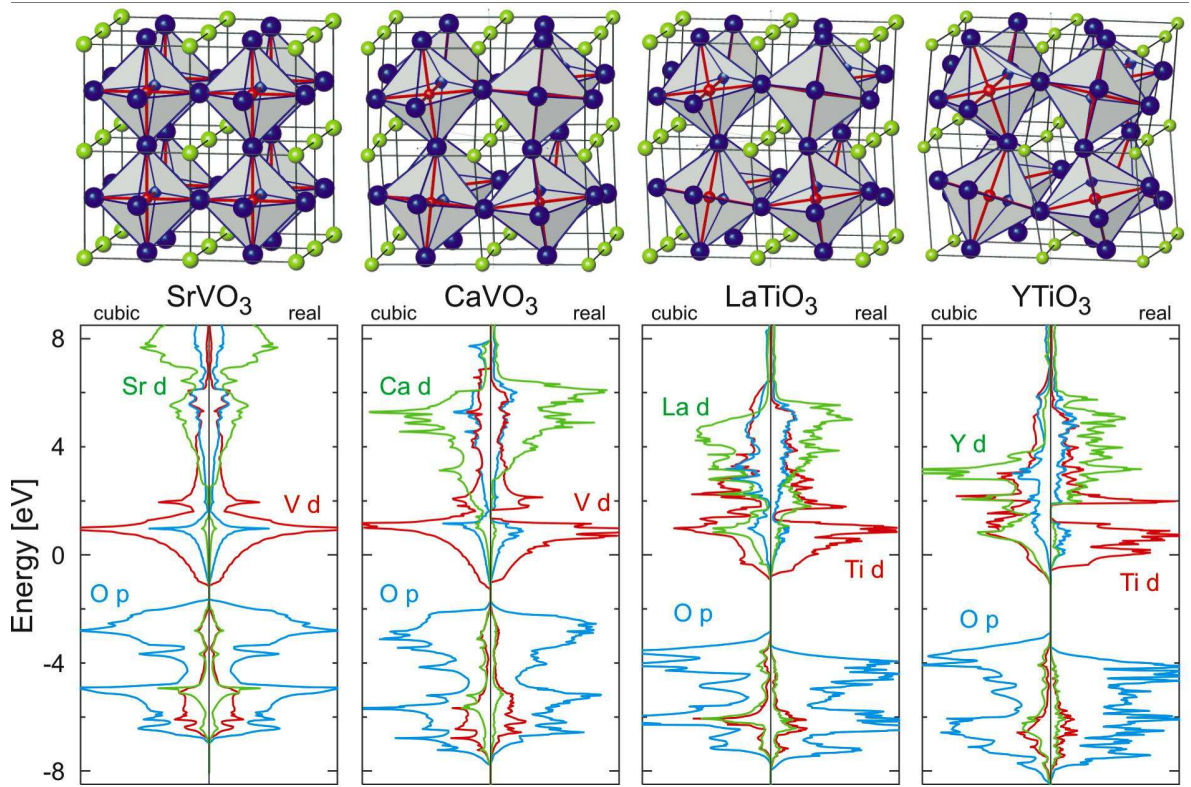}
\end{center}
\caption{Crystal structures and electronic bandstructures for the series of 3%
$d\left( t_{2g}\right) ^{1}$ orthorhombic ABO$_{3}$ perovskites considered
in this paper; A (green) B (red) O (blue). The bottom part shows densities
of one-electron states (DOSs) calculated in the LDA for the real structures
(right-hand panels) and for hypothetical, cubic structures with the same
volumes (left-hand panels). The green, red, and blue DOSs are projected onto,
respectively, A\thinspace $d,$ B\thinspace 3$d,$ and O\thinspace 2$p$
orthonormal orbitals \protect\cite{LMTO84}. The B\thinspace 3$d\left(
t_{2g}\right) $ bands are positioned around the Fermi level (zero of energy)
and their widths, $W,$ decrease from $\sim $3 to $\sim $2\thinspace eV along
the series. The much wider B\thinspace 3$d\left( e_{g}\right) $ bands are at
higher energies.
This figure resulted from linear muffin-tin orbitals (LMTO) calculations in
which the energies, $\epsilon _{\nu Rl},$ of the linear, partial-wave
expansions were chosen at the centres of gravity of the \emph{occupied,}
partial DOS. Since those energies are in the O $p$ band, the LMTO errors
proportional to $\left( \varepsilon -\epsilon _{\nu Rl}\right) ^{4}$
slightly distort the \emph{un}occupied parts of the DOS.}
\label{fig1}
\end{figure}

Transition-metal perovskites have been studied for half a century, and most
intensively during the last decade, for their fascinating electronic and
magnetic properties arising from narrow 3$d$ bands and strong Coulomb
correlations \cite{Goodenough, imada, MochizukiNJP}.  
The 3$d\left(
t_{2g}\right) ^{1}$ series SrVO$_{3}$--CaVO$_{3}$--LaTiO$_{3}$--YTiO$_{3}$
is a paradigm because it has no complicating multiplet effects, a
progressing structural distortion illustrated at the top of figure \ref{fig1}%
, and greatly varying electronic properties: while Sr and Ca vanadate are
correlated metals, with optical mass enhancements of respectively $\sim $3
and $\sim $4 \cite{optics,inoueold,aiura}, La and Y titanate are Mott
insulators, with gaps of respectively 0.2 and 1\thinspace eV \cite{mottgap}.
These two Mott insulators, which essentially mark the end points of a series
of rare-earth titanates RTiO$_{3},$\cite{MochizukiNJP,Katsufuji} have very
different metallization pressures, 11 GPa for LaTiO$_{3}$ and much larger
for YTiO$_{3}$ \cite{loa}. Moreover, they exhibit different orbital physics 
\cite{Keim} and, at low temperature, LaTiO$_{3}$ is a 3-dimensional (G-type)
antiferromagnet with $T_{N}$=150\thinspace K \cite{Goral} and a small moment
of 0.57$\,\mu _{B}$ \cite{Cwik03}, while YTiO$_{3}$ is a ferromagnet with a
low Curie temperature of $T_{C}$=30\thinspace K and a good-sized moment of
0.8$\mu _{B}\mathbf{\ }$\cite{Goral,Garrett}.

In the Mott-Hubbard picture the metal-insulator transition occurs when the
ratio of the on-site Coulomb repulsion to the one-electron bandwidth exceeds
a critical value, $\left( U/W\right) _{c}.$ As figure \ref{fig1} shows, in
the ABO$_{3}$ perovskites the B\thinspace 3$d$ ions are on a nearly cubic
(orthorhombic) lattice and at the centres of corner-sharing O$_{6}$
octahedra. The 3$d$ band thus splits into covalent O$\,p$ B\thinspace $d$ $%
\pi $ antibonding $t_{2g}$ bands and covalent O$\,p$ B$\,d$ $\sigma $
antibonding $e_{g}$\thinspace bands, of which the former lie lower, have
less O character, and couple less to the octahedra than the latter. Simple
theories for the $d^{1}$ perovskites\thinspace \cite{imada}\ are based on a
Hubbard model with three independent, 
two-dimensional, \emph{degenerate}, $\frac{%
1}{6}$-filled $t_{2g}$\thinspace bands per B ion, and the variation of the
electronic properties along the series is ascribed to a progressive
reduction of $W$ due to the increased bending of the $pd\pi $ hopping paths,
the B-O-B bonds seen in figure \ref{fig1}.

But this need not be the whole story, because the value of $(U/W)_{c}$ is
expected \cite{Olle} to decrease with decreasing degeneracy. Such a decrease
of degeneracy can be achieved by splitting the $t_{2g}\,$levels by merely $%
ZW,$ the reduced bandwidth associated with quasiparticle excitations in the
correlated metal \cite{Manini}. As a consequence, if the level-splitting
increased along the series, this could significantly influence the Mott
transition.

However, unlike in $e_{g}$ band perovskites, where large (10\%) cooperative
Jahn-Teller (JT) distortions of the oxygen octahedron indicate that the
orbitals are not degenerate but spatially ordered, the octahedron in the $%
t_{2g}$ band perovskites is nearly perfect. For that reason the $t_{2g}$
orbitals have often been assumed to be degenerate. If that is the case,
quantum fluctuations will lead to an orbital \emph{liquid} \cite%
{Keim,Khal1,Khal2} rather than orbital ordering in the Mott insulating
phase. The observation of an isotropic, small gap spin-wave spectrum both in
antiferromagnetic LaTiO$_{3}$ and also in ferromagnetic YTiO$_{3}$\thinspace 
\cite{Keim} has lent support to this orbital-liquid scenario, because if the
orbital moments were quenched, such a spectrum would seem accidental. On the
other hand, the predicted contribution to the specific heat from the orbital
liquid has not been observed in LaTiO$_{3}$ \cite{Fritsch}. Moreover, a 3\%
JT stretch of one of the basal O squares into a rectangle was recently
discovered in LaTiO$_{3}$ \cite{Cwik03,Hemberger}. This is of similar
magnitude as the JT distortion known to exist in YTiO$_{3}$ where, however,
the square is stretched into a rhomb \cite{ystr}.

\begin{figure}[t]
\par
\begin{center}
\includegraphics[width=0.74\textwidth]{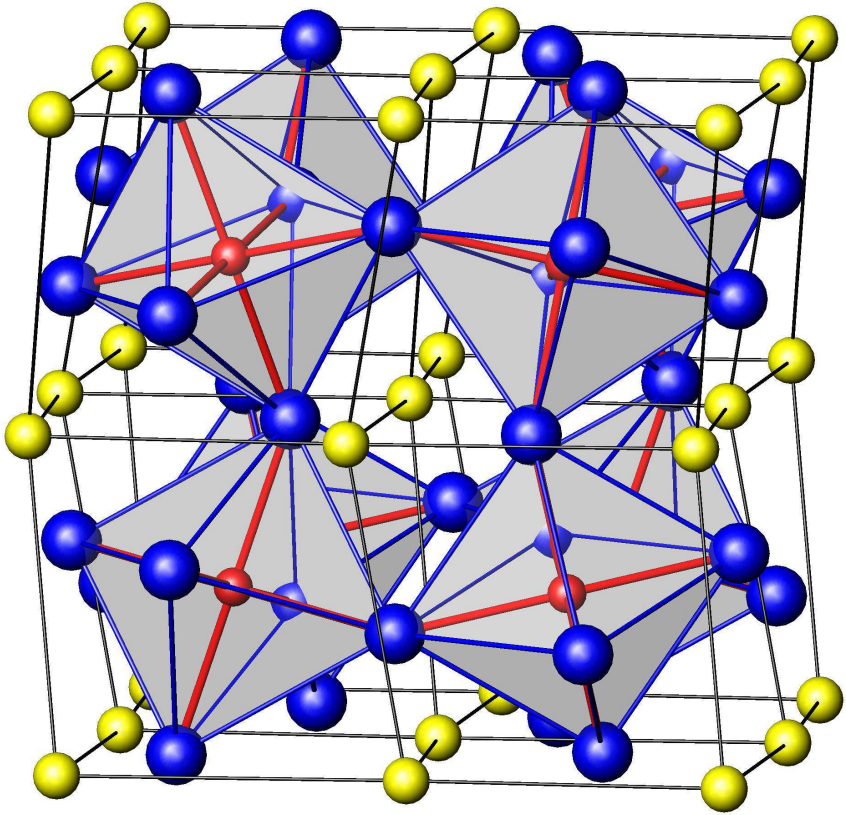}
\end{center}
\caption{Crystal structure of a GdFeO$_{3}$-distorted ABO$_{3}$ perovskite
(YTiO$_{3})$ in \textit{Pbnm} symmetry. 
The horizontal AO-planes perpendicular to the $z$-axis are mirrors. The
global $x$-axis points outward in the front and the $y$-axis to the right.
The B-sites (red) form a primitive monoclinic, nearly cubic lattice with
the following
translation vectors: 
$\mathbf{R}_{x}=\left[ 1+\left( \protect\beta +\protect%
\alpha \right) /2\right] \mathbf{\hat{x}}+\left[ \left( \protect\beta -%
\protect\alpha \right) /2\right] \mathbf{\hat{y}},$ 
$\mathbf{R}_{y}=\left[
\left( \protect\beta -\protect\alpha \right) /2\right] \mathbf{\hat{x}}+%
\left[ 1+\left( \protect\beta +\protect\alpha \right) /2\right] \mathbf{\hat{%
y}},$ and 
$\mathbf{R}_{z}=\left( 1+\protect\gamma \right) \mathbf{\hat{z}}$ 
with $\protect\alpha $, $\protect\beta $ and $\protect\gamma $ small. The
latter vector is orthogonal to the two former, which have the same length,
but need not be quite orthogonal to each other. For LaTiO$_{3}$ (YTiO$_{3}),$
$\left\vert \mathbf{\hat{x}}\right\vert =\left\vert \mathbf{\hat{y}}%
\right\vert =\left\vert \mathbf{\hat{z}}\right\vert =397.1~(385.8)$~pm and $%
\protect\alpha =33~(-258),\,\protect\beta =1~(407),$ $\protect\gamma %
=-33~(-137)~\times 10^{-4}.$ The position of a B-site is: $\mathbf{R}=x%
\mathbf{R}_{x}+y\mathbf{R}_{y}+z\mathbf{R}_{z}$ with $x$, $y$, and $z$
integers. The structure is orthorhombic with 4 ABO$_{3}$ units per cell, 
\textit{e.g.}, the four in the front plane, which we shall label 1 (bottom
left, $ xyz \mathrm{=} 000 $), 2 (bottom
right, $010$), 3 (top left, $001$) and 4 (top right, $011$). 
The orthorhombic translation vectors are: $\mathbf{a}%
=\left( \mathbf{\hat{x}}-\mathbf{\hat{y}}\right) \left( 1+\protect\alpha %
\right) =\mathbf{R}_{x}-\mathbf{R}_{y},\ \mathbf{b}=\left( \mathbf{\hat{x}}+%
\mathbf{\hat{y}}\right) \left( 1+\protect\beta \right) =\mathbf{R}_{x}+%
\mathbf{R}_{y},$ and $\mathbf{c}=2\mathbf{\hat{z}}\left( 1+\protect\gamma %
\right) =2\mathbf{R}_{z}.$ Reflection in a vertical $bc$-plane containing a
B-site ($a\leftrightarrow -a,$ or equivalently, $x\leftrightarrow y)$,
followed by a translation $\mathbf{R}_{y}=\frac{1}{2}\left( \mathbf{b}+%
\mathbf{a}\right) \mathbf{,}$ takes the crystal into itself (glide plane).
All A-ions (yellow) and all B-ions (red) are equivalent, but there are two
kinds of oxygen (blue): O1 in a mirror A-plane, and O2 in a vertical,
buckled A-plane. Proceeding along the series in figure \protect\ref{fig1},
the GdFeO$_{3}$-type distortion tilts the corner-sharing octahedra by 0, 9,
13 (12), and 20$^{\circ }$ around the $b$-axes in alternating directions,
and rotates them around the $c$-axis by 0, 7, 10 (9), and 13$^{\circ }$ in
alternating directions \protect\cite{Cwik03,ystr,srstr,castr}. Here, the
values in parentheses are from the older data \protect\cite{lastr}.}
\label{fig1e}
\end{figure}

By \textit{ab initio} calculation of the Wannier functions of the LDA $%
t_{2g} $ band, it was recently found \cite{Eva03} that in LaTiO$_{3}$ and
YTiO$_{3}$ the $t_{2g}$ degeneracy is lifted at the classical level. This is 
\emph{not} due to the small JT distortions, but to the GdFeO$_{3}$-type
distortion which tilts and rotates the corner-sharing octahedra as
illustrated in figure \ref{fig1e}. As we shall see in the present paper,
this distortion is partly driven by the covalency between occupied oxygen $p$
states and empty A-cation $d$ states, which pulls each O1 (O2) closer to 
\emph{one} (\emph{two}) of its four nearest A neighbours. As a result, each
A cation has 4 of its 12 near oxygens pulled closer in. In addition, the A
cube gets distorted so that \emph{one }diagonal becomes the shortest. The $%
t_{2g}$ degeneracy is now lifted, essentially by residual covalent
interactions between empty A\thinspace $d$ orbitals and full B\thinspace $%
t_{2g}$ and O $p$ orbitals, the details of which are different in LaTiO$_{3}$
and YTiO$_{3}.$ These residual covalent interactions not only perturb the
on-site, but also the hopping matrix elements of the Hamiltonian.

Already forty years ago, Goodenough \cite{Goodenough71} speculated that
covalency between occupied O\thinspace $p$ and empty A\thinspace $d$
orbitals may be a driving force behind the GdFeO$_{3}$-type distortion, and
this hypothesis was recently supported in an extensive series of
semiempirical simulations \cite{Woodward}.

That the GdFeO$_{3}$-type distortion lifts the $t_{2g}$ degeneracy in the
titanates had been realized slightly earlier by Mochizuki and Imada \cite%
{Mochizuki03}, but only after a long search by them and other groups for a
model Hamiltonian which could reproduce all observed magnetic and orbital
orderings \cite%
{Mizokawa96,Mizokawa99,Mochizuki01,Mochizuki01Lett,Mochizuki04}. These model
Hartree-Fock and strong-coupling studies for the entire family of 3$d^{n}$
perovskites mapped out the roles played JT distortions, spin-orbit coupling, 
$e_{g}$ degrees of freedom, orbital misalignment caused by the GdFeO$_{3}$
distortion, and, finally, by the electrostatic field and hybridization from
the A cations which enter via the GdFeO$_{3}$-type distortion.

What enabled the single, independent study in reference \cite{Eva03} to
reach the same conclusion concerning the role of the GdFeO$_{3}$-type
distortion, was the use of parameter-free density-functional (LDA) theory to
generate $t_{2g}$ Wannier functions for a representative series of real
materials by means of a new technique \cite{nmto}. The point is that it is
virtually impossible to know \emph{a priori} which on-site and hopping
matrix elements are important, and what their values are. To extract them
with the required accuracy from experiments and/or from LDA bandstructures
is often impossible, and in this respect the $t_{2g}$ perovskites are
particularly nasty: at first sight, the $t_{2g}$ Hamiltonian of these nearly
cubic materials is the simplest possible, but cation covalency in the
presence of GdFeO$_{3}$-type distortion makes it far more complicated than, 
\textit{e.g.}, the $e_{g}$ Hamiltonian relevant for the $t_{2g}^{3}e_{g}^{1}$
manganites, which have similar GdFeO$_{3}$ distortions. The reason is that
the $t_{2g}$ orbitals interact with the \emph{same} oxygen $p$ orbitals as
the cation orbitals do. This is not the case for $e_{g}$ orbitals.

Now the problem that relevant parameters might be overlooked in model
calculations occurs less frequently with the LDA+$U$ method, because this
method uses a complete basis set. In fact, at about the time when Mizokawa
and Fujimori \cite{Mizokawa96} carried out their pioneering model
calculations, Solovyev, Hamada, Sawada, and Terakura \cite%
{Solovyev96,Sawada98} performed LDA+$U$ calculations in which the value of
the on-site Coulomb repulsion, $U,$ was adjusted to the observed optical
gaps. Those calculations yielded the correct magnetic orders in LaTiO$_{3}$
and YTiO$_{3}$, but the underlying mechanism was not recognized.
The magnetic moment and orbital
order predicted for YTiO$_{3}$ were subsequently confirmed by NMR \cite{Itho}
and neutron scattering \cite{neutrons}.

The static mean-field approximation used in Hartree-Fock and LDA+$U$
calculations cannot describe the paramagnetic-metal to
paramagnetic-insulator (Mott) transition. To be more specific, this
approximation is inadequate for strongly correlated metals and for Mott
insulators at temperatures above the magnetic ordering temperature. For such
purposes, the dynamical mean-field approximation (DMFT) \cite{dmft} has
recently been developed and applied to solve low-energy Hubbard Hamiltonians
derived from the LDA \cite{ldadmft}. The calculations reported in reference 
\cite{Eva03} employed the above-mentioned basis of LDA $t_{2g}$ Wannier
functions, and a single, adjusted value of $U$ for all four materials. This
implementation of the LDA+DMFT approach properly describes orbital
fluctuations in the multiband Hubbard model by including the \emph{off}%
-diagonal $mm^{\prime }$-matrix elements of the self-energy matrix, taken to
be local in DMFT. All earlier implementations had used a scalar self-energy,
as is appropriate for cubic systems~\cite{NekrasovLaSr2000,mergedpaper}. A
recent LDA+DMFT calculation for La$_{1-x}$Sr$_{x}$TiO$_{3}$ used the
eigenrepresentation of the on-site LDA Hamiltonian and then neglected the
off-diagonal elements of the self-energy \cite{Craco03}, apparently a
reasonable approximation for this system.

This paper is a pedagogical review of the calculations reported in reference 
\cite{Eva03} and a presentation of many new results. The use of LDA Wannier
functions to bring out the materials aspects is emphasized. The paper takes
off (section \ref{HighE}) with a discussion of the chemistry: after
analyzing the high-energy part of the LDA bandstructures displayed in figure %
\ref{fig1}, we demonstrate the role of O-A covalency in driving the
progressive GdFeO$_{3}$ distortion. In particular, we visualize the bonds by
means of the O\thinspace $2p$ Wannier functions. In section \ref{LowE}, we
zoom in on the low-energy LDA $t_{2g}$-bands and discuss their Wannier
functions, first in the cubic $xy,yz,xz$-representation and then in the
crystal-field representation. The influence of the GdFeO$_{3}$-type and JT
distortions on the orbital energies, inter-orbital couplings, and the
bandstructures is discussed in detail, and the on-site and hopping integrals
are tabulated for use in future many-body calculations. In section \ref{DMFT}
we set up the $t_{2g}$ Hubbard Hamiltonian and explain how it is solved with
the DMFT many-body technique. Using the same value of $U$ for all four
materials, the resulting high-temperature electronic structures are
presented in section \ref{HT}. They reproduce the increased localization
observed along the series, including the mass enhancements, the Mott
transition, and the gap sizes. For the Mott insulators, it turns out that
the Coulomb correlation localizes the electron almost exclusively in one
orbital, and more so for YTiO$_{3}$ than for LaTiO$_{3},$ and that this
orbital is the eigenfunction of the lowest $t_{2g}$ level in the LDA. The
roles of band-width reduction and level splittings are analyzed, and it is
concluded that these alone do not suffice to explain the strong decrease of $%
\left( U/W\right) _{c}$ found by DMFT when progressing through the series.
We then compute the metallization volumes for the titanates, compare with
most recent high-pressure experiments \cite{loa}, and discuss the role of
the GdFeO$_{3}$-type and JT distortions. In order to further elucidate the
difference in the electronic structure of the Mott insulators LaTiO$_{3}$
and YTiO$_{3}$ we present calculations of the onset of the optical
conductivity. In section \ref{M} we use our computed orbital orders and
hopping integrals to calculate the magnetic exchange couplings, $J_{se},$
within conventional super-exchange theory. Reasonable agreement with the
observed magnetic orders is obtained, but --as evidenced by the extensive
model calculations \cite%
{Mizokawa96,Mizokawa99,Mochizuki01,Mochizuki01Lett,Mochizuki04}, and as
pointed out by Ulrich \emph{et al. }\cite{Ulrich} -- such $J_{se}$ values
are extremely sensitive to detail, in particular in YTiO$_{3}$ where we find
the small JT distortion to be decisive for the ferromagnetism. In section %
\ref{One} we uncover the ingredient so far missing in our understanding of
the calculated trend in $\left( U/W\right) _{c},$ namely the formation of a
lowest subband, whose width is significantly smaller than $W.$ Specifically,
the $N$th-order muffin-tin orbitals (NMTO) method
enables us to show that the complicated orthorhombic LDA $%
t_{2g}$ bandstructures for the titanates can be approximately folded out to
a pseudo-cubic zone and that, in this representation, the lowest band is
separated from the two upper bands by a direct gap. The development of this
orbitally ordered band from the three degenerate cubic bands is explained
and the relation to the increasing tendency towards ferromagnetism with
increasing GdFeO$_{3}$-distortion is pointed out. In section \ref{Concl} we
sum up our main conclusions. Appendix A explains how downfolding
within the NMTO method \cite{nmto} is used to construct truly minimal basis
sets which pick out selected bands, such as the O\thinspace $p$ or
B\thinspace $t_{2g}$ bands. When symmetrically orthonormalized, such a truly
minimal basis set constitutes a set of atom-centred, highly localized
Wannier functions.
Finally, since a weak point of the present calculations 
is our use of the  standard LMTO-ASA method \cite{LMTO84} 
to generate the LDA potentials, we give the technical details in 
appendix B. 

\begin{figure}[t]
\par
\begin{center}
\includegraphics[width=\textwidth]{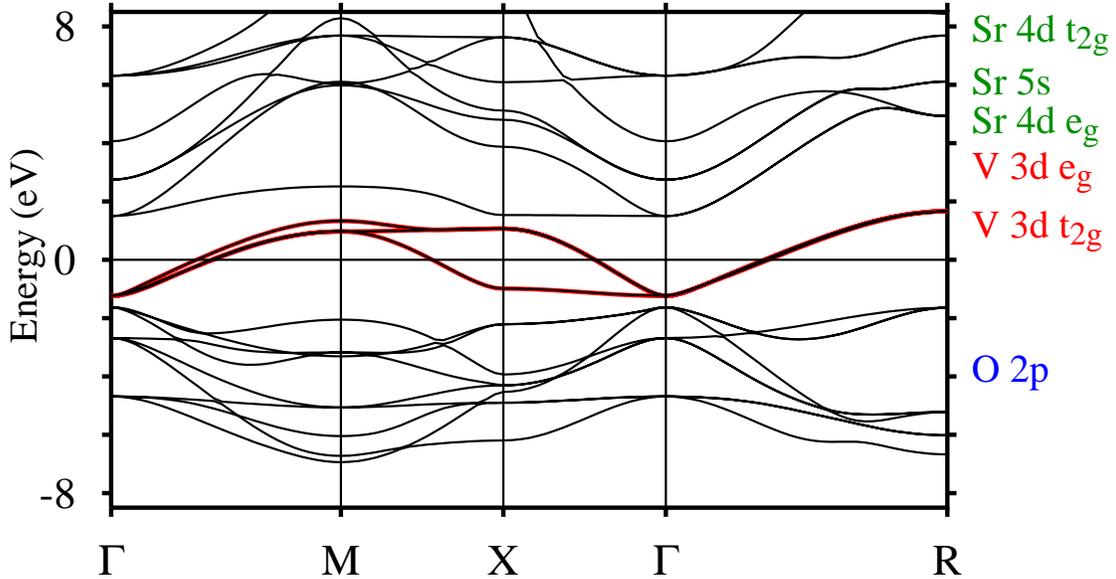}
\end{center}
\caption{LDA bandstructure of cubic SrVO$_{3}.$ The Brillouin zone (BZ) 
is shown
in blue in figure \protect\ref{fig5}. As is discussed in 
appendix A,
the bands obtained with the downfolded, truly minimal V 3$d\left(
t_{2g}\right) $ NMTO basis (red) are indistinguishable from those obtained
with the full NMTO basis (black).}
\label{fig2}
\end{figure}

\section{High-energy LDA bandstructures and O-A covalent mechanism of the
GdFeO$_{3}$-type distortion\label{HighE}}

In the present section we shall use the LDA bandstructures shown in figures %
\ref{fig1}, \ref{fig2}, and \ref{fig1a}, as well as the oxygen $p$ band
Wannier functions, to demonstrate the role of O $p\ -\ $A $d$ covalency for
the GdFeO$_{3}$-type distortion, and the role of O $p\ -\ $B $d$ covalency
for the stability of the octahedron and the splitting of the B 3$d$ band
into separate $t_{2g}$ and $e_{g}$ bands. This will set the stage for
understanding the role of covalency for the splitting of the $t_{2g}$ levels.

Apart from providing insights into the bonding, the LDA bands --except for
the $t_{2g}$ bands, which we shall treat separately with a Hubbard model--
give information about the one-electron high-energy excitations.
Specifically, we shall approximate these excitations by the LDA Hamiltonian
plus a self-energy, $\Sigma \left( \varepsilon \right) ,$ which only couples
inside $t_{2g}$ space and is independent of the crystal momentum, $\mathbf{k.%
}$ To compute this self-energy is the task of the LDA+DMFT to be
considered in section \ref{DMFT}. Future calculations might choose to
include in the many-body calculation a larger basis set of Wannier functions
than those describing merely the low-energy $t_{2g}$ bands. For that reason,
too, it makes sense to consider first the high-energy LDA band structure and
the O\thinspace $p$ Wannier functions.

\subsection{Bandstructures}

The bottom part of figure \ref{fig1} exhibits densities of states (DOSs)
projected onto various groups of orbitals, O\thinspace $p$, B\thinspace $d$,
and A\thinspace $d$, of a standard set of nearly orthonormal LMTOs \cite%
{LMTO84}. These DOS projections provide information about the mixing of
characters due to hybridization between various kinds of orbitals. Consider
for instance the panel relating to SrVO$_{3}$: in the O\thinspace $p$ bands
(mainly blue), we see V\thinspace $d$ character (red), and in the
V\thinspace $d$ bands (mainly red), we see O $p$ character (blue). This
hybridization between the O\thinspace $p$ and the V\thinspace $d$ bands has
pushed them apart, and since the oxygen bands are occupied and the
V\thinspace $d$ bands are nearly empty, band-structure energy has been
gained; this is O-B covalency. Some orbitals hybridize more than others, 
\textit{e.g.}, $\sigma $ bonds are stronger than $\pi $ bonds, and it is
therefore the lower part of the O\thinspace $p$ band and the upper, $e_{g}$
part of the V\thinspace $d$ band which have the most foreign character mixed
in. The less familiar result of this figure is that it also exhibits a large
amount of O-A (blue-green) covalency. We shall see that this is because each
oxygen has two $p$ orbitals $\sigma $ bonding with Sr, but only one $\sigma $
bonding with V. That, to some extent, compensates for the distance to Sr
being $\sqrt{2}$ longer than the distance to V.

When we move along the series SrVO$_{3}$--CaVO$_{3}$--LaTiO$_{3}$--YTiO$_{3}$
a tilt is expected if the ionic radius is such that the Goldschmidt
tolerance factor, $\left( r_{A}+r_{O}\right) \left/ \left[ \sqrt{2}\left(
r_{B}+r_{O}\right) \right] \right. ,$ is smaller than 1. The radii satisfy: $%
r_{\mathrm{Sr}^{2+}}\sim r_{\mathrm{La}^{3+}}>r_{\mathrm{Ca}^{2+}}
\sim r_{\mathrm{Y}^{3+}},$ while $%
r_{\mathrm{V}^{5+}}<r_{\mathrm{Ti}^{4+}}$ 
(in this scheme the $d$ electron is taken as
localized), and the tolerance factor decreases by about 10\% along the
series, although it is the same for CaVO$_{3}$ and LaTiO$_{3}$. The
progressive tilt is thus partly due to the fact that the size of the A
cation shrinks in relation to that of the BO$_{3}$ octahedron. However, as
pointed out by Woodward \cite{Woodward}, this does not explain why the tilt
is of the GdFeO$_{3}$-type. This type, he found by using empirical
interatomic potentials and the extended H\"{u}ckel method, is unique in
maximizing the O-A covalent bonding at the same time as minimizing the O-A
repulsive overlap. Our LDA calculations support and detail the importance of
O-A covalency for the GdFeO$_{3}$-type distortion:

%
\begin{figure}[t]
\par
\begin{center}
\rotatebox{270}{\includegraphics[height=\textwidth]{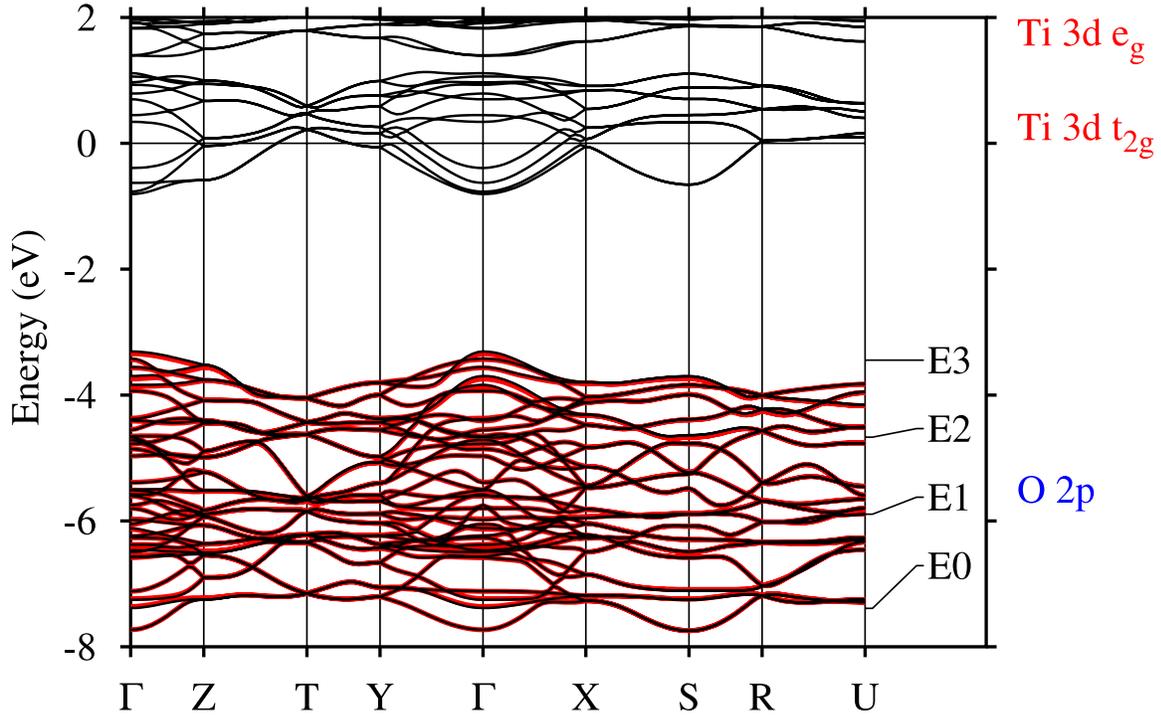}}
\end{center}
\caption{LDA bandstructure of orthorhombic LaTiO$_{3}$. The Brillouin zone
is shown in red in figure \protect\ref{fig5}. The bands obtained with the
truly minimal (downfolded) O\thinspace 2$p$ NMTO basis (red) are
indistinguishable from those obtained with the full NMTO basis (black). This
is explained in 
appendix A.
The basis functions of the truly minimal
set are shown in figures \protect\ref{fig1b} and \protect\ref{fig1c}. The
recent structural data \protect\cite{Cwik03} was used.}
\label{fig1a}
\end{figure}

It is well accepted that equilibrium crystal structures, such as those shown
at the top of figure \ref{fig1}, may be computed \emph{ab initio} with good
accuracy using the LDA. In the bottom right-hand panels we therefore show
the LDA densities of states (DOS) calculated for the real structures
--determined experimentally though \cite{ystr,srstr,castr,lastr}-- and in
the bottom left-hand panels we show the LDA DOS calculated for hypothetical,
cubic structures with the same volume. Now, the energy gain associated with
a structural distortion is approximately the gain in band-structure plus
Madelung energy, so let us consider the trend in the former:

For SrVO$_{3}$ the left- and right-hand panels are identical because the
real structure of SrVO$_{3}$ is cubic. Each Sr ion is at the corner of a
cube and has 12 nearest oxygens at the face centres. Going now to cubic CaVO$%
_{3}$ the empty 3$d$ band of Ca lies lower and thereby closer to the oxygen 2%
$p$ band than the empty 4$d$ band of Sr. It is therefore conceivable that a
GdFeO$_{3}$-type distortion which pulls some of the oxygen neighbours closer
to the A ion and thereby increases the covalency with those, is
energetically more favourable in CaVO$_{3}$ than in SrVO$_{3},$ and this is
what the figure shows: an increase of the O\thinspace 2$p$-Ca\thinspace 3$d$
gap associated with the distortion in CaVO$_{3}.$ The Ca\thinspace 3$d$
character is essentially swept out of the lower part of the V\thinspace 3$d$
band.

\begin{figure}[t]
\par
\begin{center}
\includegraphics[width=\textwidth]{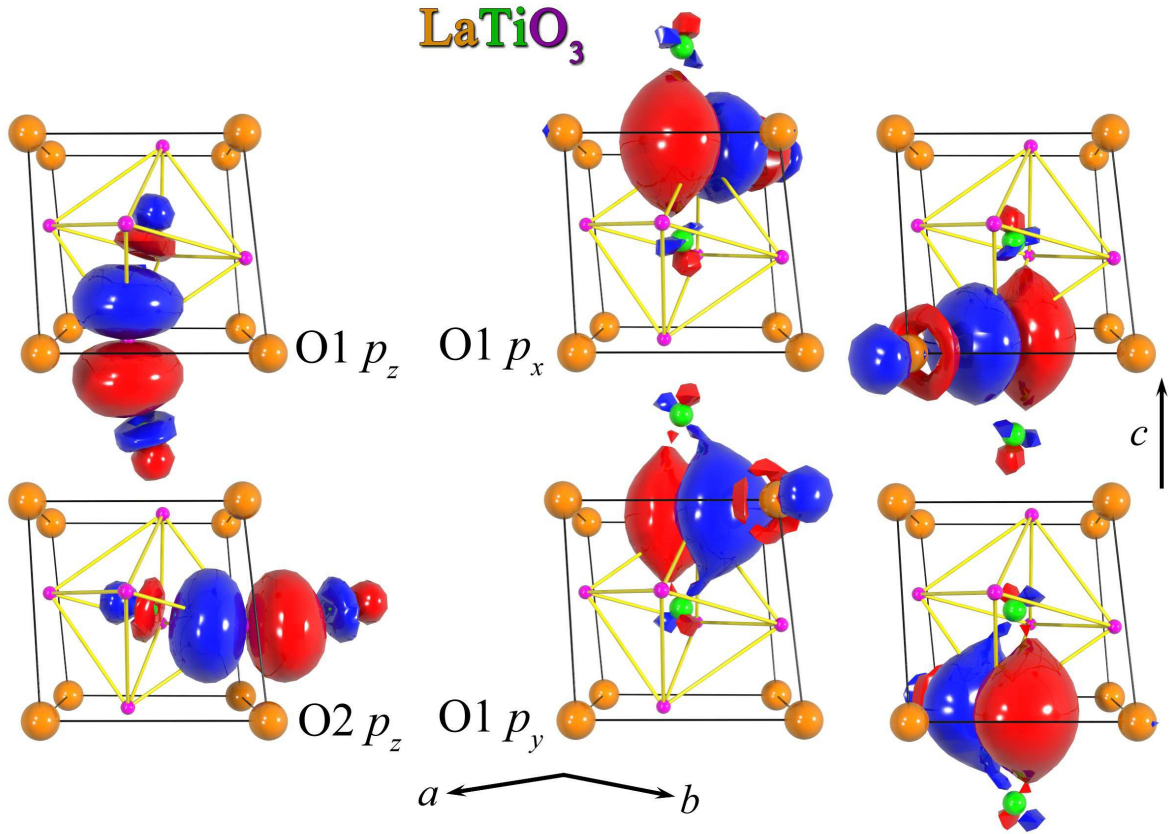}
\end{center}
\caption{Basis functions of the truly minimal set of O\thinspace 2$p$ NMTOs
calculated for LaTiO$_{3}$ with the structure of reference \protect\cite%
{Cwik03}. Shown are the orbital shapes (constant-amplitude surfaces) with
the $\pm $ signs labelled by red and blue. O1 is in a flat face of the
distorted La cube and the O2s are in the buckled faces (see caption to
figure \protect\ref{fig1e}). In column 1 we show the orbitals perpendicular
to the faces $\left( p_{z}\right) $ which exhibit symmetric Ti-O-Ti $\protect%
\sigma $-bonds. In column 2 we show the orbitals in the top horizontal face
(O1 $p_{x}$ and $p_{y}).$ Although they are equivalent to those in the
bottom face shown in column 3, we show both for the sake of clarity. The O1 $%
p_{x}$ and $p_{y}$ orbitals, as well as those for O2 shown in the following
figure \protect\ref{fig1c}, exhibit weak, symmetric Ti-O-Ti $\protect\pi $%
-bonds. Most importantly, however, the O1 $p_{x}$ and $p_{y}$ orbitals show
asymmetric O-La $\protect\sigma $-bonds. The latter, together with the
O2\thinspace $p_{x}$ - La $\protect\sigma $-bonds shown in figure  \protect
\ref{fig1c}, are responsible for the GdFeO$_{3}$-type distortion. This O-A
bonding is shown schematically in figure \protect\ref{fig1d}.}
\label{fig1b}
\end{figure}
\begin{figure}[t]
\par
\begin{center}
\includegraphics[width=\textwidth]{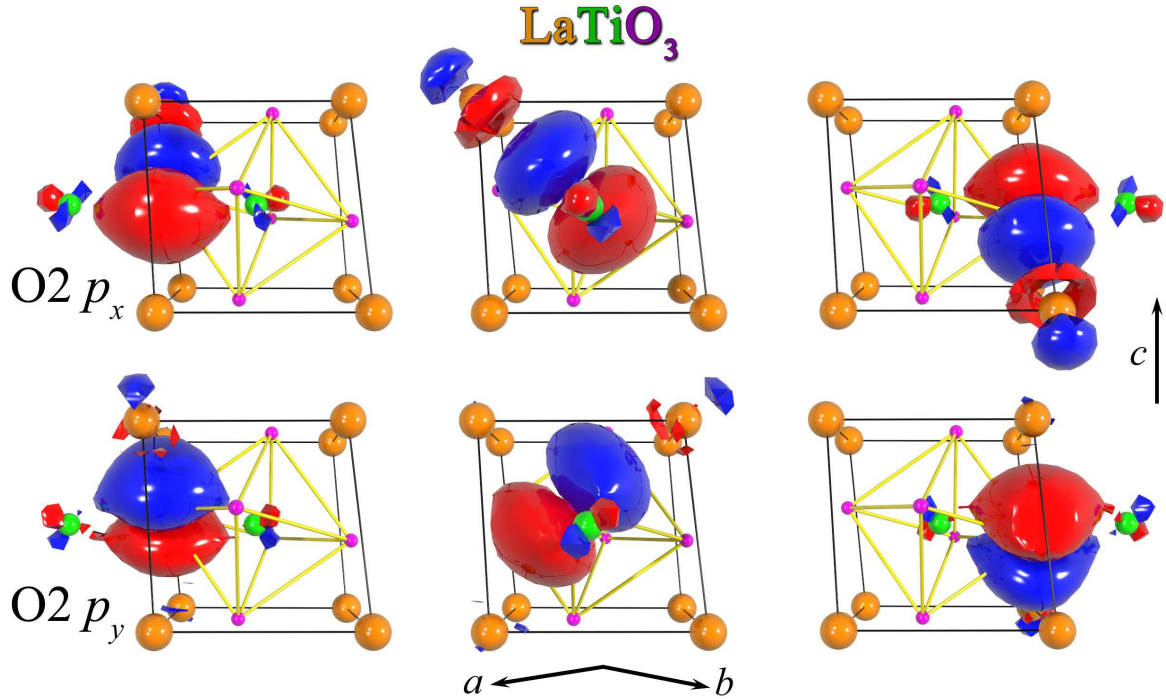}
\end{center}
\caption{Same as figure \protect\ref{fig1b}, but for the oxygen orbitals (O2 
$p_{x}$ and $p_{y})$ in the buckled faces of the distorted La cube. O2 $%
p_{x} $ is seen to bind much stronger to La than O2 $p_{y}.$ The orbitals in
the 1st and 3rd columns are equivalent, but for clarity we show both. Not
shown are the orbitals on the rear face since they are equivalent to those
in the front face shown in the 2nd column. The rear $p_{x}$ orbital binds
strongly to La at the bottom right corner. The O-A bonding is summarized in
figure \protect\ref{fig1d}.}
\label{fig1c}
\end{figure}

When proceeding to the titanates, the A and B cations become 1st- rather
than 3rd-nearest neighbours in the periodic table. The B 3$d$ band therefore
moves up and the A\thinspace $d$ band down with respect to the O\thinspace $%
p $ band. Hence, the O-B covalency decreases and the O-A covalency
increases. Most importantly, the A\thinspace $d$ band becomes nearly
degenerate with the Ti\thinspace 3$d$ band, and more so for Y\thinspace 4$d$
than for La\thinspace 5$d.$ It is only the GdFeO$_{3}$-distortion which,
through increase of the O\thinspace 2$p$-A\thinspace $d$ hybridization,
pushes the A\thinspace $d$ band above the Fermi level. This, as well as the
concomitant lowering of the O\thinspace 2$p$ band with respect to $\epsilon
_{F},$ can be seen in the figure.

In the series of trivalent rare-earth titanates, the GdFeO$_{3}$-type
distortion --and the low-temperature magnetic properties-- are known to
change gradually from those of LaTiO$_{3}$ to essentially those of YTiO$_{3}$
\cite{MochizukiNJP}. This trend follows the decrease of the ionic radius and
of the $5d$ level position.

\subsection{Oxygen $p$ bonds}

The covalent bonds can be visualized by a set of localized Wannier functions
for the \emph{occupied} bands, \textit{i.e.} the O 2$p$ bands. Here we shall
not consider the much smaller contribution from the O 2$s$ bands. In figure %
\ref{fig1a} we illustrate for the case of LaTiO$_{3}$ that we can construct
a set of O $p$ NMTOs which span the O 2$p$ bands, and no other bands. This
basis set contains as many orbitals as there are occupied bands and, hence,
it is what we call a \emph{truly minimal} basis set.\emph{\ }Its
inequivalent orbitals are shown in figures \ref{fig1b} and \ref{fig1c}. Each
oxygen has two nearest B neighbours and four near A neighbours, and we have
chosen its three $p$ orbitals such that one of them, call it $p_{z},$ points
approximately towards the B neighbours (see caption to figure \ref{fig1e}),
and the two other orbitals, labelled $p_{x}$ and $p_{y},$ point
approximately towards two of the four A neighbours.

Only the \emph{central} parts of these orbitals have respectively $p_{x},$ $%
p_{y},$ or $p_{z}$ character, because in order to describe the hybridization
of the O 2$p$ band with the B and A derived bands, the orbitals of a truly
minimal set must have those characters folded into their tails. In fact, 
\emph{all} partial-wave characters \emph{other} than oxygen $p$ are folded
into the O\thinspace $p$ set. Now, as is explained in 
appendix A,
the NMTOs
are localized by construction, but they are not quite orthogonal. Therefore,
in order to become a set of localized Wannier functions, the truly minimal
set must be symmetrically orthonormalized. The sum of the squares of these
Wannier functions is the valence charge density, and the sum of their
orbital energies is the band-structure energy (except for a small
contribution from the B\thinspace 3$d$ electron). Hence, the truly minimal
set of O\thinspace 2$p$ NMTOs shown in figures \ref{fig1b} and \ref{fig1c}
visualizes the covalent
bonds.


\begin{figure}[th]
\par
\begin{center}
\includegraphics[width=\textwidth]{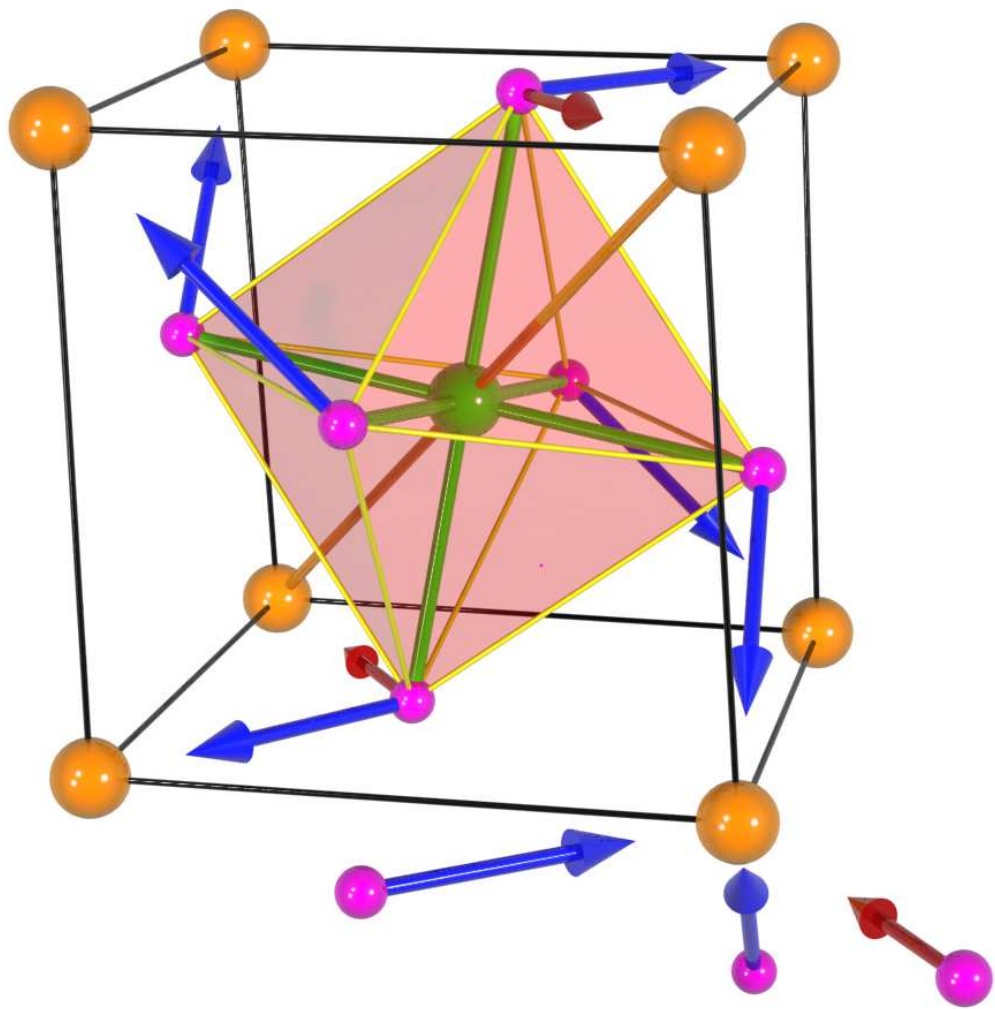}
\end{center}
\caption{Schematic representation of the O-A covalent bonds shown for LaTiO$%
_{3}$ in figures \protect\ref{fig1b} and \protect\ref{fig1c}: O1 binds to 
\emph{two} while O2 binds to \emph{one} of the four A-neighbours. One of the
two O1-A bonds is relatively weak and is indicated by a short, red arrow.
The resulting GdFeO$_{3}$-type distortion shortens the O-A bonds
correspondingly. In CaVO$_{3},$ LaTiO$_{3},$ and YTiO$_{3},$ the shortest
O1-A bond is shortened by respectively 10, 17, and 28\% with respect to the
average, the 2nd-shortest O1-A bond by respectively 4, 11, and 23\%, and the
shortest O2-A bond by respectively 12, 16, and 22\%. The oxygen coordination
of the A-ion is reduced from 12 to 4, with two of the near oxygens being in
the horizontal, flat face of the distorted A-cube, and the two others in 
\emph{one} of the vertical, buckled faces. The A-B-A diagonal (orange bar)
lying in the plane of the short, red arrows is shortened by respectively 3,
7, and 9\% of the average. The unit shown is the front bottom left one
(subcell 1) seen in \textit{e.g.} figure \protect\ref{fig1e}.}
\label{fig1d}
\end{figure}

From the first column in figure \ref{fig1b} we see that each oxygen forms 
\emph{one} strong Ti-O-Ti covalent, symmetric $\sigma $ bond with its two Ti
neighbours, the Ti character being 3$d\left( e_{g}\right) $ with a bit of 4$%
sp.$ From the right-hand side of the figure, we see that each oxygen orbital
in the horizontal, flat face of the distorted La cube (O1\thinspace $p_{x}$
or $p_{y}$) forms a weak Ti-O-Ti covalent, symmetric $\pi $ bond, the Ti
character being predominantly 3$d\left( t_{2g}\right) .$ As may be seen from
figure \ref{fig1c}, the same holds for each oxygen orbital (O2 $p_{x}$ or $%
p_{y})$ in the vertical, buckled faces. The respective O-Ti covalent
interaction pushes the anti-bonding Ti 3$d\left( e_{g}\right) $-like band
well above the Ti 3$d\left( t_{2g}\right) $-like band, as we saw in figure %
\ref{fig1}. 

Most importantly, however, figure \ref{fig1b} shows that \emph{%
each} of the two oxygen orbitals in the flat faces (O1\thinspace $p_{x}$ or 
$p_{y}$) forms an O-A covalent, asymmetric $\sigma $ bond with \emph{one} of
the two La neighbours towards which it points. It is clearly seen how the
weight of the orbital is shifted from one towards the other La ion.
Moreover, we see that the bond with O1 $p_{x}$ is somewhat stronger than
with O1\thinspace $p_{y}$. The La character of the O1 $p_{x}$ orbital is 5$%
d_{3x^{2}-1}$ with some by-mixing of 6$sp.$ This $d$ character is mostly $%
t_{2g},$ because with $x$ and $y$ along the cubic directions, it is $d_{%
\frac{3}{2}\left( x\pm y\right) ^{2}-1}=\mp \frac{\sqrt{3}}{2}d_{xy}-\frac{1%
}{2}d_{3z^{2}-1}.$ Figure \ref{fig1c} shows that for the buckled faces of
the distorted La cube, only \emph{one} of the oxygen orbitals, O2 $p_{x},$
bonds significantly to La, and that this bonding is as strong as for
O1\thinspace $p_{x}.$ As a result, we obtain the schematic picture of the
O-A covalent $\sigma $ bonds shown in figure \ref{fig1d}. The resulting GdFeO%
$_{3}$-type distortion shortens the two O1 bonds by respectively 17\% (15\%)
and 11\% (8\%) of the average of the four O1-La distances, and it shortens
the O2 bond by 16\% (14\%) of the average of the four O2-La distances \cite%
{Cwik03}. Here the numbers in parentheses are from the older data \cite%
{lastr}. For CaVO$_{3}$ the corresponding bond-length reductions are 10\%
and 4\% for O1 and 12\% for O2, while for YTiO$_{3},$ they are as large as
28\% and 23\% for O1, and 22\% for O2. For YTiO$_{3}$ the shortest O-Y
distance is, in fact, only 10\% longer than the O-Ti distance.

Also indicated in figure \ref{fig1d} is the oxygen coordination of the A
cations, which is reduced from 12-fold and cubic to 4-fold and roughly
tetrahedral. The embedding of the unit in the entire structure can be
understood by comparison with figure \ref{fig1e}, where the unit is the one
in the front bottom left corner.

Finally, the orange bar in figure \ref{fig1d} indicates that one of the four
A-B-A diagonals, [111], is shortened by respectively 3, 7 (5), and 9\% of
the average A-B-A distance in CaVO$_{3},$ LaTiO$_{3},$ and YTiO$_{3}$. The
two A ions closest to B are those which bond weakest to the oxygen
octahedron (red arrows in the figure). The corresponding distortion of the A
cube is clearly visible in figure \ref{fig1e}, and is presumably caused by
the hard-core repulsion from the three nearest oxygens (the blue arrows in
figure \ref{fig1d}). Corresponding to this shortening of the [111] diagonal
in subcell 1, is a stretching of the [11\={1}] diagonal by nearly the same
amount. In YTiO$_{3},$ the tilt of the oxygen octahedron around the $b$
axis, \textit{i.e.} towards the two Y atoms along [1\={1}1], is particularly
strong (20$%
{{}^\circ}%
,$ see figure \ref{fig1e}) and the corresponding A-B-A distance is shortened
by 5\%. None of the other three materials have such a second, short
diagonal. In the distorted structure, each A ion (all equivalent) has two
nearest B neighbours (two short diagonals). From the point of view of the A
ion in the lower right corner of figure \ref{fig1d}, \textit{i.e.} the one
at $\frac{1}{2} \frac{1}{2} \frac{-1}{2},$ these B neighbours
are at $110$ and $11\bar{1}$.

In the following we shall turn to our primary interest, the development of
the \emph{low}-energy electronic structure along the series of $d^{1}$
perovskites. We shall see that the minute A character left behind in the B 3$%
d\left( t_{2g}\right) $ band after most of this character has been swept
away by O-A covalency and the concomitant GdFeO$_{3}$-distortion, is
decisive.

\section{$t_{2g}$ Wannier functions and their Hamiltonian \label{LowE}}

The physical properties of the $d^{1}$ perovskites are determined by the
Wannier functions of the low-energy B\thinspace 3$d\left( t_{2g}\right) $
bands, such as those shown in red in figure \ref{fig2}, and their on-site
Coulomb repulsion. This, at least, is the working hypothesis of the present
paper.

In figure \ref{fig4} we show the truly minimal $t_{2g}$ NMTO basis set for
the case of LaTiO$_{3}.$ In order to generate these orbitals, it is not
necessary to choose local axes oriented after the oxygen octahedron: with $%
x, $ $y,$ and $z$ referring to the \emph{global} axes (see caption to figure %
\ref{fig1e}) and the active channels specified simply as $d_{xy},$ $d_{yz},$
and $d_{xz}$ on each of the four B sites, each orbital automatically adjusts
to its environment. In the present case, this is mainly due to the
downfolding of the on-site $e_{g}$ character, as is explained in 
appendix A.
The names and signs of the orbitals in subcells 2, 3, and 4 are thus the
ones shown in the \emph{middle} row of the figure. This is the natural
choice when the structure is nearly cubic.

\subsection{Cubic $t_{2g}$ bands\label{CubicBands}}

%
\begin{figure}[t]
\par
\begin{center}
\includegraphics[width=\textwidth]{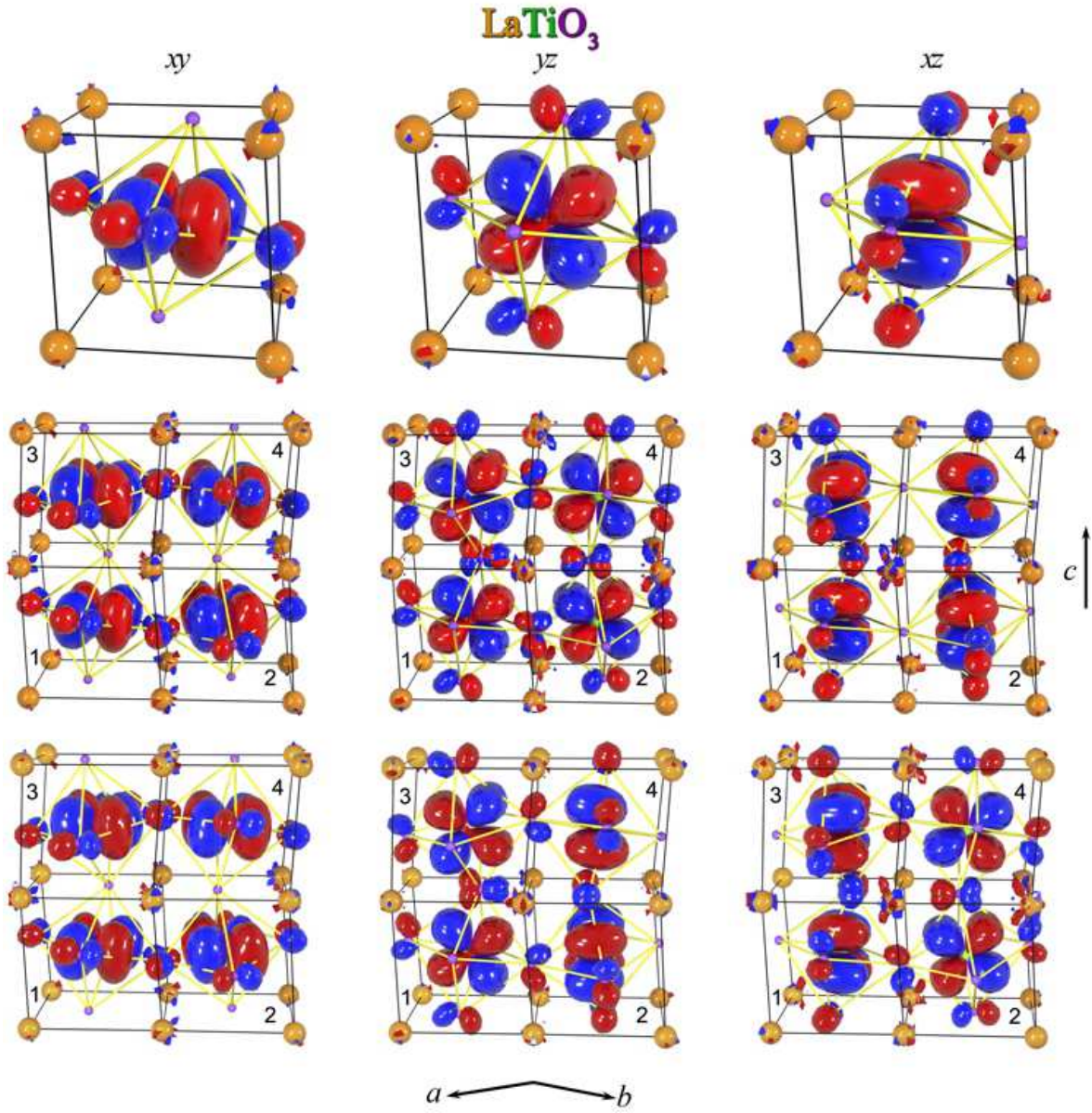}
\end{center}
\caption{B\thinspace $t_{2g}$ $xy,$ $yz,$ and $xz$ NMTOs for LaTiO$_{3}.$
See also caption to figure \protect\ref{fig1b}. In the top row, the three
different orbital shapes are in subcell 1. In the bottom row, the mirror and
the glide-mirror have been used to place them also in subcells 2, 3, and 4
(see caption to figure \protect\ref{fig1e}). In the middle row, the orbitals
have been grouped together in columns, and the signs have been chosen so as
if the orbitals in subcells 2, 3, and 4 were mere translations of the ones
in subcell 1. We shall use the middle-row convention for naming the $xy,$ $%
yz,$ and $xz$ orbitals in subcells 2, 3, and 4. This convention is the
natural one when the distortion from the cubic structure is small. If the
structure were cubic, the three orbital shapes would be identical, and the
sum over the 4 orbitals in each of the middle-row pictures would give a
Bloch wave with $\mathbf{k\mathrm{=}0.}$}
\label{fig4}
\end{figure}

In the \emph{cubic} structure, the $xy,$ $yz,$ and $xz$ Wannier orbitals are 
\emph{equivalent.} Moreover, they are nearly \emph{independent} because each
orbital is even around its own plane and odd around the two other cubic
planes. For that reason, the on-site and the 1st-nearest-neighbour couplings
between different orbitals vanish. As regards the coupling to 2nd-nearest
neighbours, the $yz$ orbital, for instance, does couple to $xy$ orbitals at
the four 101 sites and to $xz$ orbitals at the 110 sites, but the
strength is negligible (6\thinspace meV).

\begin{figure}[t]
\par
\begin{center}
\includegraphics[width=\textwidth]{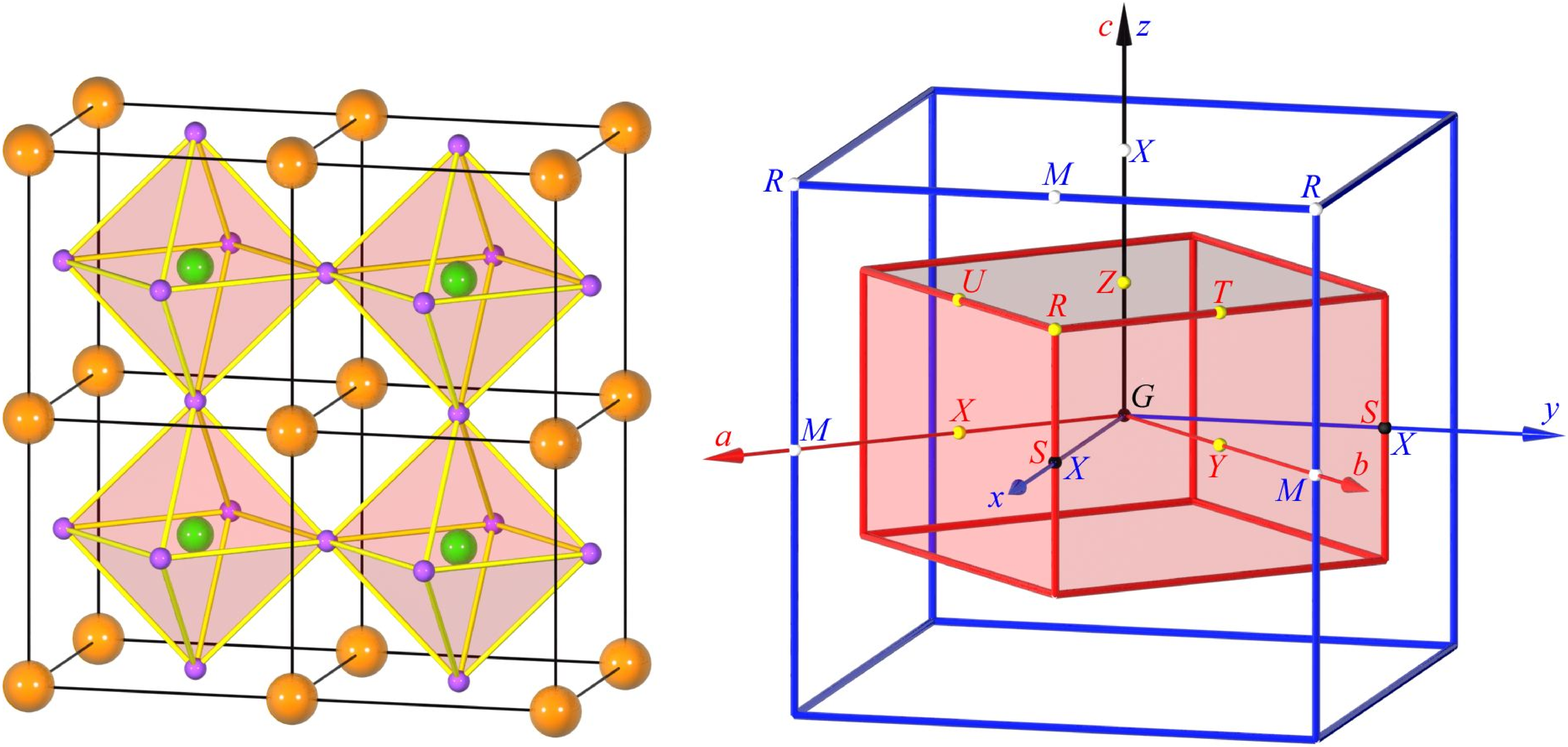}
\end{center}
\caption{Primitive cells of the monoclinic, nearly cubic and the
quadrupled orthorhombic \textit{Pbnm} structures (left), as well as their
respective (blue and red) Brillouin zones (right). See also figure 
\protect\ref{fig1e}. Taking the pseudo-cubic lattice constant, $\left\vert 
\mathbf{\hat{x}}\right\vert ,$ as unity, the primitive reciprocal-lattice
translations in the monoclinic structure are given by: $2\protect%
\pi \mathbf{G}_{x}=\left\{ \left[ 1+\left( \protect\alpha +\protect\beta %
\right) /2\right] \mathbf{\hat{x}}+\left[ \left( \protect\alpha -\protect%
\beta \right) /2\right] \mathbf{\hat{y}}\right\} /\left( 1+\protect\alpha +%
\protect\beta +\protect\alpha \protect\beta \right) ,$ $2\protect\pi \mathbf{%
G}_{y}=\left\{ \left[ \left( \protect\alpha -\protect\beta \right) /2\right] 
\mathbf{\hat{x}}+\left[ 1+\left( \protect\alpha +\protect\beta \right) /2%
\right] \mathbf{\hat{y}}\right\} /\left( 1+\protect\alpha +\protect\beta +%
\protect\alpha \protect\beta \right) ,$ and $2\protect\pi \mathbf{G}_{z}=%
\mathbf{\hat{z}/}\left( 1+\protect\gamma \right) \mathbf{.}$ With $\mathbf{k}%
\equiv k_{x}\mathbf{G}_{x}+k_{y}\mathbf{G}_{y}+k_{z}\mathbf{G}_{z},$ the $%
k_{x} k_{y} k_{z} $ coordinates of the points marked in
the blue, primitive monoclinic BZ are: G$\left( \Gamma \right)$ $000$, X
$\pi00 ,$ M $\pi\pi0 ,$ R $\pi\pi\pi ,$ and the equivalent ones. 
The high-symmetry points of the red, folded-in
orthorhombic BZ are: G$_{o}\left( \Gamma _{o}\right)$ $000 ,$ Z%
$_{o}$ $00\frac{\pi}{2} ,$ X$_{o}$ $\frac{-\pi}{2}\frac{\pi}{2}0 ,$ 
Y$_{o}$ $\frac{\pi}{2}\frac{\pi}{2}0 ,$ S$_{o}$ $\pi00 ,$ 
R$_{o}$ $\pi0\frac{\pi}{2} ,$ 
U$_{o}$ $\frac{-\pi}{2}\frac{\pi}{2}\frac{\pi}{2} ,$ 
T$_{o}$ $\frac{\pi}{2}\frac{\pi}{2}\frac{\pi}{2} ,$ 
and the equivalent ones. The points explicitly listed are
those used in figure \protect\ref{fig3b}.}
\label{fig5}
\end{figure}

Before using the Wannier functions in figure \ref{fig4}, let us start simply
by explaining the red $t_{2g}$ band structure, $\varepsilon
_{x_{i}x_{j}}\left( \mathbf{k}\right) ,$ of \emph{cubic} SrVO$_{3}$ in
figure \ref{fig2} in same way --but now for each Bloch vector-- as we
discussed the gross features of the densities of states in figure \ref{fig1}%
; that is, in terms of a large basis set, the one giving rise to the black
bands. In terms of Bloch waves, hybridization with oxygen states pushes $%
t_{2g}$ band states \emph{up} in energy and hybridization with A=Sr states
pushes them \emph{down. }In the cubic structure there is only one ABO$_{3}$
unit per primitive cell and the Brillouin zone (BZ) is the blue one shown in
figure \ref{fig5}. We may use the middle row in figure \ref{fig4} to
illustrate the three degenerate Bloch waves: with $\mathbf{R}$ being the
positions of the B atoms (cells), each cell should be decorated with a phase
factor $\exp i\mathbf{k\cdot R;}$ \textit{i.e.} $\exp ik_{y}$ in cell 2, $%
\exp ik_{z}$ in cell 3, $\exp i\left( k_{y}+k_{z}\right) $ in cell 4, etc.

At $\Gamma $ $\left( \mathbf{k\mathrm{=}0}\right) $, there is no coupling to
oxygen $p$ because each $t_{2g}$ Bloch wave is odd around the OA planes
perpendicular to the plane of the wave and the O $p\pi $ character is even.
The direct $dd\pi $ coupling is seen to be antibonding and it therefore
tends to increase the energy at $\Gamma ,$ but since $\Gamma $ marks the
bottom of the band, the direct coupling cannot be the dominant source of the
band dispersion. Finally, we realize that at $\Gamma $ all A characters
vanish, except the same A\thinspace $d\left( t_{2g}\right) $ character as
that of the Bloch wave. Hence, $\varepsilon _{x_{i}x_{j}}\left( \mathbf{0}%
\right) $ is pushed down by interaction with A\thinspace $d_{x_{i}x_{j}}.$

Going now from $\Gamma $ to X $00\pi$ in figure \ref{fig2}%
, $\exp i\mathbf{k\cdot R}$ changes sign on sites 3 and 4, whereby the $yz$
and $xz$ waves become even around the horizontal AO mirror plane and can
couple $pd\pi $ antibonding along the $z$ direction. This raises their band
energy by $4t_{pd\pi }^{2}/\left( \varepsilon -\epsilon _{p}\right) \sim 2$
eV, minus the direct $dd\pi $ contribution, and plus the lost contribution
from the bonding interaction with respectively A\thinspace $d_{yz}$ and
A\thinspace $d_{xz},$ which is now forbidden and is not substituted by any
other coupling to A\thinspace $s$ or $d$. The $xy$ wave becomes odd around
the horizontal AO mirror plane, whereby its energy increases by merely a few
tenths of an eV, due to direct $dd\delta $ hopping counteracted by the
increased bonding interaction arising from the A\thinspace $d_{xy}$
character being substituted by A\thinspace $d_{xz}$ and A\thinspace $d_{yz}.$

Proceeding then from X $00\pi$ to M $\pi0\pi$ 
in figure \ref{fig2}, the sign on sites 3 and 4 remains opposite
to that on sites 1 and 2, but the sign alternates on the sheets which are
perpendicular to $x$ and not shown in figure \ref{fig4}. The energy of the $%
yz$ band decreases by some tenths of an eV because the bonding interaction
with A\thinspace $d_{xy}$ is now allowed, but this is counteracted by $%
dd\delta $. The energy of the $xz$ band increases by less than 0.5 eV as the
result of a further increase by $4t_{pd\pi }^{2}/\left( \varepsilon
-\epsilon _{p}\right) ,$ and of reductions due to oxygen $pp$ hopping, $%
dd\pi $ hopping, and coupling to A\thinspace $s$. The energy of the $xy$
band finally goes up by essentially $4t_{pd\pi }^{2}/\left( \varepsilon
-\epsilon _{p}\right) $ and becomes degenerate with the $yz$ band.

At R $\pi\pi\pi$ the sign on all 6 nearest B neighbours
is reversed from that of the middle row in figure \ref{fig4}, and the energy
of the three degenerate bands is slightly higher than that of the $xz$ band
at M $\pi0\pi .$ This is because of direct $dd\delta $
hopping and because the coupling to A\thinspace $s$ is lost. Going finally
from R back to $\Gamma ,$ the three bands stay degenerate.

If we neglect the very weak inter-orbital coupling (6 meV in SrVO$_{3}$),
the dispersion of a cubic $t_{2g}$ band can be written as a Fourier series: 
\begin{equation}
\varepsilon _{m}\left( \mathbf{k}\right) =H_{m,m}^{LDA}\left( \mathbf{k}%
\right) =t_{m,m}^{000}+\sum_{xyz\neq \bar{x}\bar{y}\bar{z}%
}2t_{m,m}^{xyz}~\cos \left( xk_{x}+yk_{y}+zk_{z}\right) ,  \label{eq8}
\end{equation}%
where $x,$ $y,$ and $z$ run over all integers and where we have used the
inversion symmetry of the orbitals and the lattice to combine the complex
exponentials into cosines. In table \ref{tableSr} we give the energy of the
three degenerate $t_{2g}$ Wannier functions, $%
t_{xy,xy}^{000}=t_{yz,yz}^{000}=t_{xz,xz}^{000}\equiv \epsilon
_{t_{2g}}-\varepsilon _{F},$ and their transfer- or hopping integrals, such
as the most important 1st-nearest-neighbour integrals, $%
t_{xy,xy}^{100}=t_{yz,yz}^{001}=t_{xz,xz}^{001}\equiv t_{\pi }$ and $%
t_{xy,xy}^{001}=t_{yz,yz}^{100}=t_{xz,xz}^{010}\equiv t_{\delta },$ and the
most important 2nd-nearest-neighbour integral, $%
t_{xy,xy}^{110}=t_{yz,yz}^{011}=t_{xz,xz}^{101}\equiv t_{\sigma }^{\prime }.$
The description in terms of Wannier functions will be considered in detail
in the following subsection, where we shall also explain why all the
above-mentioned, most important hopping integrals are negative. Neglecting
all other hopping integrals, the Fourier series (\ref{eq8}) simplifies to:%
\begin{equation}
\varepsilon _{x_{i}x_{j}}\left( \mathbf{k}\right) =\epsilon
_{t_{2g}}+2t_{\pi }\left( c_{i}+c_{j}\right) +2t_{\delta }c_{k}+4t_{\sigma
}^{\prime }c_{i}c_{j},  \label{eq9}
\end{equation}%
where $x_{i}=x,~y,$ or $z,$ and $i,$ $j,$ and $k$ are all different.
Moreover, $c_{i}\equiv \cos k_{x_{i}}.$ In this approximation the $\Gamma -$%
R bandwidth is $-8t_{\pi }-4t_{\delta },$ which for SrVO$_{3}$ is $2.4$ eV.
Hops to farther neighbours give another 0.5 eV.

Had it not been for the direct and A-mediated \emph{effective} $dd\delta $
couplings between $t_{2g}$ Wannier functions expressed by $t_{\delta },$ the
dispersion in figure \ref{fig2} would have been \emph{two dimensional}, with
the bottom at $\Gamma ,$ top at M, and saddle-points at X. The latter would
give rise to a logarithmic van Hove singularity in the density of states.
The strong reminiscence of this 2D singularity for SrVO$_{3}$ is clearly
seen at the bottom left-hand side of figure \ref{fig1}. However, as we go
along the series --but stay in the cubic structure-- this sharp peak is
gradually washed out due to increase of the A-mediated effective $dd\delta $
coupling and, in the titanates, to A\thinspace $d$ bands overlapping the $%
t_{2g}$ bands.

\begin{table}[t]
\caption{$H_{t_{2g}}^{LDA}$ {in meV for cubic SrVO}$_{{3}}.$}
\label{tableSr}
\par
\begin{center}
\begin{tabular}{crrrrrrrrrrr}
\br $xyz$ & $000$ & $001$ & $010$ & $100$ & $011$ & $101$ & $110$ & $111$ & $%
002$ & $020$ & $200$ \\ 
$m^{\prime },m$ &  &  &  &  &  &  &  &  &  &  &  \\ 
$yz,yz$ & $625$ & $-281$ & $-281$ & $-33$ & $-96$ & $9$ & $9$ & $-11$ & $10$
& $10$ & $0$ \\ 
\br &  &  &  &  &  &  &  &  &  &  & 
\end{tabular}%
\end{center}
\end{table}

\subsection{The $\left( xy,yz,xz\right) $ set of Wannier orbitals in the
presence of GdFeO$_{3}$-type distortion\label{xyyzzx}}

Distorting now the structures to the \emph{real }ones, the density-of-states
profiles are seen to sharpen in figure \ref{fig1}. This is because the
increased O-A covalency pushes the A\thinspace $d$ bands up above the $%
t_{2g} $ band. But new structures arise, and these are caused by lifting the
degeneracy between the $xy,$ $yz,$ and $xz$ Bloch waves, by coupling between
the waves, and by quadrupling the primitive cell. Let us now explain this
--without getting into the details of the 4$\times $3 orthorhombic bands--
in terms of the $\left( xy,yz,xz\right) $ set of Wannier functions for LaTiO$%
_{3}$ shown (before orthogonalization) in figure \ref{fig4}.

In the top row of this figure we see that the $t_{2g}$ orbitals have $\pi $
antibonding oxygen $p$ character. This is the partner to the $\pi $ bonding
B\thinspace $t_{2g}$ character of the oxygen $p$ orbitals in the cube faces
seen in figures \ref{fig1b} and \ref{fig1c}. Since those are the oxygen
orbitals with which the A\thinspace $d$ orbitals interact strongly ($\sigma $
interaction) and thereby cause the GdFeO$_{3}$-type distortion, they mix
less with the B $t_{2g}$ orbitals and the oxygen $p$ character of the
B\thinspace $t_{2g}$ orbitals in figure \ref{fig4} is correspondingly weak.
The oxygen character of the $t_{2g}$ orbitals therefore decreases with
increasing O-A covalency and the concomitant decrease of the shortest O-A
distances (figure \ref{fig1d}). Adding to this trend comes the reduction of
oxygen character, $\propto t_{pd\pi }^{2}/\left( \epsilon _{d}-\epsilon
_{p}\right) ^{2},$ caused by the fact that the 3$d$ level of the earlier
transition-metal ion, Ti, is higher above the O\thinspace 2$p$ level than
that of the later transition-metal ion, V. This increase of $\epsilon
_{d}-\epsilon _{p}$ seen in figure \ref{fig1} is partly compensated by an
increase of $t_{pd\pi }$ caused by the expansion of the 3$d$ orbital when
going from V to Ti.

In addition to the oxygen $pd\pi $ antibonding character, the $t_{2g}$
orbitals have some A character and, in the distorted titanates, some $e_{g}$
character on neighbouring B sites where it bonds to the appropriate $p$ lobe
of a displaced oxygen. Such characters are barely seen in figure \ref{fig4},
but they are seen in figure \ref{fig12} where a lower contour was chosen --
albeit in subcell 2 and for a particular linear combination of $t_{2g}$
orbitals.

The oxygen character is decisive for the \emph{width} of the $t_{2g}$ band,
that is, for the overall size of the matrix elements of the Hamiltonian
between $t_{2g}$ Wannier orbitals on different sites (hopping integrals).
The A character is important for \emph{lifting} the cubic symmetry of on-
and off-site matrix elements, and the B\thinspace $e_{g}$ character induces
on-site coupling between the orbitals. In order to demonstrate the specifics
of this, we must first explain how the shapes of the Wannier orbitals are
modified by the A $spd$ and B\thinspace $e_{g}$ characters:

For \emph{all three }$t_{2g}$ orbitals in the top row of figure \ref{fig4}
(subcell 1), the red lobes attain some bonding $sp$ and $d_{3z_{111}^{2}-1}=%
\left( d_{xy}+d_{yz}+d_{xz}\right) /\sqrt{3}$ character on the \emph{nearest}
A neighbours, the ones at approximately $\frac{1}{2}\frac{1}{2}%
\frac{1}{2} .$ The red lobes thereby stretch along the short A-B-A
diagonal (orange bar in figure \ref{fig1d}). This is not merely an effect of
A-B covalency, but the oxygens set the stage as follows: the antibonding
oxygen characters of the $t_{2g}$ orbitals are deformed by bonding
interaction with their nearest A neighbour (blue arrows in figure \ref{fig1d}%
), and those A neighbours are thereby prevented from bonding to $t_{2g}$. As
a result, the blue oxygen $p$ lobes push the red $t_{2g}$ lobes towards
those two A ions which are \emph{not} nearest neighbour to any oxygen, 
\textit{i.e.}\thinspace those along the [111] diagonal. The latter A ions
are the ones free to bond with the red $t_{2g}$ lobes, and they do this by
contributing A\thinspace $sp~d_{3z_{111}^{2}-1}$ character. This stretching
of the $t_{2g}$ orbitals along the shortest A-B-A diagonal is less
pronounced in YTiO$_{3}$ than in LaTiO$_{3},$ because in YTiO$_{3},$ O1 is
unusually close (red arrow in figure \ref{fig1d}) to the A ion at 
$\frac{1}{2}\frac{1}{2}\frac{1}{2}$ and thereby partly blocks the A-B
covalency. We shall later see how this difference between the two titanates
influences their physical properties.

The $yz$ orbital is special in that its blue lobes attain bonding $sp$ and $%
d_{xy}$ character on the two 2nd-nearest A neighbours, 
the ones at $\frac{1}{2}\frac{-1}{2}\frac{1}{2}$ in subcell 1. This is because
the $yz$ orbital has no $p$ character on those oxygens, $\frac{1}{2}%
00, $ which have the $\frac{1}{2}\frac{-1}{2}\frac{1}{2}$
ions as their closest A neighbours, and therefore cannot prevent the blue $%
yz$ lobes from hybridizing with them. At the same time, a blue $yz$ lobe is
pushed towards the ion at $\frac{1}{2}\frac{-1}{2}\frac{1}{2}$
by the red O2 $p_{z}$ lobe tending to bond with \emph{its} nearest A
neighbour, the one at $\frac{1}{2}\frac{-1}{2}\frac{1}{2} .$
The hybridization is with A\thinspace $d_{xy}$ because the tilt has moved
the blue $yz$ lobe towards the $c$ direction, but the red O1 $p_{y}$ lobe
then repels the blue $yz$ lobe, which finally ends up running parallel to
the flat O1-A face (mirror plane). The strong displacement of O1 leaves
space for the blue $yz$ lobe to bend over and become $d_{xy}$-like at 
A $\frac{1}{2}\frac{-1}{2}\frac{1}{2} .$ This effect is stronger
in YTiO$_{3}$ than in LaTiO$_{3},$ where also the 2nd-shortest A-B distance
is 5\% shorter than the average, rather than merely 1\%. As a consequence,
there is more A\thinspace $d_{xy}$ character in YTiO$_{3}$ than in LaTiO$%
_{3}.$ This difference between the two titanates will also turn out to be
important for their physical properties.

The $xy$ orbital is the one whose orientation is most influenced by the GdFeO%
$_{3}$-type distortion, simply because the axes of tilt and rotation are
perpendicular to its lobe axes. The tilt moves the blue lobes towards the
2nd-nearest A ions at $\frac{1}{2}\frac{-1}{2}\frac{1}{2} $, where
the $xy$ orbital attains some bonding A\thinspace $sp~d_{xy}$ character, but
this is blocked by the attraction of the red O2 $p_{y}$ lobe to its closest
A neighbour, as was explained above. For small GdFeO$_{3}$-type distortions, 
\textit{i.e.} in CaVO$_{3},$ this reshaping of the $t_{2g}$ orbital by O-A
covalency can be neglected, and the pure tilt of the $xy$ orbital and its
hybridization with A $sp~d_{xy}$ are the most important effects of the
distortion.

Finally, we restate that an oxygen $p$ lobe which points opposite to the
direction of its displacement in the cube face, may attain bonding $e_{g}$
character on the B neighbour. This enhances the $p$ lobe and bends it
outwards. This effect increases with the displacement of oxygen, \textit{i.e.%
} with the degree of GdFeO$_{3}$ distortion.

\subsection{Effective hopping integrals\label{hopping}}

In table \ref{tableSr} above, and in tables \ref{tableCa}--\ref{tableY} 
we give the matrix
elements of the LDA Hamiltonian (relatively to $\varepsilon _{F}$), 
\[
H_{\mathbf{0}m^{\prime },\mathbf{R}m}^{LDA}\equiv \left\langle \chi _{%
\mathbf{0}\,t_{2g}\,m^{\prime }}^{\perp }\left\vert \mathcal{H}%
^{LDA}-\varepsilon _{F}\right\vert \chi _{\mathbf{R}\,t_{2g}\,m}^{\perp
}\right\rangle \equiv t_{m^{\prime },m}^{xyz}~, 
\]%
in a $t_{2g}$ Wannier representation (\ref{eq3}). Here, $\mathbf{R}=x\mathbf{%
R}_{x}+y\mathbf{R}_{y}+z\mathbf{R}_{z}$ is a B site (see caption to figure %
\ref{fig1e}) and $x,$ $y,$ and $z$ are integers. The matrix element between
orbitals $m^{\prime }$ and $m,$ both on site 000, is $t_{m^{\prime
},m}^{000},$ and the hopping integral from orbital $m^{\prime }$ on site 000
to orbital $m$ on site $xyz$ is $t_{m^{\prime },m}^{xyz}.$ Having the
symmetry of $d$ functions, these orbitals are \emph{even }with respect to
inversion in their own B site and, as a consequence, 
\begin{equation}
t_{m^{\prime },m}^{xyz}=t_{m^{\prime },m}^{\bar{x}\bar{y}\bar{z}}.
\label{invers}
\end{equation}%
In this set of tables, the cubic basis $\left( m=yz,xz,xy\right) $ is
used, with orbitals named as in the middle row of figure \ref{fig4}. The
values of the matrix elements,%
\[
H_{\mathbf{R}^{\prime }m^{\prime },\mathbf{R}m}^{LDA}\equiv \left\langle
\chi _{\mathbf{R}^{\prime }\,t_{2g}\,m^{\prime }}^{\perp }\left\vert 
\mathcal{H}^{LDA}-\varepsilon _{F}\right\vert \chi _{\mathbf{R}%
\,t_{2g}\,m}^{\perp }\right\rangle \equiv t_{m^{\prime },m}^{\mathbf{R}%
^{\prime },\mathbf{R}}~ 
\]%
with $\mathbf{R}^{\prime }\neq \mathbf{0,}$ are given by the following
rules, obtained by use of the mirror $\left( z\leftrightarrow -z\right) $
and the glide-mirror $\left( x\leftrightarrow y\right) :$%
\begin{equation}
\fl 
\begin{array}{lll}
t_{m^{\prime},m}^{\left( 0,0,1\right), \left( x,y,z+1\right) }\!=\!
t_{m^{\prime}\left( z\rightarrow -z\right), m\left( z\rightarrow -z\right)
}^{xy\bar{z}} & \  & t_{m^{\prime},m}^{\left(0,1,1\right),
\left(x,y+1,z+1\right) }\!\!=\!t_{m^{\prime }\left( x\leftrightarrow y,
z\rightarrow -z\right), m\left( x\leftrightarrow y,~z\rightarrow -z\right)
}^{yx\bar{z}} \\ 
&  &  \\[-5pt] 
t_{m^{\prime},m}^{\left( 0,0,0\right) ,\left( x,y,z\right) }\equiv
t_{m^{\prime},m}^{xyz} &  & t_{m^{\prime },m}^{\left( 0,1,0\right) ,\left(
x,y+1,z\right) }=t_{m^{\prime }\left( x\leftrightarrow y\right) ,~m\left(
x\leftrightarrow y\right) }^{yxz}%
\end{array}
\label{eqCubicHop}
\end{equation}



Before discussing these results, we stress that our downfolding to $t_{2g}$%
Wannier functions and evaluation of the Hamiltonian matrix is numerically
exact. That is, our Wannier functions span the Kohn-Sham $t_{2g}$
eigenfunctions exactly. Specifically, \emph{all} partial waves other than B $%
t_{2g}$ are downfolded, and their dependence on energy over the range of the 
$t_{2g}$ band is properly represented. This is explained in 
appendix A and
is illustrated in figures \ref{fig2} and \ref{fig3}. The downfolding
procedure used in model calculations such as \cite{Mizokawa96,Mochizuki04}
is different, and not exact in the above-mentioned sense. In model
calculations, only orbitals such as O 2$p,$ and recently A 5$d,$ expected to
be relevant are downfolded, and their hopping integrals are assumed to
follow the Slater-Koster rules. The latter may be a serious approximation,
because, \textit{e.g.}, the O $p$ orbitals have A and B $s$ and $p$ tails
unless \emph{those} orbitals are also being downfolded explicitly. Moreover,
model calculations use L\"{o}wdin downfolding with $\varepsilon \equiv
\epsilon _{t_{2g}}$ (see equation (\ref{eq6}))$.$ In the NMTO scheme, this
corresponds to taking $N=0,$ that is, to using merely a single energy point.
Once the Coulomb correlations are taken into account, not all the hopping
integrals we tabulate may be relevant, of course, but the dominant ones are,
and their values matter.

We now discuss our results and start by considering various contributions to
a dominant $t_{2g}$ hopping integral, namely the effective,
1st-nearest-neighbour $dd\pi $ hopping integral $t_{\pi }.$ A particularly
important example will turn out to be $t_{yz,yz}^{001}.$ This is the matrix
element of the LDA Hamiltonian between the $yz$ Wannier orbital in subcell 1
and the similar orbital in subcell 3, defined as in the middle row of figure %
\ref{fig4}. Its value is seen to be \emph{negative }(bonding), and to take
the values $-281$\thinspace meV in cubic SrVO$_{3},$ $-240$\thinspace meV in
CaVO$_{3},$ $-193$\thinspace meV in LaTiO$_{3},$ and a mere $-65$\thinspace
meV in YTiO$_{3}.$ This hopping integral between Wannier orbitals may be
interpreted in terms of the energies of, and the hopping integrals between
the simpler orbitals of a large basis --like the one giving rise to the
black bands in figure \ref{fig2} and like the ones discussed in for instance
references \cite{Mizokawa96,Mochizuki04,Mahedavan}-- containing O\thinspace $%
p,$ the B\thinspace $d$ and A\thinspace $s$ and A\thinspace $d$ orbitals,
all assumed to be orthonormal and to have interactions between 1st-nearest
neighbours only. If we perform L\"{o}wdin downfolding (\ref{eq6}) of the
Hamiltonian in the large basis and use the middle row of figure \ref{fig4}
to keep track of the geometry and the signs, we obtain the following kind of
result:%
\begin{eqnarray}
\fl 
t_{\pi}: \quad
t_{yz,yz}^{001}\ &\sim &\ t_{dd\pi }-\frac{t_{pd\pi }^{2}}{\varepsilon
-\epsilon _{p}}-\frac{4t_{Byz,As}^{2}}{\varepsilon -\epsilon _{As}}+\frac{%
4t_{Byz,Ayz}^{2}}{\varepsilon -\epsilon _{Ad}}+\frac{4t_{Byz,Axz}^{2}}{%
\varepsilon -\epsilon _{Ad}}-\frac{4t_{Byz,Axy}^{2}}{\varepsilon -\epsilon
_{Ad}}  \nonumber \\* 
\fl &=&t_{dd\pi }-\frac{t_{pd\pi }^{2}}{\varepsilon -\epsilon _{p}}+\frac{%
4t_{Byz,As}^{2}}{\epsilon _{As}-\varepsilon }-\frac{4t_{Byz,Ayz}^{2}}{%
\epsilon _{Ad}-\varepsilon }-\frac{4t_{Byz,Axz}^{2}}{\epsilon
_{Ad}-\varepsilon }+\frac{4t_{Byz,Axy}^{2}}{\epsilon _{Ad}-\varepsilon }.
\label{eq7}
\end{eqnarray}%
In the second line all signs are \emph{explicit} because the \emph{direct}
hopping integral $t_{dd\pi }$ is \emph{positive }(anti-bonding), and $%
\epsilon _{p}<\varepsilon <\epsilon _{As}\sim \epsilon _{Ad}$ when $%
\varepsilon $ is in the $t_{2g}$\thinspace band. The dominant term is the
second one, the $pd\pi $ hopping via oxygen. This term is \emph{bonding}
(negative) and is \emph{weakened} by the direct $dd\pi $ term, as well as by
hops via A orbitals which are \emph{even} with respect to the mirror plane
between the two $yz$ orbitals. The oxygen-mediated $pd\pi $ hopping is \emph{%
strengthened} by hops via A orbitals which are \emph{odd} with respect to
the mirror plane.

That the first two terms in equation\thinspace (\ref{eq7}) have opposite
signs, and that the second term dominates, should 
warn against
interpreting hopping integrals between Wannier orbitals as overlap
integrals. 
Moreover, characters barely visible in figures like \ref{fig4}
and \ref{fig1}, can contribute significantly to hopping if their energy is
high, because the contribution to a hopping integral by a character of
magnitude $t_{BA}^{2}/\left( \epsilon _{A}-\varepsilon \right) ^{2}$ is $%
t_{BA}^{2}/\left( \epsilon _{A}-\varepsilon \right) ,$ that is, $\epsilon
_{A}-\varepsilon $ times larger.

Now, the contributions from the four nearest A neighbours to an A term in
expression (\ref{eq7}) are identical only if the structure is cubic. In this
case, actually, the two last terms in equation (\ref{eq7}) cancel. Moreover,
only if the structure is cubic do hopping integrals like $t_{dd\pi }$ and $%
t_{pd\pi }$ have their full value. Otherwise, they are reduced due to
misalignment.

\begin{table}[t]
\caption{$H_{t_{2g}}^{LDA}$ {in meV for CaVO}$_{{3}}$~\cite{castr}.}
\label{tableCa}
\begin{center}
\smallskip {\setlength{\tabcolsep}{3pt} 
\begin{tabular}{crrrrrrrrrrrrrr}
\br $xyz $ & $000$ & $001$ & $010$ & $100$ & $011$ & $01\bar{1}$ & $101$ & $%
10\bar{1}$ & $110$ & $1\bar{1}0$ & $111$ & $11\bar{1}$ & $\bar{1}11$ & $1%
\bar{1}1$ \\ 
$m^{\prime },m$ &  &  &  &  &  &  &  &  &  &  &  &  &  &  \\ 
$yz,yz$ & $620$ & $-240$ & $-223$ & $-17$ & $-91$ & $-90$ & $6$ & $10$ & $2$
& $10$ & $-8$ & $-8$ & $-3$ & $-3$ \\ 
$xz,xz$ & $612$ & $-249$ & $-17$ & $-223$ & $10$ & $6$ & $-90$ & $-91$ & $3$
& $3$ & $-2$ & $-2$ & $-7$ & $-7$ \\ 
$xy,xy$ & $542$ & $-23$ & $-231$ & $-231$ & $6$ & $7$ & $7$ & $6$ & $-98$ & $%
-85$ & $-2$ & $-2$ & $-3$ & $-3$ \\ 
$yz,xz$ & $4$ & $-21$ & $45$ & $45$ & $6$ & $22$ & $22$ & $6$ & $8$ & $-2$ & 
$-1$ & $0$ & $0$ & $0$ \\ 
$xz,yz$ & $4$ & $-21$ & $-33$ & $-33$ & $-20$ & $-1$ & $-1$ & $-20$ & $8$ & $%
-2$ & $0$ & $-1$ & $0$ & $0$ \\ 
$yz,xy$ & $-17$ & $-36$ & $-17$ & $-25$ & $-2$ & $13$ & $8$ & $-4$ & $-17$ & 
$5$ & $-5$ & $-4$ & $-2$ & $-5$ \\ 
$xy,yz$ & $-17$ & $36$ & $-34$ & $33$ & $11$ & $-5$ & $0$ & $-11$ & $-17$ & $%
5$ & $4$ & $5$ & $5$ & $2$ \\ 
$xz,xy$ & $-4$ & $25$ & $33$ & $-34$ & $11$ & $0$ & $5$ & $-11$ & $15$ & $-1$
& $8$ & $0$ & $4$ & $2$ \\ 
$xy,xz$ & $-4$ & $-25$ & $-25$ & $-17$ & $4$ & $-8$ & $-13$ & $2$ & $15$ & $%
-1$ & $0$ & $-8$ & $-2$ & $-4$ \\ 
\br &  &  &  &  &  &  &  &  &  &  &  &  &  & 
\end{tabular}%
}%
\end{center}
\bigskip
\par
\caption{$H_{t_{2g}}^{LDA}$ in meV{\ for LaTiO$_{3}$ using the older data 
\protect\cite{lastr}.}}
\label{tableLaMac}
\begin{center}
\smallskip {\setlength{\tabcolsep}{3pt} 
\begin{tabular}{crrrrrrrrrrrrrr}
\br $xyz$ & $000$ & $001$ & $010$ & $100$ & $011$ & $01\bar{1}$ & $101$ & $10%
\bar{1}$ & $110$ & $1\bar{1}0$ & $111$ & $11\bar{1}$ & $\bar{1}11$ & $1\bar{1%
}1$ \\ 
$m^{\prime },m$ &  &  &  &  &  &  &  &  &  &  &  &  &  &  \\ 
$yz,yz$ & $474$ & $-202$ & $-195$ & $-8$ & $-57$ & $-67$ & $-8$ & $18$ & $4$
& $-3$ & $-9$ & $-9$ & $2$ & $2$ \\ 
$xz,xz$ & $543$ & $-221$ & $-8$ & $-195$ & $18$ & $-8$ & $-67$ & $-57$ & $%
-10 $ & $14$ & $-1$ & $-1$ & $-7$ & $-7$ \\ 
$xy,xy$ & $525$ & $-30$ & $-197$ & $-197$ & $-3$ & $13$ & $13$ & $-3$ & $-60$
& $-56$ & $3$ & $3$ & $4$ & $4$ \\ 
$yz,xz$ & $-21$ & $-36$ & $77$ & $77$ & $20$ & $16$ & $16$ & $20$ & $-2$ & $%
8 $ & $-6$ & $11$ & $-10$ & $-11$ \\ 
$xz,yz$ & $-21$ & $-36$ & $-38$ & $-38$ & $-24$ & $-13$ & $-13$ & $-24$ & $%
-2 $ & $8$ & $11$ & $-6$ & $-11$ & $-10$ \\ 
$yz,xy$ & $-78$ & $-58$ & $-27$ & $-43$ & $19$ & $11$ & $-3$ & $4$ & $-22$ & 
$-3$ & $0$ & $-6$ & $-21$ & $4$ \\ 
$xy,yz$ & $-78$ & $58$ & $-49$ & $72$ & $-14$ & $10$ & $-12$ & $-1$ & $-22$
& $-3$ & $6$ & $0$ & $-4$ & $21$ \\ 
$xz,xy$ & $-60$ & $54$ & $72$ & $-49$ & $1$ & $12$ & $-10$ & $14$ & $12$ & $%
6 $ & $22$ & $-6$ & $11$ & $0$ \\ 
$xy,xz$ & $-60$ & $-54$ & $-43$ & $-27$ & $-4$ & $3$ & $-11$ & $-19$ & $12$
& $6$ & $6$ & $-22$ & $0$ & $-11$ \\ 
\br &  &  &  &  &  &  &  &  &  &  &  &  &  & 
\end{tabular}
} 
\end{center}
\end{table}

\begin{table}[t]
\caption{$H_{t_{2g}}^{LDA}$ in meV for LaTiO$_{3}$ 
\protect\cite{Cwik03}.$^{\rm *}$}
\label{tableLaCwik}
\begin{center}
\smallskip {\setlength{\tabcolsep}{3pt} 
\begin{tabular}{crrrrrrrrrrrrrr}
\br $xyz$ & $000$ & $001$ & $010$ & $100$ & $011$ & $01\bar{1}$ & $101$ & $10%
\bar{1}$ & $110$ & $1\bar{1}0$ & $111$ & $11\bar{1}$ & $\bar{1}11$ & $1\bar{1%
}1$ \\ 
$m^{\prime },m$ &  &  &  &  &  &  &  &  &  &  &  &  &  &  \\ 
$yz,yz$ & $445$ & $-193$ & $-185$ & $1$ & $-46$ & $-49$ & $-6$ & $21$ & $-1$
& $-3$ & $-6$ & $-6$ & $0$ & $0$ \\ 
$xz,xz$ & $530$ & $-208$ & $1$ & $-185$ & $21$ & $-6$ & $-49$ & $-46$ & $-17$
& $2$ & $5$ & $5$ & $-1$ & $-1$ \\ 
$xy,xy$ & $486$ & $-22$ & $-183$ & $-183$ & $2$ & $8$ & $8$ & $2$ & $-47$ & $%
-46$ & $5$ & $5$ & $0$ & $0$ \\ 
$yz,xz$ & $-42$ & $-42$ & $75$ & $75$ & $21$ & $20$ & $20$ & $21$ & $-2$ & $%
7 $ & $-2$ & $7$ & $-6$ & $-7$ \\ 
$xz,yz$ & $-42$ & $-42$ & $-43$ & $-43$ & $-21$ & $-4$ & $-4$ & $-21$ & $-2$
& $7$ & $7$ & $-2$ & $-7$ & $-6$ \\ 
$yz,xy$ & $-89$ & $-59$ & $-34$ & $-33$ & $17$ & $11$ & $-2$ & $1$ & $-23$ & 
$-2$ & $-2$ & $-8$ & $-12$ & $2$ \\ 
$xy,yz$ & $-89$ & $59$ & $-54$ & $73$ & $-14$ & $10$ & $-13$ & $-5$ & $-23$
& $-2$ & $8$ & $2$ & $-2$ & $12$ \\ 
$xz,xy$ & $-42$ & $52$ & $73$ & $-54$ & $5$ & $13$ & $-10$ & $14$ & $9$ & $%
11 $ & $19$ & $-3$ & $11$ & $0$ \\ 
$xy,xz$ & $-42$ & $-52$ & $-33$ & $-34$ & $-1$ & $2$ & $-11$ & $-17$ & $9$ & 
$11$ & $3$ & $-19$ & $0$ & $-11$ \\ 
\br 
\end{tabular}
} %
\end{center}
$^{\rm *}$ We used their room-temperature data. 
For calculation of exchange constants, we also used the 8\thinspace K
data for LaTiO$_{3}.$ The result is shown in table \ref{tableLa8noJT16G}.
\bigskip 
\par
\caption{$H_{t_{2g}}^{LDA}$ in meV for{\ YTiO$_{3}$~\cite{ystr}.}}
\label{tableY}
\begin{center}
\smallskip {\setlength{\tabcolsep}{3pt} 
\begin{tabular}{crrrrrrrrrrrrrr}
\br $xyz$ & $000$ & $001$ & $010$ & $100$ & $011$ & $01\bar{1}$ & $101$ & $10%
\bar{1}$ & $110$ & $1\bar{1}0$ & $111$ & $11\bar{1}$ & $\bar{1}11$ & $1\bar{1%
}1$ \\ 
$m^{\prime },m$ &  &  &  &  &  &  &  &  &  &  &  &  &  &  \\ 
$yz,yz$ & $375$ & $-65$ & $-184$ & $28$ & $-34$ & $-26$ & $-7$ & $18$ & $14$
& $14$ & $-3$ & $-3$ & $-13$ & $-13$ \\ 
$xz,xz$ & $605$ & $-178$ & $28$ & $-184$ & $18$ & $-7$ & $-26$ & $-34$ & $-5$
& $-39$ & $8$ & $8$ & $11$ & $11$ \\ 
$xy,xy$ & $417$ & $-8$ & $-162$ & $-162$ & $9$ & $-1$ & $-1$ & $9$ & $-50$ & 
$8$ & $14$ & $14$ & $0$ & $0$ \\ 
$yz,xz$ & $43$ & $-63$ & $70$ & $70$ & $11$ & $34$ & $34$ & $11$ & $19$ & $%
13 $ & $-2$ & $1$ & $2$ & $-9$ \\ 
$xz,yz$ & $43$ & $-63$ & $-41$ & $-41$ & $-19$ & $1$ & $1$ & $-19$ & $19$ & $%
13$ & $1$ & $-2$ & $-9$ & $2$ \\ 
$yz,xy$ & $-103$ & $-64$ & $-54$ & $-22$ & $15$ & $0$ & $-7$ & $2$ & $-18$ & 
$10$ & $9$ & $-10$ & $15$ & $1$ \\ 
$xy,yz$ & $-103$ & $64$ & $-64$ & $65$ & $-13$ & $7$ & $-23$ & $-12$ & $-18$
& $10$ & $10$ & $-9$ & $-1$ & $-15$ \\ 
$xz,xy$ & $-17$ & $83$ & $65$ & $-64$ & $12$ & $23$ & $-7$ & $13$ & $16$ & $%
16$ & $24$ & $-10$ & $9$ & $4$ \\ 
$xy,xz$ & $-17$ & $-83$ & $-22$ & $-54$ & $-2$ & $7$ & $0$ & $-15$ & $16$ & $%
16$ & $10$ & $-24$ & $-4$ & $-9$ \\ 
\br &  &  &  &  &  &  &  &  &  &  &  &  &  & 
\end{tabular}%
} 
\end{center}
\end{table}

In SrVO$_{3}$ the contribution from hops via A ions is small, as evidenced
in figure \ref{fig1} by the weak Sr\thinspace $d$ character in the
V\thinspace $t_{2g}$ band. But in the titanates, the $t_{2g}$ band has
almost as much A\thinspace $d$ as O\thinspace $p$ character. This is not so
evident from the appearance of a single $t_{2g}$ orbital in figure \ref{fig4}%
, because for a B\thinspace $t_{2g}$ orbital, B:A=1:8 rather than 1:1, as in
the formula unit, ABO$_{3}.$ Similarly, for a B\thinspace $t_{2g}$ orbital,
B:O=1:4 rather than 1:3. In the heavily distorted YTiO$_{3},$ the hopping
integral $t_{Byz,Axy}$ to the 2nd-nearest Y, the one at position 
$\frac{1}{2}\frac{-1}{2}\frac{1}{2}$ in subcell 1, is large, and
that is the main reason for the \emph{anomalously small value} \emph{of} $%
t_{yz,yz}^{001}$ \emph{in} YTiO$_{3}$. The values of the five other $t_{\pi
} $ integrals$,$ $t_{xz,xz}^{001}$, $t_{yz,yz}^{010}=t_{xz,xz}^{100}$%
,\thinspace\ and $t_{xy}^{100}=t_{xy}^{010},$ are normal and approximately $%
-170$ meV in this material.

When estimating the value of $t_{yz,yz}^{010}$ it should be remembered that,
with the notation of the middle row of figure \ref{fig4}, the $yz$ orbital
in subcell 2 has the shape of the $xz$ orbital in subcell 1; this is the
reason why $t_{yz,yz}^{010}$ does not have an abnormal A\thinspace $d_{xy}$
contribution, and also the reason why $t_{yz,yz}^{010}=t_{xz,xz}^{100}.$

The \emph{average} of the $t_{\pi }$ integrals in the tables decreases by a
surprisingly constant ratio of 22\% for every step we proceed along the
series SrVO$_{3}$--CaVO$_{3}$--LaTiO$_{3}$--YTiO$_{3}.$ This holds for the
recent LaTiO$_{3}$ data \cite{Cwik03}, while the older data \cite{lastr} 
yield a 16\% drop from CaVO$_{3}$ and a 29\% drop to YTiO$_{3}.$

When deriving expression (\ref{eq7}) from equation\thinspace (\ref{eq6}) in
configuration-space representation we have used low-order perturbation
theory and have not properly inverted $\left\langle P\left\vert \varepsilon -%
\mathcal{H}\right\vert P\right\rangle .$ This means that the reduction of
oxygen character in the B$\,t_{2g}$ Wannier orbitals due to O-A covalency,
and also the direct oxygen $pp$ hopping, are not included in expression (\ref%
{eq7}). But effectively, they both reduce the magnitude of all terms, except
the first. Moreover, higher-order hopping processes such as B-O-A-B and
B-O-A-O-B are neglected. Finally, the enhancement and outward bending of a $%
p $ lobe on a displaced oxygen due to acquisition of bonding $e_{g}$
character on the B neighbour is also neglected, but it effectively increases 
$t_{pd\pi }$ towards that B neighbour.

The second largest intra-orbital hopping integral is the effective
2nd-nearest-neighbour $dd\sigma $ integral, $t_{\sigma }^{\prime }$. Like $%
t_{\pi }$ in (\ref{eq7}), it is \emph{bonding }(negative), but smaller
because the distance is longer and because there is no oxygen-mediated
hopping, except in case of the heavily distorted YTiO$_{3}$. The typical
integrals $t_{xy,xy}^{110}$ and $t_{xy,xy}^{1\bar{1}0}$ are both $-96$ meV
in cubic SrVO$_{3},$ $-98$ and $-85$ meV in CaVO$_{3},$ $-47$ and $-46$ meV
in LaTiO$_{3},$ and $-50$ and $+8$ meV in YTiO$_{3}.$ From the middle left
part of figure \ref{fig4}, we gather that the integral $t_{xy,xy}^{110}$ $%
\left( t_{xy,xy}^{1\bar{1}0}\right) $ from subcell 1 to the same subcell
translated by $\mathbf{b~}\left( \mathbf{a}\right) $ has contributions from
direct $dd\sigma $ like hopping --which is now bonding and therefore
strengthening,-- from A mediated hopping, and from $dp\pi $-$pp$-$pd\pi $
hops. For the two former, we get:%
\begin{eqnarray}
\fl 
t_{\sigma}^{\prime}: \quad
t_{xy,xy}^{110}\left( t_{xy,xy}^{1\bar{1}0}\right) &\sim &t_{dd\sigma }+%
\frac{2t_{Bxy,As}^{2}}{\varepsilon -\epsilon _{As}}-\frac{2t_{Bxy,Ayz}^{2}}{%
\varepsilon -\epsilon _{Ad}}-\frac{2t_{Bxy,Axz}^{2}}{\varepsilon -\epsilon
_{Ad}}+\frac{2t_{Bxy,Axy}^{2}}{\varepsilon -\epsilon _{Ad}}  \nonumber \\* 
\fl &=&-\left\vert t_{dd\sigma }\right\vert -\frac{2t_{Bxy,As}^{2}}{%
\epsilon _{As}-\varepsilon }+\frac{2t_{Bxy,Ayz}^{2}}{\epsilon
_{Ad}-\varepsilon }+\frac{2t_{Bxy,Axz}^{2}}{\epsilon _{Ad}-\varepsilon }-%
\frac{2t_{Bxy,Axy}^{2}}{\epsilon _{Ad}-\varepsilon }.  \label{eq7a}
\end{eqnarray}%
For $t_{xy,xy}^{110},$ the two important A ions are $\frac{1}{2}\frac{%
1}{2}\frac{\pm 1}{2}$ when seen from subcell 1, and $\frac{-1%
}{2}\frac{-1}{2}\frac{\pm 1}{2}$ when seen from $\mathbf{b.}$ For $%
t_{xy,xy}^{1\bar{1}0},$ the two important A-ions are $\frac{1}{2}%
\frac{-1}{2}\frac{\pm 1}{2} $ and $\frac{-1}{2}\frac{1}{2}%
\frac{\pm 1}{2}$ as seen from respectively the origin and $\mathbf{a}
$. Keeping in mind the A characters of the $xy$ orbital, we realize that
although the intermediate A ions are different in cases $\mathbf{b}$ and $%
\mathbf{a},$ strong A\thinspace $d_{xy}$ or $d_{yx/xz}$ character does not
exist on the \emph{same} A ion; hence, A $d$ mediated coupling is weak in
all cases. The two first terms in expression (\ref{eq7a}) seem to dominate,
except in YTiO$_{3}$ where the rotation of the $xy$ orbital and the
deformation of its $d$ and $p$ lobes are so strong that there is hopping, $%
t_{pd},$ from the red $p_{y}$ lobe on O2 to the blue $xy$ lobe translated by 
$\mathbf{a.}$ This gives a \emph{positive} term, $t_{pd\pi }t_{pd}/\left(
\varepsilon -\epsilon _{p}\right) ,$ which diminishes $t_{xy,xy}^{1\bar{1}%
0}. $ Due to the different deformation of the $xy$ lobe, no such term exists
for $t_{xy,xy}^{110}.$

The effective 1st-nearest-neighbour $dd\delta $-hopping, $t_{\delta },$ 
\textit{e.g.} $t_{yz,yz}^{100},$ is an order of magnitude smaller than $%
t_{\pi }$. It consists of the direct term $-\left\vert t_{dd\delta
}\right\vert ,$ plus A mediated contributions. It is negative for SrVO$_{3},$
changes sign near LaTiO$_{3},$ and is positive for YTiO$_{3}.$

If the semi-quantitative expressions (\ref{eq7}) and (\ref{eq7a}) for the
effective hopping integrals are inserted in equation (\ref{eq8}) for the
dispersion of a cubic $t_{2g}$ band, the results of the discussion at the
beginning of this section will of course be reproduced.

We may define the root-mean-square value, $t_{rms},$ of the hopping
integrals by $\sum t^{2}\equiv 12\left( t_{rms}\right) ^{2},$ where the sum
runs over all neighbours and all three orbitals, and where 12 is the number
of orbitals times the number of strong $\left( t_{\pi }\right) $ hopping
integrals in the cubic structure. These rms values are listed in table \ref%
{tabletrms} and are seen to decrease with 19\% from SrVO$_{3}$ to CaVO$_{3},$
with 15\% from CaVO$_{3}$ to LaTiO$_{3}$ (using the recent data \cite{Cwik03}%
), and by 11\% from LaTiO$_{3}$ to YTiO$_{3}.$ Even though the hopping
integrals of type $t_{\delta }$ and $t_{\sigma }^{\prime },$ which decrease
rapidly through the series, are included in the $t^{2}$ sum$,$ the
19--15--11\% decrease of $t_{rms}$ is significantly smaller than the
22--22--22\% decrease of the average $t_{\pi }$. The reason is that the
hopping \emph{between }orbitals increases; without including inter-orbital
hopping in the $t^{2}$-sum, $t_{rms}$ would be 11\% smaller in LaTiO$_{3}$ (%
\cite{Cwik03}) and 19\% smaller in YTiO$_{3}.$

\begin{table}[h]
\caption{Rms values of the hopping integrals in meV}
\label{tabletrms}
\begin{center}
\begin{tabular}{cccccc}
\br & SrVO$_{3}$\cite{srstr} & CaVO$_{3}$\cite{castr} & LaTiO$_{3}$\cite%
{lastr} & LaTiO$_{3}$\cite{Cwik03} & YTiO$_{3}$\cite{ystr} \\ 
$t_{rms}$ & 298 & 250 & 228 & 217 & 196 \\ 
\br &  &  &  &  & 
\end{tabular}%
%
%
%
%
%
%
%
%
%
\end{center}
\end{table}

\subsection{Effective on-site matrix elements\label{on-site}}

The Hamiltonian matrix elements most influenced by the lowering of the
symmetry in the distorted perovskites are not the hopping integrals, but the 
\emph{on-site} elements. The change of a hopping integral, however, usually
perturbs the LDA bandstructure more than the same change of an on-site
element, because the perturbation via the former is multiplied by the number
of neighbours reached by that kind of hop. On the other hand, Coulomb
correlations will effectively enhance the on-site elements and reduce the
hopping integrals.

As seen in table~\ref{tableCa} for CaVO$_{3}$, 
the energy of the $xy$ orbital is 75 meV
below that of the nearly degenerate $yx$ and $xz$ orbitals. One reason is
that, for weak GdFeO$_{3}$-distortion, the $xy$ orbital is the one whose
orientation is most optimized with respect to the A anions, specifically the
two along [1\={1}1]. The \emph{electrostatic field} from these positive ions
lowers the energy of orbitals pointing towards them. And so does the \emph{%
ligand field} caused by hybridization with A orbitals, because for a \emph{%
diagonal} matrix element, all interactions with characters of higher energy
are bonding and push the energy of the Wannier orbital down. Conversely, all
interactions with characters of lower energy are antibonding and push the
energy up. As mentioned in connection with equation (\ref{eq7}), $\epsilon
_{A}>\varepsilon >\epsilon _{p}.$ An additional reason for the energy of the 
$xy$ orbital being the lowest is, therefore, that it has the least
antibonding oxygen character. This comes about because the $xy$ orbital
interacts with four O2 ions, while each of the $yz$ and $xz$ orbitals
interact with two O1 ions and two O2 ions. In addition, since Ca bonds more
to O2 than to O1 (the distance is 2\% shorter), less O2 character is left
for the V $xy$ orbital. This is a pure ligand-field effect, since from
oxygen there is no electrostatic effect due to the lack of JT distortion in
CaVO$_{3}$.

\begin{table}[tbp]
\caption{$H_{t_{2g}}^{LDA}$ in meV{\ for LaTiO$_{3}$.
}}
\label{tableLa8noJT16G}
\begin{center}
\smallskip {\setlength{\tabcolsep}{2pt} 
\begin{tabular}{crrrrrrrrrrrrrrr}
\br &  & \multicolumn{4}{c}{LaTiO$_3$ at 8K~\cite{Cwik03}} &  & 
\multicolumn{4}{c}{YTiO$_3$ without JT} &  & 
\multicolumn{4}{c}{YTiO$_3$ at 15.9 GPa~\cite{loa}} \\ 
\cline{3-6}\cline{8-11}\cline{13-16}
$xyz$ & \hspace{0.3cm} & $000$ & $001$ & $010$ & $100$ & \hspace{0.3cm} & $%
000$ & $001$ & $010$ & $100$ & \hspace{0.3cm} & $000$ & $001$ & $010$ & $100$
\\ 
$m^{\prime},m$ &  &  &  &  &  &  &  &  &  &  &  &  &  &  &  \\ 
$yz,yz$ &  & $465$ & $-195$ & $-185$ & $3$ &  & $420$ & $-68$ & $-168$ & $21$
&  & $471$ & $-41$ & $-211$ & $41$ \\ 
$xz,xz$ &  & $540$ & $-208$ & $3$ & $-185$ &  & $540$ & $-148$ & $21$ & $%
-168 $ &  & $638$ & $-184$ & $41$ & $-211$ \\ 
$xy,xy$ &  & $500$ & $-21$ & $-185$ & $-185$ &  & $457$ & $-7$ & $-207$ & $%
-207$ &  & $566$ & $-10$ & $-167$ & $-167$ \\ 
$yz,xz$ &  & $-52$ & $-45$ & $76$ & $76$ &  & $36$ & $-89$ & $53$ & $53$ & 
& $94$ & $-79$ & $73$ & $73$ \\ 
$xz,yz$ &  & $-52$ & $-45$ & $-40$ & $-40$ &  & $36$ & $-89$ & $5$ & $5$ & 
& $94$ & $-79$ & $-35$ & $-35$ \\ 
$yz,xy$ &  & $-93$ & $-61$ & $-36$ & $-32$ &  & $-96$ & $-66$ & $-40$ & $-57$
&  & $-129$ & $-82$ & $-61$ & $-13$ \\ 
$xy,yz$ &  & $-93$ & $61$ & $-56$ & $73$ &  & $-96$ & $66$ & $-46$ & $95$ & 
& $-129$ & $82$ & $-75$ & $58$ \\ 
$xz,xy$ &  & $-51$ & $50$ & $73$ & $-56$ &  & $-50$ & $82$ & $95$ & $-46$ & 
& $-23$ & $91$ & $58$ & $-75$ \\ 
$xy,xz$ &  & $-51$ & $-50$ & $-32$ & $-36$ &  & $-50$ & $-82$ & $-57$ & $-40$
&  & $-23$ & $-91$ & $-13$ & $-61$ \\ 
\br &  &  &  &  &  &  &  &  &  &  &  &  &  &  & 
\end{tabular}%
}%
\end{center}
\end{table}

In the \emph{titanates}, the energy of the $yz$ orbital at sites 000 and 001
is $\sim $40 meV below that of the $xy$ orbital. This is so because the
orientation of the $yz$ orbital towards the 1st- and 2nd-nearest A ions,
those along the [111] and [1\={1}1] diagonals, respectively, exploits both
the electrostatic and the ligand fields. An additional ligand-field effect
is that, in the titanates the bonding of the A ion to O1 is stronger than to
O2. Therefore, less antibonding O1 character is available for the Ti $yz$ and 
$xz$ Wannier functions. This can actually be seen in the top row of figure %
\ref{fig4}. Finally, the B neighbour $e_{g}$ character which binds to the
red back lobe of the O1$\,p_{y}$ orbital --the one which is displaced most
along its own direction-- also contributes to lower the energy of the $yz$
Wannier orbital. The orbital with the highest energy at sites 000 and 001 is 
$xz.$ In LaTiO$_{3},$ its energy is 85 meV above that of the $yz$ orbital,
and in YTiO$_{3},$ it is as much as 230 meV above. The main reason is that
the $xz$ orbital is the one of the three which is least favourably oriented
with respect to the A ions, particularly in YTiO$_{3}$. In addition, the $xz$
orbital has little B $e_{g}$ character because its oxygen $p$ lobes are not
directed towards the face centres. A final reason why, particularly in YTiO$%
_{3},$ the energy of the $xz$ orbital is very far above that of the $yz$
orbital is the JT distortion: the distance to Ti 000 of the oxygens along $%
z$ and $x$ is 3\% shorter than the distance along $y,$ and the $pd\pi $
character of the $t_{2g}$ functions is antibonding at the same time as the
electrostatic field from the oxygen ions repel the near $t_{2g}$ lobes.

There is a strong on-site coupling \emph{between} the two \emph{lowest}
orbitals, $xy$ and $yz,$ in the titanates. It is $-89$ meV in LaTiO$_{3}$
and $-103$ meV in YTiO$_{3}.$ This is due to the deformations of the $xy$
and $yz$ orbitals towards the 1st and 2nd-nearest A ions, those along [111]
and [1\={1}1]. With the signs chosen for the two orbitals, the lobes have
the same sign when they point towards a near A ion, and the opposite sign
when they point towards a far A ion. For this reason, both electrostatics
and hybridization with all A characters enforce each other to make this
on-site inter-orbital interaction strong and negative.

In all cases considered, the electrostatic and ligand-field effects work in
the same direction.  In the following, we shall therefore refer to
anisotropy of the on-site Hamiltonian matrix for the $t_{2g}$ Wannier
functions as a \emph{crystal-field effect,} regardless of whether its origin
is more electrostatic or covalent. 

\subsection{Influence of the Jahn-Teller distortion in YTiO$_{3}$\label{JT}}

Although it is now generally recognized that the GdFeO$_{3}$-type distortion
sets up a surprisingly large crystal field in the titanates, it is not
agreed upon to what extent the JT distortion is important for the physical
properties of YTiO$_{3}$ \cite{Eva03,Mochizuki03,Mochizuki04}. In table \ref%
{tableLa8noJT16G}, we have included $H_{t_{2g}}^{LDA}$ calculated for
a hypothetical structure of YTiO$_{3}$ in which the octahedra are perfect,
i.e., they lack the 3\% 
JT-elongation of the Ti-O2 bonds in the $y$($x$) direction 
in subcells 1 and 3 (2 and 4), 
and the orthorhombic lattice constants are as in the real structure \cite%
{ystr}. 
As expected, the energy of the $xz$ orbital at site 000 is lowered, but
remains 120 meV above that of the $yz$ orbital at the same site. But the
on-site \emph{couplings} have not decreased, and the coupling between the $%
xz $ and $xy$ orbitals has even increased. Undoing the JT-distortion also
influences the hopping integrals, but that does not make them more
cubic: $t_{\pi }$ decreases between $xz$ orbitals and increases between $xy$
orbitals. With other plausible hypothetical structures without JT
distortion, we obtained similar results.

Recent high-pressure measurements \cite{loa} have shown that the 3\% JT
distortion essentially disappears in the pressure region between 9 and 14
GPa. At the same time, the Y ions move even further away from their cubic
positions. At 16 GPa, the Ti-O2 distances in the $x$ and $y$ directions have
become nearly equal, with the Ti-O1 distance being just 1\% smaller.
Moreover the
GdFeO$_{3}$-type tilt has increased from $\sim $20 to 21$%
{{}^\circ}%
,$ while the rotation is unchanged. The on-site Hamiltonian and
nearest-neighbour hopping integrals calculated for this structure are
included in table \ref{tableLa8noJT16G}. We see that the ordering of the
orbital energies is the \emph{same} as for the JT distorted and undistorted
structures at normal pressure, and that the energy of the $xz$ orbital is
\textquotedblleft merely\textquotedblright\ 167 meV above that of the $yz$
orbital, while that of the $xy$ orbital is 95 meV above. These results are
similar to the ones obtained for the hypothetical 0 GPa structure without JT
distortion, and so is the result that the on-site coupling between the $yz$
and $xy$ orbitals increases upon removal of the JT distortion and
readjustment of the Y positions.

In conclusion, the JT distortion influences the crystal field in YTiO$_{3},$
but is not its source. We shall return to this subject at the end of section %
\ref{DistBS}, as well as in sections \ref{XtalFieldBasis}, \ref{Pres}, and %
\ref{M}.

\subsection{Influence of GdFeO$_{3}$-type and JT distortions on the $t_{2g}$
bandstructures\label{DistBS}}

In the preceding subsections, we have clarified how the on-site and hopping
matrix elements of the LDA Hamiltonian develop along the series. Now we need
to understand how the GdFeO$_{3}$-type distortion perturbs the spatial
coherence of the simple, cubic Bloch waves.

\begin{figure}[t]
\par
\begin{center}
\includegraphics[width=\textwidth]{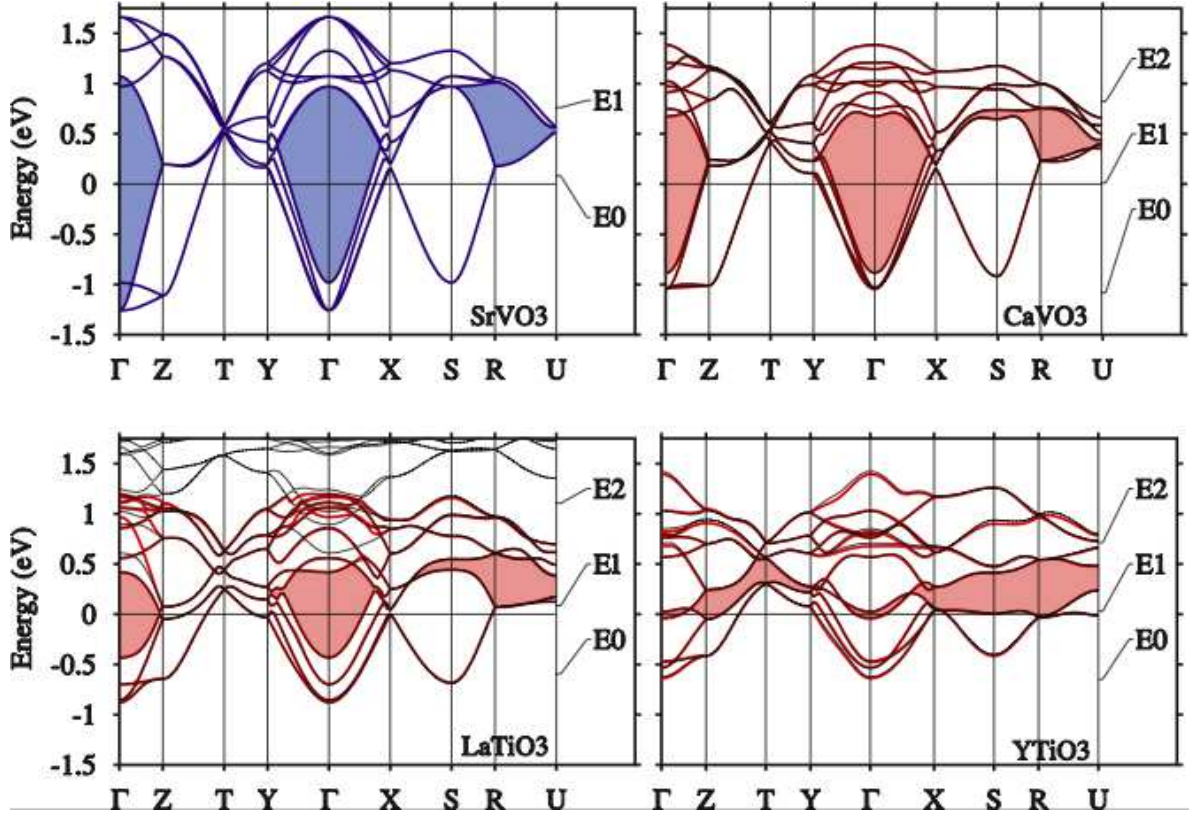}
\end{center}
\caption{Low-energy LDA bandstructures of SrVO$_{3},$ CaVO$_{3},$ LaTiO$_{3}$%
, and YTiO$_{3}.$ The cubic bands for SrVO$_{3}$ (figure \protect\ref{fig2})
have been folded into the orthorhombic BZ (figure \protect\ref%
{fig5}). The black bands were obtained with a large NMTO basis, while the
coloured ones were obtained with the truly-minimal B $d\left( t_{2g}\right) $
basis. The two are indistinguishable, except for LaTiO$_{3}$ where the
bottom of the La 5$d$-band overlaps the top of the Ti $t_{2g}$-band because
the older structural data \protect\cite{lastr} was used here. With the
recent, slightly more distorted structure\ \protect\cite{Cwik03}, O-A
covalency removes this overlap as discussed in section \protect\ref{HighE}
and figure \protect\ref{fig1}, and as shown in figures \protect\ref{fig1a}, 
\protect\ref{fig11}, and \protect\ref{fig3b}. In order to emphasize the
development of the bandstructure along the series, the gap between the lower
1/3 and the upper 2/3 of the bands has been shaded.}
\label{fig3}
\end{figure}

\begin{table}[h]
\caption{$t_{2g}$ edge-to-edge $\left( W_{t_{2g}}\right) $ and rms $\left(
W\right) $ bandwidths in eV.}
\label{tableW}
\begin{center}
\begin{tabular}{cccccc}
\br & SrVO$_{3}$\cite{srstr} & CaVO$_{3}$\cite{castr} & LaTiO$_{3}$\cite%
{lastr} & LaTiO$_{3}$\cite{Cwik03} & YTiO$_{3}$\cite{ystr} \\ 
$W_{t_{2g}}$ & 2.85 & 2.45 & 2.09 & 1.92 & 2.05 \\ 
$W$ & 2.85 & 2.39 & 2.18 & 2.08 & 1.87 \\ 
\br &  &  &  &  & 
\end{tabular}%
%
%
%
%
%
%
%
%
%
\end{center}
\end{table}
The LDA bandstructures in the region of the $t_{2g}$ bands are shown in
figure \ref{fig3} for all four materials. Like in figure \ref{fig2}, the red
bands, $\varepsilon _{i}\left( \mathbf{k}\right) ,$ are obtained from the $%
t_{2g}$ Wannier functions, \textit{e.g.} by forming Bloch sums (see equation
(\ref{Bloch})) of the Wannier orbitals and diagonalizing the Bloch
transformed Hamiltonian,%
\begin{equation}
H_{\mathbf{R}^{\prime }m^{\prime },\,\mathbf{R}m}^{LDA}\left( \mathbf{k}%
\right) \equiv \sum_{\mathbf{T}}t_{m^{\prime },m}^{\mathbf{R}^{\prime },%
\mathbf{R}+\mathbf{T}}\,e^{i\mathbf{k\cdot }\left( \mathbf{T}+\mathbf{R}-%
\mathbf{R}^{\prime }\right) }.  \label{Hk}
\end{equation}%
We thus obtain the eigenvalues, $\varepsilon _{i}\left( \mathbf{k}\right) ,$
and eigenvectors, $\mathbf{u}_{i}\left( \mathbf{k}\right) .$ Since the
orthorhombic structure has four (equivalent) subcells per translational
cell, there are 4$\times $3=12 orthorhombic $t_{2g}$ bands. In equation (\ref%
{Hk}), $\mathbf{T}$ are the orthorhombic translations, and $\mathbf{R}$ and $%
\mathbf{R}^{\prime }$ run over the four B sites, 000, 010, 001, and 011. The
structure is specified in the caption to figure \ref{fig1e}. The Hamiltonian
is a periodic function of $\mathbf{k}$ in the lattice reciprocal to the
orthorhombic lattice. The corresponding BZ is shown in red
in figure \ref{fig5}, whose caption specifies the reciprocal-space
structure. The 12 bands are doubly degenerate on the faces of the
orthorhombic BZ due to the presence of the mirror and the glide mirror in
the space group.

As we proceed along the series, the $t_{2g}$ edge-to-edge \emph{bandwidth}, $%
W_{t_{2g}},$ is clearly seen to decrease. It is tabulated in the top row of
table \ref{tableW}. In the second row, we give an rms bandwidth, $W,$ which
we define to be proportional to $t_{rms}$ (table \ref{tabletrms}) with the
prefactor chosen such that $W=W_{t_{2g}}$ in case of cubic SrVO$_{3},$ i.e.: 
$W\equiv \left( W_{t_{2g}}/t_{rms}\right) _{\mathrm{SrVO}_{3}}\times
t_{rms}. $ We see that the decrease of the edge-to-edge bandwidth does \emph{%
not} follow the trend of $t_{rms}.$ So, clearly, band \emph{shapes} change
along the series. 

In view of the extreme simplicity of the cubic bandstructure of SrVO$_{3}$
(subsection \ref{CubicBands}), the orthorhombic ones are bewilderingly
complicated. They do not immediately tell us what the progressing GdFeO$_{3}$%
-type distortion actually does. Simple theories have assumed that all
hopping integrals scale uniformly with the bandwidth, \textit{i.e.}, that
the bandstructures of all four materials look the same; clearly, they do not.

The next simplest case is that of a pure, cooperative JT distortion, \textit{%
e.g.} the one in which the energy of the $yz$ orbital on site 000 is $\Delta 
$ lower than that of the $xz$ orbital on the same site. The glide mirror,
which exchanges $yz$ and $xz$ on the nearest neighbours in the $x$ and $y$
directions, will then make a $yz/xz$ orbital
feel a spatially alternating potential, $\mp \frac{1}{2}\Delta \exp i\pi
\left( x+y\right) =\mp \frac{1}{2}\Delta \exp [i\left( \pi ,\pi ,0\right)
\cdot \mathbf{R}].$ This makes the cubic $\varepsilon _{yz}\left( \mathbf{k}%
\right) $ and $\varepsilon _{yz}\left( \mathbf{k-} \pi \pi 0
\right) $ bands split by $\Delta $ at the surface in $\mathbf{k}$ space
where they would have crossed had there been no JT distortion. The same
holds for the $xz$ band. In this case, the $xy$ band is unaffected, and
there is no need for folding the bands along the $z$ direction, \textit{i.e.}
by the reciprocal lattice vectors $00\pi$ and $\pi\pi\pi .$

For the $t_{2g}^{1}$ materials, however, the dominant crystal-field
splitting, $\Delta =t_{yz,yz}^{000}-t_{xz,xz}^{000},$ is at most a tenth of
the bandwidth, and yet, individual bands are seen to shift by up to half the
bandwidth when we proceed along the series in figure \ref{fig3}. The
following shifts (splittings) are large and important, because they move
bands which are near the Fermi level: 
the 4th band at $\Gamma _{o}$ moves up, and
in YTiO$_{3}$ it almost empties$.$ At the same time, the 5th band at $\Gamma
_{o}$ and the degenerate 3rd and 4th bands at S$_{o}$ move down, so that in
the Stoner model for ferromagnetism in YTiO$_{3}$ they become occupied
(see also figure 18). This
downward shift of the 3rd and 4th bands near S$_{o}$ is the cause for the
gradual development of a \emph{pseudo-gap} between the lower 1/3 and the
upper 2/3 of the $t_{2g}$ band. In comparison, the $\Delta $-gapping of the $%
yz$ and $xz$ bands along the lines Y$_{o}$T$_{o}$ and X$_{o}$U$_{o}$ $
\frac{\pm \pi }{2}\frac{\pi }{2}k_{z} $ is less spectacular.

Coulomb correlations will of course change all of this, but in a fairly simple
way, at least in the DMFT as we shall see in section \ref%
{OrbPol}: the one-electron excitations can be described as the LDA
Hamiltonian plus a self-energy, which in the single-site DMFT is simply an
energy-dependent, complex on-site matrix, $\Sigma _{mm^{\prime }}\left( \omega
\right) .$ It is therefore important to understand the LDA bandstructure
and, even better, to have an analytical model of it, to which a given $%
\Sigma _{mm^{\prime }}\left( \omega \right) $ can be added or even
determined, say, by a model DMFT calculation. Such an analytical model
should for instance be useful for understanding why recent
de Haas--van Alphen (dHvA)
measurements
for CaVO$_{3}$ gave a Fermi surface more cubic than the orthorhombic one
obtained from the LDA \cite{inoue}.

An analytical bandstructure model may be obtained by transforming from the
mixed $\left( \mathbf{k,R}\right) $ representation (\ref{Hk}), where $%
\mathbf{R}$ are the four sites in the orthorhombic cell, to a pure
wave-vector $\mathbf{k}-\mathbf{Q}$ representation, where $%
\mathbf{Q}$ are the four smallest, inequivalent vectors,%
\begin{equation}
\mathbf{Q}=000 \mathbf{,\;} \pi\pi0
,\; 00\pi ,\;\mathrm{and\;} \pi\pi\pi,
\label{4Q}
\end{equation}%
of the orthorhombic reciprocal lattice ($\mathbf{Q}\cdot \mathbf{T}=2\pi
\times $integer). In figure \ref{fig5}, these are the blue points $\Gamma ,$
M$_{xy},$ X$_{z},$ and R of high symmetry in the cubic (actually primitive
monoclinic) BZ. The transformed Bloch-waves are:%
\begin{eqnarray}
\frac{1}{\sqrt{4}}\sum_{\mathbf{R}}^{4}\left\vert \mathbf{R},m,\mathbf{k}%
\right\rangle e^{-i\mathbf{Q}\cdot \mathbf{R}} &=&\sum_{\mathbf{R}}^{4}\frac{%
1}{\sqrt{4L_{o}}}\sum_{\mathbf{T}}^{o}\left\vert \mathbf{R},m\right\rangle
e^{i\left( \mathbf{k-Q}\right) \mathbf{\cdot }\left( \mathbf{R}+\mathbf{T}%
\right) }  \nonumber \\* 
&=&\frac{1}{\sqrt{L_{c}}}\sum_{\mathbf{R}}^{c}\left\vert \mathbf{R,}%
m\right\rangle e^{i\left( \mathbf{k-Q}\right) \mathbf{\cdot R}}=\left\vert m,%
\mathbf{k-Q}\right\rangle ,  \label{Blochcub}
\end{eqnarray}%
where, in the first expressions, $\mathbf{R}$ runs over the 4 sites and, in
the last expression, over all $\left( L_{c}\rightarrow \infty \right) $
translations of the cubic lattice. $\mathbf{T}$ runs over all $\left(
L_{o}\right) $ translations of the orthorhombic lattice (see figure \ref%
{fig1e}). $\left\vert \mathbf{R},m\right\rangle $ denotes a $yz,$ $xz,$ or $%
xy$ Wannier orbital with the convention of the middle row in figure \ref%
{fig4}, so that the energy and shape of the $m$ orbital depends on which of
the four types of sites it is centred on. In the $\mathbf{Q}$
representation, the expression for the Hamiltonian matrix becomes:%
\begin{equation}
\left\langle m^{\prime },\mathbf{k-Q}^{\prime }\left\vert \mathcal{H}%
\right\vert m,\mathbf{k-Q}\right\rangle =\sum_{\mathbf{R}}^{c}t_{m^{\prime
},m}^{\mathbf{Q}^{\prime }\mathbf{-Q},\mathbf{R}}\,e^{i\left( \mathbf{k-Q}%
\right) \cdot \mathbf{R}},  \label{kmQ}
\end{equation}%
with%
\[
t_{m^{\prime },m}^{\mathbf{Q},\mathbf{R}}\equiv \frac{1}{4}\sum_{\mathbf{R}%
^{\prime }}^{4}t_{m^{\prime },m}^{\mathbf{R}^{\prime },\mathbf{R}^{\prime }+%
\mathbf{R}}\,e^{i\mathbf{Q\cdot R}^{\prime }}, 
\]%
and where for simplicity of notation
we have dropped the superscript $LDA$ on $\mathcal{H}.$ Note that
the difference between any of the four $\mathbf{Q}$ vectors also belongs to
set (\ref{4Q}). If $t_{m^{\prime },m}^{\mathbf{R}^{\prime },\mathbf{R}%
^{\prime }+\mathbf{R}}$ were independent of $\mathbf{R}^{\prime },$ as would
be the case if the energy and shape of each $m$ orbital were independent of
its position, we would have: $t_{m^{\prime },m}^{\mathbf{Q},\mathbf{R}%
}=\delta _{\mathbf{Q},\mathbf{0}}\,t_{m^{\prime },m}^{\mathbf{R}},$ and the
Hamiltonian would have cubic translational symmetry. But the GdFeO$_{3}$%
-type distortion of the on-site energies and shapes of the Wannier orbitals
introduces coupling between the four orthorhombic subdivisions of the cubic
BZ. Each subdivision is denoted by a specific $\mathbf{Q}$
vector in the set (\ref{4Q}), or one of their equivalents.

We must now calculate $t_{m^{\prime },m}^{\mathbf{Q},\mathbf{R}}$ using the
unitary matrix,%
\[
\frac{1}{\sqrt{4}}e^{i\mathbf{Q}\cdot \mathbf{R}}=\frac{1}{\sqrt{4}}\times 
\fbox{$%
\begin{array}{crrrrr}
\mathbf{Q} & \mathbf{R} & 000 & 010 & 001 & 011 \\ 
000 &  & 1 & 1 & 1 & 1 \\ 
\pi \pi 0 &  & 1 & -1 & 1 & -1 \\ 
00\pi &  & 1 & 1 & -1 & -1 \\ 
\pi \pi \pi &  & 1 & -1 & -1 & 1%
\end{array}%
$}, 
\]%
and the space group $\left( Pbnm\right) $ and orbital symmetries expressed
by equations (\ref{eqCubicHop}) and (\ref{invers}). In order to obtain
simple, explicit expressions, let us limit the lattice sum in (\ref{kmQ}) to
include merely the on-site, the six 1st-nearest, and the six $t_{\sigma
}^{\prime }$-type 2nd-nearest-neighbour hoppings. 
For the latter, we neglect the
tiny differences between the three numbers, $t_{yz,yz}^{011}=t_{xz,xz}^{10%
\bar{1}},$ $t_{xz,xz}^{101}=t_{yz,yz}^{01\bar{1}}$ and $\frac{1}{2}\left(
t_{xy,xy}^{1\bar{1}0}+t_{xy,xy}^{110}\right) .$ The 
$\mathbf{Q}^{\prime}\mathbf{Q}$ matrix element of the Hamiltonian can then be
expressed as:%
\begin{eqnarray}
\fl \left\langle \mathbf{k\!-\!Q}^{\prime }\left\vert \mathcal{H}\right\vert 
\mathbf{k\!-\!Q}\right\rangle &\!\!=\!& t^{\mathbf{Q}^{\prime }\!-\!\mathbf{Q%
},000}\!+\!2t^{\mathbf{Q}^{\prime }\!-\!\mathbf{Q},001}c_{z}\!+\!2t^{\mathbf{%
Q}^{\prime }\!-\! \mathbf{Q},010}c_{y}\!+\!2t^{\mathbf{Q}^{\prime }\!-\!%
\mathbf{Q},100}c_{x}  \nonumber \\* 
&+\!&4t^{\mathbf{Q}^{\prime }\!-\!\mathbf{Q},110}c_{x}c_{y}\!+\! 4t^{%
\mathbf{Q}^{\prime }\!-\!\mathbf{Q},101}c_{x}c_{z}\!+\!4t^{\mathbf{Q}%
^{\prime }\!-\!\mathbf{Q},011}c_{y}c_{z}\!+\!4\tilde{t}^{\mathbf{Q}^{\prime
}\!-\! \mathbf{Q},110}s_{x}s_{y},  \label{kmQ1}
\end{eqnarray}%
where $c_{x}\equiv \cos \left( k_{x}-Q_{x}\right) $ and $s_{x}\equiv \sin
\left( k_{x}-Q_{x}\right) ,$ etc. In order to specify the $t^{\mathbf{Q},%
\mathbf{R}}$ matrices in orbital space, it will finally prove convenient to
define the following unit, symmetric, and anti-symmetric matrices:%
\begin{equation}
\begin{array}{c}
E^{yz}\equiv \fbox{$%
\begin{array}{rrrr}
& yz & xz & xy \\ 
yz & 1 & 0 & 0 \\ 
xz & 0 & 0 & 0 \\ 
xy & 0 & 0 & 0%
\end{array}%
$},\qquad S^{yz,xz}\equiv \fbox{$%
\begin{array}{rrrr}
& yz & xz & xy \\ 
yz & 0 & 1 & 0 \\ 
xz & 1 & 0 & 0 \\ 
xy & 0 & 0 & 0%
\end{array}%
$}, \\ 
\mbox{} \\[-5pt] 
A^{yz,xz}\equiv A^{xz,yz}\equiv \fbox{$%
\begin{array}{rrrr}
& yz & xz & xy \\ 
yz & 0 & 1 & 0 \\ 
xz & -1 & 0 & 0 \\ 
xy & 0 & 0 & 0%
\end{array}%
$},\;\mathrm{etc.}%
\end{array}
\label{ESA}
\end{equation}%
Note that the A matrices are defined to have 1 or 0 in their upper triangle,
and $-1$ or 0 in their lower triangle.

\begin{figure}[t]
\begin{center}
\includegraphics[width=0.89\textwidth]{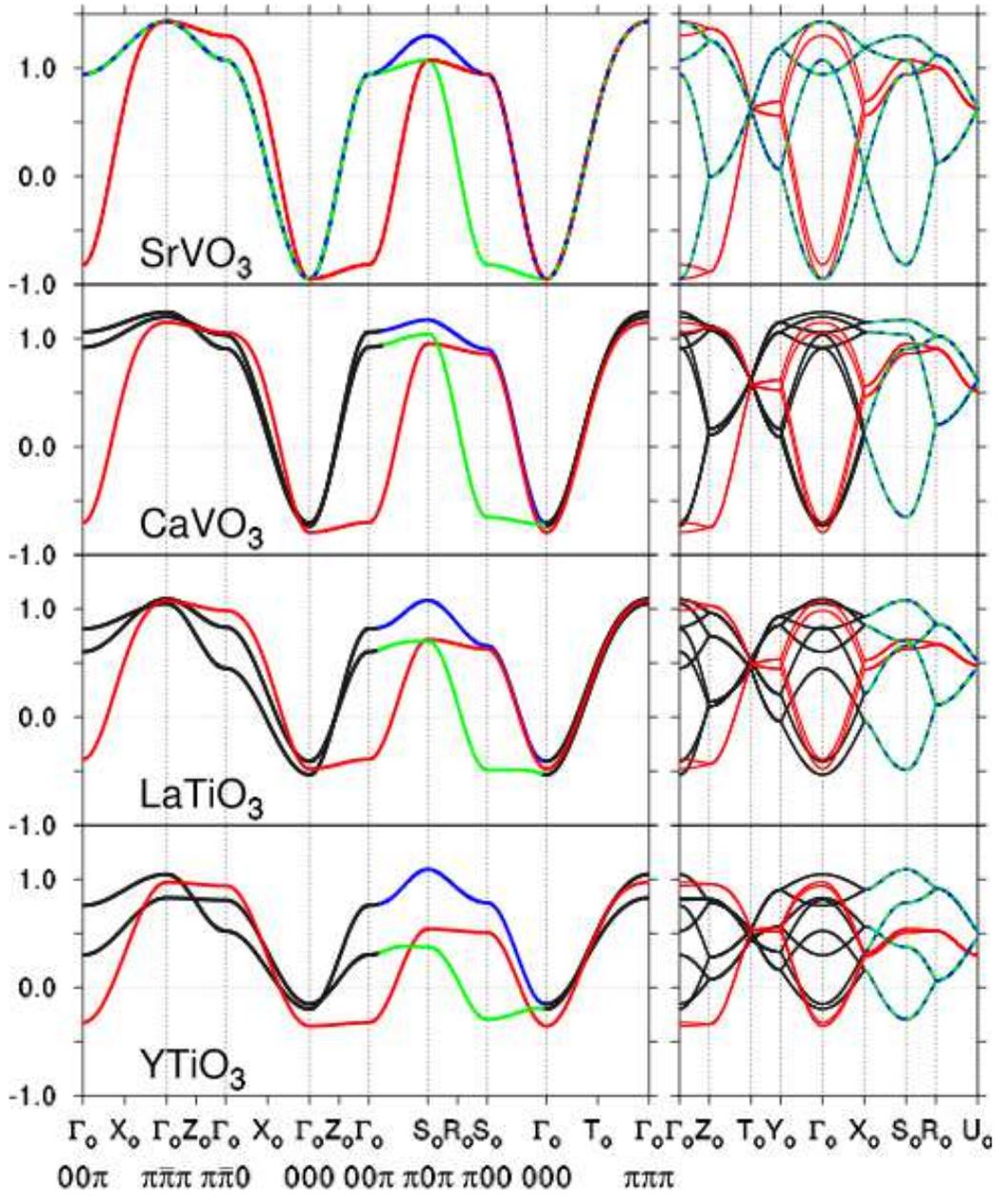}
\end{center}
\caption{Cubically averaged $\left( \mathbf{Q}=\mathbf{0}\right) $ $t_{2g}$
LDA bandstructures in eV, in the primitive monoclinic BZ (left), and folded
into the orthorhombic BZ (right). The letters above the $k_{x}k_{y}k_{z}$
coordinates denote the orthorhombic high-symmetry points (red in figure 
\protect\ref{fig5}), $\mathbf{k}+\mathbf{Q},$ into which the corresponding
cubic point is folded on the right-hand side. 
Bands with predominant $xy,$ $yz,$ and $zx$
character are respectively red, green, and blue. Black bands are of strongly
mixed $yz$ and $zx$ character. The definitions of characters 
are those of the middle row
in figure \protect\ref{fig4}. The bands were obtained 
with the parameters for $\mathbf{Q}$=$000$
in table \protect\ref{tableQ}. In
order to obtain the bandstructures in figure \protect\ref{fig3}, albeit in the
2nd-nearest neighbour approximation, the folded-in bands should be coupled
using the matrix elements (\protect\ref{tQ110})--(\protect\ref{tQ111}). }
\label{fig3a}
\end{figure}

\begin{table}[t]
\caption{Coefficients of the on-site energies and hopping integrals $t^{%
\mathbf{Q,R}}$ in meV (see equations (\protect\ref{kmQ1})--(\protect\ref%
{tQ111})) }
\label{tableQ}
\begin{center}
{\small {\setlength{\tabcolsep}{0.2pt} 
\begin{tabular}{cccrrrrrrrrrrcrrrr}
\br $\mathbf{R}$ & \ \ \  &  & \multicolumn{4}{c}{$000$} & \ \  & 
\multicolumn{4}{c}{$001$} & \ \ \  &  & \multicolumn{4}{c}{$010/100$} \\ 
&  &  & Sr & Ca & La & Y &  & Sr & Ca & La & Y &  &  & Sr & Ca & La & Y \\ 
$\mathbf{Q}$ &  &  &  &  &  &  &  &  &  &  &  &  &  &  &  &  &  \\ 
$000$ &  & $\left\langle t_{\llcorner z,\llcorner z}\right\rangle $ & 
$\!625$ & $616$ & $488$
& $490$ &  & $-281$ & $-245$ & $-201$ & $-122$ &  & $t_{\shortparallel
z,\shortparallel z}$ & $\!\!\!\!-281$ & $-223$ & $-185$ & $-184$ \\ 
&  & $t_{xy,xy}$ & $\!625$ & $542$ & $486$ & $417$ &  & $-33$ & $-23$ & $-22$
& $-8$ &  & $t_{xy,xy}$ & $\!\!\!\!-281$ & $-231$ & $-183$ & $-162$ \\ 
&  &  &  &  &  &  &  &  &  &  &  &  & $t_{\llcorner z,\llcorner z}$ & $-33$
& $-17$ & $1$ & $28$ \\ 
&  & $t_{yz,xz}$ & $0$ & $4$ & $-42$ & $43$ &  & $0$ & $-21$ & $-42$ & $-63$
&  & $\left\langle t_{\shortparallel z,\llcorner z}\right\rangle $ & $0$ & $%
6 $ & $16$ & $15$ \\ 
&  &  &  &  &  &  &  &  &  &  &  &  &  &  &  &  &  \\ 
$\pi \pi 0$ &  & $t_{yz,yz}\!\!-\!\!t_{xz,xz}$ & $0$ & $8$ & $-85$ & $-230$
&  & $0$ & $9$ & $15$ & $113$ &  & $t_{yz,xz}\!\!-\!\!t_{xz,yz}$ & $0$ & $78$
& $118$ & $111$ \\ 
&  &  &  &  &  &  &  &  &  &  &  &  &  &  &  &  &  \\ 
$00\pi$ &  & $t_{yz,xy}\!\!+\!\!t_{xz,xy}$ & $0$ & $-21$ & $-131$ & $-120$ & 
& $0$ & $-11$ & $-7$ & $19$ &  & $t_{\shortparallel
z,xy}\!\!+\!\!t_{xy,\shortparallel z}$ & $0$ & $-51$ & $-88$ & $-118$ \\ 
&  &  &  &  &  &  &  &  &  &  &  &  & $t_{\llcorner
z,xy}\!\!+\!\!t_{xy,\llcorner z}$ & $0$ & $8$ & $40$ & $43$ \\ 
&  &  &  &  &  &  &  &  &  &  &  &  &  &  &  &  &  \\ 
$\pi\pi\pi$ &  & $t_{yz,xy}\!\!-\!\!t_{xz,xy}$ & $0$ & $-13$ & $-47$ & $-86$
&  & $0$ & $-61$ & $-111$ & $-147$ &  & $t_{yz,xy}^{010}\!\!-\!%
\!t_{xy,yz}^{010}$ & $0$ & $17$ & $20$ & $10$ \\[5pt] 
&  &  &  &  &  &  &  &  &  &  &  &  & $t_{xz,xy}^{010}\!\!-\!%
\!t_{xy,xz}^{010}$ & $0$ & $58$ & $106$ & $87$ \\ 
&  & & & & &  &  &  &  &  &  &  &  & &  &  &  \\ 
$\mathbf{R}$ &  &  & \multicolumn{4}{c}{$110/011/101$} &  &  & 
&  &  &  &  &  &  &  &  \\ 
 &  &  & Sr & Ca & La & Y &  &  &  &  &  &  &  &  &  &  &  \\ 
$\mathbf{Q}$ &  &  & & & & &  &  &  &  &  &  &  &  &  &  &  \\ 
$000$ &  & $t_{\sigma}^{\prime}$ & $-96$ & $-91$ & $-47$ & $-27$ &  &  &  & 
&  &  &  &  &  &  &  \\ 
&  & $s_{\sigma}^{\prime}$ & $0$ & $7$ & $1$ & $29$ &  &  &  &  &  &  &  & 
&  &  &  \\ 
\br &  &  &  &  &  &  &  &  &  &  &  &  &  &  &  &  & 
\end{tabular}%
}}
\end{center}
\end{table}

The $\mathbf{Q=Q}^{\prime }=\mathbf{0}$ element, $\left\langle \mathbf{k}%
\left\vert \mathcal{H}\right\vert \mathbf{k}\right\rangle ,$ is a
``cubically'' averaged $3\times 3$ Hamiltonian. This is a generalization of
the non-interacting cubic bands given by (\ref{eq9}). Its on-site 
and 001-hopping matrices are given by:%
\begin{equation}
t^{\mathbf{Q}=000,\mathbf{R}=00z}=\left\langle 
t_{\llcorner z,\llcorner z}^{00z}\right\rangle
\left( E^{yz}+E^{xz}\right) +t_{xy,xy}^{00z}E^{xy}+t_{yz,xz}^{00z}S^{yz,xz},
\label{tQ000}
\end{equation}%
if $z$ takes respectively the value 0 and 1. Here, the coefficients are linear
combinations of the basic Hamiltonian matrix elements given in tables \ref%
{tableSr}-\ref{tableY}. For instance is%
\[
\left\langle t_{\llcorner z,\llcorner z}^{00z}
\right\rangle \equiv \frac{1}{2}\left(
t_{yz,yz}^{00z}+t_{xz,xz}^{00z}\right) 
\]%
the on-site energy or 001 $t_{\pi }$ hopping integral,
averaged over the $yz$ and $xz$ orbitals. The matrices for hopping parallel
to the mirror plane ($\mathbf{R}=010$ or $100$) have the very similar form:%
\[
t^{\mathbf{Q}=000,\mathbf{R}=010/100}=t_{\shortparallel z,\shortparallel
z}E^{\shortparallel z}+t_{\llcorner z,\llcorner z}E^{\llcorner
z}+t_{xy,xy}E^{xy}+\left\langle t_{\shortparallel z,\llcorner
z}\right\rangle S^{\shortparallel z,\llcorner z}, 
\]%
where $\shortparallel $ and $\llcorner ,$ respectively, denote the directions
parallel and perpendicular to the direction of the hop, \textit{i.e.}, for $%
\mathbf{R}$ = 010, $\shortparallel =y$ and $\llcorner =x,$ whereas for $%
\mathbf{R}$ = 100, $\shortparallel =x$ and $\llcorner =y.$ So the first
three terms are the $t_{\pi }$ or $t_{\delta }$ hoppings in the plane, and
the last term is the average of the hopping \emph{between} the $yz$ and $xz$
orbitals along 010, or 100:%
\[
\left\langle t_{\shortparallel z,\llcorner z}\right\rangle \equiv \frac{1}{2}%
\left( t_{\shortparallel z,\llcorner z}+t_{\llcorner z,\shortparallel
z}\right) =\frac{1}{2}\left( t_{yz,xz}^{010}+t_{xz,yz}^{010}\right) =\frac{1%
}{2}\left( t_{xz,yz}^{100}+t_{yz,xz}^{100}\right) . 
\]%
Remember that $t_{yz,xz}^{100}=t_{yz,xz}^{010}\neq
t_{xz,yz}^{010}=t_{xz,yz}^{100}.$ 
The hopping to the 2nd-nearest neighbours will be \emph{diagonal} in
both orbital and $\mathbf{Q}$ spaces:%
\[
\fl
t^{\mathbf{Q},110}\!=\delta_{\mathbf{Q,0}}\,t_{\sigma}^{\prime}\,E^{xy},\quad 
t^{\mathbf{Q},101}\!=\delta_{\mathbf{Q,0}}\,t_{\sigma}^{\prime}\,E^{xz},\quad 
t^{\mathbf{Q},011}\!=\delta_{\mathbf{Q,0}}\,t_{\sigma}^{\prime}\,E^{yz},\quad  
\tilde{t}^{\mathbf{Q},110}\!=
\delta_{\mathbf{Q,0}}\,s_{\sigma}^{\prime}\,E^{xy},
\]
as a consequence of the approximation made above. 
Here,%
\[
t_{\sigma }^{\prime }=\frac{1}{6}\left[ t_{xy,xy}^{1\bar{1}%
0}+t_{xy,xy}^{110}+2\left( t_{yz,yz}^{011}+t_{xz,xz}^{101}\right) \right] ,\;%
\mathrm{and\;}s_{\sigma }^{\prime }=\frac{1}{2}\left( t_{xy,xy}^{1\bar{1}%
0}-t_{xy,xy}^{110}\right) . 
\]%
Note that in these expressions, and in (\ref{tQ110})--(\ref{tQ111}) below,
the $t^{\mathbf{Q},\mathbf{R}},$ $E,$ $S,$ and $A$ matrices in orbital space
are the \emph{only }such matrices; the prefactors, such as $t_{xy,xz}^{001},$
are parameters and their subscripts do not label elements of $3\times 3$
matrices.

The values of these parameters are displayed in table \ref{tableQ}, whose
top four rows, denoted $\mathbf{Q}$=$000$, 
determine the cubically averaged bands. We see that in SrVO$%
_{3}$ and LaTiO$_{3}$ the average (on-site) energy of the $xy$ band is the
same as the common, average energy of the $yz$ and $xz$ bands, whereas in
CaVO$_{3}$ and YTiO$_{3},$ it is $\sim $75 meV lower. The values of the
hopping integrals confirm that the cubically averaged bands narrow along the
series (due to misalignment and increased theft of oxygen character), and
that the hopping in the $z$ direction is anomalously small in YTiO$_{3},$
which therefore has the most anisotropic $t_{2g}$ band. The reason for that
is the
smallness of $t_{yz,yz}^{001}$ due to hybridization with Y 4$%
d_{xy},$ as was discussed in connection with equation (\ref{eq7}).

The cubically averaged bandstructures are shown in figure \ref{fig3a}. We
clearly see how these bands develop from the cubic, non-interacting, nearly
two-dimensional $xy$, $yz$, and $xz$ bands of SrVO$_{3}$ into three
distorted, monoclinic bands in which nearly degenerate $yz$ and $xz$ levels
split by $\pm \left[ t_{yz,xz}^{000}+2t_{yz,xz}^{001}\cos
k_{z}+2\left\langle t_{\shortparallel z,\llcorner z}\right\rangle \left(
\cos k_{x}+\cos k_{y}\right) \right] .$ The $xy$ band (red) stays pure and
nearly two-dimensional. For YTiO$_{3},$ the weak $k_{z}$ dispersion of the $%
yz$ band (green) is clearly seen in the (010) plane where there is no mixing
with $xz.$ The $xz$ band (blue) will have the same behaviour in the (100)
plane in this cubically averaged bandstructure. The letters $\Gamma _{o},$ T$%
_{o},$ S$_{o},$ R$_{o}$, Z$_{o}$, and X$_{o}$ on the abscissa refer to the
orthorhombic high-symmetry points, $\mathbf{k}+\mathbf{Q}$ (red in figure %
\ref{fig5}), into which the monoclinic $\mathbf{k}$ points (blue in figure %
\ref{fig5}) should be folded prior to $\mathbf{Q}$-coupling. The folded-in
bands are shown on the right-hand side, and comparison with the real $t_{2g}$
bands in figure \ref{fig3} reveals that, in fact, the cubically averaged
bandstructures go ``halfway'' towards the real bandstructures.


%


The \emph{hybridization} between two cubically averaged bandstructures
displaced in $\mathbf{k}$ space by the orthorhombic reciprocal lattice
vector $\mathbf{Q}$=$\pi\pi0$ is given by the
matrix (\ref{kmQ1}) with:%
\begin{eqnarray}
2t^{\mathbf{Q=}\pi \pi 0,\mathbf{R}=00z} &=&\left(
t_{yz,yz}^{00z}-t_{xz,xz}^{00z}\right) \left( E^{yz}-E^{xz}\right)
\label{tQ110} \\* 
2t^{\mathbf{Q}=\pi \pi 0,\mathbf{R}=010/100} &=&\left(
t_{yz,xz}-t_{xz,yz}\right) A^{yz,xz},  \nonumber
\end{eqnarray}%
and is seen to express the asymmetry between the $yz$ and $xz$ orbitals, as
well as the coupling between them. Here, the top line provides the $\Delta $
splitting by $\left( t_{yz,yz}^{000}-t_{xz,xz}^{000}\right) +2\left(
t_{yz,yz}^{001}-t_{xz,xz}^{001}\right) c_{z}$ of the $\varepsilon
_{yz}\left( \mathbf{k}\right) $ and $\varepsilon _{yz}\left( \mathbf{k-}%
\pi \pi 0 \right) $ bands at their crossings. The same holds
for the splitting of the $\varepsilon _{xz}\left( \mathbf{k}\right) $ and $%
\varepsilon _{xz}\left( \mathbf{k-} \pi \pi 0 \right) $
bands. Note that since $t_{yz,yz}^{00z}-t_{xz,xz}^{00z}$ is a parameter, 
it should 
\emph{not} be substituted by $t_{xz,xz}^{00z}-t_{yz,yz}^{00z}$ when
describing the $xz$ band, because that change of sign is accounted for as a
prefactor to $E^{xz}.$ Note also that $%
t_{yz,xz}^{010}-t_{xz,yz}^{010}=t_{yz,xz}^{100}-t_{xz,yz}^{100}$ because $%
t_{yz,xz}^{010}=t_{yz,xz}^{100}\neq t_{xz,yz}^{100}=t_{xz,yz}^{010}.$

Similarly, the hybridization between two cubically averaged bandstructures
displaced by $\mathbf{Q}$=$00\pi$ is specified by:%
\begin{eqnarray}
2t^{\mathbf{Q}=00\pi ,\mathbf{R}=000} &=&\left(
t_{yz,xy}^{000}+t_{xz,xy}^{000}\right) \left( S^{yz,xy}+S^{xz,xy}\right) 
\nonumber \\* 
2t^{\mathbf{Q=}00\pi ,\mathbf{R}=001} &=&\left(
t_{yz,xy}^{001}+t_{xz,xy}^{001}\right) \left( A^{yz,xy}+A^{xz,xy}\right)
\label{tQ001} \\* 
2t^{\mathbf{Q}=00\pi ,\mathbf{R}=010/100} &=&\left( t_{\shortparallel
z,xy}+t_{xy,\shortparallel z}\right) S^{\shortparallel z,xy}+\left(
t_{\llcorner z,xy}+t_{xy,\llcorner z}\right) S^{\llcorner z,xy},  \nonumber
\end{eqnarray}%
and is seen to express the average of the coupling from the $xy$ orbital to
the $yz$ and $xz$ orbitals. As above, $\shortparallel $ and $\llcorner $
denote the direction parallel and perpendicular to the direction of the hop, 
\textit{i.e.}, for $\mathbf{R}$ = 010, $\shortparallel =y$ and $\llcorner
=x, $ whereas for $\mathbf{R}$ = 100, $\shortparallel =x$ and $\llcorner =y.$
Note that $t_{iz,xy}^{001}=-t_{xy,iz}^{001}$, that 
$t_{yz,xy}^{010}=t_{xy,xz}^{100}%
\neq t_{xz,xy}^{100}=t_{xy,yz}^{010}$, and that 
$t_{xz,xy}^{010}=t_{xy,yz}^{100}%
\neq t_{yz,xy}^{100}=t_{xy,xz}^{010}.$

Finally, two cubic bands displaced by $\mathbf{Q}$=$ \pi \pi \pi$ 
hybridize by means of a matrix (\ref{kmQ1}) specified by:%
\begin{eqnarray}
2t^{\mathbf{Q}=\pi \pi \pi ,\mathbf{R}=000} &=&\left(
t_{yz,xy}^{000}-t_{xz,xy}^{000}\right) \left( S^{yz,xy}-S^{xz,xy}\right) 
\nonumber \\* 
2t^{\mathbf{Q=}\pi \pi \pi ,\mathbf{R}=001} &=&\left(
t_{yz,xy}^{001}-t_{xz,xy}^{001}\right) \left( A^{yz,xy}-A^{xz,xy}\right)
\label{tQ111} \\* 
2t^{\mathbf{Q}=\pi \pi \pi ,\mathbf{R}=010} &=&\left(
t_{yz,xy}^{010}-t_{xy,yz}^{010}\right) A^{yz,xy}+\left(
t_{xz,xy}^{010}-t_{xy,xz}^{010}\right) A^{xz,xy}  \nonumber \\* 
2t^{\mathbf{Q}=\pi \pi \pi ,\mathbf{R}=100} &=&-\left(
t_{xz,xy}^{010}-t_{xy,xz}^{010}\right) A^{yz,xy}-\left(
t_{yz,xy}^{010}-t_{xy,yz}^{010}\right) A^{xz,xy}  \nonumber \\* 
&=&\left( t_{yz,xy}^{100}-t_{xy,yz}^{100}\right) A^{yz,xy}+\left(
t_{xz,xy}^{100}-t_{xy,xz}^{100}\right) A^{xz,xy}  \nonumber
\end{eqnarray}%
This matrix expresses the difference between the couplings from $xy$ to $yz$
and to $xz.$
We emphasize that there is no $\mathbf{Q}$ coupling between $xy$ bands.

The values of the coefficients given in table \ref{tableQ} exhibit the
general trend that the non-cubic perturbations increase along the series.
We also see that \emph{details} differ: which ones of the many non-cubic
couplings dominate depends on the material. Moreover, the on-site,\emph{\ }$%
\mathbf{k}$-averaged splittings, $t_{yz,yz}^{000}-t_{xz,xz}^{000},$ and
inter-orbital couplings, $t_{yz,xy}^{000}\pm t_{xz,xy}^{000},$ are not much
larger than their modulations given by the corresponding hopping integrals.
This is very different from the situation in materials with strong JT
distortions, where $\mathbf{Q}$ coupling due to on-site terms dominate. So
according to the present conventional description of the $t_{2g}$
perovskites in terms of $\mathbf{Q}$ couplings, the distortion from the
cubically averaged to the real bandstructure is the sum of several terms
with varying signs.

As a most important example, let us discuss the \emph{pseudo-gap.} More
specifically, let us estimate the levels at S$_{o}~\pi00$ 
with intermediate energy by using
expressions (\ref{kmQ1})--(\ref{tQ111}), table \ref{tableQ}, 
and figure \ref{fig3a}:

The relevant cubically averaged $yz$ levels are those at $\mathbf{k}-\mathbf{%
Q}= \pi00 \,-\, 00\pi = \pi0\bar{\pi}
$ and $ \pi00 \,-\, \pi\pi0 = 0\bar{\pi}0$. Their
energies are:%
\begin{eqnarray*}
&& \varepsilon _{yz}\left( \pi 0\pi \right) ~\left/ ~\varepsilon _{yz}\left(
0\pi 0\right) \right. =\left\langle t_{\llcorner z,\llcorner z}^{000}
\right\rangle -4t_{\sigma
}^{\prime }\mp 2\left( \left\langle t_{\llcorner z,\llcorner z}^{001}
\right\rangle
-t_{\shortparallel z,\shortparallel z}+t_{\llcorner z,\llcorner z}\right) \\* 
&& =\left( 
\begin{array}{rrrr}
1009 & 980 & 676 & 598%
\end{array}%
\right) \mp \left( 
\begin{array}{rrrr}
-66 & -58 & -30 & 180%
\end{array}%
\right) \\* 
&& =\left( 
\begin{array}{cccc}
1075 & 1038 & 706 & 418%
\end{array}%
\right) \left/ \left( 
\begin{array}{cccc}
943 & 922 & 646 & 778%
\end{array}%
\right) \right. ~\mathrm{meV}
\end{eqnarray*}%
for SrVO$_{3}$ through YTiO$_{3}.$ Note that in YTiO$_{3},$ the $\varepsilon
_{yz}( \pi0\pi ) $ energy is anomalously low, as is also seen
in figure \ref{fig3a}. This is one reason for the pronounced pseudo-gap in
that material. The $yz$ levels are separated by $\mathbf{Q}^{\prime }-%
\mathbf{Q}$=$\pi\pi\bar{\pi}$ and, according to (\ref{tQ111}),
or table \ref{tableQ}, do therefore not couple. Degenerate with these $yz$
levels are the cubically averaged $xz$ levels,%
\[
\varepsilon _{xz}\left( 0\pi \pi \right) \left/ \varepsilon _{xz}\left( \pi
00\right) \right. =\varepsilon _{yz}\left( \pi 0\pi \right) \left/
\varepsilon _{yz}\left( 0\pi 0\right) \right. , 
\]%
at $\mathbf{k}-\mathbf{Q} = \pi00 \,-\, \pi\pi\pi
 = 0\bar{\pi}\bar{\pi}\,$ and $\,\pi00 \,-\,
000 = \pi00.$ Also these $xz$ levels are separated
by $\pi\pi\pi$, and do therefore not couple with each
other. The $yz$ level at $\pi0\bar{\pi}$ and the $xz$ level at 
$0\bar{\pi}\bar{\pi}$ are separated by $\pi\pi0$
and therefore couple via a matrix element obtained from equation (\ref{tQ110}%
), which is seen to vanish, however, because $c_{x}+c_{y}=0.$ The same holds
for the $yz$ level at $0\bar{\pi}0$ and the $xz$ level at $%
\pi00.$ Finally, the $yz$ level at $\pi0\bar{\pi}
$ and the $xz$ level at $\pi00$ are separated by $%
00\pi$ so that, according to (\ref{tQ001}) or table \ref%
{tableQ}, they do not couple. In conclusion, the four $yz$ and $xz$ states
at S$_{o}$ with intermediate energy can only couple via $xy$ states, which
we consider next: 

When folded into S$_{o},$ all four cubically averaged $xy$
bands (the red ones in figure \ref{fig3a}) are nearly degenerate. One pair
of degenerate levels come from $\mathbf{k}-\mathbf{Q}^{\prime }= \pi
00 \,-\, 00\pi  = \pi0\bar{\pi}\,$ and $\,%
\pi00 \,-\, \pi\pi\pi = 0\bar{\pi}\bar{\pi}
$, and the other from $\mathbf{k}-\mathbf{Q}^{\prime }= \pi
00 \,-\,  000 = \pi00\,$ and $\,\pi
00 \,-\, \pi\pi0 = 0\bar{\pi}0.$ As for
the $yz$ and $xz$ levels, the two $xy$ energies are those of $\pi0\pi$ and
$\pi00$
points, specifically:%
\begin{eqnarray*}
\fl 
\left[ \varepsilon _{xy}\left( \pi 0\pi \right) \mathrm{=}
\varepsilon _{xy}\left(
0\pi \pi \right) \right] &&~\left/ ~\left[ \varepsilon _{xy}\left( \pi
00\right) \mathrm{=}\varepsilon _{xy}\left( 0\pi 0\right) \right] \right.
=t_{xy,xy}^{000}-4t_{\sigma }^{\prime }\mp 2t_{xy,xy}^{001} \\* 
\fl &&\hspace{-10mm}=\left( 
\begin{array}{rrrr}
1009 & 906 & 674 & 525%
\end{array}%
\right) \mp \left( 
\begin{array}{rrrr}
-66 & -46 & -44 & -16%
\end{array}%
\right) \\* 
\fl &&\hspace{-10mm}=\left( 
\begin{array}{cccc}
1075 & 952 & 718 & 541%
\end{array}%
\right) \left/ \left( 
\begin{array}{cccc}
943 & 860 & 630 & 509%
\end{array}%
\right) \right. ~\mathrm{meV.}
\end{eqnarray*}

Since there is no $\mathbf{Q}$-coupling between $xy$ bands, we now only have
to couple $xy$ to $yz$ and $xz.$

The $yz$ level at $\pi0\bar{\pi}$ couples to the lowest,
degenerate $xy$ level, that is the level at $\pi00$ and $%
0\bar{\pi}0$, because for those, $\mathbf{Q}^{\prime }-\mathbf{Q%
}= \pi0\bar{\pi} \,-\, \pi00 =  00\bar{\pi}\,$ and 
$\pi0\bar{\pi} \,-\, 0\bar{\pi}0 =
\pi\pi\bar{\pi} ,$ respectively$.$ It does,
however, not couple to the $xy$ level at $\pi0\pi$, 
because for those states, $\mathbf{Q%
}^{\prime }-\mathbf{Q}= \pi0\bar{\pi} \,-\, \pi0\bar{\pi}
= 000\,$ and $\, \pi0\bar{\pi} \,-\,
0\bar{\pi}\bar{\pi} = \pi\pi0 .$ Similarly, the $xz$
level at $0\bar{\pi}\bar{\pi} ,$ which is degenerate with the $yz$
level at $\bar{\pi}0\bar{\pi}$, 
also couples to the $xy$ level at $\pi00$ and $%
0\bar{\pi}0$, because, here, $\mathbf{Q}^{\prime }-\mathbf{Q}%
= 0\bar{\pi}\bar{\pi} \,-\, \pi00 = \bar{\pi}\bar{\pi}
\bar{\pi}\,$ and $\, 0\bar{\pi}\bar{\pi} \,-\, 0\bar{\pi}0
= 00\bar{\pi} .$ As for the $yz$ level, the $xz$ level does not
couple to the $xy$ level at $0\pi\pi$.

The $yz$ level at $0\bar{\pi}0 ,$ which is lower than the one
at $\pi0\bar{\pi} ,$ except in YTiO$_{3},$ couples to the
highest, degenerate $xy$ level, the one at $\pi0\bar{\pi}$
and $0\bar{\pi}\bar{\pi}$, because for those, $\mathbf{Q}^{\prime }-%
\mathbf{Q}= 0\bar{\pi}0 \,-\, \pi0\bar{\pi} =
\bar{\pi}\bar{\pi}\pi\,$ and $\, 0\bar{\pi}0 \,-\, 0\bar{\pi}\bar{\pi}
= 00\pi .$ It does not couple to the $xy$ level at $0\bar{\pi}0$.
Similarly, the $xz$ level at $\pi00$ couples to the $xy$
level at $\pi0\bar{\pi}$ and $0\bar{\pi}\bar{\pi} ,$
because $\mathbf{Q}^{\prime }-\mathbf{Q}= \pi00 \,-\,
\pi0\bar{\pi} = 00\pi\,$ and $\, \pi00
\,-\, 0\bar{\pi}\bar{\pi} = \pi\pi\pi .$ Again, there
is no coupling between $xz$ and $xy$ levels at $\pi00$.

All couplings relevant for the pseudo-gap at S$_{o}$ thus have $\mathbf{Q}%
^{\prime }-\mathbf{Q}= 00\pi\,$ and $\, \pi\pi\pi
 ,$ and are therefore described by the five hopping parameters listed
in the corresponding rows of table \ref{tableQ}. We emphasize that the
dominant crystal-field coupling, caused by $t_{yz,yz}^{000}-t_{xz,xz}^{000},$
has $\mathbf{Q}^{\prime }-\mathbf{Q}= \pi\pi0 ,$ and is
therefore not relevant for the pseudo-gap at S$_{o}.$ For the couplings via $%
\mathbf{Q}^{\prime }-\mathbf{Q}= 00\pi ,$ we find from
equations (\ref{kmQ1}) and (\ref{tQ001}):%
\begin{eqnarray*}
\fl \left\langle xy,\pi 00\left\vert \mathcal{H}\right\vert \pi 0\pi
,yz\right\rangle &&=\left\langle xy,0\pi 0\left\vert \mathcal{H}\right\vert
0\pi \pi ,xz\right\rangle \\* 
\fl &&\hspace{-25mm} =\frac{1}{2}\left(
t_{yz,xy}^{000}+t_{xz,xy}^{000}\right) +\left(
t_{yz,xy}^{001}+t_{xz,xy}^{001}\right) +\left( t_{\shortparallel
z,xy}+t_{xy,\shortparallel z}\right) -\left( t_{\llcorner
z,xy}+t_{xy,\llcorner z}\right) \\* 
\fl &&\hspace{-25mm} =\frac{1}{2}\left( {\setlength{\arraycolsep}{1pt} 
\begin{array}{rrr}
-21 & -131 & -120%
\end{array}%
} \right) +\left( {\setlength{\arraycolsep}{2pt} 
\begin{array}{rrr}
-11 & -7 & 19%
\end{array}%
} \right) +\left( {\setlength{\arraycolsep}{2pt} 
\begin{array}{rrr}
-51 & -88 & -118%
\end{array}%
} \right) -\left( {\setlength{\arraycolsep}{2pt} 
\begin{array}{rrr}
8 & 40 & 43%
\end{array}%
} \right) \\* 
\fl &&\hspace{-25mm} =\left( {\setlength{\arraycolsep}{2pt} 
\begin{array}{ccc}
-81 & -201 & -202%
\end{array}%
} \right) ~\mathrm{meV,}
\end{eqnarray*}%
for CaVO$_{3}$ through YTiO$_{3}.$ These couplings are numerically large for
the titanates because terms add up. The other $00\pi$
couplings are:%
\begin{eqnarray*}
\fl \left\langle xy,0\pi \pi \left\vert \mathcal{H}\right\vert 0\pi
0,yz\right\rangle &&=\left\langle xy,\pi 0\pi \left\vert \mathcal{H}%
\right\vert \pi 00,xz\right\rangle \\* 
\fl &&\hspace{-25mm} =\frac{1}{2}\left(
t_{yz,xy}^{000}+t_{xz,xy}^{000}\right) -\left(
t_{yz,xy}^{001}+t_{xz,xy}^{001}\right) -\left( t_{\shortparallel
z,xy}+t_{xy,\shortparallel z}\right) +\left( t_{\llcorner
z,xy}+t_{xy,\llcorner z}\right) \\* 
\fl &&\hspace{-25mm} =\frac{1}{2}\left( {\setlength{\arraycolsep}{2pt} 
\begin{array}{rrr}
-21 & -131 & -120%
\end{array}%
} \right) -\left( {\setlength{\arraycolsep}{2pt} 
\begin{array}{rrr}
-11 & -7 & 19%
\end{array}%
} \right) -\left( {\setlength{\arraycolsep}{2pt} 
\begin{array}{rrr}
-51 & -88 & -118%
\end{array}%
} \right) +\left( {\setlength{\arraycolsep}{2pt} 
\begin{array}{rrr}
8 & 40 & 43%
\end{array}%
} \right) \\* 
\fl &&\hspace{-25mm} =\left( {\setlength{\arraycolsep}{2pt} 
\begin{array}{ccc}
60 & 70 & 82%
\end{array}%
} \right) ~\mathrm{meV,}
\end{eqnarray*}%
which are small for the titanates due to cancellation of terms. For the
couplings via $\mathbf{Q}^{\prime }-\mathbf{Q}= \pi\pi\pi$
we find from expressions (\ref{tQ111}):%
\begin{eqnarray*}
\fl \left\langle xy,0\pi 0\left\vert \mathcal{H}\right\vert \pi 0\pi
,yz\right\rangle &&=-\left\langle xy,\pi 00\left\vert \mathcal{H}\right\vert
0\pi \pi ,xz\right\rangle \\* 
\fl &&\hspace{-25mm} =\frac{1}{2}\left(
t_{yz,xy}^{000}-t_{xz,xy}^{000}\right) +\left(
t_{yz,xy}^{001}-t_{xz,xy}^{001}\right) -\left(
t_{yz,xy}^{010}-t_{xy,yz}^{010}\right) -\left(
t_{xz,xy}^{010}-t_{xy,xz}^{010}\right) \\* 
\fl &&\hspace{-25mm} =\frac{1}{2}\left( {\setlength{\arraycolsep}{2pt} 
\begin{array}{rrr}
-13 & -47 & -86%
\end{array}%
} \right) +\left( {\setlength{\arraycolsep}{2pt} 
\begin{array}{rrr}
-61 & -111 & -147%
\end{array}%
} \right) -\left( {\setlength{\arraycolsep}{2pt} 
\begin{array}{rrr}
17 & 20 & 10%
\end{array}%
} \right) -\left( {\setlength{\arraycolsep}{2pt} 
\begin{array}{rrr}
58 & 106 & 87%
\end{array}%
} \right) \\* 
\fl &&\hspace{-25mm} =\left( {\setlength{\arraycolsep}{2pt} 
\begin{array}{ccc}
-143 & -261 & -287%
\end{array}%
} \right) ~\mathrm{meV,}
\end{eqnarray*}%
which are large for all three materials, and 
\begin{eqnarray*}
\fl \left\langle xy,\pi 0\pi \left\vert \mathcal{H}\right\vert 0\pi
0,yz\right\rangle &&=-\left\langle xy,0\pi \pi \left\vert \mathcal{H}%
\right\vert \pi 00,xz\right\rangle \\* 
\fl &&\hspace{-25mm} =\frac{1}{2}\left(
t_{yz,xy}^{000}-t_{xz,xy}^{000}\right) -\left(
t_{yz,xy}^{001}-t_{xz,xy}^{001}\right) +\left(
t_{yz,xy}^{010}-t_{xy,yz}^{010}\right) +\left(
t_{xz,xy}^{010}-t_{xy,xz}^{010}\right) \\* 
\fl &&\hspace{-25mm} =\frac{1}{2}\left( {\setlength{\arraycolsep}{2pt} 
\begin{array}{rrr}
-13 & -47 & -86%
\end{array}%
} \right) -\left( {\setlength{\arraycolsep}{2pt} 
\begin{array}{rrr}
-61 & -111 & -147%
\end{array}%
} \right) +\left( {\setlength{\arraycolsep}{2pt} 
\begin{array}{rrr}
17 & 20 & 10%
\end{array}%
} \right) +\left( {\setlength{\arraycolsep}{2pt} 
\begin{array}{rrr}
58 & 106 & 87%
\end{array}%
} \right) \\* 
\fl &&\hspace{-25mm} =\left( {\setlength{\arraycolsep}{2pt} 
\begin{array}{ccc}
130 & 214 & 201%
\end{array}%
} \right) ~\mathrm{meV,}
\end{eqnarray*}%
which are large as well.

Finally, we can form the matrix $\left\langle \pi00 \mathbf{%
-Q}^{\prime }\left\vert \mathcal{H}\right\vert \pi00 
\mathbf{-Q}\right\rangle $ for the 8 states of intermediate energy at S$%
_{o}. $ This matrix blocks into two. The first block involves the states: $%
\left\vert xy,\pi00 \mathrm{-} 000 \right\rangle ,$ $%
\left\vert xy, \pi00 \mathrm{-} \pi\pi0 \right\rangle ,$ 
$\left\vert yz, \pi00 \mathrm{-} 00\pi \right\rangle ,$
and $\left\vert xz, \pi00 \mathrm{-} \pi\pi\pi
\right\rangle .$ Along its diagonal, it has the degenerate levels $%
\varepsilon _{xy}\left( \pi 00\right) $ and $\varepsilon _{xy}\left( 0\pi
0\right) ,$ followed by the degenerate levels $\varepsilon _{yz}\left( \pi
0\pi \right) $ and $\varepsilon _{xz}\left( 0\pi \pi \right) .$ The
off-diagonal elements are: $\left\langle xy,\pi 00\left\vert \mathcal{H}%
\right\vert \pi 0\pi ,yz\right\rangle =\left\langle xy,0\pi 0\left\vert 
\mathcal{H}\right\vert 0\pi \pi ,xz\right\rangle $ and $\left\langle xy,0\pi
0\left\vert \mathcal{H}\right\vert \pi 0\pi ,yz\right\rangle =-\left\langle
xy,\pi 00\left\vert \mathcal{H}\right\vert 0\pi \pi ,xz\right\rangle .$ The
states contributing to the second block are: $\left\vert zx, \pi
00 \mathrm{-} 000 \right\rangle ,$ $\left\vert yz, \pi
00 \mathrm{-} \pi \pi 0 \right\rangle ,$ $\left\vert xy,
\pi 00 \mathrm{-} 00\pi  \right\rangle ,$ and $\left\vert
xy, \pi 00 \mathrm{-} \pi \pi \pi  \right\rangle .$ Along
its diagonal, there are the degenerate levels $\varepsilon _{xz}\left( \pi
00\right) $ and $\varepsilon _{yz}\left( 0\pi 0\right) ,$ followed by the
degenerate $\varepsilon _{xy}\left( \pi 0\pi \right) $ and $\varepsilon
_{xy}\left( 0\pi \pi \right) $ levels. Its off-diagonal elements are: $%
\left\langle xy,0\pi \pi \left\vert \mathcal{H}\right\vert 0\pi
0,yz\right\rangle =\left\langle xy,\pi 0\pi \left\vert \mathcal{H}%
\right\vert \pi 00,xz\right\rangle $ and $\left\langle xy,\pi 0\pi
\left\vert \mathcal{H}\right\vert 0\pi 0,yz\right\rangle =-\left\langle
xy,0\pi \pi \left\vert \mathcal{H}\right\vert \pi 00,xz\right\rangle .$ Each
of these 4$\times 4$ blocks can now be transformed into two degenerate 2$%
\times 2$ blocks. For the first set of states, this 2$\times 2$ block is
simply:%
\[
\fl         
\begin{array}{c}
\left( 
\begin{array}{ccc}
\varepsilon _{xy}\left( \pi 00\right) =\varepsilon _{xy}\left( 0\pi 0\right)
& &
\sqrt{\left\langle xy,\pi 00\left\vert \mathcal{H}\right\vert \pi 0\pi
,yz\right\rangle ^{2}+\left\langle xy,0\pi 0\left\vert \mathcal{H}%
\right\vert \pi 0\pi ,yz\right\rangle ^{2}} \\ 
\mathrm{hc} && 
\varepsilon _{yz}\left( \pi 0\pi \right) =\varepsilon _{xz}\left(
0\pi \pi \right)%
\end{array}%
\right) = \\ 
\\[-5pt] 
{\setlength{\arraycolsep}{2pt} 
\begin{array}{rrr}
\left( 
\begin{array}{cc}
860 & \sqrt{81^{2}+143^{2}} \\ 
\mathrm{hc} & 1038%
\end{array}%
\right) , & \left( 
\begin{array}{cc}
630 & \sqrt{201^{2}+261^{2}} \\ 
\mathrm{hc} & 706%
\end{array}%
\right) , & \left( 
\begin{array}{cc}
509 & \sqrt{202^{2}+287^{2}} \\ 
\mathrm{hc} & 418%
\end{array}%
\right)%
\end{array}%
} ~\mathrm{meV,}%
\end{array}%
\]
where we have inserted the values for CaVO$_{3}$ through YTiO$_{3}.$ Note
that the size of the hybridization is $\sqrt{2}$ larger than a typical
matrix element. Diagonalization yields the corresponding S$_{o}$ levels:%
\[
\varepsilon _{1}\left( \mathrm{S}_{o}\right) =%
\begin{array}{rrr}
\left( 
\begin{array}{r}
762 \\ 
1136%
\end{array}%
\right) , & \left( 
\begin{array}{r}
336 \\ 
1000%
\end{array}%
\right) , & \left( 
\begin{array}{r}
817 \\ 
110%
\end{array}%
\right)%
\end{array}%
~\mathrm{meV.} 
\]%
The 2$\times 2$ matrix and eigenvalues for the second set of states are
respectively:%
\[
\fl         
\begin{array}{c}
\left( 
\begin{array}{ccc}
\varepsilon _{xy}\left( \pi 0\pi \right) =\varepsilon _{xy}\left( 0\pi \pi
\right) && 
\sqrt{\left\langle xy,0\pi \pi \left\vert \mathcal{H}\right\vert
0\pi 0,yz\right\rangle ^{2}+\left\langle xy,\pi 0\pi \left\vert \mathcal{H}%
\right\vert 0\pi 0,yz\right\rangle ^{2}} \\ 
\mathrm{hc} &&
 \varepsilon _{yz}\left( 0\pi 0\right) =\varepsilon _{xz}\left( \pi
00\right)%
\end{array}%
\right) = \\ 
\\[-5pt] 
{\setlength{\arraycolsep}{2pt} 
\begin{array}{rrr}
\left( 
\begin{array}{cc}
952 & \sqrt{60^{2}+130^{2}} \\ 
\mathrm{hc} & 922%
\end{array}%
\right) , & \left( 
\begin{array}{cc}
718 & \sqrt{70^{2}+214^{2}} \\ 
\mathrm{hc} & 646%
\end{array}%
\right) , & \left( 
\begin{array}{cc}
541 & \sqrt{82^{2}+201^{2}} \\ 
\mathrm{hc} & 778%
\end{array}%
\right)%
\end{array}%
} ~\mathrm{meV}%
\end{array}%
\]
and%
\[
\varepsilon _{2}\left( \mathrm{S}_{o}\right) =%
\begin{array}{rrr}
\left( 
\begin{array}{r}
1081 \\ 
793%
\end{array}%
\right) , & \left( 
\begin{array}{r}
910 \\ 
454%
\end{array}%
\right) , & \left( 
\begin{array}{r}
412 \\ 
907%
\end{array}%
\right)%
\end{array}%
~\mathrm{meV} 
\]%
We realize that these eigenvalues give the trends well, although accurate
agreement with the bands in figure \ref{fig3} requires inclusion also of the
remaining 4 states at S$_{o},$ as well as inclusion of longer-range
hoppings. The decrease through the series of the lowest, degenerate
eigenvalue, that is, of the 3rd and 4th bands at S$_{o},$ from 943 meV in
SrVO$_{3}$ and 762 meV in CaVO$_{3}$ to 336 meV in LaTiO$_{3}$ and, finally,
to 110 meV in YTiO$_{3}$ is spectacular. Note also the inversion of the
cubically averaged $\varepsilon _{yz}\left( \pi 0\pi \right) \mathrm{=}
\varepsilon
_{xz}\left( 0\pi \pi \right) $ and $\varepsilon _{xy}\left( \pi 00\right)
\mathrm{=}\varepsilon _{xy}\left( 0\pi 0\right) $ 
levels in YTiO$_{3}.$ In \emph{both}
titanates, the couplings between the $xy$ and the $yz$ and $xz$ Bloch waves
are strong, but the anomalously low-lying $\varepsilon _{yz}\left( \pi 0\pi
\right) \mathrm{=}
\varepsilon _{xz}\left( \pi 00\right) $ in YTiO$_{3}$ makes the
difference.

\begin{figure}[t]
\par
\begin{center}
\rotatebox{270}{\includegraphics[height=\textwidth]{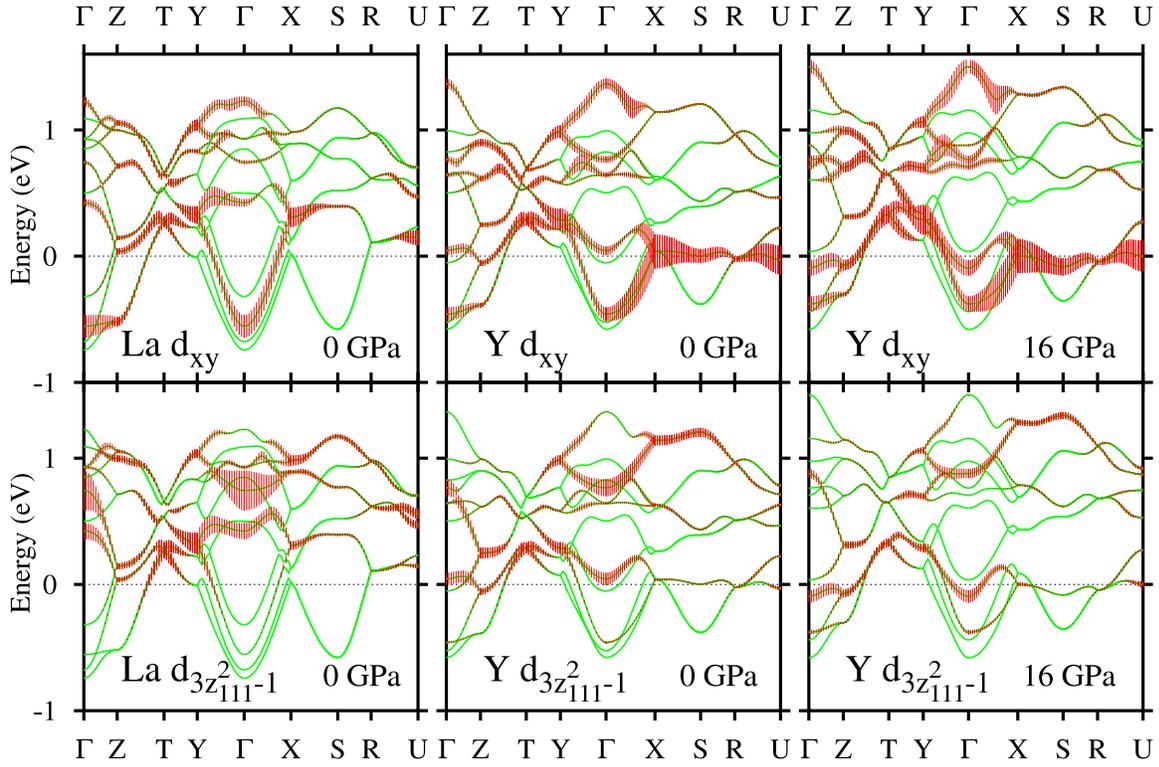}}
\end{center}
\caption{LDA LMTO bandstructures of LaTiO$_{3}$ \protect\cite{Cwik03} and
YTiO$_{3}$ at normal and 16 GPa pressure \protect\cite{loa}. The bands have
been decorated with A $d_{xy}$ and A $\left( d_{xy}+d_{yx}+d_{xz}\right) /%
\protect\sqrt{3}$ partial-wave characters, averaged over all A sites. These
characters are seen in figure \protect\ref{fig12} to be characteristic for
the lowest crystal-field orbital.}
\label{fig11}
\end{figure}

The chemical reason for the development of the pseudo-gap is the increasing
residual A character in the $t_{2g}$ band. This was mentioned before and is
demonstrated for the titanates in figure \ref{fig11}, where we have
projected the bands onto the A\thinspace $d_{3z_{111}^{2}-1}$ and A $d_{xy}$
partial waves. The lowering of the 5th band near $\Gamma _{o}$ and the 3rd
and 4th bands near X is clearly correlated with, respectively, the
A\thinspace $d_{3z_{111}^{2}-1}$ and A $d_{xy}$ characters. The cause for
this lowering is therefore likely to be hybridization with the above-lying A
bands. We saw in section \ref{HighE} that the GdFeO$_{3}$-type distortion is
larger in YTiO$_{3}$ than in LaTiO$_{3},$ because Y$^{3+}$ is smaller than La%
$^{3+},$ and because the Y 4$d$ bands are lower and more narrow than the La 5%
$d$ bands (figure \ref{fig1}). As a result, the A $d$ character left behind
to hybridize with the Ti $t_{2g}$ band is different in the two cases, and
this together with the electrostatics is the reason why the shape of these
bands, and, hence, the \emph{orbital orders,} to be considered below, are
very different. 

In the far right-hand side of figure \ref{fig11} we have
shown the bandstructure calculated for the structure measured at 16 GPa, in
which the JT distortion is nearly absent, the GdFeO$_{3}$-type distortion is
slightly increased, and the A positions have moved most \cite{loa}. As seen
from table \ref{tableLa8noJT16G} and as mentioned at the end of section \ref%
{JT}, at 16 GPa the crystal-field splitting $t_{yz,yz}^{000}-t_{xz,xz}^{000}$
is reduced from $-230$ to $-167$ meV, the crystal-field coupling $%
t_{yz,xy}^{000}-t_{xz,xy}^{000}$ is increased from $-86$ (see table \ref%
{tableQ}) to $-106$ meV, and the anomalously small hopping integral, $%
t_{yz,yz}^{001},$ is decreased from $-65$ to $-41$ meV. As expected, the
pseudo-gap \emph{and} the A $d_{xy}$ character have both increased. So at 16
GPa, the bandstructure of YTiO$_{3}$ is distorted more, rather than less
away from the cubic bandstructure of SrVO$_{3}$.

Still, one may ask: is it the displacement or the scattering properties of
the A ion by which the band shape (the pseudo-gap) is determined. To answer
this, we have performed LMTO calculations of the bandstructure of LaTiO$_{3}$
using the crystal structure of YTiO$_{3},$ both with the cell volume of YTiO$%
_{3}$ and with that of LaTiO$_{3}.$ We have also calculated the
bandstructure of YTiO$_{3}$ using the crystal structure of LaTiO$_{3},$ both
with the cell volume of LaTiO$_{3}$ and with that of YTiO$_{3}.$ In all
cases, the result was that, given the structure, it is the 
GdFeO$_{3}$-distortion
rather than the nature of the A ion which determines the shape of the 
$t_{2g}$ band.

\begin{figure}[t]
\par
\begin{center}
\rotatebox{270}{\includegraphics[height=\textwidth]{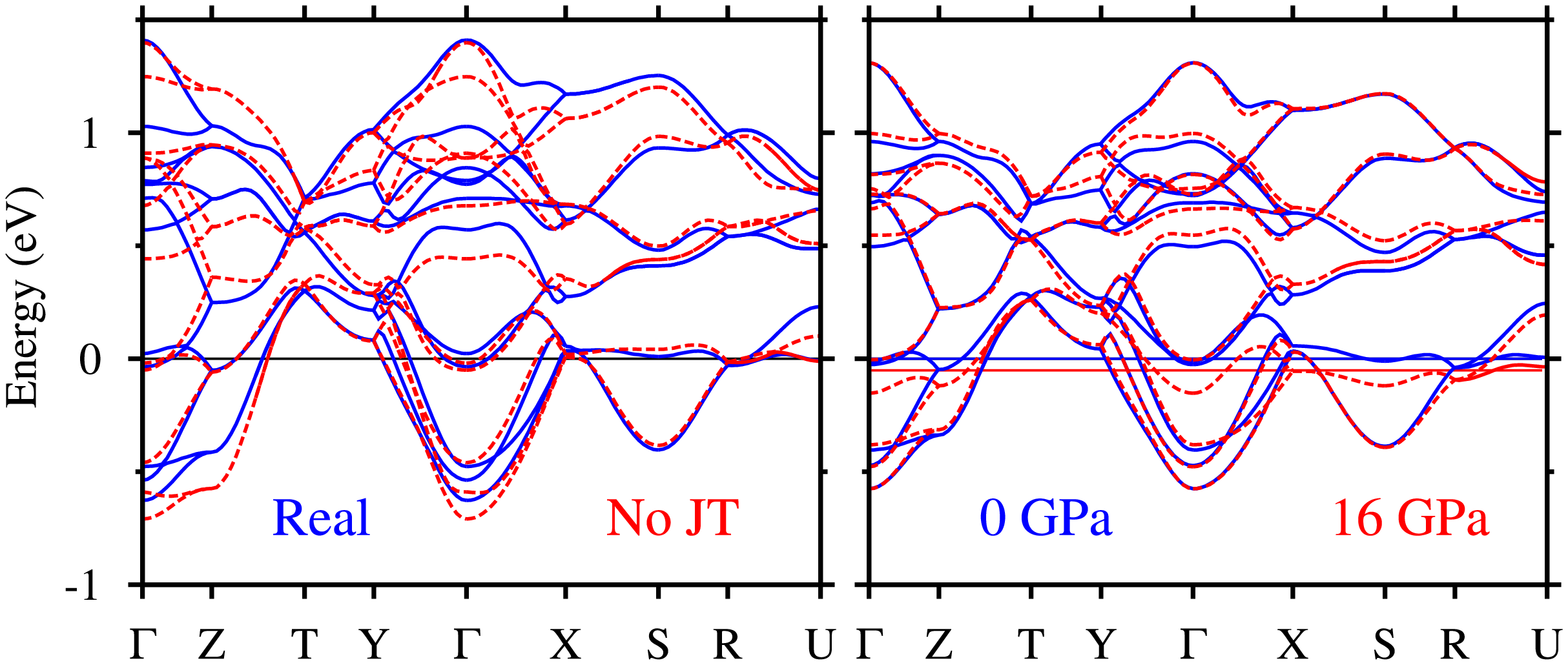}}
\end{center}
\caption{\textit{Left}: YTiO$_{3}$ $t_{2g}$ bands (blue) for the real
structure \protect\cite{ystr} and those obtained from the hypothetical
structure without JT-distortion used also for producing table \protect\ref%
{tableLa8noJT16G} (dashed red). \textit{Right}: YTiO$_{3}$ $t_{2g}$ bands
for the real structure at normal pressure \protect\cite{loa} (blue) and
those calculated for the structure at 16 GPa \protect\cite{loa}, where the
JT-distortion has essentially disappeared. The energy axis of the
high-pressure bandstructure has been scaled such as to have the bands at
normal and high pressure match well. The true scale may be found in figure 
\protect\ref{fig11}.}
\label{fig13}
\end{figure}

Let us finally demonstrate that the JT distortion is not the reason for the
development of the pseudo-gap. The blue $t_{2g}$ bands in figure \ref{fig13}
are those of YTiO$_{3}$ with the real, JT-distorted structure, which amounts
to a 3\% stretch of the O2 square into a rhomb in the $y$ direction in
subcells 1 and 3, and in the $x$ direction in subcells 2 and 4. The dashed
red bands in the left-hand figure were obtained for the hypothetical
structure with perfect TiO$_{6}$ octahedra discussed in connection with
table \ref{tableLa8noJT16G}. We see that the band shapes, in particular
those of the pseudo-gap, are \emph{the same,} compared for instance with the
huge difference in band shapes between LaTiO$_{3}$ and YTiO$_{3}.$ The
results for the 16 GPa structure, in which the JT distortion is minimal,
confirm this conclusion.

\subsection{Crystal-field representation and orbital order\label%
{XtalFieldBasis}}

Since for the titanates the calculated crystal-field splittings are an order
of magnitude larger than the spin-orbit splitting (20 meV) and the
magnetic ordering temperatures, it will prove useful to transform from the $%
yz,$ $zx,$ and $xy$ Wannier orbitals to those linear combinations, $%
\left\vert 1\right\rangle ,$ $\left\vert 2\right\rangle ,$ and $\left\vert
3\right\rangle ,$ which diagonalize the on-site LDA $t_{2g}$ Hamiltonian, $%
t^{000}$, given in the first column of tables \ref{tableSr}--\ref{tableY} and 
\ref{tableLa8noJT16G}. Of
these so-called crystal-field orbitals, $\left\vert 1\right\rangle $ has the
lowest energy and $\left\vert 3\right\rangle $ the highest. In CaVO$_{3}$
the lowest orbital remains almost purely $xy,$ in LaTiO$_{3}$ it is a fairly
equal mixture of all three orbitals, and in YTiO$_{3}$ it is a mixture of $%
yz $ and $xy$ only.

Specifically for CaVO$_{3}$, the eigenvalues relatively to $\varepsilon _{F}$
and the eigenvectors at site 000 are respectively:%
\begin{equation}
\begin{array}{ll}
\left( 
\begin{array}{ccc}
\epsilon _{1} & \epsilon _{2} & \epsilon _{3}%
\end{array}%
\right) & =\left( 
\begin{array}{ccc}
538 & 610 & 625%
\end{array}%
\right) \;\mathrm{meV}\qquad \mathrm{and} \\ 
\left( {\setlength{\arraycolsep}{3pt}%
\begin{array}{ccc}
\left\vert 1\right\rangle & \left\vert 2\right\rangle & \left\vert
3\right\rangle%
\end{array}%
}\right) & =\left( \setlength{\arraycolsep}{3pt}%
\begin{array}{ccc}
\left\vert yz\right\rangle & \left\vert xz\right\rangle & \left\vert
xy\right\rangle%
\end{array}%
\right) \left( {\setlength{\arraycolsep}{3pt}%
\begin{array}{rrr}
.202\, & -.340 & .918 \\ 
.042 & .940 & .339\, \\ 
.979 & .030 & -.204\,%
\end{array}%
}\right) .%
\end{array}
\label{eq10}
\end{equation}%
For LaTiO$_{3}$ with the older data \cite{lastr}, we get:%
\begin{equation}
\begin{array}{ll}
\left( 
\begin{array}{ccc}
\epsilon _{1} & \epsilon _{2} & \epsilon _{3}%
\end{array}%
\right) & =\left( 
\begin{array}{ccc}
398 & 537 & 608%
\end{array}%
\right) \;\mathrm{meV\qquad \mathrm{and}} \\ 
\left( {\setlength{\arraycolsep}{3pt}%
\begin{array}{ccc}
\left\vert 1\right\rangle & \left\vert 2\right\rangle & \left\vert
3\right\rangle%
\end{array}%
}\right) & =\left( \setlength{\arraycolsep}{3pt}%
\begin{array}{ccc}
\left\vert yz\right\rangle & \left\vert xz\right\rangle & \left\vert
xy\right\rangle%
\end{array}%
\right) \left( {\setlength{\arraycolsep}{3pt}%
\begin{array}{rrr}
.715 & -.609 & .344 \\ 
.353 & .739 & .574 \\ 
.604 & .289 & -.743%
\end{array}%
}\right) ,%
\end{array}
\label{eq11}
\end{equation}%
whereas with the recent LaTiO$_{3}$ data \cite{Cwik03}, results are somewhat
different: 
\begin{equation}
\begin{array}{ll}
\left( 
\begin{array}{ccc}
\epsilon _{1} & \epsilon _{2} & \epsilon _{3}%
\end{array}%
\right) & =\left( 
\begin{array}{ccc}
354 & 546 & 561%
\end{array}%
\right) \;\mathrm{meV} \\ 
\left( {\setlength{\arraycolsep}{3pt}%
\begin{array}{ccc}
\left\vert 1\right\rangle & \left\vert 2\right\rangle & \left\vert
3\right\rangle%
\end{array}%
}\right) & =\left( \setlength{\arraycolsep}{3pt}%
\begin{array}{ccc}
\left\vert yz\right\rangle & \left\vert xz\right\rangle & \left\vert
xy\right\rangle%
\end{array}%
\right) \left( {\setlength{\arraycolsep}{3pt}%
\begin{array}{rrr}
.735 & -.538 & .413 \\ 
.320 & .812 & .489 \\ 
.599 & .227 & -.768%
\end{array}%
}\right) ,%
\end{array}
\label{eq13}
\end{equation}%
in particular for states $\left\vert 2\right\rangle $ and $\left\vert
3\right\rangle .$ For YTiO$_{3}$ the eigenvalues and eigenvectors are:%
\begin{equation}
\begin{array}{ll}
\left( 
\begin{array}{ccc}
\epsilon _{1} & \epsilon _{2} & \epsilon _{3}%
\end{array}%
\right) & =\left( 
\begin{array}{ccc}
289 & 488 & 620%
\end{array}%
\right) \;\mathrm{meV} \\ 
\left( {\setlength{\arraycolsep}{3pt}%
\begin{array}{ccc}
\left\vert 1\right\rangle & \left\vert 2\right\rangle & \left\vert
3\right\rangle%
\end{array}%
}\right) & =\left( \setlength{\arraycolsep}{3pt}%
\begin{array}{ccc}
\left\vert yz\right\rangle & \left\vert xz\right\rangle & \left\vert
xy\right\rangle%
\end{array}%
\right) \left( {\setlength{\arraycolsep}{3pt}%
\begin{array}{rrr}
.781 & -.571 & .253 \\ 
-.073 & .319 & .945 \\ 
.620 & .757 & -.207%
\end{array}%
}\right) .%
\end{array}
\label{eq15}
\end{equation}
These crystal-field splittings are much larger than the spin-orbit splitting
($\sim $20 meV) and $kT,$ and they agree with what was deduced (0.12-0.30
eV) from spin-polarized x-ray scattering for LaTiO$_{3}$ \cite{Haverkort}.


%
\begin{figure}[t]
\par
\begin{center}
\includegraphics[width=\textwidth]{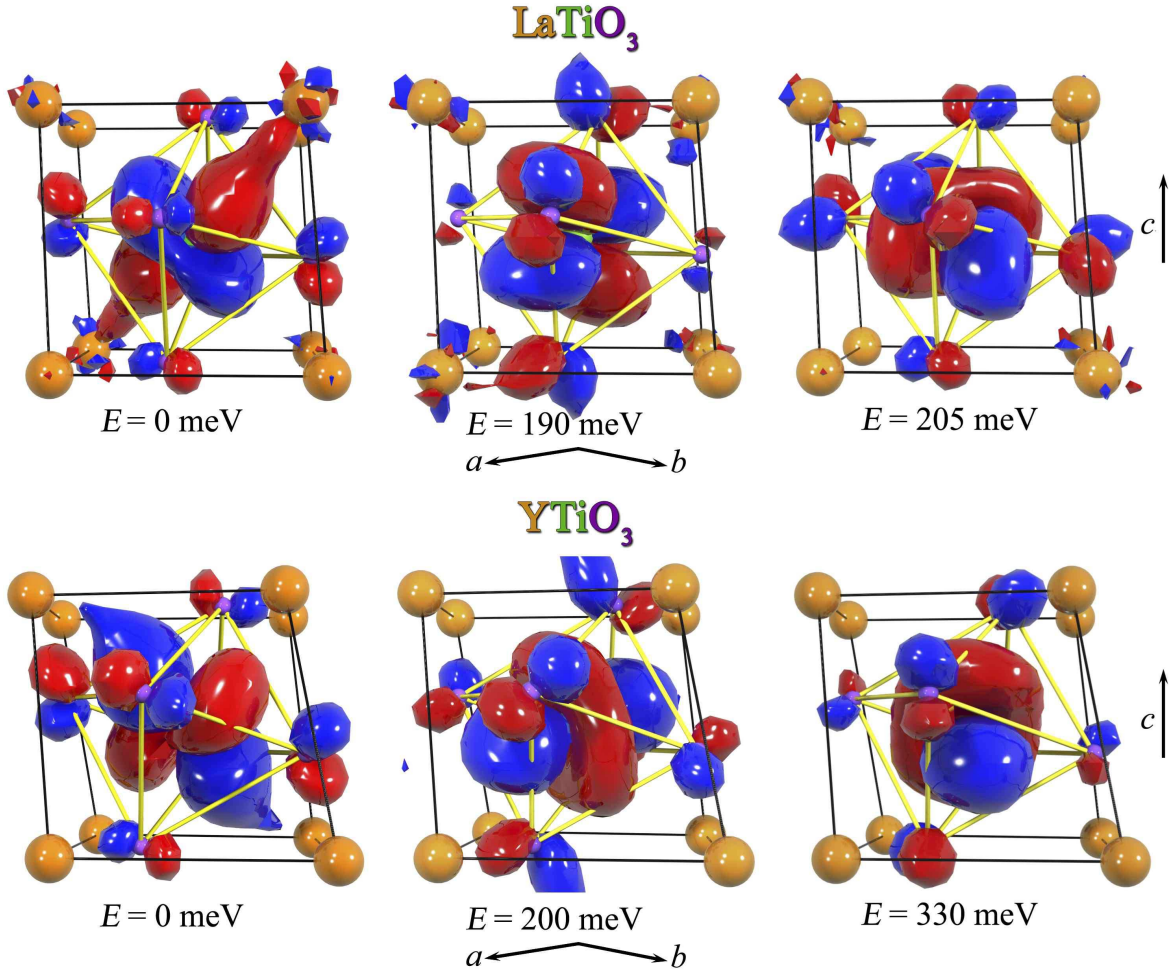}
\end{center}
\caption{Ti $t_{2g}$ crystal-field splittings and NMTO orbitals $\left\vert
1\right\rangle ,$ $\left\vert 2\right\rangle ,$ and $\left\vert
3\right\rangle $ at site 000 (subcell 1) for LaTiO$_{3}$ \protect\cite%
{Cwik03} and YTiO$_{3}$ \protect\cite{ystr} obtained by diagonalization of
the on-site LDA Hamiltonian in the orthonormalized $t_{2g}$-NMTO basis. With
the older LaTiO$_{3}$ data \protect\cite{lastr}, we get crystal-field levels
of 0, 140, and 210 meV, whereas using the recent \protect\cite{Cwik03}
8\thinspace K data yields 0, 207, and 221 meV. For the hypothetical
structure of YTiO$_{3}$ without JT-distortion used in table \protect\ref%
{tableLa8noJT16G}, the crystal field levels are 0, 136, and 259 meV. Orbital 
$\left\vert 1\right\rangle $ has 97 (94) \% of its charge density inside the
central La$_{8}$TiO$_{6}$ (Y$_{8}$TiO$_{6})$ unit. The 3\% difference is due
to $e_{g}$-character on the Ti 010 neighbours in YTiO$_{3}.$ The
same contour was chosen as in figure \protect\ref{fig4}. The strong $e_{g}$
character on Ti 001 for orbital $\left\vert 2\right\rangle $ in YTiO$_{3}$
has been cropped in this figure.}
\label{fig7}
\end{figure}

All three eigenfunctions and the two level splittings are shown in figure %
\ref{fig7} for the titanates. The orbital of lowest energy, $\left\vert
1\right\rangle ,$ is of course the one for which the electrostatic
attraction and the bonding hybridization with the A ions are maximized at
the same time as the electrostatic repulsion and the antibonding
hybridization with the oxygens is minimized. In LaTiO$_{3},$ this orbital is
directed towards the two La cations closest to Ti, the ones along the
shortest diagonal. Also in YTiO$_{3}$ do the nearest cations attract the red
lobes of the lowest orbital, and for the same reasons, but in YTiO$_{3}$
there is also competition from the 2nd-nearest cations and, as a result, the
blue lobes bond along the [1\={1}1] diagonal with the Y 4$d_{xy}$ orbitals.
The details of the hybridizations were explained for the constituent $yz,$ $%
xy,$ and $xz$ orbitals in the preceding subsection.

\begin{figure}[t]
\par
\begin{center}
\includegraphics[width=\textwidth]{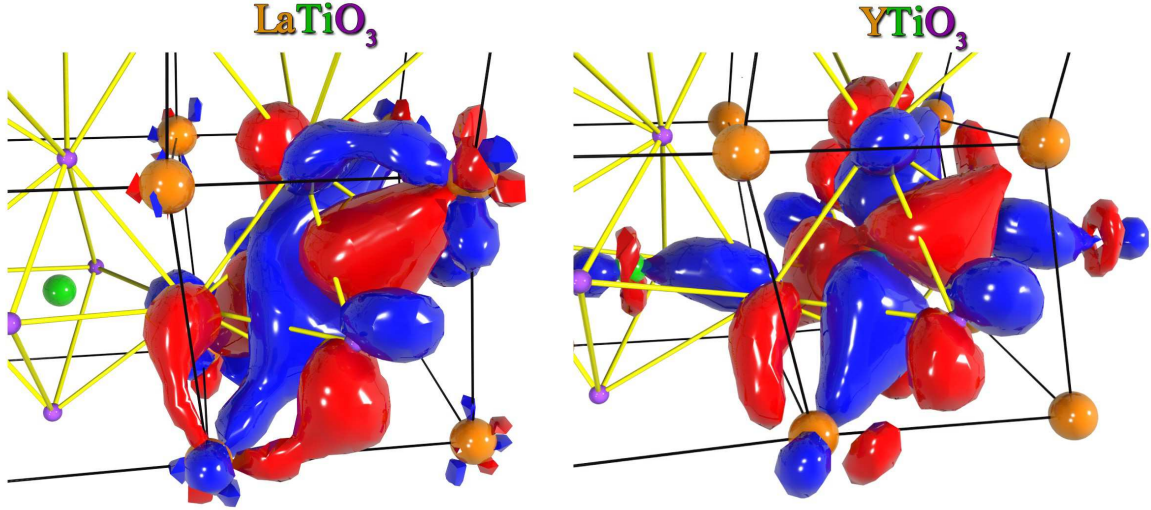}
\end{center}
\caption{Crystal-field orbital with the lowest energy, $\left\vert
1\right\rangle ,$ at site 010 (subcell 2). A contour value 2/3 of the one
used in the previous figures was chosen in order to exhibit the
hybridization with the O $p$ and A $d_{3z_{111}^{2}-1}$ and $d_{xy}$ states.
In LaTiO$_{3},$ the hybridization with $d_{3z_{111}^{2}-1}$ on the two
1st-nearest A ions dominates over the hybridization with $d_{xy}$ on the two
2nd-nearest A-ions towards which the octahedron tilts. In YTiO$_{3},$ the
opposite is true. For the heavily distorted YTiO$_{3},$ there is also
bonding from O2 $p$ to $e_{g}$ on two Ti-neighbours. The A $%
d_{3z_{111}^{2}-1}$ and $d_{xy}$ hybridizations of the $t_{2g}$ bands were
shown in figure \protect\ref{fig11}.}
\label{fig12}
\end{figure}

\begin{figure}[t]
\par
\begin{center}
\includegraphics[width=\textwidth]{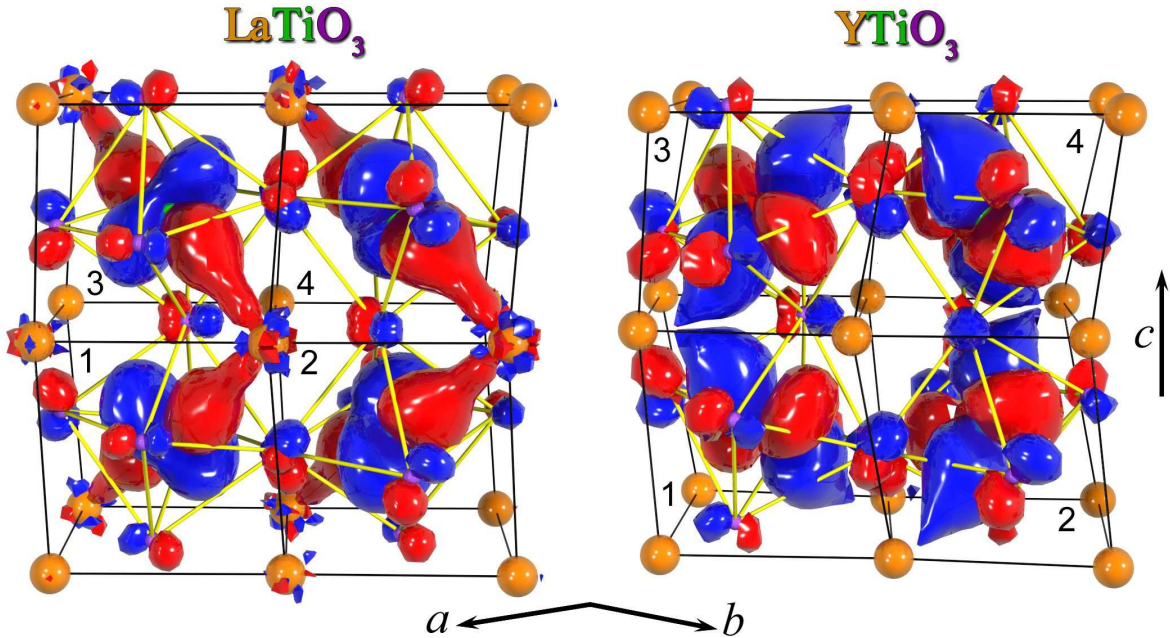}
\end{center}
\caption{Orbital of the lowest energy, $\left\vert 1\right\rangle ,$ placed
in the 4 subcells (1=000, 2=010, 3=001, and 4=011) according to the mirror
and glide-mirror symmetries, \textit{i.e.} like in the bottom row of figure 
\protect\ref{fig4}. This illustrates the orbital order which will crystalize
out due to the Coulomb correlations (see subsection \protect\ref{OrbPol}).}
\label{fig8}
\end{figure}

The difference between the lowest orbital in LaTiO$_{3}$ and YTiO$_{3}$ is
brought out clearly in figure \ref{fig12} where we have chosen a lower
contour than in figure \ref{fig7}, as well as the orbital at site 010 (see
also figure \ref{fig8}). In both materials does each red lobe \emph{bond} to
the closest-cation $d_{3z_{111}^{2}-1}$
orbital directed along the shortest A-B-A diagonal, and in both
materials does each blue lobe form a complex 5-centre, B\thinspace
--\thinspace 2$\times $O2\thinspace --\thinspace A\thinspace $d_{xy}$%
\thinspace --\thinspace O1 \emph{bond} around the 2nd-nearest A-ion towards
which the octahedron tilts. But the former mechanism prevails in La and the
latter in Y titanate. The effect of A $d_{3z_{111}^{2}-1}$ and A $d_{xy}$
hybridizations on the $t_{2g}$ bandstructures were shown in figure \ref%
{fig11}. Finally, for the YTiO$_{3}$ orbital, we see that two oxygen $p$
lobes which point opposite to the direction of their GdFeO$_{2}$-type
displacement, attain bonding $e_{g}$ character on their Ti neighbour. As was
mentioned in subsection \ref{xyyzzx}, this enhances the $p$ lobe and bends
it outwards. This hybridization --allowed only by the strong tilt--
contributes as well to lowering the energy of orbital $\left\vert
1\right\rangle .$

These quantitative differences in the shape of the lowest orbital in La and
Y titanate are \emph{not} caused by the small JT distortions. As an example,
for YTiO$_{3}$ without JT distortion the on-site Hamiltonian in table \ref%
{tableLa8noJT16G} yields the following eigenvalues and eigenvectors:%
\begin{equation}
\begin{array}{ll}
\left( 
\begin{array}{ccc}
\epsilon _{1} & \epsilon _{2} & \epsilon _{3}%
\end{array}%
\right) & =\left( 
\begin{array}{ccc}
341 & 477 & 600%
\end{array}%
\right) \;\mathrm{meV} \\ 
\left( {\setlength{\arraycolsep}{3pt}%
\begin{array}{ccc}
\left\vert 1\right\rangle & \left\vert 2\right\rangle & \left\vert
3\right\rangle%
\end{array}%
}\right) & =\left( \setlength{\arraycolsep}{3pt}%
\begin{array}{ccc}
\left\vert yz\right\rangle & \left\vert xz\right\rangle & \left\vert
xy\right\rangle%
\end{array}%
\right) \left( {\setlength{\arraycolsep}{3pt}%
\begin{array}{rrr}
.766 & -.473 & .435 \\ 
-.023 & .696 & .718 \\ 
.642 & .540 & -.544%
\end{array}%
}\right) .%
\end{array}
\label{eq17}
\end{equation}%
For the lowest orbital, this is nearly identical with the results for the
JT-distorted structure in equation (\ref{eq15}), but it differs
substantially for the higher orbitals, and that will prove important for the
stability of ferromagnetic order in YTiO$_{3}$ (section \ref{M}). For the
structure measured at 16 GPa, table \ref{tableLa8noJT16G} yields%
\begin{equation}
\begin{array}{ll}
\left( 
\begin{array}{ccc}
\epsilon _{1} & \epsilon _{2} & \epsilon _{3}%
\end{array}%
\right) & =\left( 
\begin{array}{ccc}
365 & 584 & 726%
\end{array}%
\right) \;\mathrm{meV}\\ 
\left( \setlength{\arraycolsep}{3pt}%
\begin{array}{ccc}
\left\vert 1\right\rangle & \left\vert 2\right\rangle & \left\vert
3\right\rangle%
\end{array}%
\right) & =\left( \setlength{\arraycolsep}{3pt}%
\begin{array}{ccc}
\left\vert yz\right\rangle & \left\vert xz\right\rangle & \left\vert
xy\right\rangle%
\end{array}%
\right) \left( {\setlength{\arraycolsep}{3pt}%
\begin{array}{rrr}
.829 & -.221 & .514 \\ 
-.243 & .685 & .689 \\ 
.504 & .693 & -.514%
\end{array}%
}\right) .%
\end{array}
\label{eq18}
\end{equation}%
Here again, the shape of the lowest orbital is nearly the same as at normal
pressure, although \emph{less} cubic and \emph{more} YTiO$_{3}$'ish. This
proves that the shape of the lowest orbital is determined mainly
by the GdFeO$_{3}$%
-type rather than by the JT distortion.

The shape of the orbital becomes crucial when it is placed in the crystal.
As seen in figure \ref{fig8}, the \emph{orbital order} is very \emph{%
different} in the two titanates. In LaTiO$_{3},$ the lowest orbital roughly
has the $bc$ plane as mirror $\left( x\leftrightarrow y\right) ,$ and this
means that the glide-mirror operation from site 1 to 2 is roughly a
translation, and so are therefore all cubic translations in the $xy$ plane.
In YTiO$_{3},$ the lowest orbital does not have this symmetry, and the
orbitals in the $xy$ plane therefore avoid each other even more than in LaTiO%
$_{3}.$ As a result, the dominating integral, $t_{11}^{100}=t_{11}^{010},$
for hopping between two lowest orbitals on nearest neighbours parallel to
the mirror plane is only half as large in YTiO$_{3}$ ($-48$ meV) as in LaTiO$%
_{3}$ ($-98$ meV). In both cases, this hopping integral is mainly via $%
d_{xy} $ character on the two A ions which are 1st-nearest neighbours to one
Ti and 2nd-nearest neighbours to the other. But whereas both large hopping
integrals, $t_{yz,yz}^{010}$ and $t_{xy,xy}^{010},$ contribute in LaTiO$_{3}$%
, only the latter does so in YTiO$_{3}$. This may be seen in detail from the
matrix transformation (\ref{eqLa010}) and (\ref{eqY010}) given below. Both
for La and Y titanate, there is only 
little O2-mediated hopping because the $p$
character of one orbital hardly couples to the other orbital.

The 1st-nearest-neighbour hopping in the $z$ direction, $t_{11}^{001},$ is
antibonding (105 meV) in LaTiO$_{3},$ with the sign-convention dictated by
the mirror plane, and it is mainly via oxygen $p.$ In fact, since according
to the transformation (\ref{eq13}), orbital $\left\vert 1\right\rangle $ in
LaTiO$_{3}$ is mainly composed of $\left\vert yz\right\rangle $ and $%
\left\vert xy\right\rangle ,$ and since the latter hardly couples in the $z$
direction, $t_{11}^{001}\sim \ -0.5t_{yz,yz}^{001}$ and $%
t_{yz,yz}^{001}=-193 $ meV, as was discussed in connection with
equation (\ref{eq7}). In YTiO$_{3},$ orbital $\left\vert 1\right\rangle $ is
almost exclusively composed of $\left\vert yz\right\rangle $ and $\left\vert
xy\right\rangle ,$ so that here we should have: $t_{11}^{001}\sim
-0.6t_{yz,yz}^{001}$. With $t_{yz,yz}^{001}$ anomalously small ($-65$ meV)
due to coupling via Y\textbf{\thinspace }4$d_{xy},$ $t_{11}^{001}$ becomes
not only anomalously small, but even bonding ($-38$ meV) because now the
small coupling ($\pm $64 meV) between $\left\vert yz\right\rangle $ and $%
\left\vert xy\right\rangle $ in the $z$ direction cannot be neglected. This
can be followed explicitly in equations (\ref{eqLa001}) and (\ref{eqY001})
given below.

Hence for the lowest crystal-field orbital the paths for hopping between
nearest Ti neighbours in the $xy$ plane are very different from those for
hopping along $z.$ Since exchange couplings are proportional to hopping
integrals \emph{squared} in conventional super-exchange theory, and since
the spin-wave spectra in antiferromagnetic La as well as in ferromagnetic Y
titanate were both measured to be isotropic, this could be taken as an
argument against the applicability of conventional theory for the
low-temperature properties of these systems~\cite{Ulrich}. We shall return
to this in section \ref{M}.

Hopping integrals squared are very sensitive quantities, and since the
residual cation covalencies in the materials with large GdFeO$_{3}$-type
distortions have completely ruined the cubic symmetry of 
the LDA $t_{2g}$ Hamiltonians, let us
explicitly perform the transformation of the dominating $t_{\pi }$ hopping
integrals, $t^{010},~t^{100},$ and $t^{001},$ from the cubic to the
crystal-field representation in LaTiO$_{3}$ and YTiO$_{3}.$ We start from
the cubic representation 
in tables \ref{tableLaCwik} and \ref{tableY}, and use the
transformation valid for site 000 given by equations (\ref{eq13}) and (\ref%
{eq15}). When transforming the hopping integrals, we should remember that in
the cubic representation (middle row in figure \ref{fig4}) those $t_{2g}$
orbitals which would couple strongly if the crystal were cubic, carry the
same name in all four subcells, whereas in the crystal-field representation,
the orbitals are named and have signs following the space-group symmetry, 
that is, the bottom row in figure \ref{fig4} and in figure \ref{fig8}.

The $t^{010}$ matrix for hopping between crystal-field orbitals $\left\vert
1\right\rangle ,$ $\left\vert 2\right\rangle,$ and $\left\vert
3\right\rangle $ at sites 000 and 010, which are related by the 
glide mirror $\left( x\leftrightarrow y\right) ,$ is for LaTiO$_{3}$:%
\begin{eqnarray} &&\nonumber
\fl 
\left( \begin{array}{rrr} 
.735 & .320 & .599 \\ 
-.538 & .812 & .227 \\ 
.413 & .489 & -.768\end{array}\right) 
\left(\begin{array}{rrr}
 -185 & 75 & -34 \\ 
-43 & 1 & 73 \\ 
-54 & -33 &
-183\end{array}\right)
 \left( \begin{array}{rrr}
 .320 &  .812 &  .489 \\ 
 .735 & -.538 &  .413 \\ 
 .599 &  .227 & -.768\end{array}\right) 
\mathrm{meV} \\ &&
 =\left(
\begin{array}{rrr}
 -98 & -192 & 12 \\ 
4 & 76 & -22 \\ 
120 & -40 & -128\end{array}\right) \mathrm{meV}, 
\label{eqLa010} 
\end{eqnarray}%
%
%
with the order of rows and columns being like in
equations (\ref{eq10})--(\ref{eq18}), \textit{i.e.}: $yz,$ $xz,$ $xy$ for
the cubic and $1,$ 2, 3 for the crystal-field representation. In the cubic
representation, the elements along the diagonal of the hopping matrix are
the well-known $t_{\pi },$ $t_{\delta },$ $t_{\pi }.$ It may be realized
that the hopping integrals, $t_{mm^{\prime }}^{010},$ resulting from this
transformation are sums of contributions with varying signs. For YTiO$_{3}:$%
\begin{eqnarray} \nonumber &&
\fl  
\left( \begin{array}{rrr} .781 & -.073 &
.620 \\ -.571 & .319 & .757 \\ .253 & .945 & -.207\end{array}\right) \!\!
\left( \begin{array}{rrr} -184 & 70 & -54 \\ -41 & 28 & 65 \\ -64 & -22 &
-162\end{array}\right) \!\! \left( \begin{array}{rrr} -.073 & .319 & .945 \\
.781 & -.571 & .253 \\ .620 & .757 & -.207\end{array}\right) 
\mathrm{meV} \\  &&
= \left(
\begin{array}{rrr} -48 & -191 & -130 \\ -84 & -13 & 44 \\ 94 & 11 &
-73\end{array}\right) \mathrm{meV}. \label{eqY010} 
\end{eqnarray}%
Note that at the second site 010, we must exchange the $yz$ and $xz$
orbitals. For the $t^{100}$ matrix of hopping integrals between the orbitals
at sites 000 and site $100= 010 +\mathbf{a}$ the
transformation is the same, and we simply get:%
\[
t_{mm^{\prime }}^{100}=t_{m^{\prime }m}^{010}. 
\]%
For these hoppings parallel to the mirror plane, the
differences between La and Y-titanate are rooted more in the different
crystal-field eigenvectors than in the different hoppings between the $yz,$ $%
xz,$ and $xy$ orbitals.

The $t^{001}$ matrix for hopping between crystal-field orbitals $\left\vert
1\right\rangle ,$ $\left\vert 2\right\rangle ,$ and $\left\vert
3\right\rangle $ at sites 000 and 001, which are related by the A-O1
mirror plane perpendicular to the $z$ axis $\left( z\rightarrow -z\right) $,
is for LaTiO$_{3}$:%
\begin{eqnarray}
\nonumber && \fl 
\left( \begin{array}{rrr} .735 & .320 &
.599 \\ -.538 & .812 & .227 \\ .413 & .489 & -.768\end{array}\right) \left(
\begin{array}{rrr} -193 & -42 & -59 \\ -42 & -208 & 52 \\ 59 & -52 &
-22\end{array}\right) \left( \begin{array}{rrr} -.735 & .538 & -.413 \\
-.320 & -.812 & -.489 \\ .599 & .227 & -.768\end{array}\right) 
\mathrm{meV} \\ && =\left(
\begin{array}{rrr} 105 & 31 & 143 \\ 31 & 188 & -10 \\ 143 & -10 &
85\end{array}\right) \mathrm{meV} , \label{eqLa001} 
\end{eqnarray}%
and for YTiO$_{3}$ it is:%
\begin{eqnarray} \nonumber &&
\fl \left( \begin{array}{rrr} .781 & -.073 &
.620 \\ -.571 & .319 & .757 \\ .253 & .945 & -.207\end{array}\right) \left(
\begin{array}{rrr} -65 & -63 & -64 \\ -63 & -178 & 83 \\ 64 & -83 &
-8\end{array}\right) \left( \begin{array}{rrr} -.781 & .571 & -.253 \\ .073
& -.319 & -.945 \\ .620 & .757 & -.207\end{array}\right) 
\mathrm{meV} \\ && =\left(
\begin{array}{rrr} -38 & -21 & 97 \\ -21 & 107 & 51 \\ 97 & 51 &
167\end{array}\right) \mathrm{meV}. \label{eqY001} 
\end{eqnarray}%
Here we must flip the sign of the $yz$ and $xz$ orbitals at site 001. In
case of hopping perpendicular to the mirror plane, not only the different
crystal-field eigenvectors matter, but also the large non-cubic
perturbations of the hopping integrals in YTiO$_{3}.$

In the cubic representation, the matrices of
1st-nearest-neighbour hopping integrals have two large diagonal elements, $%
t_{\pi },$ and all other (inter-orbital) elements are small, except in YTiO$%
_{3},$ where $t_{yz,yz}^{001}=-65$ meV is anomalously small and the
inter-orbital elements are of similar size. In the crystal-field
representation, the matrices of 1st-nearest-neighbour hopping
integrals do not have this form at all. In particular in YTiO$_{3},$
the hopping between orbitals $\left\vert 1\right\rangle $ is anomalously
small and, except in one case, smaller than the hopping from orbital $%
\left\vert 1\right\rangle $ to orbitals $\left\vert 2\right\rangle $ and $%
\left\vert 3\right\rangle .$ As we shall see in section \ref{M}, this is what
makes YTiO$_{3}$ ferromagnetic at low temparature.

\begin{figure}[t]
\par
\begin{center}
\includegraphics[width=0.88\textwidth]{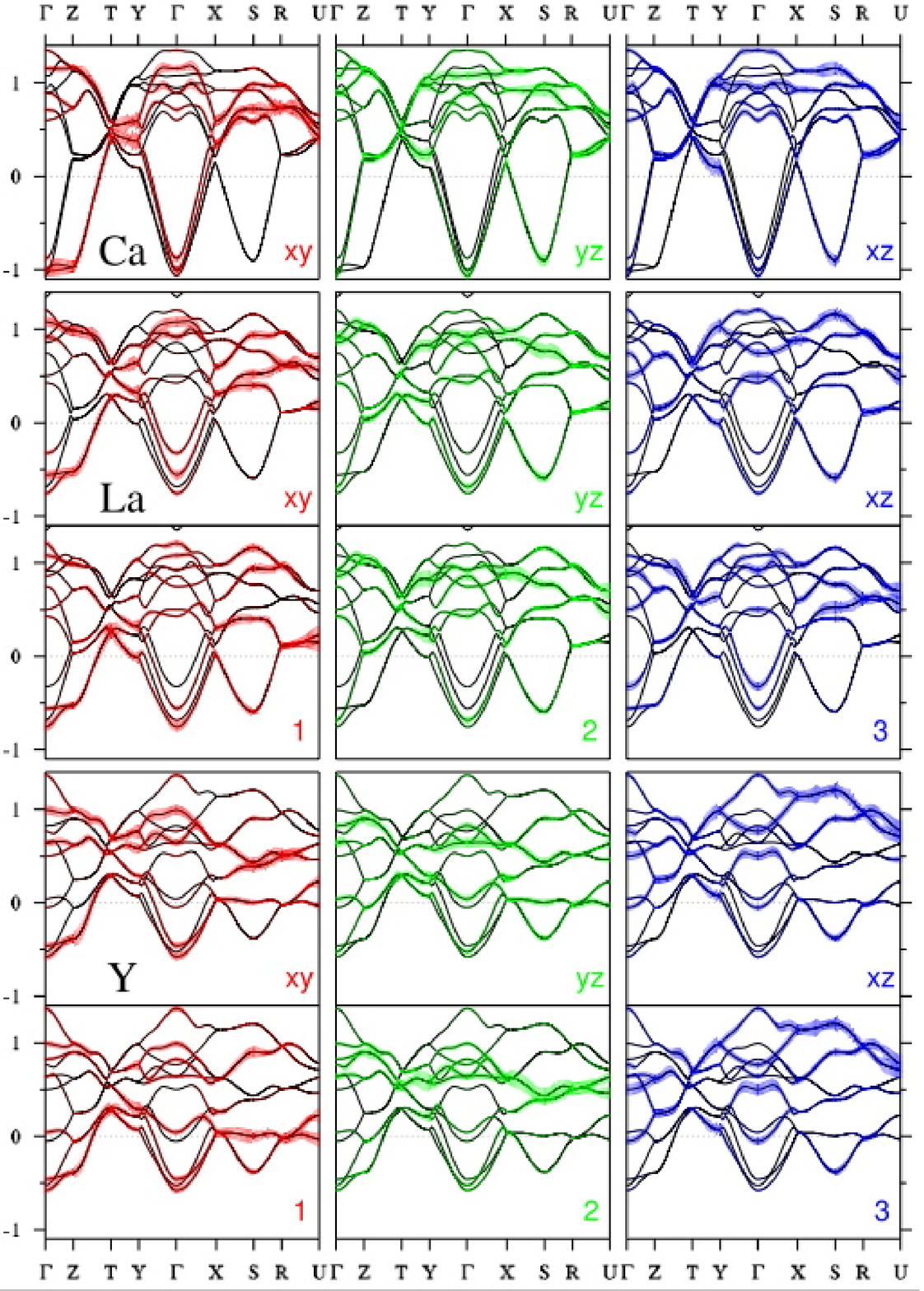}
\end{center}
\caption{LDA LMTO bandstructures of the orthorhombic perovskites decorated
with B partial-wave characters at site 000 in the cubic and the
crystal-field representations, or equivalently, according to the \emph{bottom%
} row of figure \protect\ref{fig4} and figure \protect\ref{fig8}. For CaVO$%
_{3}$ $\left\vert 1\right\rangle \approx \left\vert xy\right\rangle ,$ $%
\left\vert 2\right\rangle \approx \left\vert xz\right\rangle ,$ and $%
\left\vert 3\right\rangle \approx \left\vert yz\right\rangle .$ The points
of high symmetry are those mentioned explicitly in the caption to figure 
\protect\ref{fig5}, and not any of their equivalents.}
\label{fig3b}
\end{figure}


\begin{figure}[t]
\begin{center}
\includegraphics[width=\textwidth]{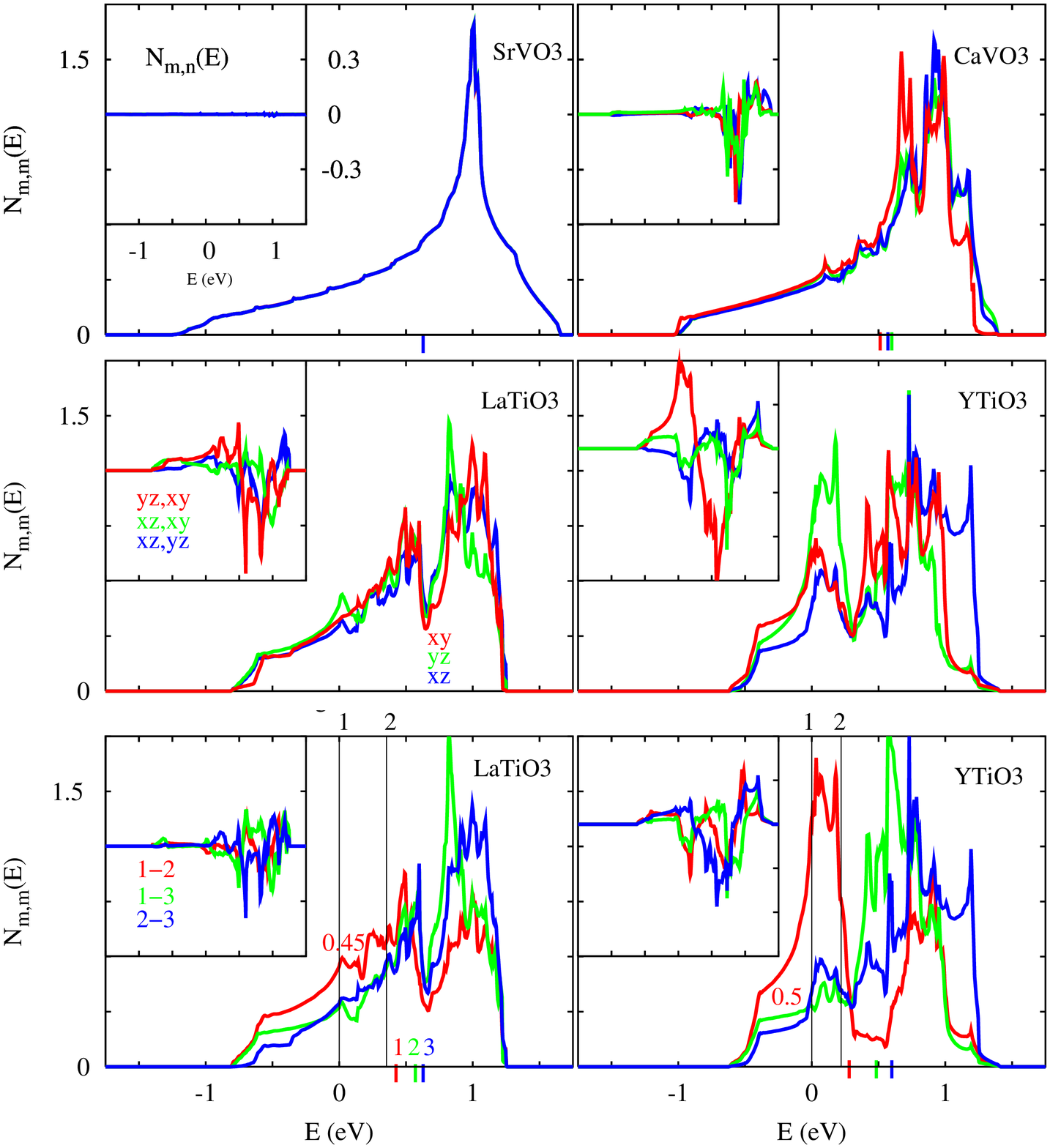}
\end{center}
\caption{The LDA on-site DOS matrices $N_{mm^{\prime }}\left( \protect%
\varepsilon \right) $ for site 000 (subcell 1) in the representation of the $%
t_{2g}$ Wannier functions. The zero of energy is the Fermi level. The DOS
unit is states/(eV$\cdot $spin$\cdot $ABO$_{3}).$ The insets show the
off-diagonal elements on a reduced energy scale, but the same DOS scale. 
\textit{Top two rows:} cubic basis as defined in the bottom row of figure 
\protect\ref{fig4}. \textit{Bottom row:} crystal-field basis as defined in
figure \protect\ref{fig7} and equations (\protect\ref{eq10})--(\protect\ref%
{eq15}). The tick marks indicate the positions of the $t_{2g}$ crystal-field
levels. The occupancy of the lowest crystal-field orbital is 0.45 in LaTiO$%
_{3}$ and 0.50 in YTiO$_{3}$. The Fermi levels for occupations with one and
two electrons are marked; the latter is the majority-spin Fermi level for a $%
d^{1}$ ferromagnet in the Stoner model. The older data \protect\cite{lastr}
was used for LaTiO$_{3}$.}
\label{fig6}
\end{figure}

In the crystal-field representation, we do \emph{not} mirror the orbitals
like in equation (\ref{eqCubicHop}) in order to obtain the values of
integrals corresponding to hops starting from another site than the first.
Here the rules are simply:%
\[
\begin{array}{lll}
t_{m^{\prime },m}^{\left( 0,0,1\right) ,\left( x,y,z+1\right) }=t_{m^{\prime
},m}^{xy\bar{z}} &  & t_{m^{\prime },m}^{\left( 0,1,1\right) ,\left(
x,y+1,z+1\right) }=t_{m^{\prime },m}^{yx\bar{z}} \\ 
&  &  \\ 
t_{m^{\prime },m}^{\left( 0,0,0\right) ,\left( x,y,z\right) }\equiv
t_{m^{\prime },m}^{xyz} &  & t_{m^{\prime },m}^{\left( 0,1,0\right) ,\left(
x,y+1,z\right) }=t_{m^{\prime },m}^{yxz}%
\end{array}%
\]

In figure \ref{fig3b} we show the bandstructures decorated with B $t_{2g}$
partial-wave characters in the cubic and the crystal-field representations
at site 000 or, equivalently, according to the \emph{bottom} row of figure %
\ref{fig4} (see also figure \ref{fig8}). In general, the mixing of
characters is considerable in both representations, although the
crystal-field splittings -- which are small on the scale of the bandwidth --
do cause a slight preference for lower-lying bands to have stronger $%
\left\vert 1\right\rangle $ character (red) and higher-lying bands to have
stronger $\left\vert 3\right\rangle $ character (blue). Nevertheless, 
there \emph{are} two
important cases of character separation: the bands at the lower edge of the
pseudo-gap, which move down and get increasingly occupied as we proceed along
the series, have strong $\left\vert 1\right\rangle $ character, and the 4th
band near $\Gamma _{o},$ which moves upwards and almost empties, has strong $%
\left\vert 3\right\rangle $ character. As we have seen in figures \ref%
{fig11} and \ref{fig12}, the $\left\vert 1\right\rangle $ bands are those
with strong A $d_{3z_{111}^{2}-1}$ and A $d_{xy}$ characters. The dominant
role of those bands in the development of the pseudo-gap is also evident from
the top right and the bottom row parts of figure \ref{fig6}, exhibiting the
on-site 000 elements,%
\[
N_{mm^{\prime }}\left( \varepsilon \right) \equiv \sum_{i\mathbf{k}}u_{%
\mathbf{R}m,i}\left( \mathbf{k}\right) \,\delta \left[ \varepsilon
-\varepsilon _{i}\left( \mathbf{k}\right) \right] \,u_{\mathbf{R}m^{\prime
},i}^{\ast }\left( \mathbf{k}\right) , 
\]%
of the density-of-states (DOS) matrix. Here, $\mathbf{u}_{i}\left( \mathbf{k}%
\right) $ is an eigenvector of the LDA Hamiltonian (\ref{Hk}) in a $t_{2g}$
Wannier-function basis.


\section{Multi-band Hubbard Hamiltonian and its solution in the dynamical
mean-field approximation\label{DMFT}}

Having discussed at length the one-electron, \textquotedblleft
chemical\textquotedblright\ part of the Hamiltonian, $H_{t_{2g}}^{LDA},$ we
now add the on-site Coulomb repulsion and obtain the multi-band Hubbard
Hamiltonian:%
\begin{equation}
\mathcal{H}_{t_{2g}}=\mathcal{H}_{t_{2g}}^{LDA}+\sum\nolimits_{R}\mathcal{U}%
_{R}~.  \label{H}
\end{equation}%
Here, $\mathcal{U}$ is such that for a doubly occupied site $\left(
t_{2g}^{2}\right) ,$ its 15 degenerate states are as follows: a 9-fold
degenerate triplet with energy $U-3J,$ a 5-fold degenerate singlet with
energy $U-J,$ and a singlet with energy $U+2J.$ For the analogous $p^{2}$%
-configuration these states would be $^{3}P,$ $^{1}D,$ and $^{1}S,$
respectively. The average $d^{2}$ energy is%
\begin{equation}
U-2J\equiv U^{\prime }.  \label{Uprime}
\end{equation}%
$U$ is the Coulomb repulsion between two electrons in the same orbital, and $%
J$ is the Hund's-rule exchange coupling.

Since $\mathcal{H}_{t_{2g}}$ involves only the correlated $t_{2g}$ orbitals,
the number of correlated electrons is fixed and the double-counting
correction therefore amounts to an irrelevant shift of the chemical
potential, which we omit in (\ref{H}).
For transitions from or to a non-correlated LDA band, such as those seen in
figures \ref{fig1}, \ref{fig2}, and \ref{fig1a}, there \emph{is} a
double-counting correction, but it is presumably $\lesssim $ 1 eV, because
the number of B\thinspace $t_{2g}$ electrons in the LDA calculation is not
far from 1. In those figures, the LDA $t_{2g}$ band should therefore simply
be substituted by the correlated $t_{2g}$ band with the Fermi levels lined
up.

In general, an LDA+$U$ Hamiltonian like (\ref{H}) depends on which
electronic degrees of freedom, \textit{i.e.} orbitals, are included, and how
the correlated orbitals are chosen. This means that in order to be able to
treat electronic correlations beyond the LDA, we have to depart from the 
\textit{ab initio} philosophy and make a system- and even property-dependent
choice of an appropriate low-energy Hubbard Hamiltonian. In the present work
for the $d^{1}$ perovskites, the correlated orbitals are taken as the set
of localized $t_{2g}$ LDA Wannier functions. The degrees of freedom from all
other bands are neglected, not only in the correlation term, but also in $%
\mathcal{H}^{LDA}.$ For instance, the B $e_{g}$ band, whose centre is 3 eV
above that of the $t_{2g}$ band (see figures \ref{fig2} and \ref{fig1a}),
is not treated as correlated, but the $e_{g}$ characters are downfolded
into the $t_{2g}$ orbitals as seen in figures \ref{fig7} and \ref{fig12}.
For the Mott transition, this seems to be the appropriate treatment, as is
also suggested by the work of Manini \textit{et al.} \cite{Manini}. For
super-exchange couplings this also seems to hold, as we shall argue in
section \ref{M}, but for high-energy optical or inverse-photoemission
spectra, it is clearly inappropriate. Here, $e_{g}$ degrees of freedom must
be taken explicitly into account in the Hubbard Hamiltonian. This can be
done by using as basis either the present $t_{2g}$ basis augmented by the
Wannier functions for the LDA $e_{g}$ band, or the set of Wannier functions
for the entire $d\left( t_{2g}+e_{g}\right) $ band. The advantage of the
former basis 
set is that it has \emph{no} single-particle coupling between the $%
t_{2g}$ and $e_{g}$ bands, and the advantage of the latter is that it is
more localized. Anyhow, in the present work we do not compute high-energy
properties.

In principle, the on-site Coulomb matrix should be calculated from the
Wannier-function basis used, but in the present case it was simpler to keep 
a single, adjustable parameter, $U$, and --
as we shall see -- this yields more insight.
The value of this parameter we took to be the \emph{same}
for all four materials. For the Hund's-rule coupling we used the following:
\[
J=0.68~\mathrm{eV}\qquad \mathrm{and}\qquad 0.64~\mathrm{eV,} 
\]
for the
vanadates and titanates, respectively. 
These are atomic Hartree-Fock values times 0.8 in order to
account for the screening in the solid \cite{Mizokawa96}.

The Hubbard Hamiltonian (\ref{H}) is solved in the DMFT~\cite{dmft}, 
\textit{i.e.}\ under the assumption that
the elements, $\Sigma _{Rm,R^{\prime }m^{\prime }}\left( \omega \right) ,$
of the self-energy matrix \emph{between} different sites can be neglected.
The self-energy is thus assumed to be an effective, energy-dependent and
complex crystal-field term. In this case, the on-site Green matrix is:%
\begin{equation}
G_{Rm,Rm^{\prime }}\left( \omega \right) =\sum_{\mathbf{k}}\left\{ \left[
\omega -\Sigma \left( \omega \right) -H^{LDA}\left( \mathbf{k}\right) \right]
^{-1}\right\} _{Rm,Rm^{\prime }},  \label{G}
\end{equation}%
and the many-body lattice problem is then mapped onto an Anderson impurity
problem in which the inverse of the bath Green function of the uncorrelated
host is required to be $G\left( \omega \right) ^{-1}+\Sigma \left( \omega
\right) .$ Solution of the Anderson impurity problem must now yield the same 
$G_{Rm,Rm^{\prime }}\left( \omega \right) $ as equation (\ref{G}), and this
is a self-consistency condition for determination of the self-energy matrix, 
$\Sigma _{Rm,Rm^{\prime }}\left( \omega \right) ,$ and the host.

The bottleneck in a DMFT calculation is to solve the Anderson impurity
problem, and only a few correlated orbitals can be handled at the moment. In
the orthorhombic perovskites all 4 B-sites are \emph{equivalent}, so the $%
t_{2g}$ impurity problem involves only 3 correlated orbitals. To solve it,
we employed the numerically exact Hirsch-Fye \cite{hirsch} quantum Monte
Carlo (QMC) method. In order to access temperatures down to 770\thinspace K,
we used up to 100 slices in imaginary time and about 10$^{6}$ QMC sweeps. To
reach convergence, 15--20 DMFT iterations were needed. Finally, the spectral
function was obtained on the real $\omega $-axis by analytical continuation
using the maximum entropy method \cite{jarrell}. Unfortunately, this does
not provide us with the self-energy matrix for real $\omega ,$ so we do not
obtain correlated bandstructures in the present applications.

In order to be able to perform the DMFT calculations we adopted the usual
approximation of keeping only the density-density terms in the on-site
Coulomb-repulsion, \textit{i.e.}%
\[
\fl\mathcal{U}_{R}\approx \frac{1}{2}\sum_{mm^{\prime }\sigma \sigma
^{\prime }}U_{m\sigma, m^{\prime }\sigma ^{\prime }}n_{Rm\sigma
}n_{Rm^{\prime }\sigma ^{\prime }},\quad U_{m\sigma, m^{\prime }\sigma
^{\prime }}=\left\{ 
\begin{array}{lllll}
U & \mathrm{if} & m=m^{\prime } & \mathrm{and} & \sigma \neq \sigma ^{\prime
} \\ 
U-2J & \mathrm{if} & m\neq m^{\prime } & \mathrm{and} & \sigma \neq \sigma
^{\prime } \\ 
U-3J & \mathrm{if} & m\neq m^{\prime } & \mathrm{and} & \sigma =\sigma
^{\prime }%
\end{array}%
\right. . 
\]%
Here, $n_{Rm\sigma }=a_{Rm\sigma }^{\dagger }a_{Rm\sigma },$ and $%
a_{Rm\sigma }^{\dagger }$ creates an electron with spin $\sigma $ in a
localized orbital $m$ at B-site $R.$ In the summation, at least two of the
indices on the operators must be different. The repulsion averaged over all
doubly occupied states remains as in equation (\ref{Uprime}) \cite{fresard}.

In previous implementations of the LDA+DMFT method \cite%
{ldadmft,NekrasovLaSr2000} it was assumed that the on-site block of the
single-particle Green function is diagonal with identical elements in the
space of the correlated orbitals, and the latter were usually taken as
orthonormal LMTOs \cite{LMTO84}, approximated by truncated and renormalized
partial waves. These approximations mean that the partial waves \emph{not}
belonging to the irreducible representation of the correlated LMTOs --and
therefore arising from the tails of LMTOs on neighbouring sites-- are
neglected. As a consequence, the self-energy in equation (\ref{G}) merely
distorts the energy scale of the LDA DOS, from which the on-site $G\left(
\varepsilon \right) $ can then be obtained by Hilbert transformation.
Although this may be appropriate for \textit{e.g.} cubic systems, it is
clearly \emph{not} appropriate for the series of materials considered in the
present paper. Here, the self-energy must be allowed to be an effective
crystal field -- enhanced, energy dependent, and complex 
through the influence of $%
\mathcal{U}_{R}$ -- which produces $\mathbf{k}$-dependent broadenings and 
distortions of the
bandstructure of the kind we have been discussing extensively in section \ref%
{DistBS}. Specifically, in table \ref{tableQ} we should perform the
substitutions:%
\begin{equation}
t_{x_{i}x_{j},x_{j}x_{k}}^{000}\rightarrow
t_{x_{i}x_{j},x_{j}x_{k}}^{000}+\Sigma _{x_{i}x_{j},x_{j}x_{k}}\left( \omega
\right)  \label{tSigma}
\end{equation}
It is \emph{not} correct to merely distort the energy scales of the
different partial DOS functions differently.

In our new implementation~\cite{Eva03} of the LDA+DMFT method, the highly
accurate NMTO method is used to create a localized set of correlated
orbitals, which then defines the Hubbard Hamiltonian. We choose the localized
orbitals to span a subset of LDA bands around the Fermi level \emph{exactly}
--the $t_{2g}$ bands in the present application,-- and we choose to
orthonormalize them symmetrically 
so that they become a set of Wannier functions. With
this set, the matrix elements are then calculated correctly (except the
matrix elements of the on-site Coulomb repulsion). Hence, we take all
components of the self-energy matrix $\Sigma _{mm^{\prime }}\left( \omega
\right) $ between different orbitals on a given site into account. From this
3$\times $3 matrix we use the \textit{Pbnm} symmetry (figure \ref{fig1e}) to
construct a 12$\times $12 block-diagonal self-energy matrix, which is then
inserted in expression (\ref{G}). The 12$\times 12$ matrix $\omega
-\Sigma \left( \omega \right) -H^{LDA}\left( \mathbf{k}\right) $ is now
inverted as a function of $\mathbf{k,}$ and, finally, the on-site 3$\times 3$
block is summed over $\mathbf{k}$ to yield the 3$\times 3$ on-site $G\left(
\varepsilon \right) $ matrix.

A recent LDA+DMFT calculation for La$_{1-x}$Sr$_{x}$TiO$_{3}$ \cite{Craco03}
used the crystal-field representation and neglected the off-diagonal
elements of the DOS and the self-energy. As seen in the bottom row of figure %
\ref{fig6}, this may be an intelligible approximation. Nevertheless, even
with the crystal-field eigenvectors frozen at the LDA values, the $\Sigma
\left( \omega \right) +H^{LDA}\left( \mathbf{k}\right) $ bandstructure
should be allowed to change as the effective crystal-field splittings change
from iteration to iteration.

Another LDA+DMFT implementation based on Wannier functions was proposed
after the completion of the present work \cite{Anisimov05}. Applications to
cubic SrVO$_{3}$ using $t_{2g}$ Wannier functions yielded low-energy
spectral functions similar to those presented in the next section. It is not
obvious to us how $t_{2g}$ Wannier functions can provide matrix elements of
the self-energy connecting to other bands, except for the trivial
double-counting shifts mentioned above. A good idea in that paper is to use
the on-site Green function (\ref{G}) to evaluate the contribution to the
charge density from the correlated orbitals and, hence, to adjust the
Wannier functions self-consistently, away from their LDA values. 
For the $t_{2g}^{1}$ titanates, a useful 
application  would be to calculate the JT
distortions from the LDA+DMFT charge density.

\section{High-temperature properties\label{HT}}

\begin{figure}[t]
\par
\begin{center}
\includegraphics[width=\textwidth]{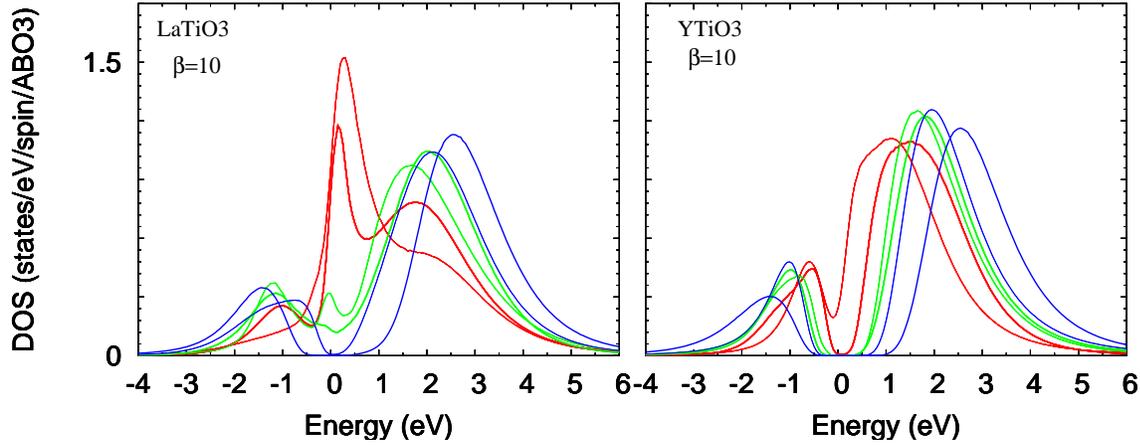}
\end{center}
\caption{DMFT spectral function for $U$=3.5 (red), 4.0 (bold red), 4.5
(green), 4.75 (bold green), 5.0 (blue), and 6.0 (bold blue) eV. The average
Coulomb repulsion is $U^{\prime }$=$U-1.28$\thinspace \textrm{eV}. $kT$=%
\textrm{\ 100\thinspace meV}$\sim $1200\thinspace K. For LaTiO$_{3},$ the
older structural data were used \protect\cite{lastr}. Extensive DMFT
calculations were not performed with the recent data \protect\cite{Cwik03}.}
\label{fig8a}
\end{figure}

\subsection{The Mott transition\label{Mott}}

With LDA+DMFT we computed the spectral functions for all four materials as a
function of $U$ and at $kT$=100~meV$\sim $1200\thinspace K, \textit{i.e.}%
\thinspace way above any magnetic ordering temperature$.$ We do not
investigate the complete phase diagram or the order of the Mott transition;
rather, we estimate the critical value,
$U_{c},$ for the high-temperature region where the system
is paramagnetic and for which we find a unique solution of the DMFT
equations.

For the titanates, the spectra are shown in figure \ref{fig8a}. We found
that the materials become insulating when $U^{\prime }\geq U_{c}^{\prime }$
with the values of $U_{c}^{\prime }$ given in the first row of table \ref%
{tableUc}. Cubic SrVO$_{3}$ remained metallic at the highest value of $%
U^{\prime }$ considered (4.6 eV), and the calculation employed the older
structural data \cite{lastr} for LaTiO$_{3}$. The decrease of the critical
value seen along the series is dramatic, and if we form the ratio with the
rms bandwidth $W$ from table \ref{tableW}, we find the numbers given 
in the central row
of table \ref{tableUc}. This ratio is \emph{not} constant, but
decreases along the series by 35\% from CaVO$_{3}$ to YTiO$_{3}$. Therefore,
the Mott transition is \emph{not} driven merely by the narrowing of the $%
t_{2g}$ band, but also by its deformation. That is unexpected. 

\begin{table}[tbp]
\caption{Critical Coulomb repulsion, $U_{c}^{\prime },$ in eV and relatively
to the rms bandwidth, $W$. The crystal-field splitting between the lowest
levels is $\Delta _{12}$.}
\label{tableUc}
\begin{center}
\begin{tabular}{cccccc}
\br & SrVO$_{3}$\cite{srstr} & CaVO$_{3}$\cite{castr} & LaTiO$_{3}$\cite%
{lastr} & LaTiO$_{3}$(\cite{Cwik03}) & YTiO$_{3}$\cite{ystr} \\ 
$U_{c}^{\prime }$ & $>4.6$ & 4.4 & 3.6 & -- & 2.5 \\ 
$U_{c}^{\prime }/W$ & (1.96) & 1.84 & 1.65 & -- & 1.34 \\ 
$\Delta _{12}/W$ & 0 & 0.031 & 0.064 & 0.093 & 0.106 \\ 
\br &  &  &  &  & 
\end{tabular}%
\end{center}
\end{table}

The importance of orbital degeneracy for the Mott transition was first
pointed out by Gunnarsson \textit{et al.}\thinspace \cite{Olle}, who argued
that due to the increase in the number of hopping processes in many-body
theory compared with band theory, there is an enhancement of the hopping,
which for a half-full band is approximately proportional to the square root
of its degeneracy. The value, $U_{c}^{\prime },$ necessary to cause a Mott
transition for a given bare bandwidth, $W,$ therefore increases with the
degeneracy. Imagine now with Manini \textit{et al.}\thinspace \cite{Manini}
that this degeneracy is split by a small crystal field: upon increasing $%
U^{\prime },$ and thereby reducing the width of the quasiparticle peak\
towards zero, $ZW\sim \alpha \left[ 1-U^{\prime }/U_{c}^{\prime }\left(
N\right) \right] W,$ for some value of $U^{\prime }<U_{c}^{\prime }\left(
N\right) $ the reduced bandwidth will have reached the same size as the
crystal-field splitting. At that point the degeneracy is effectively
decreased from $N$ to $N-n,$ and herewith $U_{c}^{\prime },$ so that a Mott
transition may be triggered. One therefore expects a critical crystal-field
splitting, $\Delta _{c}\sim \alpha \left[ 1-U_{c}^{\prime }\left( N-n\right)
/U_{c}^{\prime }\left( N\right) \right] W$. Manini \textit{et al.}
specifically solved a simple Hubbard model with one electron in two
identical, non-interacting bands whose on-site energies differed by $\Delta
 $ For $\Delta =0,$ they found $U_{c}^{\prime }\left( 2\right) /W\sim 1.8,$
and for $\Delta \rightarrow \infty ,$ they found $U_{c}^{\prime }\left(
1\right) /W\sim 1.35,$ so that $U_{c}^{\prime }\left( 2\right)
/U_{c}^{\prime }\left( 1\right) \sim 1.33.$ The surprising result was that $%
U_{c}^{\prime }/W$ decreases from the $\frac{1}{4}$-filled two-band limit to
the $\frac{1}{2}$-filled one-band limit for $\Delta $ increasing from 0 to
merely 10\% of the bandwidth, \textit{i.e.} $\Delta _{c}/W\sim 0.1,$ so that 
$\alpha \sim 0.1/\left( 1-1.35/1.8\right) \approx 0.4.$ The crystal field
splitting does, however, only control $U_{c}^{\prime }/W$ for the Mott
transition within the limits from $U_{c}^{\prime }\left( 2\right) /W$ to $%
U_{c}^{\prime }\left( 1\right) /W;$ if for $\Delta >\Delta _{c},$ $U^{\prime
}/W$ is increased from below 1.35, then the system undergoes a phase
transition from a two-band metal to a one-band metal before it goes from a
one-band metal to a one-band Mott insulator at $U_{c}^{\prime }/W=1.35$.

\begin{figure}[t]
\par
\begin{center}
\rotatebox{270}{
\includegraphics[height=\textwidth]{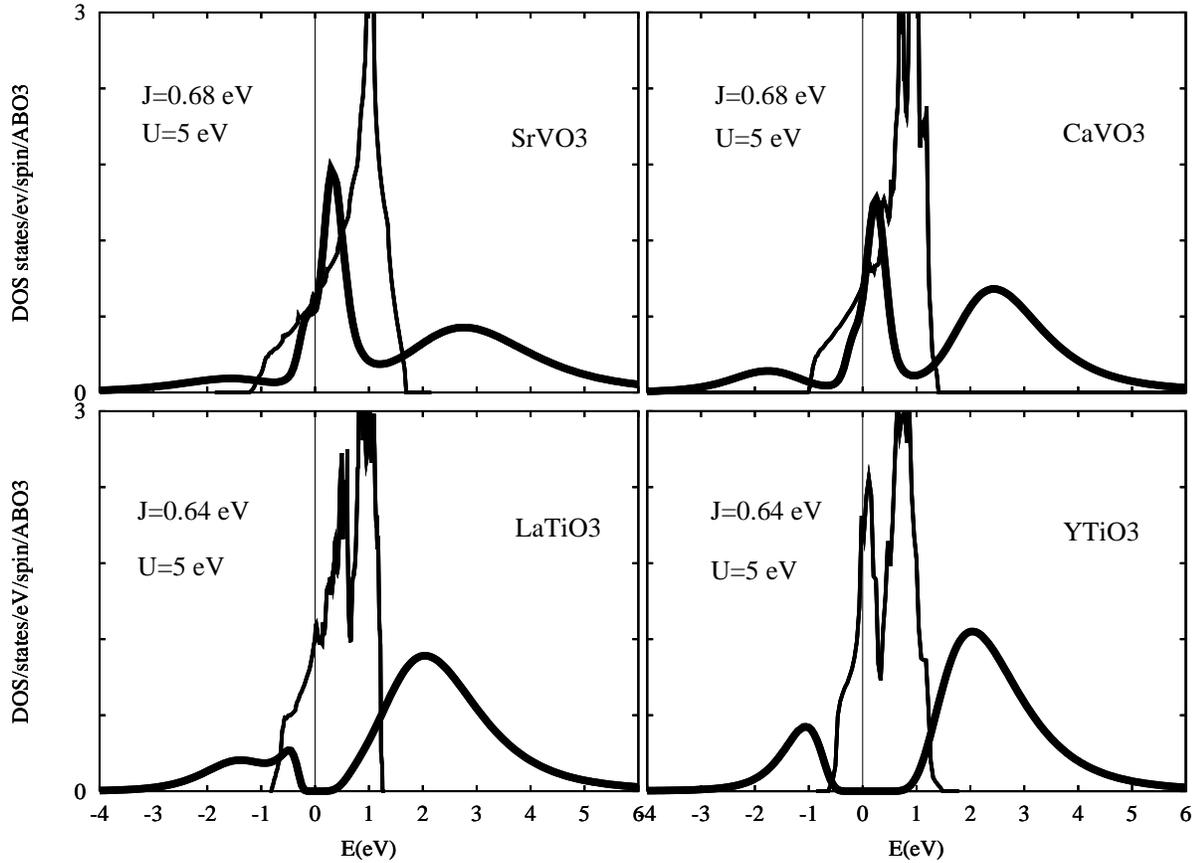}}
\end{center}
\caption{DMFT spectral functions for $U$=5 eV and $T$=770 K (bold) and LDA
total density of states (weak). For LaTiO$_{3},$ the old structure was used.
Reproduced from \protect\cite{Eva03}. }
\label{fig9}
\end{figure}

For our case of a $\frac{1}{6}$-filled three-band system, Koch \textit{et al.}%
\thinspace \cite{Erik} solved a Hubbard model for the $t_{1u}$-band in C$%
_{60}$ and found that hopping is enhanced by merely $\sim \left( 1+\sqrt{2}%
\right) /2 = 1.21.$ Thus, the degeneracy reduction expected from the
crystal-field splitting is at most 30\%. The striking fact about our values
of $U_{c}^{\prime }/W$ is that they decrease by considerably more: if we use
Manini \textit{et al.}'s $\alpha $-value, then the 3\% crystal-field
splitting in CaVO$_{3}$ has reduced $U_{c}^{\prime }/W$ by about 6\% from $%
U_{c}^{\prime }\left( 3\right) /W,$ the value appropriate for SrVO$_{3}$ and
listed in parentheses in table \ref{tableUc}. Our LDA+DMFT calculations thus
yield a decrease in $U_{c}^{\prime }/W$ of $\sim $50\% when going from cubic
SrVO$_{3}$ to YTiO$_{3}.$ That is not possible unless the width of the
lowest subband narrows more than the rms bandwidth, that is, unless the band
deforms along the series, an effect not considered previously.
We shall return to this point in section \ref{One}.

\subsection{Spectral functions for $U$=5 eV and comparison with experiments}

The main features of the photoemission spectra for all four materials, as
well as the correct values of the Mott-Hubbard gap for the insulators, are
reproduced by taking $U$ material independent and equal to 5$\,$eV. This is
satisfying because $U$ is expected to be similar for vanadates and
titanates, or maybe slightly smaller for the latter \cite%
{Mizokawa96,photoemission2}: Bocquet \textit{et al. }fitted 2$p$ core-level
photoemission spectra with a model Hamiltonian containing all O\thinspace $p$
and B\thinspace $d$ orbitals and obtained $U^{\prime }\approx 4.0$\thinspace
eV for La and Y titanate \cite{photoemission3}. This is consistent with the
value $U^{\prime }=5.0-2\times 0.64=3.7$\thinspace eV used by us for the
less localized $t_{2g}$ orbitals. Similar values of $U^{\prime }$ have been
used by other authors \cite{mergedpaper,Anisimov05}

In Figure \ref{fig9} we show the spectral functions together with the total
LDA DOS. The vanadates exhibit a quasiparticle peak and are therefore
metallic, while the titanates are Mott insulators.

For cubic SrVO$_{3}$ we reproduce the results of previous calculations \cite%
{mergedpaper,liebsch}: the lower Hubbard band is around $-1.8$\thinspace eV
and the upper Hubbard band around 3\thinspace eV. For CaVO$_{3}$, the
quasiparticle peak loses weight to the lower Hubbard band, which remains at $%
-1.8$\thinspace eV, while the upper band moves down to 2.5\thinspace eV.
These results are in very good agreement with photoemission data \thinspace 
\cite{mergedpaper,photoemission}. From the linear regime of the self-energy
at small Matsubara frequencies we estimate the quasi-particle weight to be $%
Z=0.45$ for SrVO$_{3}$ and 0.29 for CaVO$_{3}$. For a \textbf{k} independent
self-energy, as assumed in DMFT, this yields $m^{\ast }/m=1/Z=2.2$ for SrVO$%
_{3}$ and $3.5$ for CaVO$_{3}$. These results are in good agreement with the
optical-conductivity values of $2.7$ and $3.6$ \cite{optics}, as well as
with the values 2-3 obtained by thermodynamics \cite{inoueold,aiura} and 
dHvA
experiments \cite{inoue}; they show that SrVO$_{3}$ and
CaVO$_{3}$ are rather similar, with the latter slightly more correlated than
the former. Similar conclusions were drawn in \cite{mergedpaper}.

For the titanates, the lower Hubbard band is around $-1.5$\thinspace eV, in
accord with photoemission\thinspace \cite{photoemission3,photoemission2}.
Despite very similar bandwidths, the gaps are very different, approximately
0.3 eV for LaTiO$_{3}$ and approximately 1\thinspace eV for YTiO$_{3}$, and
this also appears to agree with optical experiments\thinspace \cite{mottgap}.

\subsection{Orbital polarization\label{OrbPol}}

Diagonalization of the matrix of occupation numbers,%
\[
n_{x_{i}x_{j},x_{j}x_{k}}=\frac{1}{\pi }\mathrm{Im}\int^{\mu }d\omega
G_{Rx_{i}x_{j},Rx_{j}x_{k}}\left( \omega \right) , 
\]%
obtained with LDA+DMFT for values of $U^{\prime }$ exceeding the $t_{2g}$
bandwidth, $W,$ reveals that -- within the numerical accuracy -- the \emph{%
eigenvectors, }$w_{x_{i}x_{j},m},$ are the \emph{same} as those which
diagonalize the on-site LDA Hamiltonian matrix, that is, $n$ and $t^{\mathbf{%
0}}$ commute. For the titanates, the eigenvectors are the ones given by
equations (\ref{eq11}), (\ref{eq13}), and (\ref{eq15}), and the
corresponding eigenfunctions are the crystal-field orbitals shown in subcell
1 in figure \ref{fig7}. However, upon increasing $U$ the orbital \emph{%
polarizations} increase around the metal-insulator transition --from the
small LDA values given in the top and bottom rows of figure \ref{fig6}-- and
become \emph{nearly complete} after the Mott transition. For this to be
true, in the insulators 
the DMFT must create a self-energy matrix which strongly and uniformly
enhances the non-cubic part of the on-site LDA Hamiltonian matrix, \textit{%
i.e.} for $\Sigma $ in equation (\ref{tSigma}) we must have:%
\begin{equation}
\Sigma _{x_{i}x_{j},x_{j}x_{k}}\left( \omega \right) \approx
\sum_{m}w_{x_{i}x_{j},m}n_{m}U_{m}^{\prime }\left( \omega \right)
w_{x_{j}x_{k},m},  \label{SigmanU}
\end{equation}%
so that in the crystal-field representation%
\begin{equation}
\Sigma _{mm^{\prime }}\left( \omega \right) \approx \delta _{mm^{\prime
}}\Sigma _{m}\left( \omega \right) =\delta _{mm^{\prime }}n_{m}U_{m}^{\prime
}\left( \omega \right) ,  \label{sigmaomega}
\end{equation}%
where, as $U^{\prime }$ increases beyond $U_{c}^{\prime },$\ $%
n_{m}\rightarrow \,\sim  1$ while 
$\mathrm{Re}\; U_{m}^{\prime }\left( \omega
\right) $ tends towards a function which is roughly 
$\frac{a}{\omega} -\frac{1}{2}U^{\prime } +O(\omega)$
for $\omega $ negative and 
$ \frac{a}{\omega}+\frac{1}{2}U^{\prime }  +O(\omega)$ for 
$\omega $ positive, and 
$\mathrm{Re}\; U_{2}^{\prime }\left( \omega \right) $ and 
$\mathrm{Re}\; U_{3}^{\prime }\left(
\omega \right) $ tend towards functions which are roughly 
$\frac{a}{\omega} +\frac{1}{2} U^{\prime } +O(\omega)$ 
independently of the sign of $\omega$.
Here, $a$  is a constant.
This would mean that the Mott
transition takes place in only \emph{one} band, and that this band is the one
described by the crystal-field orbital with lowest energy. Static mean-field
methods like Hartree-Fock and LDA+$U$ have no $\omega $-dependence of $%
U_{m}^{\prime }\left( \omega \right) ,$ and are therefore unable to split
the half-full, lowest band into two. Unfortunately, we cannot present figures
of $\Sigma _{m}\left( \omega \right) $ because the methods we use to solve
the Anderson impurity problem do not enable us to evaluate the self-energy
for real $\omega .$

\begin{figure}[t]
\par
\begin{center}
{\includegraphics[width=\textwidth]{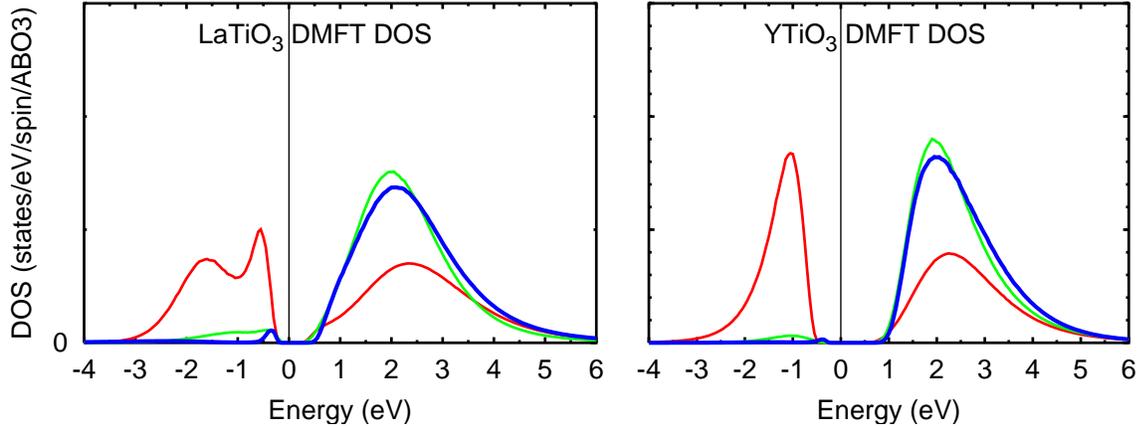}}
\end{center}
\caption{Diagonal elements of the spectral function in the crystal-field
basis calculated with LDA+DMFT ($U$=5 and $kT$=0.1 eV). The corresponding
figure for $U$=0 and $kT$=0 is the bottom part of figure \protect\ref{fig6},
which also has the same colour coding: $\left\vert 1\right\rangle $ red, $%
\left\vert 2\right\rangle $ green, and $\left\vert 3\right\rangle $ blue.
The occupation of $\left\vert 1\right\rangle $ is 0.88 for LaTiO$_{3}$
(older structure) and 0.96 for YTiO$_{3}.$ Below $\protect\varepsilon _{F}$
there is one, and above there are five electrons. The off-diagonal elements
are two orders of magnitude smaller than the diagonal ones.}
\label{fig10}
\end{figure}

Figure \ref{fig10} shows the DOS for the titanates calculated with LDA+DMFT (%
$U$=5 eV and $kT$=0.1 eV) and projected onto the common eigenvectors of the
occupation numbers and the crystal-field Hamiltonian. Hence, for LaTiO$_{3}$
the lowest crystal-field orbital contains 0.49 \cite{Cwik03} (0.45 \cite%
{lastr}) electrons when $U$=0 and 0.91 \cite{Cwik03} (0.88 \cite{lastr})
electrons when $U$= 5 eV. For YTiO$_{3},$ the respective occupations are
0.50 and 0.96. The nearly complete orbital polarization found for the two
insulators indicates that \emph{correlation effects} in the paramagnetic
Mott insulating state considerably decrease orbital fluctuations, and makes
it unlikely that YTiO$_{3}$ is a realization of an orbital liquid\thinspace 
\cite{Khal}. In LaTiO$_{3}$ some orbital fluctuations are still active,
although quite weak.

Most recent NMR \cite{Kiyama03} and x-ray scattering \cite{Tokura}
experiments point to an orbital order which is very similar to the one
obtained in our calculations. For LaTiO$_{3}$ our occupied orbital is very
similar to the one arrived at slightly earlier by Mochizuki and Imada \cite%
{Mochizuki03} through model calculations to which we shall return in more
detail in section \ref{M}. Also some of the results in the recent LDA+DMFT
calculation for La$_{1-x}$Sr$_{x}$TiO$_{3}$ by Craco \textit{et al.} \cite%
{Craco03} mentioned at the end of section \ref{DMFT} are similar to ours.

Our occupied orbital for YTiO$_{3}$ is similar to the one obtained with the
spin-unrestricted LDA (actually GGA) scheme for $T$=0\thinspace \cite%
{Sawada98}. This is understandable because the LDA properly accounts for the
covalency and the electrostatics, and because the occupancy in the
spin-unrestricted GGA is very similar to the one obtained by occupying the
LDA DOS in the bottom right of figure \ref{fig6} with a single spin-up
electron, \textit{i.e.} filling it to the Fermi level corresponding to $n$%
\textrm{=2}. The width of the band filled with one spin-up electron is seen
to be 0.8 eV so that in order to stabilize this ferromagnetic state, a
Hund's-rule coupling constant exceeding 0.4 eV is required; this is well
satisfied by our $J$=0.64 eV. In being \emph{metallic,} this ferromagnetic
ground state is, however, incorrect and the LDA+$U$ scheme was needed to
produce an insulator \cite{Sawada98}. Our orbital in equation (\ref{eq15})
agrees almost perfectly with the one deduced from NMR \cite{Itho}: $%
\left\vert 1\right\rangle =0.8\left\vert yz\right\rangle +0.6\left\vert
xy\right\rangle ,$ and well with the one deduced from neutron scattering 
\cite{neutrons} and a small correction \cite{srxr}: $\left\vert
1\right\rangle =0.7\left\vert yz\right\rangle +0.7\left\vert xy\right\rangle
.$ Also resonant $x$-ray scattering confirms the orbital order \cite%
{srxr,rxr}. Mochizuki and Imada obtained a similar orbital to which we shall
return in more detail in section \ref{M}.

To visualize the \emph{orbital order,} the nearly full orbital was placed on
each of the 4 Ti sites in figure \ref{fig8}. Despite the fact that LaTiO$%
_{3} $ and YTiO$_{3}$ have the same space group, the orbital orders look
very different. This difference is, however, quantitative rather than
qualitative; as mentioned in section \ref{XtalFieldBasis}, it reflects the
extent to which the orbital has the $bc$ plane as mirror. We emphasize once
more that the two different types of JT distortions observed in LaTiO$_{3}$
and YTiO$_{3}$ are \emph{not the cause} for the difference in the orbital
orders. This was discussed for YTiO$_{3}$ in section \ref{JT}, was clearly
shown in figures \ref{fig11} and \ref{fig13}, and was finally explicitly
brought out in section \ref{XtalFieldBasis} by the eigenvectors (\ref{eq17})
and (\ref{eq18}). It is however obvious that the oxygen octahedron will
relax to the shape of the charge density of the localized electron, and this
seems to be the reason why in YTiO$_{3}$ the O2 square is stretched by 3\%
into a rhomb along $y$ in subcells 1 and 3, and along $x$ in subcells 2 and
4 \cite{ystr}, while in LaTiO$_{3},$ it is stretched into a rectangle by 3\%
along $a$ \cite{Cwik03}. This is remarkable because in systems such as LaMnO$%
_{3},$ where the low-energy bands are of $e_{g}$ type, orbital order and
large (10\%) JT distortions occur \emph{together,} and the different types
of orbital orderings are therefore often classified according to the type
and spatial arrangement of the JT distortions. We have seen that this does
not apply when the low-energy bands are of $t_{2g}$ type, first of all
because the $t_{2g}$ orbitals have $pd\pi $ coupling to the \emph{same}
oxygen orbitals as the A ions have $pd\sigma $ coupling to and, secondly,
because this $pd\pi $ coupling of the $t_{2g}$ electrons is much weaker than
the $pd\sigma $ coupling of the $e_{g}$ electrons (see figures \ref{fig1b}
and \ref{fig1c}). In conclusion, the crystal-field splitting in the $%
t_{2g}^{1}$ perovskites is due to the GdFeO$_{3}$-type distortion.

We have thus seen that it is the Coulomb repulsion which causes the electron
to localize, but it is chemistry which sets the stage: it determines $U_{c}$
and selects the orbital to be occupied.

\subsection{Pressure-induced metallization\label{Pres}}

Loa \textit{et al.} recently performed high-pressure experiments on the
titanates, in which they monitored the structures and the optical gaps \cite%
{loa}. Some of their results are shown in the left-hand side of figure \ref%
{fig13a}. At 11\thinspace GPa, LaTiO$_3$ becomes metallic, its volume
contracts slightly, but no apparent change of internal parameters occurs.
For YTiO$_{3},$ the optical gap decreases with increasing pressure, but the
material remains insulating up to at least 17\thinspace GPa. By
extrapolation of this pressure dependence, it was estimated that
metallization will occur above 40 GPa \cite{loa}.

\begin{figure}[t]
\par
\begin{center}
\rotatebox{270}{\includegraphics[height=\textwidth]{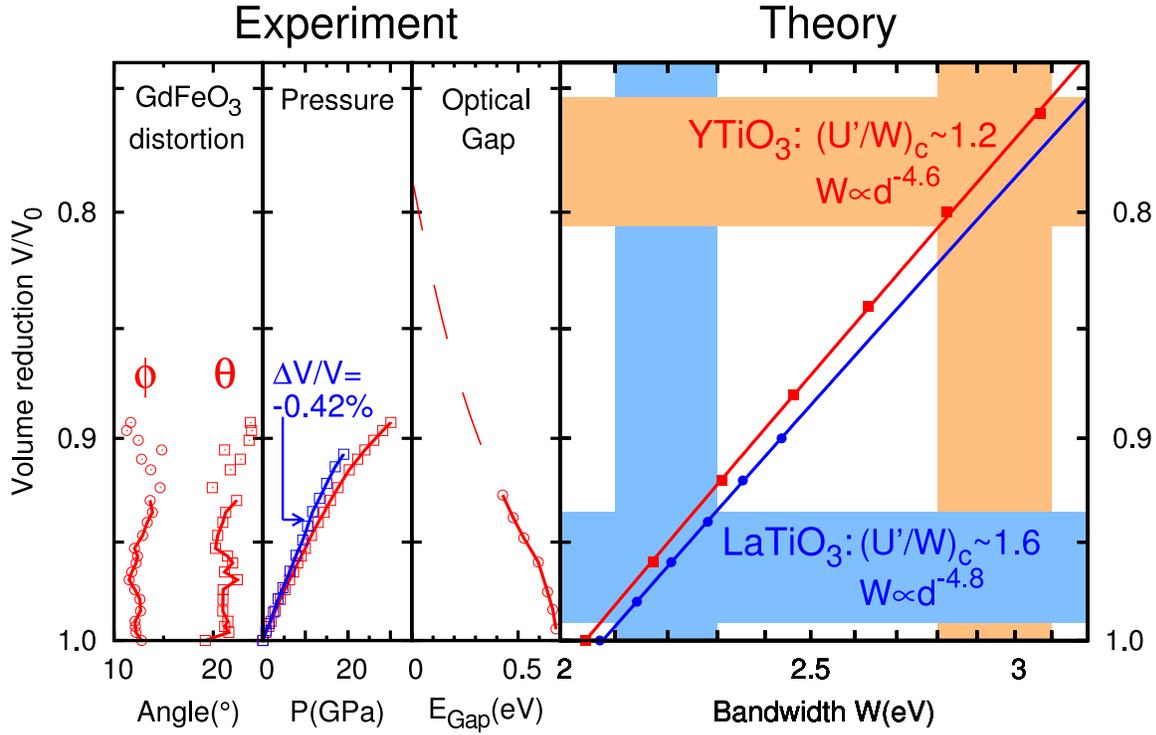}}
\end{center}
\caption{High-pressure metallization of LaTiO$_{3}$ (blue) and YTiO$_{3}$
(red). Experimental \protect\cite{loa} and theoretical data (horizontal
axes) as a function of the reduction of volume from that at normal pressure
(vertical axis, log scale). $\protect\theta $ and $\protect\phi $ are the
tilt and rotation angles. Rietveld refinement of the crystallographic data
was only possible for $V>0.93V_{0}$. For YTiO$_{3}$ the Mott transition has
not been observed within the available pressure range, and the dashed line
extrapolating the optical gap to zero is a theoretical guess from the
right-hand side of the figure. For LaTiO$_{3},$ the Mott transition takes
place at 11 GPa and $V\mathrm{=}0.94V_{0},$ where a discontinuous
contraction is observed. $W$ is the calculated LDA $t_{2g}$-bandwidth
(distance between band edges), which is seen to follow a power law in the
Ti-Ti distance $d\propto V^{1/3}.$ Since the experimental GdFeO$_{3}$-type
distortion seems to be independent of volume, at least up to $V\mathrm{=}%
0.9V_{0},$ the values $U_{c}^{\prime }$=3.6 and 2.5 eV (table \protect\ref%
{tableUc}) obtained from our LDA+DMFT study of the Mott transition (figure 
\protect\ref{fig8a}) can be used to predict the metallization volume.}
\label{fig13a}
\end{figure}

We may compute the metallization pressures using LDA+DMFT, provided that we
know the structure. Luckily, the experimental data in the left-hand side of
figure \ref{fig13a} give no reason to expect that the structures change by
anything but a uniform compression; the GdFeO$_{3}$-type distortion seems to
be fairly constant both for LaTiO$_{3}$ and for YTiO$_{3},$ at least up to a
volume reduction of 0.9. We can therefore use our previous LDA+DMFT
calculation of $U_{c}^{\prime },$ the value of the average on-site Coulomb
repulsion necessary to drive the metal through a Mott transition (figure \ref%
{fig8a} and table \ref{tableUc}), to predict the LDA $t_{2g}$ bandwidth
necessary to drive the insulator metallic. This gives:%
\[
W_{c}\sim 2.09~\mathrm{eV\times }\frac{5-2\times 0.64}{3.6}=2.2~\mathrm{%
eV\;for\;LaTiO}_{3} 
\]%
and%
\[
W_{c}\sim 2.05~\mathrm{eV\times }\frac{5-2\times 0.64}{2.5}=3.0~\mathrm{%
eV\;for\;YTiO}_{3}. 
\]%
Since only the width --not the shape-- of the band changes, we have used
here the simplest measure of the bandwidth, namely the energy distance
between the band edges $\left( W_{t_{2g}}\right) $. To find the critical
volume reduction, $V_{c}/V_{0},$ we now merely need to compute this
bandwidth as a function of volume. The result is shown in the right-hand
side of figure \ref{fig13a}. The critical volume and corresponding pressure
is then%
\[
V_{c}\sim 0.96V_{0}\quad \mathrm{and}\quad P_{c}\sim 7~\mathrm{%
GPa\;for\;LaTiO}_{3}, 
\]%
which is in reasonable accord with the experiments, and 
\[
V_{c}\sim 0.78V_{0}\quad \mathrm{and}\quad P_{c}\sim 100~\mathrm{%
GPa\;for\;YTiO}_{3}. 
\]%
For YTiO$_{3}$, $V_{c}$ is uncertain because the GdFeO$_{3}$-distortion is
only known to be constant for volumes larger than $0.9V_{0},$ but the
corresponding, sketched behaviour of the optical gap does not look
unreasonable. The critical pressure is more uncertain than the volume
because it hinges on an extrapolation of the pressure-volume relation as
well.

A most recent, refined study of YTiO$_{3}$ \cite{loa} revealed that 
between 9 and 14 GPa the JT
distortion essentially disappears, but the distortion of the Y sublattice
increases. This was mentioned at the end of section \ref{JT}, and the
16\thinspace GPa structure was used to calculate the bandstructure shown on
the right-hand sides of figures \ref{fig11} and \ref{fig13}. This
bandstructure justifies our approximation of assuming a rigid bandshape,
although the Y positions shift so as to increase the residual covalency and,
hence, to increase the pseudo-gap. A corresponding correction of $U_{c}$
would be towards a slightly lower value, and that would \emph{increase} our
estimate --to \emph{above} 100\thinspace GPa-- for the pressure where
metallization should occur if no further structural changes were to take
place. Most importantly, the fact that the JT distortion is strongly reduced
in the 16\thinspace GPa structure does \emph{not} mean that the orbital
order has changed drastically. That was explained at the end of the
preceding subsection. The reduced JT distortion will, however, change the
low-temperature magnetic order from ferromagnetic to A-type
antiferromagnetic, as well as the ordering temperature. This we shall show
in section \ref{M}.

\subsection{Onset of optical absorption in the titanates\label{Opt}}

Let us calculate how the difference between LaTiO$_{3}$ and YTiO$_{3}$ shows
up in the optical transitions between the lower and upper $t_{2g}$ Hubbard
bands seen in figure \ref{fig10}. Experimentally, the onset of the optical
conductivity is more gradual in LaTiO$_{3}$ than in YTiO$_{3}$ \cite%
{loa,Crandles94,Grueninger}.

The optical conductivity has previously been evaluated for LaTiO$_{3}$ in
LDA+$U$ \cite{Bourab96} and LDA+DMFT \cite{Oudo} calculations. However, the
GdFeO$_{3}$-type distortion, crucial to the present study, was neglected in
both calculations where, on the other hand, the optical matrix elements were
treated with more care than in our study.

\begin{figure}[t]
\par
\begin{center}
\rotatebox{270}{\includegraphics[height=\textwidth]{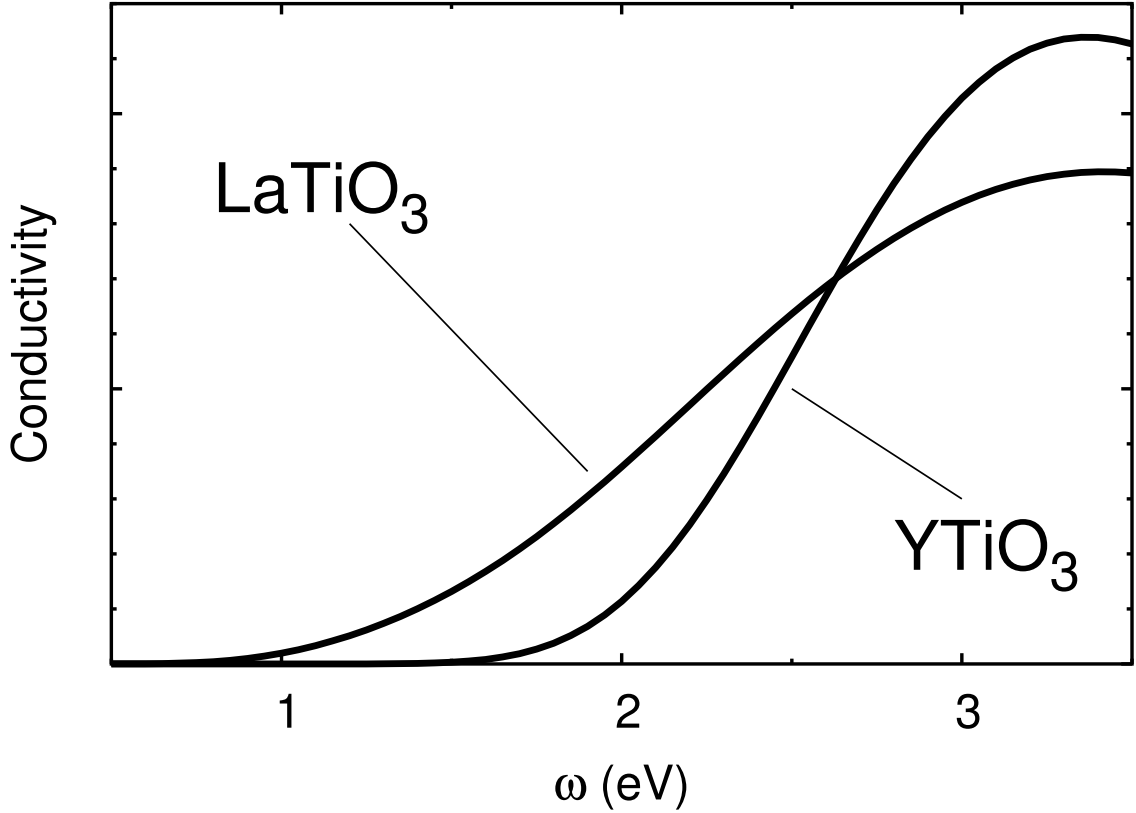}}
\end{center}
\caption{Real part of the optical conductivity obtained by LDA+DMFT with $U$%
=5 eV, $kT$=0.1 eV, and treating the matrix elements as explained in the
text. The recent structural data for LaTiO$_{3}$ was used \protect\cite%
{Cwik03}.}
\label{fig18}
\end{figure}

The optical conductivity can be expressed in terms of the current-current
correlation function \cite{Mahn}. In the dynamical mean-field approximation
there are no vertex corrections and the conductivity can therefore be
written as \cite{Khurana,Pruschke,Bluemer}:%
\begin{equation}
\fl\sigma (\omega )\propto \frac{1}{\omega }\sum_{\mathbf{k}}\!\int
\!d\omega ^{\prime }\left[ n(\omega ^{\prime })-n(\omega +\omega ^{\prime })%
\right] \sum_{ijkl}M_{ij}\left( \mathbf{k}\right) A_{jk}(\mathbf{k},\omega
^{\prime })M_{kl}\left( \mathbf{k}\right) A_{li}(\mathbf{k},\omega ^{\prime
}+\omega ).  \label{sigma}
\end{equation}%
Here, $n(\omega )$ is the Fermi distribution function, $M_{ij}\left( \mathbf{%
k}\right) $ the optical matrix element, and%
\[
A_{ij}\left( \mathbf{k},\omega \right) =\frac{1}{\pi }\mathrm{Im}\left\{ %
\left[ \omega -\Sigma \left( \omega \right) -H^{LDA}\left( \mathbf{k}\right) %
\right] ^{-1}\right\} _{ij} 
\]%
the spectral function at real energy. The latter can be obtained by
analytical continuation from the QMC data created in the course of an
LDA+DMFT calculation. We shall assume that all transitions between $t_{2g}$
crystal-field orbitals (see figure \ref{fig10}), and allowed by the Fermi
functions, have \emph{equal} probability. Compared with charge-transfer
transitions (O\thinspace 2$p$--Ti\thinspace 3$d$ and O\thinspace 2$p$%
--A\thinspace $d),$ these transitions are all weak, and we do not expect
some to be stronger than others. With this approximation, equation (\ref%
{sigma}) reduces to:%
\[
\fl\sigma (\omega )\;\propto \;\frac{1}{\omega }\sum_{\mathbf{k}}\int
d\omega ^{\prime }\left[ n(\omega ^{\prime })-n(\omega +\omega ^{\prime })%
\right] \sum_{ij}A_{ij}(\mathbf{k},\omega ^{\prime })\sum_{ij}A_{ij}(\mathbf{%
k},\omega ^{\prime }+\omega ), 
\]%
where $i$ and $j$ now run over the crystal-field orbitals.

The real part of the optical conductivity obtained from our LDA+DMFT ($U$=5
eV, $kT$=0.1 eV) calculations are given in figure \ref{fig18}. By going from
the indirect DOS gaps in figures \ref{fig9} and \ref{fig10} to the direct
gaps in figure \ref{fig18}, the latter are seen to be increased by
approximately 0.5 eV, to 0.8 eV in LaTiO$_{3}$ and to 1.5 eV YTiO$_{3}$. We
observe that, even without detailed matrix-element effects, the onset of
inter-band transitions is more gradual in LaTiO$_{3}$ than in YTiO$_{3}.$
That is presumably the same trend as seen for the upper Hubbard band in
figure \ref{fig10}, where the lower edge of bands $\left\vert 2\right\rangle 
$ and $\left\vert 3\right\rangle $ rises more gradually in LaTiO$_{3}$ than
in YTiO$_{3};$ in LaTiO$_{3}$ the peak is reached after 1.5 eV, but in YTiO$%
_{3}$ after 1.0 eV.

\section{Low-temperature, magnetic properties of the titanates\label{M}}

\subsection{Introduction}

The origin of the magnetic phases of LaTiO$_{3}$ and YTiO$_{3}$ have been a
puzzle for decades \cite{imada,MochizukiNJP}. Whereas LaTiO$_{3}$ is a
G-type antiferromagnet with $T_{N}$=150\thinspace K and a small moment of
0.57$\,\mu _{B}$, YTiO$_{3}$ is a ferromagnet with a low Curie temperature, $%
T_{C}$=30\thinspace K, and a good-sized moment, 0.8$\mu _{B}\mathbf{\ }$\cite%
{Goral,Cwik03}.

The early idea was that in Mott insulators, the magnetic structure is rooted
in orbital order. Hence, the different magnetic ground states in LaTiO$_{3}$
and YTiO$_{3}$ ought to arise from different orbital orders. Orbital order,
presumably caused by electron correlation, should be accompanied by a JT
elongation of the TiO$_{6}$ octahedron, and in perovskites the stacking
along the $c$ axis was found to be either parallel (d type) or perpendicular
(a type). Since the $d^{1}$ titanates were found to have a mirror plane
perpendicular to the $c$ axis, they have d type orbital order if they are JT
distorted. In YTiO$_{3},$ a small (3$\%$) JT distortion was observed, but
hardly any in LaTiO$_{3}$. In their series of model Hartree-Fock
calculations, Mizokawa and Fujimori could obtain an antiferromagnetic ground
state for LaTiO$_{3}$ --but only by virtue of the spin-orbit coupling-- as
well as the ferromagnetic ground state for YTiO$_{3}.$ However, they could
not explain why d- rather than a-type orbital order is realized in YTiO$_{3}$
\cite{Mizokawa96}. This problem was later solved by Mizokawa, Khomskii, and
Sawatzky by including A $d$ orbitals together with the GdFeO$_{3}$-type
distortion in their model. However, these authors also noted that, given the
smallness of any JT distortion in LaTiO$_{3},$ this mechanism would make
LaTiO$_{3}$ ferro- rather than antiferromagnetic \cite{Mizokawa99}.

Early on in the game, the correct magnetic orders had actually been computed
by Solovyev, Hamada, and Terakura for LaTiO$_{3}$ \cite{Solovyev96}
and by Sawada and Terakura for YTiO$_{3}$ \cite{Sawada98} 
with the LDA+\textit{U} method. They adjusted 
$U$ to the experimentally measured Mott gap.
Their \emph{%
orbital} order for YTiO$_{3}$ was subsequently confirmed by NMR \cite{Itho},
neutron scattering \cite{neutrons}, and resonant x-ray scattering \cite{rxr}.

The spin-wave spectra were then measured and found to be essentially
isotropic and gapless in both titanates \cite{Keim,Ulrich}. This lead
Khaliullin and collaborators to question the existence of orbital order in
the titanates. They demonstrated that with the accepted orbital order in YTiO%
$_{3},$ conventional super-exchange theory, as used in the model studies,
leads to an \emph{an}isotropic spin-wave spectrum and, even worse, to an
antiferromagnetic ground state. If, on the other hand, the $t_{2g}$ levels
were degenerate, the isotropy of the spin-wave spectra could easily be
explained by strong quantum fluctuations in orbital space (orbital liquid
theory \cite{Khal1,Khal}). The assumption that the $t_{2g}$ levels are
nearly degenerate was justified by the smallness of the observed JT
distortions.

Mochizuki and Imada \cite{Mochizuki01Lett} then pointed out that in order to
obtain the observed G-type antiferromagnetism, as well as exchange-coupling
constants in accord with the measured spin-wave spectra ($%
J_{se}^{001}=J_{se}^{010/100}=15$\thinspace meV), it would suffice if the Ti 
$3d_{3z_{111}^{2}-1}$ level were lower than the two other $t_{2g}$ levels by
an amount exceeding the spin-orbit splitting and $kT,$ \textit{i.e.} by $%
\gtrsim $40 meV. A tiny deformation of the octahedron along [111] would
enable this. They also gave arguments why such an orbital order had not been
observed in resonant x-ray scattering \cite{Keim}. At the same time \cite%
{Mochizuki01}, Mochizuki and Imada published strong-coupling (in $U/t$, as
in super-exchange theory) studies of the antiferromagnetic-to-ferromagnetic
transition, which takes place as the angle of the O-Ti-O bond ($\sim $
degree of GdFeO$_{3}$-type distortion) decreases beyond that in LaTiO$_{3}.$
In these studies, the JT distortion was assumed to be d-like, that is, with
one level ($xz$ on site 1) \emph{above} a doubly-degenerate level ($yz$ and $%
xy$ on site 1). The dominant effect was found to be the $t_{2g}$-to-$e_{g}$
hopping via oxygen which increases with distortion, as we have seen in
figures \ref{fig7} and \ref{fig12}. Such hopping processes make it possible
to reach virtual high-spin states with configuration $t_{2g}^{\uparrow
}e_{g}^{\uparrow }$ and, hence, favour ferromagnetism. This study \cite%
{Mochizuki01} therefore predicted that the orbital order is not influenced
by the increasing GdFeO$_{3}$-type distortion, but that the exchange coupling
along the $z$ direction, where the $t_{2g}$-$p$-$e_{g}$ hopping is
strongest, changes sign from A-type antiferromagnetic (\textit{i.e.,}
ferromagnetic parallel to the mirror plane) to ferromagnetic. None of these
studies \cite{Mochizuki01,Mochizuki01Lett} included the A $d$ orbitals 
which had been found crucial
by Mizokawa, Khomskii, and Sawatzky \cite{Mizokawa99}.

Now, orbital-liquid theory had predicted that there be a significant
contribution to the specific heat at low temperature, but this was not 
observed \cite{Fritsch,Hemberger}. Moreover, a 3\% JT stretch of one of the
basal O squares into a rectangle was recently discovered in LaTiO$_{3}$ \cite%
{Cwik03}, thus bringing strong evidence for non-degenerate $t_{2g}$ levels
and a JT distortion of a different type than the one in YTiO$_{3}$.
Mochizuki and Imada finally \cite{Mochizuki03,Mochizuki04} included the A $d$
orbitals and realized that what stabilizes a single level with $%
d_{3z_{111}^{2}-1}$-character, and therefore G-type antiferromagnetism, is
not JT distortion, but the GdFeO$_{3}$-type distortion which makes the A
ions produce a crystal field. Its electrostatic and covalent contributions
were estimated to give about the same lowest state, namely $\left\vert
1\right\rangle = 0.68\left\vert yz\right\rangle +0.41\left\vert
zx\right\rangle +0.61\left\vert xy\right\rangle .$ With this orbital order,
they obtained
super-exchange coupling constants ($J_{se}^{001}=19.7$\thinspace
meV and $J_{se}^{010/100}=18.5$\thinspace meV) in almost perfect agreement
with the experimental values \cite{Keim}. Proceeding along the series of
rare-earth titanates with decreasing ionic radius from LaTiO$_{3}$ to SmTiO$%
_{3},$ the $J_{se}$ values were found to decrease by 20\%, 
and the lowest orbital to
become $\left\vert 1\right\rangle = 0.73\left\vert yz\right\rangle
+0.24\left\vert zx\right\rangle +0.63\left\vert xy\right\rangle .$ From this
continuous transition towards the case of YTiO$_{3},$ it was concluded that
beyond Sm the JT distortion should be \emph{the} factor which controls the
magnetism, \textit{i.e.} the JT distortion should be responsible for the
ferromagnetism of YTiO$_{3}$~\cite{Mochizuki04}. 
This work did not mention including the
coupling to the $e_{g}$ degrees of freedom.

Essentially the same conclusions were reached independently, and in one
shot, by using a new, parameter-free density-functional approach \cite{Eva03}%
. The present paper tells the story as it is seen from that side. The fact
that very similar conclusions concerning the splitting of the orbital
degeneracy and its universal origin were reached in such a short time, not
only by different theory groups, but by many experimental groups as well,
strengthens the case for those conclusions. Nevertheless, we, too, must
demonstrate that the orbital order obtained is consistent with the magnetic
structures and the spin-wave spectra \cite{Keim,Ulrich}. We thus move a step
backwards and repeat Khaliullin's calculation \cite{Ulrich} of the
inter-atomic exchange integrals by 2nd-order perturbation theory in $t/U,$
this time using the crystal-fields and hopping integrals obtained from
our LDA Wannier functions:

\subsection{Super-exchange couplings}

The super-exchange 
Hamiltonian is obtained from the $t_{2g}$ Hubbard Hamiltonian, $%
\mathcal{H}_{t_{2g}}$, given by expression (\ref{H}), including now the
spin-flip terms neglected for the purpose of the DMFT calculations.
Following the standard approach, we work in the many-electron representation
which diagonalizes the \emph{on-site} terms, $\mathcal{H}_{t_{2g};R=R^{%
\prime }}^{LDA}+\mathcal{U}_{R},$ of $\mathcal{H}_{t_{2g}}.$ The remaining,
non-diagonal part of $\mathcal{H}_{t_{2g}}$ is the LDA hopping, $\mathcal{T}%
. $ The super-exchange Hamiltonian, $\mathcal{H}_{se},$ is then obtained by L%
\"{o}wdin downfolding (\ref{eq6}) of the subspace which has one or more
sites doubly occupied:%
\begin{equation}
\mathcal{H}_{se}=\mathcal{P}_{s}\mathcal{T}\left\{ \left( 1-\mathcal{P}%
_{s}\right) \left( E-\mathcal{H}_{t_{2g}}\right) \left( 1-\mathcal{P}%
_{s}\right) \right\} ^{-1}\mathcal{TP}_{s}.  \label{Hexc}
\end{equation}%
Here, $\mathcal{P}_{s}$ projects onto the subspace of singly-occupied sites.
To order $t/U,$ the projection 
$1-\mathcal{P}_{s}$ can be substituted by $\mathcal{P}_{d},$
which projects onto the space of \emph{one} doubly-occupied site, 
and $\mathcal{H}%
_{t_{2g}}$ can be substituted by the on-site part of the Hamiltonian.
Expression (\ref{Hexc}) then reduces to the 2nd-order perturbation expansion
for $\mathcal{H}_{se}$ used by Khaliullin \textit{et al.}

We now specialize to the cases of LaTiO$_{3}$ and YTiO$_{3}$. As explained
in the previous section, the main result of our many-body LDA+DMFT
calculations was that the density matrix is diagonalized by the
crystal-field orbitals, $\left\vert m\right\rangle ,$ and that the orbital
with the lowest energy, $\left\vert 1\right\rangle ,$ is occupied with 0.91
electron in LaTiO$_{3}$ and with 0.96 in YTiO$_{3}.$ In the following, we
shall assume that $\left\vert 1\right\rangle $ is \emph{completely}
occupied. For the \emph{two} sites, $R$ and $R^{\prime },$ between which we
want to compute the super-exchange coupling, the states with \emph{no} double
occupancy, \textit{i.e.} $t_{2g}^{1}+t_{2g}^{1},$ are then the Slater
determinants $|1)$ through $|4)$ in the pictogram below, and the states with 
\emph{one} double occupancy with which they couple, line by line
in the pictogram, are the 15 
$t_{2g}^{2}$ states $|5)$ through $|19)$ (soft bras and kets denote
two-electron states):{\setlength{\arraycolsep}{4pt} 
\begin{eqnarray}
\fl 
\begin{array}{rcccrcccrcccrccc}
|1) & \!\!=\!\! & \fbox{$%
\begin{array}{ccc}
- &  & - \\ 
- &  & - \\ 
\uparrow &  & \downarrow%
\end{array}%
$} & \  & |5) & \!\!=\!\! & \fbox{$%
\begin{array}{ccc}
- &  & - \\ 
- &  & - \\ 
- &  & \uparrow \downarrow%
\end{array}%
$} & \  & |6) & \!\!=\!\! & \fbox{$%
\begin{array}{ccc}
- &  & - \\ 
- &  & \uparrow \\ 
- &  & \downarrow%
\end{array}%
$} & \  & |7) & \!\!=\!\! & \fbox{$%
\begin{array}{ccc}
- &  & \uparrow \\ 
- &  & - \\ 
- &  & \downarrow%
\end{array}%
$} & \mbox{} \\[-5pt] 
&  &  &  &  &  &  &  &  &  &  &  &  &  &  &  \\ 
|2) & \!\!=\!\! & \fbox{$%
\begin{array}{ccc}
- &  & - \\ 
- &  & - \\ 
\downarrow &  & \uparrow%
\end{array}%
$} &  & |5) & \!\!=\!\! & \fbox{$%
\begin{array}{ccc}
- &  & - \\ 
- &  & - \\ 
- &  & \uparrow \downarrow%
\end{array}%
$} &  & |8) & \!\!=\!\! & \fbox{$%
\begin{array}{ccc}
- &  & - \\ 
- &  & \downarrow \\ 
- &  & \uparrow%
\end{array}%
$} &  & |9) & \!\!=\!\! & \fbox{$%
\begin{array}{ccc}
- &  & \downarrow \\ 
- &  & - \\ 
- &  & \uparrow%
\end{array}%
$} & \mbox{} \\[-5pt] 
&  &  &  &  &  &  &  &  &  &  &  &  &  &  &  \\ 
|3) & \!\!=\!\! & \fbox{$%
\begin{array}{ccc}
- &  & - \\ 
- &  & - \\ 
\uparrow &  & \uparrow%
\end{array}%
$} &  & |10) & \!\!=\!\! & \fbox{$%
\begin{array}{ccc}
- &  & - \\ 
- &  & \uparrow \\ 
- &  & \uparrow%
\end{array}%
$} &  & |11) & \!\!=\!\! & \fbox{$%
\begin{array}{ccc}
- &  & \uparrow \\ 
- &  & - \\ 
- &  & \uparrow%
\end{array}%
$} &  & |14) & \!\!=\!\! & \fbox{$%
\begin{array}{ccc}
- &  & - \\ 
- &  & \uparrow \downarrow \\ 
- &  & -%
\end{array}%
$} & \mbox{} \\[-5pt] 
&  &  &  &  &  &  &  &  &  &  &  &  &  &  &  \\ 
|4) & \!\!=\!\! & \fbox{$%
\begin{array}{ccc}
- &  & - \\ 
- &  & - \\ 
\downarrow &  & \downarrow%
\end{array}%
$} &  & |12) & \!\!=\!\! & \fbox{$%
\begin{array}{ccc}
- &  & - \\ 
- &  & \downarrow \\ 
- &  & \downarrow%
\end{array}%
$} &  & |13) & \!\!=\!\! & \fbox{$%
\begin{array}{ccc}
- &  & \downarrow \\ 
- &  & - \\ 
- &  & \downarrow%
\end{array}%
$} &  & |15) & \!\!=\!\! & \fbox{$%
\begin{array}{ccc}
- &  & \uparrow \downarrow \\ 
- &  & - \\ 
- &  & -%
\end{array}%
$} & \mbox{} \\[-5pt] 
&  &  &  &  &  &  &  &  &  &  &  &  &  &  &  \\ 
|16) & \!\!=\!\! & \fbox{$%
\begin{array}{ccc}
- &  & \downarrow \\ 
- &  & \uparrow \\ 
- &  & -%
\end{array}%
$} &  & |17) & \!\!=\!\! & \fbox{$%
\begin{array}{ccc}
- &  & \uparrow \\ 
- &  & \downarrow \\ 
- &  & -%
\end{array}%
$} &  & |18) & \!\!=\!\! & \fbox{$%
\begin{array}{ccc}
- &  & \uparrow \\ 
- &  & \uparrow \\ 
- &  & -%
\end{array}%
$} &  & |19) & \!\!=\!\! & \fbox{$%
\begin{array}{ccc}
- &  & \downarrow \\ 
- &  & \downarrow \\ 
- &  & -%
\end{array}%
$} & \;.%
\end{array}
\nonumber \\* 
\fl  \label{d2}
\end{eqnarray}%
}In this space of 19 two-electron states, the matrix element of the
Hamiltonian $\mathcal{H}_{t_{2g}}$ between states $(1|$ through $(4|$ and $%
|5)$ through $|13)$ is a hopping integral, such as $(1R|\mathcal{H}%
_{t_{2g}}|7R^{\prime })=\left\langle 1R\left\vert \mathcal{H}%
_{t_{2g}}^{LDA}\right\vert 3R^{\prime }\right\rangle \equiv t_{13}$ between
crystal-field orbital $\left\vert 1\right\rangle $ on site $R$ and
crystal-field orbital $\left\vert 3\right\rangle $ on site $R^{\prime }.$
The matrix elements in the $t_{2g}^{2}$-subspace are the eigenfunctions of $%
\mathcal{U}$ plus the crystal-field splittings, $\Delta _{m}\equiv \epsilon
_{m}-\epsilon _{1}.$ Now, the orders of magnitude produced by our LDA
calculations for the titanates are:
$
t\sim \Delta \sim 0.3J\sim 0.04U, 
$
and we may therefore neglect $\Delta $ when evaluating the $t_{2g}^{2}$
eigenfunctions (although this is not done in our numerical calculations).
The $t_{2g}^{2}$ eigenfunctions are then simply the $p^{2}$-states listed
following expression (\ref{H}). The transformation from the two-electron
basis (\ref{d2}) to the one which diagonalizes $\mathcal{U}$ is very simple:
it merely involves combining states $|6)$ and $|8)$ into a singlet and a
triplet (belonging to respectively $^{1}D$ and $^{3}P)$, the same for $|7)$
and $|9),$ and combining states $|5),$ $|14),$ and $|15)$ into the $^{1}S$
singlet and the two singlets belonging to $^{1}D.$ One of the latter does
not couple to the singly-occupied states $|1)$ through $|4),$ and the same
holds for the three triplets formed from $|16)$ through $|19).$ The singlet
formed from states $|16)$ and $|17)$ belonging to $^{1}D$ does not couple
either. As a result, the off-diagonal block of $H_{t_{2g}}-2\epsilon _{1}$
is then:{\setlength{\arraycolsep}{1.15pt} 
\begin{eqnarray}
\fl \left[ 
\begin{array}{ccccccccccc}
& \frac{1}{\sqrt{3}}\times & \frac{1}{\sqrt{6}}\times & \frac{1}{\sqrt{2}}%
\times & \frac{1}{\sqrt{2}}\times & \frac{1}{\sqrt{2}}\times & \frac{1}{%
\sqrt{2}}\times &  &  &  &  \\ 
& \left[ 
\begin{array}{r}
|5) \\ 
+|14) \\ 
+|15)%
\end{array}%
\right] & \left[ 
\begin{array}{c}
2|5) \\ 
-|14) \\ 
-|15)%
\end{array}%
\right] & \left[ 
\begin{array}{c}
|6) \\ 
+|8)\!%
\end{array}%
\right] & \left[ 
\begin{array}{c}
|7) \\ 
+|9)\!%
\end{array}%
\right] & \left[ 
\begin{array}{c}
|6) \\ 
-|8)\!%
\end{array}%
\right] & \left[ 
\begin{array}{c}
|7) \\ 
-|9)\!%
\end{array}%
\right] & |10) & |11) & |12) & |13) \\ 
(1| & \frac{t_{11}}{\sqrt{3}} & \sqrt{\frac{2}{3}}t_{11} & \frac{t_{12}}{%
\sqrt{2}} & \frac{t_{13}}{\sqrt{2}} & \frac{t_{12}}{\sqrt{2}} & \frac{t_{13}%
}{\sqrt{2}} & 0 & 0 & 0 & 0 \\ 
(2| & \frac{t_{11}}{\sqrt{3}} & \sqrt{\frac{2}{3}}t_{11} & \frac{t_{12}}{%
\sqrt{2}} & \frac{t_{13}}{\sqrt{2}} & -\frac{t_{12}}{\sqrt{2}} & -\frac{%
t_{13}}{\sqrt{2}} & 0 & 0 & 0 & 0 \\ 
(3| & 0 & 0 & 0 & 0 & 0 & 0 & t_{12} & t_{13} & 0 & 0 \\ 
(4| & 0 & 0 & 0 & 0 & 0 & 0 & 0 & 0 & t_{12} & t_{13} \\ 
. & 
\begin{array}{c}
U\!+\!2J\!+ \\ 
\frac{2}{3}(\Delta _{2}\!\!+\!\!\Delta _{3})%
\end{array}
& 0 & 0 & 0 & 0 & 0 & 0 & 0 & 0 & 0 \\ 
. & 0 & 
\begin{array}{c}
U\!-\!J\!+ \\ 
\frac{1}{3}(\Delta _{2}\!\!+\!\!\Delta _{3})%
\end{array}
& 0 & 0 & 0 & 0 & 0 & 0 & 0 & 0 \\ 
. & 0 & 0 & 
\begin{array}{c}
U\!\!-\!\!J \\ 
+\!\Delta _{2}%
\end{array}
& 0 & 0 & 0 & 0 & 0 & 0 & 0 \\ 
. & 0 & 0 & 0 & 
\begin{array}{c}
U\!\!-\!\!J \\ 
+\!\Delta _{3}%
\end{array}
& 0 & 0 & 0 & 0 & 0 & 0 \\ 
. & 0 & 0 & 0 & 0 & 
\begin{array}{c}
U\!\!-\!\!3J \\ 
+\!\Delta _{2}%
\end{array}
& 0 & 0 & 0 & 0 & 0 \\ 
. & 0 & 0 & 0 & 0 & 0 & 
\begin{array}{c}
U\!\!-\!\!3J \\ 
+\!\Delta _{3}%
\end{array}
& 0 & 0 & 0 & 0 \\ 
(10| & 0 & 0 & 0 & 0 & 0 & 0 & 
\begin{array}{c}
U\!\!-\!\!3J \\ 
+\!\Delta _{2}%
\end{array}
& 0 & 0 & 0 \\ 
(11| & 0 & 0 & 0 & 0 & 0 & 0 & 0 & 
\begin{array}{c}
U\!\!-\!\!3J \\ 
+\!\Delta _{3}%
\end{array}
& 0 & 0 \\ 
(12| & 0 & 0 & 0 & 0 & 0 & 0 & 0 & 0 & 
\begin{array}{c}
U\!\!-\!\!3J \\ 
+\!\Delta _{2}%
\end{array}
& 0 \\ 
(13| & 0 & 0 & 0 & 0 & 0 & 0 & 0 & 0 & 0 & 
\begin{array}{c}
U\!\!-\!\!3J \\ 
+\!\Delta _{3}%
\end{array}%
\end{array}%
\right]  \nonumber \\*
\fl  \label{H12}
\end{eqnarray}%
}Running along the rows, we have the following $t_{2g}^{2}$-states: the $%
^{1}S $-state, three of the five $^{1}D$-states, and six of the nine $^{3}P$%
-states. With the high-energy subspace diagonalized, it is now trivial to
downfold to the low-energy subspace, $|1)$ through $|4),$ obtaining the
super-exchange Hamiltonian (\ref{Hexc}) with $E=2\epsilon _{1}$. For the
energy difference between the N\'{e}el state and the ferromagnetic state, we
thus obtain:%
\begin{equation}
\fl(3|\mathcal{H}_{se}|3)\!-\!(1|\mathcal{H}_{se}|1)=t_{11}^{2}\left( \frac{%
\frac{1}{3}}{U+2J}+\frac{\frac{2}{3}}{U-J}\right) +\frac{%
t_{12}^{2}+t_{13}^{2}}{2}\left( \frac{1}{U-J}-\frac{1}{U-3J}\right) ,
\label{Ja}
\end{equation}%
which for spin $\frac{1}{2},$ is $\frac{1}{4}$ times the super-exchange
coupling,\emph{\ }$J_{se},$\emph{\ }between the two sites considered. For
simplicity, and because it is a good approximation, we have left out the
crystal-field splittings in the denominators, but they are trivial to
include by returning to the full Hamiltonian (\ref{H12}).

The super-exchange coupling (\ref{Ja}) is the sum of terms caused by intra-
and inter-level hops. As is well known, the first term is positive, that is,
it favours antiferromagnetic alignment, and the latter is negative.
Ferromagnetic alignment is thus favoured when the hopping to excited levels
exceeds the hopping in the ground state by a critical ratio, $\tau _{0}:$%
\begin{equation}
\frac{\sqrt{\frac{1}{2}\left( t_{12}^{2}+t_{13}^{2}\right) }}{\left\vert
t_{11}\right\vert }\equiv \tau >\tau _{0}\equiv \sqrt{\frac{\left(
1+J/U\right) \left( 1-3J/U\right) }{\left( 2J/U\right) \left( 1+2J/U\right) }%
}.  \label{tau}
\end{equation}%
For the actual value of the ratio between the Hund's-rule coupling to the
Coulomb repulsion, $J/U=0.64/5\approx 0.13,$ this equation gives $\tau
_{0}=1.47.$ Similarly, for a given ratio of inter- to intra-level hopping
integrals, one may find a critical value of the Hund's-rule coupling to be
exceeded in order to produce ferromagnetic coupling:%
\begin{equation}
\frac{J}{U}>\left( \frac{J}{U}\right) _{0}=\frac{\sqrt{4+6\tau ^{2}+\tau ^{4}%
}-1-\tau ^{2}}{3+4\tau ^{2}}.  \label{JU}
\end{equation}

We now consider the super-exchange coupling between two nearest neighbours on
either side of the mirror plane, \textit{i.e.} those separated by 001, and
between two nearest neighbours in the $xy$-plane, \textit{i.e.} those
separated by 010 or 100. In the latter case, we must average over the two
sites related by a glide mirror, and this amounts to taking the average of
the 010 and 100 exchange couplings. Couplings between farther neighbours are
very small because the squares of the corresponding hopping integrals
determine $J_{se},$ so we neglect them. In table \ref{tableLaYXJ}, we have
extracted $t_{1m}^{010}$ and $t_{1m}^{100}$ for LaTiO$_{3}$ from equation (%
\ref{eqLa010}), and $t_{1m}^{001}$ from equation (\ref{eqLa001}). For YTiO$%
_{3},$ we use equations (\ref{eqY010}) and (\ref{eqY001}) in the same way.
From these hopping integrals and the values of $U$ and $J,$ we can now
evaluate the various contributions in (\ref{Ja}) to the super-exchange
coupling, and the corresponding rows have been inserted in the tables. The
summed up, total $J_{se}$ may be found in an inserted row. Finally, $J_{se}$
evaluated numerically without neglecting the crystal-field splittings has
been included in a row marked $J_{se}^{\Delta }$.

\begin{table}[tbp]
\caption{Hopping integrals, $t_{1m}^{xyz}$, and superexchange couplings, $%
J_{se}$, in meV}
\label{tableLaYXJ}
\begin{center}
{\setlength{\tabcolsep}{1.pt} 
\begin{tabular}{crrrrrrrrrrrrrrrrrr}
\br &  & \multicolumn{8}{c}{LaTiO$_{3}$} &  & \multicolumn{8}{c}{YTiO$_{3}$}
\\ \cline{3-10}\cline{12-19}
$xyz$ & \hspace{0.3cm} & \multicolumn{2}{c}{$001$} & \hspace{0.2cm} & 
\multicolumn{2}{c}{$010$} & \multicolumn{2}{c}{$100$} & 
& \hspace{0.3cm} & \multicolumn{2}{c}{$001$} & \hspace{0.2cm} & 
\multicolumn{2}{c}{$010$} & \multicolumn{2}{c}{$100$} & 
\\ 
$1,m$ &  & $t_{1m}$ & $J_{se}$ &  & $t_{1m}$ & $J_{se}$ & $t_{1m}$ & $J_{se}$
&  &  & $t_{1m}$ & $J_{se}$ &  & $t_{1m}$ & $J_{se}$ & $t_{1m}$ & $J_{se}$ & 
\\[-3pt] 
&  &  &  &  &  &  &  &  &  &  &  &  &  &  &  &  &  &  \\ 
$1,1$ &  & $105$ & $9.1$ &  & $-98$ & $7.9$ & $-98$ & $7.9$ &  &  & $-38$ & $%
1.2$ &  & $-48$ & $1.9$ & $-48$ & $1.9$ &  \\ 
$1,2$ &  & $31$ & $-0.2$ &  & $-192$ & $-7.0$ & $4$ & $-0.0$ &  &  & $-21$ & 
$-0.1$ &  & $-191$ & $-7.0$ & $-84$ & $-1.3$ &  \\ 
$1,3$ &  & $143$ & $-3.9$ &  & $12$ & $0.0$ & $120$ & $-2.7$ &  &  & $97$ & $%
-1.8$ &  & $-130$ & $-3.2$ & $94$ & $-1.7$ &  \\ 
$J_{se}$ &  &  & $5.0$ &  &  & $0.9$ &  & $5.2$ & $3.0$ &  &  & $-0.7$ &  & 
& $-8.3$ &  & $-1.1$ & $-4.7$ \\ 
$J_{se}^{\Delta} $ &  &  & $5.0$ &  &  &  &  &  & $3.2$ &  &  & $-0.5$ &  & 
&  &  &  & $-4.0$ \\ 
$\tau $ &  &  & $0.99$ &  &  &  &  &  & $1.16$ &  &  & $1.85$ &  &  &  &  & 
& $2.74$ \\ 
$\left( \frac{J}{U}\right) _{0}$ &  &  & $0.19$ &  &  &  &  &  & $0.16$ &  & 
& $0.10$ &  &  &  &  &  & $0.05$ \\ 
\br &  &  &  &  &  &  &  &  &  &  &  &  &  &  &  &  &  & 
\end{tabular}%
} 
\end{center}

\caption{$t_{1m}^{xyz}$ and $J_{se}$ in meV for LaTiO$_{3}$ at $T$=$8$ K.}
\label{tableLa8KXJ}
\begin{center}
\begin{tabular}{crrrrrrrrr}
\br $xyz$ & \hspace{0.3cm} & \multicolumn{2}{c}{$001$} & \hspace{0.2cm} & 
\multicolumn{2}{c}{$010$} & \multicolumn{2}{c}{$100$} & 
\\ 
$1,m$ &  & $t_{1m}$ & $J_{se}$ &  & $t_{1m}$ & $J_{se}$ & $t_{1m}$ & $J_{se}$
&  \\[-3pt] 
&  &  &  &  &  &  &  &  &  \\ 
$1,1$ &  & $114$ & $10.7$ &  & $-108$ & $9.6$ & $-108$ & $9.6$ &  \\ 
$1,2$ &  & $43$ & $-0.4$ &  & $-182$ & $-6.3$ & $23$ & $-0.1$ &  \\ 
$1,3$ &  & $142$ & $-3.8$ &  & $23$ & $-0.1$ & $119$ & $-2.7$ &  \\ 
$J_{se}$ &  &  & $6.5$ &  &  & $3.2$ &  & $6.8$ & $5.0$ \\ 
$\tau$ &  &  & $0.92$ &  &  &  &  &  & $1.02$ \\ 
$\left( \frac{J}{U}\right) _{0}$ &  &  & $0.20$ &  &  &  &  &  & $0.18$ \\ 
\br &  &  &  &  &  &  &  &  & 
\end{tabular}%
%
%
%
%
%
%
%
%
%
\end{center}
\end{table}

For LaTiO$_{3}$ and the coupling to 001, we see that the antiferromagnetic
intra-level hopping dominates and that the total $J_{se}^{001}$ is
5.0\thinspace meV. Including the crystal-field splittings makes no
difference in this case. Parallel to the mirror plane, a couple of large
inter-level hopping integrals make the total exchange constant smaller, $%
J_{se}^{100/010}=3.2$\thinspace meV, but still antiferromagnetic. These
values are much smaller than the values obtained 
from the neutron scattering \cite%
{Keim} and Raman experiments \cite{raman}, and also than the values obtained
with the model Hamiltonian \cite{Mochizuki03,Mochizuki04}. With the older
structural data \cite{lastr}, we find the perpendicular and the parallel
exchange-coupling constants to be both 6 meV. Using the low-temperature
structural data \cite{Cwik03}, we compute the LDA Hamiltonian given in the
left-hand side of table \ref{tableLa8noJT16G}, and with that, we can
calculate the crystal-field eigenvalues and vectors like in equation (\ref%
{eq13}), and then transform the hopping integrals to the crystal-field
representation like in equations (\ref{eqLa010}) and (\ref{eqLa001}). The
result is given in table \ref{tableLa8KXJ}: 
due to a 10\%
increase of $t_{11}^{010}=t_{11}^{010}$ 
the isotropy is increased, $%
J_{se}^{001}=6.5 $ and $J_{se}^{100/010}=5.0$\thinspace meV, 
but our \emph{ab initio} exchange coupling
remains 2--3 times smaller than the experimental value. Had we, instead of
the computed crystal-field eigenvectors (\ref{eq13}), used the simplified
ones:%
\begin{equation}
\left( 
\begin{array}{ccc}
\left\vert 1\right\rangle & \left\vert 2\right\rangle & \left\vert
3\right\rangle%
\end{array}%
\right) =\left( 
\begin{array}{ccc}
\left\vert yz\right\rangle & \left\vert xz\right\rangle & \left\vert
xy\right\rangle%
\end{array}%
\right) \left( 
\begin{array}{rrr}
\frac{1}{\sqrt{3}} & -\frac{1}{\sqrt{2}} & \frac{1}{\sqrt{6}} \\ 
\frac{1}{\sqrt{3}} & \frac{1}{\sqrt{2}} & \frac{1}{\sqrt{6}} \\ 
\frac{1}{\sqrt{3}} & 0 & -\frac{2}{\sqrt{6}}%
\end{array}%
\right) ,  \label{eqLasimple}
\end{equation}%
where $\left\vert 1\right\rangle =$ $3z_{111}^{2}-1,$ we would have obtained 
$J_{se}^{001}=15$ and $J_{se}^{100/010}=10$\thinspace meV, as shown in table %
\ref{tableLaYsimpXJ} below. This demonstrates the extreme sensitivity of the
exchange coupling constants to the crystal-field eigenvectors.

\begin{table}[tbp]
\caption{$t_{1m}^{xyz}$ and $J_{se}$ in meV using the simplified
eigenvectors (\protect\ref{eqLasimple}) and (\protect\ref{eqYsimple}).}
\label{tableLaYsimpXJ}
\begin{center}
\smallskip 
{\setlength{\tabcolsep}{1.pt} 
\begin{tabular}{crrrrrrrrrrrrrrrrrr}
\br &  & \multicolumn{8}{c}{LaTiO$_{3}$} &  & \multicolumn{8}{c}{YTiO$_{3}$}
\\ \cline{3-10}\cline{12-19}
$xyz$ & \hspace{0.3cm} & \multicolumn{2}{c}{$001$} & \hspace{0.2cm} & 
\multicolumn{2}{c}{$010$} & \multicolumn{2}{c}{$100$} & 
& \hspace{0.3cm} & \multicolumn{2}{c}{$001$} & \hspace{0.2cm} & 
\multicolumn{2}{c}{$010$} & \multicolumn{2}{c}{$100$} & 
\\ 
$1,m$ &  & $t_{1m}$ & $J_{se}$ &  & $t_{1m}$ & $J_{se}$ & $t_{1m}$ & $J_{se}$
&  &  & $t_{1m}$ & $J_{se}$ &  & $t_{1m}$ & $J_{se}$ & $t_{1m}$ & $J_{se}$ & 
\\[-3pt] 
&  &  &  &  &  &  &  &  &  &  &  &  &  &  &  &  &  &  \\ 
$1,1$ &  & $150$ & $18.5$ &  & $-128$ & $13.5$ & $-128$ & $13.5$ &  &  & $%
-35 $ & $1.0$ &  & $-84$ & $5.8$ & $-84$ & $5.8$ &  \\ 
$1,2$ &  & $51$ & $-0.5$ &  & $-133$ & $-3.4$ & $71$ & $-1.0$ &  &  & $-36$
& $-0.2$ &  & $-132$ & $-3.3$ & $-100$ & $-1.9$ &  \\ 
$1,3$ &  & $126$ & $-3.0$ &  & $12$ & $-0.0$ & $101$ & $-1.9$ &  &  & $103$
& $-2.0$ &  & $-175$ & $-5.8$ & $65$ & $-0.8$ &  \\ 
$J_{se}$ &  &  & $15.0$ &  &  & $10.1$ &  & $10.6$ & $10.3$ &  &  & $-1.3$ & 
&  & $-3.3$ &  & $3.1$ & $-0.1$ \\ 
$\tau$ &  &  & $0.64$ &  &  &  &  &  & $0.71$ &  &  & $2.20$ &  &  &  &  & 
& $1.49$ \\ 
$\left( \frac{J}{U}\right) _{0}$ &  &  & $0.25$ &  &  &  &  &  & $0.24$ &  & 
& $0.07$ &  &  &  &  &  & $0.13$ \\ 
\br &  &  &  &  &  &  &  &  &  &  &  &  &  &  &  &  &  & 
\end{tabular}%
} \bigskip
\end{center}
\end{table}

For YTiO$_{3},$ the small intra-level 001 hopping makes the ferromagnetic
inter-level coupling dominate so that $J_{se}^{001}=-0.5$ meV, as seen in
table \ref{tableLaYXJ}. Parallel to the mirror-plane the large inter-orbital
hopping integrals, however, make the calculated exchange coupling robustly
ferromagnetic: $J_{se}^{100/010}=-4.0$\thinspace meV. This is consistent
with the measured Curie temperature and the spin-wave spectra, which give $%
-3 $ meV \cite{Ulrich}. Had we, like for LaTiO$_{3},$ used simplified
crystal-field eigenvectors for YTiO$_{3},$ namely%
\begin{equation}
\left( 
\begin{array}{ccc}
\left\vert 1\right\rangle & \left\vert 2\right\rangle & \left\vert
3\right\rangle%
\end{array}%
\right) =\left( 
\begin{array}{ccc}
\left\vert yz\right\rangle & \left\vert xz\right\rangle & \left\vert
xy\right\rangle%
\end{array}%
\right) \left( 
\begin{array}{ccc}
\frac{1}{\sqrt{2}} & -\frac{1}{\sqrt{2}} & 0 \\ 
0 & 0 & 1 \\ 
\frac{1}{\sqrt{2}} & \frac{1}{\sqrt{2}} & 0%
\end{array}%
\right) ,  \label{eqYsimple}
\end{equation}%
we would have obtained $J_{se}^{001}=-1.3$ and $J_{se}^{100/010}=-0.1$%
\thinspace meV, as demonstrated in the right-hand side of table \ref%
{tableLaYsimpXJ}. So YTiO$_{3}$ does stay ferromagnetic, but just barely.

From the hopping integrals, we can also calculate the ratio, $\tau ,$
between the intra- and inter-level hoppings using (\ref{tau}).
The corresponding lines have been added at the bottom of 
tables~\ref{tableLaYXJ}--\ref{tablenoJTYXJ}. 
We see that for both titanates, $\tau$ is \emph{nearly isotropic}
and, for LaTiO$_{3}$, it is 
significantly \emph{smaller} than the critical ratio, $%
\tau _{0}$=1.47, whereas for YTiO$_{3},$ it is significantly larger. This
agrees with the experimental facts. Obviously, the small $t_{11}$ and the
large $t_{12}$ and $t_{13}$ in YTiO$_{3},$ make this material profit from
the Hund's-rule coupling in the $t_{2g}^{2}$ state and, hence, have a
ferromagnetic ground state.

\begin{figure}[t]
\par
\begin{center}
\includegraphics[width=\textwidth]{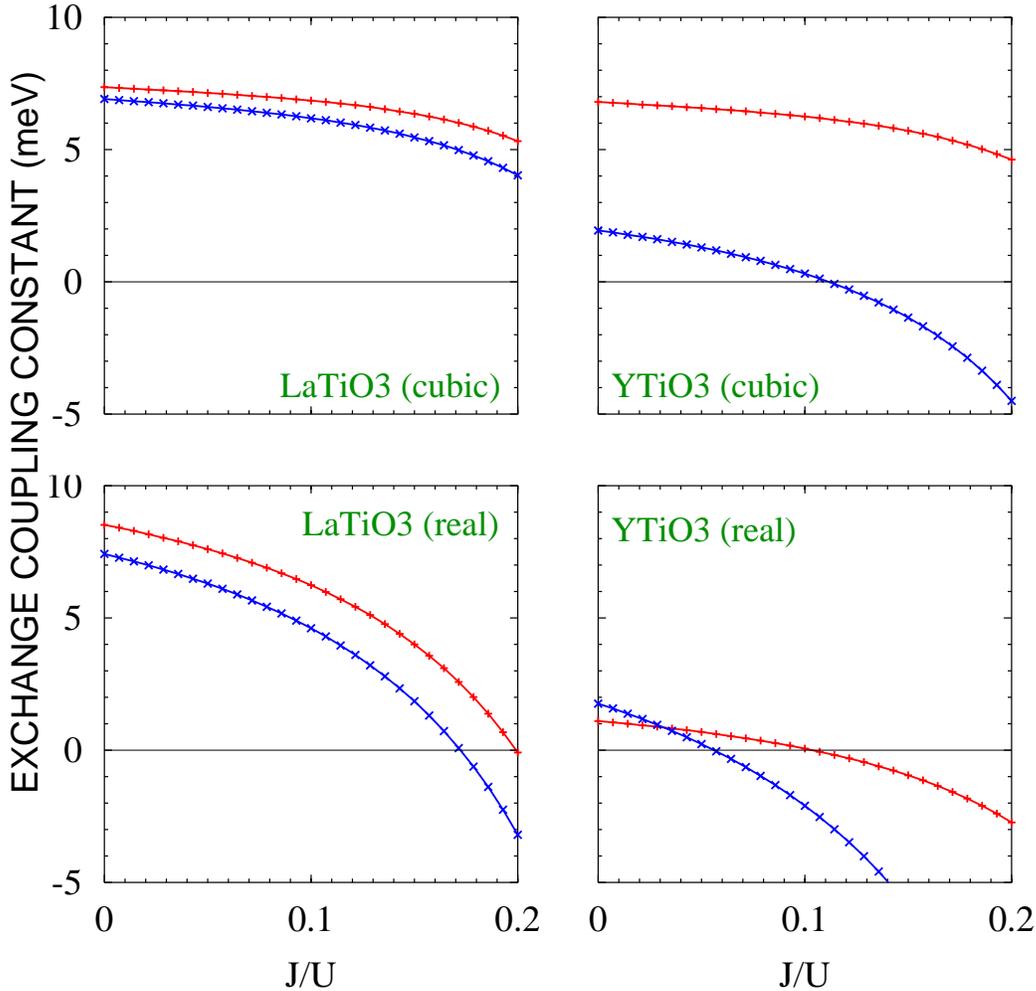}
\end{center}
\caption{Nearest-neighbour exchange coupling constants perpendicular (red)
and parallel (blue) to the mirror planes as functions of the ratio between
the Hund's-rule coupling and the on-site repulsion, $J/U$. See figure 4
in reference \protect\cite{Ulrich}. Antiferro- and ferromagnetic couplings
are respectively positive and negative. Input were $U=5$ eV and the
orbitally ordered states given by equations (\protect\ref{eq13}) and (%
\protect\ref{eq15}). In the upper row, degenerate $t_{2g}$-levels and cubic
hopping with $t_{\pi }$=$-150$ meV were used, like in table  \protect
\ref{tableCubicXJ}. The values of the exchange-coupling constants for $%
J/U=0.64/5=0.13$ are those listed in table \protect\ref{tableLaYXJ} as $%
J_{se}^{\Delta }.$}
\label{fig14}
\end{figure}

Using (\ref{JU}) we can calculate the critical $J/U$ ratio needed for the
coupling to be ferromagnetic. $\left( J/U\right) _{0}$ is seen to be
significantly larger than the actual value (0.13) for LaTiO$_{3},$ and
significantly smaller in YTiO$_{3}.$ In the bottom panels of figure \ref{fig14}
we show the perpendicular (red) and parallel (blue) exchange couplings as
functions of $J/U,$ computed with inclusion of the crystal-field splittings.
The latter are seen to have only minor effects, at least on the scale of the
figure, but not on the scale of experiments, \textit{e.g.} the largest
deviation of the critical $J/U$ values obtained without crystal-field
splitting is found for the in-plane exchange coupling in LaTiO$_{3}.$ Here,
the full calculation in the figure gives $\left( J/U\right) _{0}$=0.175,
while the approximate value quoted in the table is 0.19.

\begin{table}[tbp]
\caption{$t_{1m}^{xyz}$ and $J_{se}$ in meV for cubic hopping.}
\label{tableCubicXJ}
\begin{center}
{\setlength{\tabcolsep}{1.pt} 
\begin{tabular}{crrrrrrrrrrrrrrrrrr}
\br &  & \multicolumn{8}{c}{LaTiO$_{3}$} &  & \multicolumn{8}{c}{YTiO$_{3}$}
\\ \cline{3-10}\cline{12-19}
$xyz$ & \hspace{0.3cm} & \multicolumn{2}{c}{$001$} & \hspace{0.2cm} & 
\multicolumn{2}{c}{$010$} & \multicolumn{2}{c}{$100$} & 
& \hspace{0.5cm} & \multicolumn{2}{c}{$001$} & \hspace{0.2cm} & 
\multicolumn{2}{c}{$010$} & \multicolumn{2}{c}{$100$} & 
\\ 
$1,m$ &  & $t_{1m}$ & $J_{se}$ &  & $t_{1m}$ & $J_{se}$ & $t_{1m}$ & $J_{se}$
&  &  & $t_{1m}$ & $J_{se}$ &  & $t_{1m}$ & $J_{se}$ & $t_{1m}$ & $J_{se}$ & 
\\[-3pt] 
&  &  &  &  &  &  &  &  &  &  &  &  &  &  &  &  &  &  \\ 
$1,1$ &  & $96$ & $7.6$ &  & $-89$ & $6.5$ & $-89$ & $6.5$ &  &  & $92$ & $%
7.0$ &  & $-49$ & $2.0$ & $-49$ & $2.0$ &  \\ 
$1,2$ &  & $-20$ & $-0.1$ &  & $-110$ & $-2.3$ & $5$ & $-0.0$ &  &  & $-70$
& $-0.9$ &  & $-108$ & $-2.2$ & $-76$ & $-1.1$ &  \\ 
$1,3$ &  & $69$ & $-0.9$ &  & $15$ & $-0.0$ & $49$ & $-0.5$ &  &  & $19$ & $%
-0.1$ &  & $-91$ & $-1.6$ & $22$ & $-0.1$ &  \\ 
$J_{se}$ &  &  & $6.6$ &  &  & $4.2$ &  & $6.1$ & $5.2$ &  &  & $6.0$ &  & 
& $-1.8$ &  & $0.8$ & $-0.5$ \\ 
$\tau $ &  &  & $0.53$ &  &  &  &  &  & $0.68$ &  &  & $0.56$ &  &  &  &  & 
& $1.65$ \\ 
$\left( \frac{J}{U}\right) _{0}$ &  &  & $0.27$ &  &  &  &  &  & $0.24$ &  & 
& $0.27$ &  &  &  &  &  & $0.11$ \\ 
\br &  &  &  &  &  &  &  &  &  &  &  &  &  &  &  &  &  & 
\end{tabular}%
} 
\end{center}
\par
\caption{$t_{1m}^{xyz}$ and $J_{se}$ in meV for simplified eigenvectors and
cubic hopping.}
\label{tableLaYcubsimpJ}
\begin{center}
{\setlength{\tabcolsep}{1.pt} 
\begin{tabular}{crrrrrrrrrrrrrrrrrr}
\br &  & \multicolumn{8}{c}{LaTiO$_{3}$} &  & \multicolumn{8}{c}{YTiO$_{3}$}
\\ \cline{3-10}\cline{12-19}
$xyz$ & \hspace{0.3cm} & \multicolumn{2}{c}{$001$} & \hspace{0.2cm} & 
\multicolumn{2}{c}{$010$} & \multicolumn{2}{c}{$100$} & 
& \hspace{0.3cm} & \multicolumn{2}{c}{$001$} & \hspace{0.2cm} & 
\multicolumn{2}{c}{$010$} & \multicolumn{2}{c}{$100$} & 
\\ 
$1,m$ &  & $t_{1m}$ & $J_{se}$ &  & $t_{1m}$ & $J_{se}$ & $t_{1m}$ & $J_{se}$
&  &  & $t_{1m}$ & $J_{se}$ &  & $t_{1m}$ & $J_{se}$ & $t_{1m}$ & $J_{se}$ & 
\\[-3pt] 
&  &  &  &  &  &  &  &  &  &  &  &  &  &  &  &  &  &  \\ 
$1,1$ &  & $100$ & $8.2$ &  & $-100$ & $8.2$ & $-100$ & $8.2$ &  &  & $75$ & 
$4.6$ &  & $-75$ & $4.6$ & $-75$ & $4.6$ &  \\ 
$1,2$ &  & $0$ & $0.0$ &  & $-61$ & $-0.7$ & $61$ & $-0.7$ &  &  & $-75$ & $%
-1.1$ &  & $-75$ & $-1.1$ & $-75$ & $-1.1$ &  \\ 
$1,3$ &  & $71$ & $-1.0$ &  & $35$ & $-0.2$ & $35$ & $-0.2$ &  &  & $0$ & $%
0.0$ &  & $-106$ & $-2.1$ & $0$ & $0.0$ &  \\ 
$J_{se}$ &  &  & $7.3$ &  &  & $7.3$ &  & $7.3$ & $7.3$ &  &  & $3.6$ &  & 
& $1.4$ &  & $3.6$ & $2.5$ \\ 
$\tau$ &  &  & $0.50$ &  &  &  &  &  & $0.50$ &  &  & $0.71$ &  &  &  &  & 
& $1.00$ \\ 
$\left( \frac{J}{U}\right) _{0}$ &  &  & $0.28$ &  &  &  &  &  & $0.28$ &  & 
& $0.24$ &  &  &  &  &  & $0.19$ \\ 
\br &  &  &  &  &  &  &  &  &  &  &  &  &  &  &  &  &  & 
\end{tabular}%
} \bigskip
\end{center}
\end{table}

Our LDA-NMTO results should be seen on the background of the claim \cite%
{Ulrich} that orbital order is inconsistent with the observed magnetic
ground states and isotropic spin-wave spectra in YTiO$_{3}$. This claim was
based on the assumption that the \emph{hopping} is as in the \emph{cubic}
structure, where the $xy,$ $yz,$ and $xz$ bands are identical, independent,
and 2D. In that case, there is only \emph{one}
nearest-neighbour hopping integral, the effective $t_{\pi }$ between two
similar orbitals in the same plane (in case of SrVO$_{3},$ this is the $-281$%
\thinspace meV quoted in table \ref{tableSr}). As we have seen and explained
in great detail in section \ref{LowE}, the hopping between the $xy, $ $yz,$
and $xz$ LDA Wannier functions is, however, \emph{very different} due to the
GdFeO$_{3}$-type distortion in the titanates. In order to demonstrate that
this is the reason for the discrepancy between our calculation and that of
Khaliullin \cite{Ulrich}, we first of all note that our expression (\ref{Ja}%
) reduces to their equation (3) if one assumes cubic hopping and uses the
eigenvector of the lowest crystal-field state $\left\vert 1\right\rangle $
in YTiO$_{3}$. Since equation (\ref{Ja}) is general, we can use it also for
``cubic'' LaTiO$_{3}$. In table \ref{tableCubicXJ} we have therefore listed
the hopping integrals obtained by applying our crystal-field transformations
(\ref{eq13}) and (\ref{eq15}) to \emph{cubic hopping} with $t_{\pi }\equiv
-150$ meV. In this case, the difference between LaTiO$_{3}$ and YTiO$_{3}$
stems exclusively from the different crystal-field eigenvectors. The
differences with the \textit{ab initio} 
values of the hopping integrals listed in table \ref{tableLaYXJ} 
are seen to be large, although some of the
trends are captured by the \textit{ab initio} 
crystal-field eigenvectors. From the values of $%
\tau $ and $\left( J/U\right) _{0},$ we realize that, although both ``cubic
materials'' are antiferromagnetic, as claimed in
\cite{Ulrich}, YTiO$_{3}$ does have a tendency to couple
ferromagnetically in the plane. The calculation for the cubic titanates is
included as the top panel of figure \ref{fig14}.

If we finally take cubic hoppings, as well as the simplified eigenvector (%
\ref{eqLasimple}) for LaTiO$_{3}$ and (\ref{eqYsimple}) for YTiO$_{3},$
these then being the only difference between the two materials, we obtain
the results shown in table \ref{tableLaYcubsimpJ}. Both ``materials'' are
fairly isotropic, G-type antiferromagnets with the exchange coupling two
times larger in LaTiO$_{3}$ than in YTiO$_{3}.$

In conclusion, the sign of the super-exchange coupling that we calculate
using conventional theory and LDA $t_{2g}$ Wannier functions is consistent
with the observed magnetic ground states of LaTiO$_{3}$ and YTiO$_{3}.$ The
mechanism by which the different magnetic orders come out of the
calculations is through orbital order caused by the Coulomb repulsion and a
strong crystal-field, and through distortion of the cubic hopping integrals.
Both the crystal field and the modification of the hopping integrals are
caused by the GdFeO$_{3}$-type distortion. The fact that we can explain 
\emph{both} the Mott gap \emph{and} the observed magnetic order from the
orbital order and hopping integrals calculated without adjustable parameters
with the LDA, seems to rule out the orbital-liquid scenario for the $%
t_{2g}^{1}$ titanates.

The discrepancies between experiment and computation do deserve further
investigation, however.

The size of our $J^{001}_{se}$ exchange-coupling for YTiO$_{3}$ agrees with the
experimental values, but the anisotropy is too large. Due to the
near-cancellation of terms, the calculated exchange coupling constants are
very small, as they should be in order to account for the low Curie
temperature, but this also makes them sensitive to detail. For instance, the
values of our calculated 2nd-nearest-neighbour constants, $%
J_{se}^{011/101}=+0.2$ meV, $J_{se}^{01\bar{1}/10\bar{1}}=+0.3$ meV, $%
J_{se}^{110}=+0.5$ meV, and $J_{se}^{1\bar{1}0}=+0.2$ meV, are \emph{anti}%
ferromagnetic and \emph{not} negligible compared with the
1st-nearest-neighbour values in table \ref{tableLaYXJ}. We shall see a
further example of this sensitivity in section \ref{JTM}, where we consider
the influence of the JT distortion.
Moreover, the computational inaccuracies mentioned in appendix B may influence these delicate exchange-coupling constants. 

For LaTiO$_{3},$ our exchange couplings in table \ref{tableLa8KXJ} are
isotropic, but three times smaller than the experimental values. In this
case, our calculated 2nd-nearest-neighbour constants, $J_{se}^{011/101}=-0.2$
meV, $J_{se}^{01\bar{1}/10\bar{1}}=-0.1$ meV, $J_{se}^{110}=+0.9$ meV, and $%
J_{se}^{1\bar{1}0}=0$ meV, are small in comparison with the
1st-nearest-neighbour values.
Also the computational inaccuracies mentioned in appendix B are 
presumably too small to explain this discrepancy. 

In general, one suspects that the LDA overestimates covalency. If that would
cause the scale of our hopping integrals to be too large, our value of $U$
chosen to account for the Mott transition would also be too large in order
to compensate for this, but then our $t^{2}/U$ should be too large -- but, in
fact, the opposite seems to be true for LaTiO$_{3}$. Nevertheless, since our
hopping integrals result from a delicate balance between Ti-O, Ti-Ti, A-O,
and Ti-A covalencies as seen in sections \ref{hopping}, \ref{on-site}, and %
\ref{XtalFieldBasis}, the degree to which the LDA gets this balance right is
decisive for the accuracy of the calculated exchange couplings. 
To check this further,
calculations with other density functionals are needed, and so are
experiments to test the implications of the strong cation covalency
predicted by our \cite{Eva03} and earlier model calculations \cite%
{Mochizuki03,Mizokawa99}.

Another reason for the factor-three discrepancy in LaTiO$_{3}$ could be that
the $e_{g}$ degrees of freedom should have been treated explicitly because,
as was pointed out by Mochizuki and Imada \cite{Mochizuki01}, the energy of
the $e_{g}$ band is merely a few times $2J$ above that of the $t_{2g}$ band
(see figure 4). However, not fully treating the $%
e_{g}$ degrees of freedom should lead us to underestimate the tendency
towards ferromagnetism -- a problem that we do not have for LaTiO$_{3}$.
Besides, if we were to include the $e_{g}$ degrees of freedom in the Hubbard
model, we could do so by adding the space of the Wannier functions for
the $e_{g}$ band to the space of Wannier functions for the $t_{2g}$ band,
because with this choice the Wannier orbitals and, hence, the parameter
values for the $t_{2g}$ band would be unchanged. But in this $d$ basis there
is no single-particle coupling between the $t_{2g}$ and $e_{g}$ subspaces,
and therefore no coupling in the 2nd-order perturbation expression (\ref%
{Hexc}). As a result, our super-exchange calculation in which only the $%
t_{2g}$ degrees of freedom are treated and the $e_{g}$- as well as all other
single-particle channels are downfolded, is correct.

The factor-three discrepancy might be related to the fact that the
magnetic moment in LaTiO$_{3}$ is anomalously small (0.57 $\mu _{B}).$ Our
assumption of complete orbital order and subsequent use of 2nd-order
perturbation theory are clearly insufficient if the orbitally-ordered state
changes at low temperature. A future spin-polarized LDA+DMFT calculation
should allow us to calculate the magnetic moment and to keep full account of
the inter-site coupling between spin and orbital degrees of freedom. For a
general discussion of the why the magnetic moment is so small in LaTiO$_{3},$
we refer to reference \cite{Mochizuki04}.

Finally, it is possible that in these $t_{2g}$ materials with strong 
A covalency the eigenvectors of the density matrix
calculated more accurately than in the single-site DMFT
 differ sufficiently from
those of the crystal-field Hamiltonian to influence the exchange-coupling
constants.

\subsection{Influence of the JT distortion in YTiO$_{3}\label{JTM}$}

We have found that the JT distortion in YTiO$_{3}$ does not create, but
merely reflects the orbital order driven by the Coulomb correlations and
controlled by the GdFeO$_{3}$-type distortion through the crystal field set
up by the A ions. Mochizuki and Imada \cite{Mochizuki04} now
speculated that for strongly GdFeO$_{3}$-distorted materials like YTiO$_{3},$
where the magnetism is weak, the JT distortion could be a controlling factor
for the \emph{magnetic} order. This we now investigate.

\begin{table}[tbp]
\caption{$t_{1m}^{xyz}$ and $J_{se}$ in meV.}
\label{tablenoJTYXJ}
\begin{center}
{\setlength{\tabcolsep}{1.pt} 
\begin{tabular}{crrrrrrrrrrrrrrrrrr}
\br &  & \multicolumn{8}{c}{YTiO$_{3}$ without JT distortion} &  & 
\multicolumn{8}{c}{YTiO$_{3}$ at 15.9 GPa} \\ \cline{3-10}\cline{12-19}
$xyz$ & \hspace{0.3cm} & \multicolumn{2}{c}{$001$} & \hspace{0.2cm} & 
\multicolumn{2}{c}{$010$} & \multicolumn{2}{c}{$100$} & 
& \hspace{0.3cm} & \multicolumn{2}{c}{$001$} & \hspace{0.2cm} & 
\multicolumn{2}{c}{$010$} & \multicolumn{2}{c}{$100$} & 
\\ 
$1,m$ &  & $t_{1m}$ & $J_{se}$ &  & $t_{1m}$ & $J_{se}$ & $t_{1m}$ & $J_{se}$
&  &  & $t_{1m}$ & $J_{se}$ &  & $t_{1m}$ & $J_{se}$ & $t_{1m}$ & $J_{se}$ & 
\\[-3pt] 
&  &  &  &  &  &  &  &  &  &  &  &  &  &  &  &  &  &  \\ 
$1,1$ &  & $-23$ & $0.4$ &  & $-104$ & $8.9$ & $-104$ & $8.9$ &  &  & $-87$
& $6.2$ &  & $11$ & $0.1$ & $11$ & $0.1$ &  \\ 
$1,2$ &  & $52$ & $-0.5$ &  & $-200$ & $-7.6$ & $-47$ & $-0.4$ &  &  & $-15$
& $0.0$ &  & $-253$ & $-12.2$ & $-22$ & $-0.1$ &  \\ 
$1,3$ &  & $123$ & $-2.9$ &  & $-24$ & $-0.1$ & $157$ & $-4.7$ &  &  & $81$
& $-1.3$ &  & $-41$ & $-0.3$ & $130$ & $-3.2$ &  \\ 
$J_{se}$ &  &  & $-3.0$ &  &  & $1.2$ &  & $3.8$ & $2.5$ &  &  & $4.9$ &  & 
& $-12.4$ &  & $-3.2$ & $-7.8$ \\ 
$\tau$ &  &  & $4.11$ &  &  &  &  &  & $1.25$ &  &  & $0.67$ &  &  &  &  & 
& $13.1$ \\ 
$\left( \frac{J}{U}\right) _{0}$ &  &  & $0.03$ &  &  &  &  &  & $0.15$ &  & 
& $0.25$ &  &  &  &  &  & $0.00$ \\ 
\br &  &  &  &  &  &  &  &  &  &  &  &  &  &  &  &  &  & 
\end{tabular}%
} \bigskip
\end{center}
\end{table}

Using the LDA Hamiltonian (table \ref{tableLa8noJT16G}) computed for our
hypothetical YTiO$_{3}$ without JT distortion, we obtain the hopping
integrals and exchange couplings given in table \ref{tablenoJTYXJ}: $%
J_{se}^{001}=-3.0$ meV and $J_{se}^{100/010}=+2.5$\thinspace meV. This gives
ferromagnetic order along $z$ and antiferromagnetic order in the plane, that
is, C-type antiferromagnetism and, hence, a completely different result than
obtained for the proper structure.

For the high-pressure phase observed by Loa \textit{et al.} to have strongly
reduced JT distortion, but slightly increased GdFeO$_{3}$-type distortion, we
find the results given in the right-hand side of table \ref{tablenoJTYXJ}: $%
J_{se}^{001}=+4.9$ meV and $J_{se}^{100/010}=-7.8$\thinspace meV. This
causes a third kind of ground state with robust antiferromagnetic order
along $z$ and ferromagnetic order in the plane: a type-A antiferromagnet.

In conclusion, our \textit{ab initio} calculations support the speculation
of Mochizuki and Imada \cite{Mochizuki04} that, although not
very important for the orbital order, the JT distortion \emph{is} of crucial
importance for the magnetic order in YTiO$_{3}.$ This should be tested
experimentally.

\section{Unfolding the orthorhombic band\label{One}}

In this paper we have first seen three degenerate, independent, and nearly
two-dimensional cubic bands develop into twelve inequivalent and coupled
orthorhombic bands, which narrow down and develop a pseudo-gap under
increasing GdFeO$_{3}$-type distortion (figures \ref{fig2}, \ref{fig3}, \ref%
{fig3b}, and \ref{fig6}). In order to treat the Coulomb correlations, we
have constructed a set of highly localized $t_{2g}$ Wannier functions for
these bands and have defined the corresponding Hubbard Hamiltonian. With
this set of Wannier functions, the splitting of the lower $t_{2g}$ levels
reaches a mere 10\% of the bandwidth, but this, together with the 30\%
decrease of the entire bandwidth through the series (table \ref{tableW}) is
important for determining where in the series the Mott transition occurs
(table \ref{tableUc}). However, $U_{c}^{\prime }/W$ decreases by 50\% when
going from SrVO$_{3}$ to YTiO$_{3},$ and that is significantly more that the 
$\sim $30\% which can be gained by a crystal-field induced decrease of the
effective degeneracy from 3 to 1. Therefore, the width of the lowest subband
must decrease faster than the width of the entire $t_{2g}$ band.

A second surprising result of the LDA+DMFT calculations was that once $%
U^{\prime }$ exceeds $\sim W,$ the eigenvectors of the density matrix are
essentially the same as those of the on-site LDA Hamiltonian, and as $%
U^{\prime }\approx U_{c}^{\prime },$ the orbital fluctuations become
strongly suppressed. As a result, only the orbital of lowest energy is
occupied in LaTiO$_{3}$ and YTiO$_{3}$. As seen in figure \ref{fig10}, the $%
\left\langle 1\left\vert \varepsilon \right\vert 1\right\rangle $ element of
the spectral function is divided into a lower occupied and a higher unoccupied
Hubbard band, while the diagonal elements of the higher orbitals, $%
\left\langle 2\left\vert \varepsilon \right\vert 2\right\rangle $ and $%
\left\langle 3\left\vert \varepsilon \right\vert 3\right\rangle ,$ have
weight almost exclusively in the upper, unoccupied Hubbard band. The
off-diagonal elements, $\left\langle 1\left\vert \varepsilon \right\vert
2\right\rangle ,$ $\left\langle 2\left\vert \varepsilon \right\vert
3\right\rangle ,$ and $\left\langle 1\left\vert \varepsilon \right\vert
3\right\rangle ,$ are completely negligible. We thus found orbital order,
particularly in YTiO$_{3}.$ Our conclusion of those two findings was that
the self-energy must behave in the way explained around equations (\ref%
{SigmanU}) and (\ref{sigmaomega}).

In section \ref{M} we found that the increasing tendency towards
ferromagnetic coupling is due to the increasing strength of the hopping in
the orbitally ordered state from the lowest crystal-field orbital to the
higher orbitals on the neighbouring sites, compared with that of the hopping
between the lowest orbitals. This is what the ratio $\tau $ given in tables %
\ref{tableLaYXJ}--\ref{tablenoJTYXJ} measures. Increasing $\tau ,$ of
course, tends to decrease the width of band 1.

What remains unexplained is the 50\% effect brought out by table \ref%
{tableUc} or, in other words, why for merely $U^{\prime }\geq 1.6W$ in LaTiO$%
_{3}$ and $1.3W$ in YTiO$_{3}$ the lower half of \emph{one} band (per Ti) --
considerably more narrow than $W$ -- lies below the bottom of the others,
and why this band has pure LDA crystal-field character $\left\vert
1\right\rangle .$ In figure \ref{fig3b} it is hard to identify such four
orthorhombic LDA bands with $\left\vert
1\right\rangle $ character. That is, the orthorhombic LDA bands display
little tendency towards symmetry breaking. Nevertheless, we shall now
demonstrate that behind the LDA pseudo-gap in the orthorhombic GdFeO$_{3}$%
-distorted perovskites there \emph{is} a real gap. This gap is a direct one
in a \emph{pseudo-cubic} $\mathbf{k}$ space where it splits off the lowest $%
t_{2g}$ band.

\begin{table}[tbp]
\caption{$H^{LDA}$ in meV for LaTiO$_{3}$ \protect\cite{Cwik03}.}
\label{tableLaLong}
\begin{center}
{\setlength{\tabcolsep}{3pt} 
\begin{tabular}{crrrrrrrrrrrrrr}
\br $xyz$ & $000$ & $001$ & $010$ & $100$ & $011$ & $10\bar{1}$ & $01\bar{1}$
& $101$ & $110$ & $1\bar{1}0$ & $111$ & $11\bar{1}$ & $\bar{1}11$ & $1\bar{1}%
1$ \\ 
$m^{\prime },m$ &  &  &  &  &  &  &  &  &  &  &  &  &  &  \\ 
I,I & $9$ & $156$ & $-60$ & $-60$ & $-31$ & $-31$ & $-25$ & $-25$ & $-46$ & $%
-38$ & $-10$ & $-10$ & $-15$ & $-15$ \\ 
II,II & $748$ & $116$ & $31$ & $31$ & $-2$ & $-2$ & $-16$ & $-16$ & $6$ & $%
16 $ & $7$ & $7$ & $-1$ & $-1$ \\ 
III,III & $716$ & $102$ & $-101$ & $-101$ & $29$ & $29$ & $38$ & $38$ & $-24$
& $-23$ & $9$ & $9$ & $12$ & $12$ \\ 
II,III & $14$ & $11$ & $-21$ & $10$ & $5$ & $-26$ & $-10$ & $-26$ & $-55$ & $%
-34$ & $-10$ & $14$ & $3$ & $14$ \\ 
III,II & $14$ & $11$ & $10$ & $-21$ & $-26$ & $5$ & $-26$ & $-10$ & $-55$ & $%
-34$ & $14$ & $-10$ & $14$ & $3$ \\[10pt] 
\multicolumn{1}{l}{} &  &  &  &  &  &  &  &  &  &  &  &  &  &  \\ 
\multicolumn{1}{r}{$002$} & $020$ & $200$ & $2\bar{1}0$ & $\bar{1}20$ & $102$
& $012$ & $0\bar{1}2$ & $\bar{1}02$ & $201$ & $\bar{2}01$ & $02\bar{1}$ & $%
021$ & $210$ & $120$ \\ 
&  &  &  &  &  &  &  &  &  &  &  &  &  &  \\ 
\multicolumn{1}{r}{$-33$} & $-55$ & $-30$ & $4$ & $4$ & $-2$ & $-2$ & $-7$ & 
$-7$ & $-12$ & $-12$ & $42$ & $42$ & $-5$ & $-5$ \\ 
\multicolumn{1}{r}{$0$} & $-1$ & $82$ & $-18$ & $-18$ & $-9$ & $-9$ & $0$ & $%
0$ & $-24$ & $-24$ & $-3$ & $-3$ & $-14$ & $-14$ \\ 
\multicolumn{1}{r}{$31$} & $5$ & $-13$ & $2$ & $2$ & $-2$ & $-2$ & $1$ & $1$
& $10$ & $10$ & $-5$ & $-5$ & $5$ & $5$ \\ 
\multicolumn{1}{r}{$10$} & $-1$ & $-6$ & $0$ & $-7$ & $1$ & $1$ & $5$ & $13$
& $8$ & $4$ & $8$ & $2$ & $0$ & $-4$ \\ 
\multicolumn{1}{r}{$10$} & $-1$ & $-6$ & $-7$ & $0$ & $1$ & $1$ & $13$ & $5$
& $4$ & $8$ & $2$ & $8$ & $-4$ & $0$ \\ 
\br &  &  &  &  &  &  &  &  &  &  &  &  &  & 
\end{tabular}%
} 
\end{center}

\caption{$H^{LDA}$ in meV{\ for YTiO$_{3}$.} }
\label{tableYLong}
\begin{center}
{\setlength{\tabcolsep}{3pt} 
\begin{tabular}{crrrrrrrrrrrrrr}
\br $xyz$ & $000$ & $001$ & $010$ & $100$ & $011$ & $10\bar{1}$ & $01\bar{1}$
& $101$ & $110$ & $1\bar{1}0$ & $111$ & $11\bar{1}$ & $\bar{1}11$ & $1\bar{1}%
1$ \\ 
$m^{\prime },m$ &  &  &  &  &  &  &  &  &  &  &  &  &  &  \\ 
I,I & $-38$ & $62$ & $-39$ & $-39$ & $-7$ & $-7$ & $-24$ & $-24$ & $-39$ & $%
-11$ & $-9$ & $-9$ & $7$ & $7$ \\ 
II,II & $674$ & $44$ & $-20$ & $-20$ & $7$ & $7$ & $1$ & $1$ & $46$ & $54$ & 
$0$ & $0$ & $-12$ & $-12$ \\ 
III,III & $706$ & $127$ & $-58$ & $-58$ & $10$ & $10$ & $-9$ & $-9$ & $-48$
& $-60$ & $17$ & $17$ & $6$ & $6$ \\ 
II,III & $59$ & $22$ & $5$ & $30$ & $-3$ & $-22$ & $-14$ & $-20$ & $-9$ & $%
-12$ & $-20$ & $10$ & $2$ & $0$ \\ 
III,II & $59$ & $22$ & $30$ & $5$ & $-22$ & $-3$ & $-20$ & $-14$ & $-9$ & $%
-12$ & $10$ & $-20$ & $0$ & $2$ \\[10pt] 
\multicolumn{1}{r}{} &  &  &  &  &  &  &  &  &  &  &  &  &  &  \\ 
\multicolumn{1}{r}{$002$} & $020$ & $200$ & $2\bar{1}0$ & $\bar{1}20$ & $102$
& $012$ & $0\bar{1}2$ & $\bar{1}02$ & $201$ & $\bar{2}01$ & $02\bar{1}$ & $%
021$ & $210$ & $120$ \\ 
&  &  &  &  &  &  &  &  &  &  &  &  &  &  \\ 
\multicolumn{1}{r}{$-35$} & $-60$ & $-32$ & $5$ & $5$ & $1$ & $1$ & $0$ & $0$
& $-14$ & $-14$ & $32$ & $32$ & $-14$ & $-14$ \\ 
\multicolumn{1}{r}{$-15$} & $-11$ & $59$ & $-4$ & $-4$ & $-9$ & $-9$ & $-11$
& $-11$ & $-10$ & $-10$ & $-8$ & $-8$ & $9$ & $9$ \\ 
\multicolumn{1}{r}{$25$} & $-8$ & $6$ & $-9$ & $-9$ & $-1$ & $-1$ & $8$ & $8$
& $0$ & $0$ & $1$ & $1$ & $-7$ & $-7$ \\ 
\multicolumn{1}{r}{$13$} & $-4$ & $49$ & $-8$ & $-19$ & $8$ & $-5$ & $-3$ & $%
13$ & $-17$ & $-14$ & $12$ & $5$ & $-20$ & $-10$ \\ 
\multicolumn{1}{r}{$13$} & $-4$ & $49$ & $-19$ & $-8$ & $-5$ & $8$ & $13$ & $%
-3$ & $-14$ & $-17$ & $5$ & $12$ & $-10$ & $-20$ \\ 
\br &  &  &  &  &  &  &  &  &  &  &  &  &  & 
\end{tabular}
} 
\end{center}
\end{table}

\begin{figure}[t]
\par
\begin{center}
\rotatebox{270}
{\includegraphics[height=\textwidth]{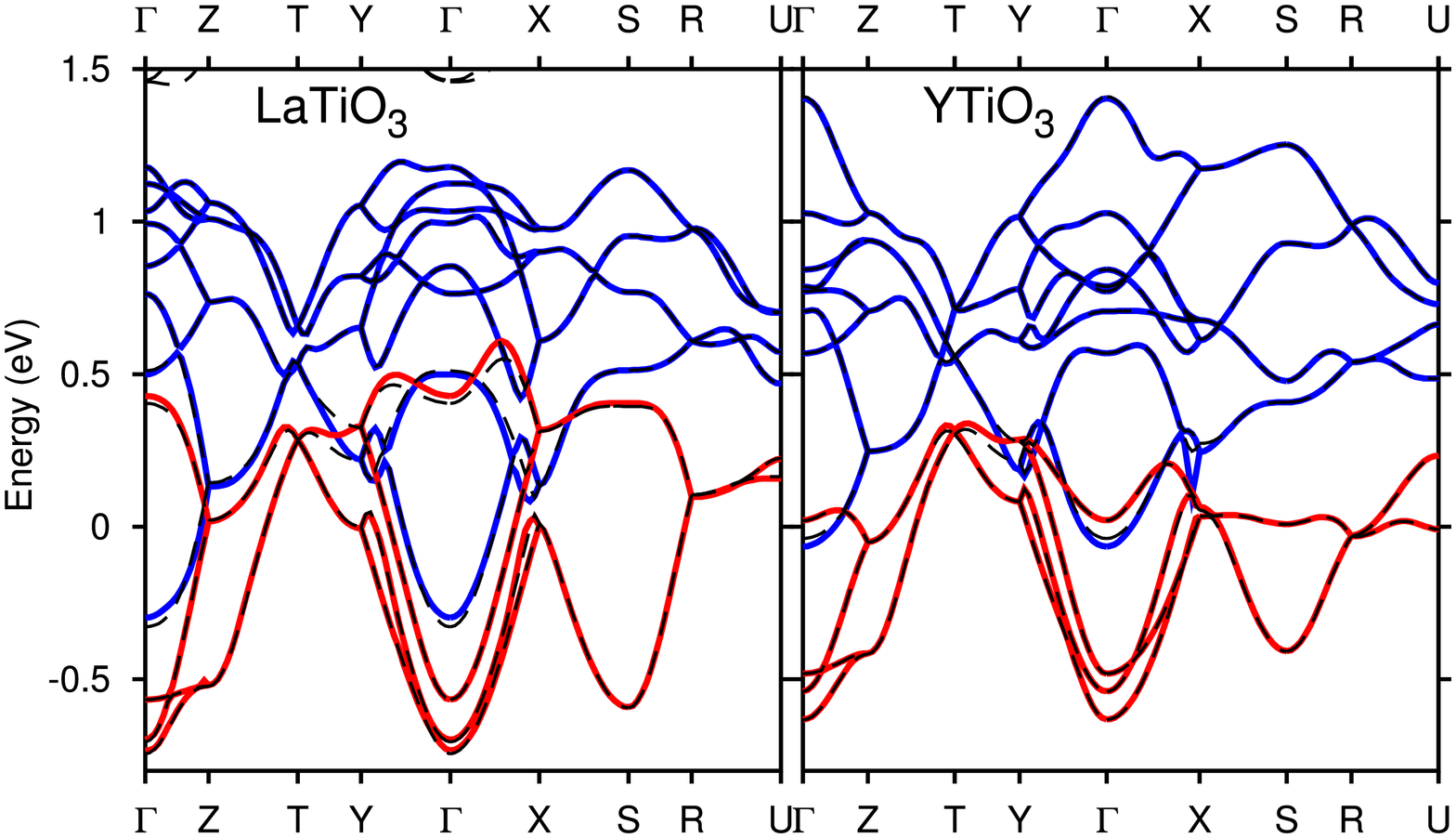}}
\end{center}
\caption{LDA bands obtained with a large NMTO basis set (dashed black) and
with truly minimal sets, downfolded to respectively the lowest $t_{2g}$
crystal-field orbital $\left\vert \mathrm{I}\right\rangle $ (red) and the
two highest crystal-field orbitals $\left\vert \mathrm{II}\right\rangle $
and $\left\vert \mathrm{III}\right\rangle $ (blue). The following NMTO
energy meshes were used: for the LaTiO$_{3}$ red bands $\protect\epsilon %
_{n}=$$-0.55$, $-0.45$, and 0.04 eV, and for the blue bands $\protect%
\epsilon _{n}=$$-0.94$, 0.70, 1.11 and 1.17 eV. For YTiO$_{3},$ $\protect%
\epsilon _{n}=$$-0.25$, $-0.12$, and 0.40 eV for the red bands, and $\protect%
\epsilon _{n}=$$-0.80$, 0.70, 1.04, 1.31, and 1.65 eV for the blue bands.}
\label{fig15}
\end{figure}

\begin{figure}[t]
\par
\begin{center}
{\includegraphics[width=.9\textwidth]{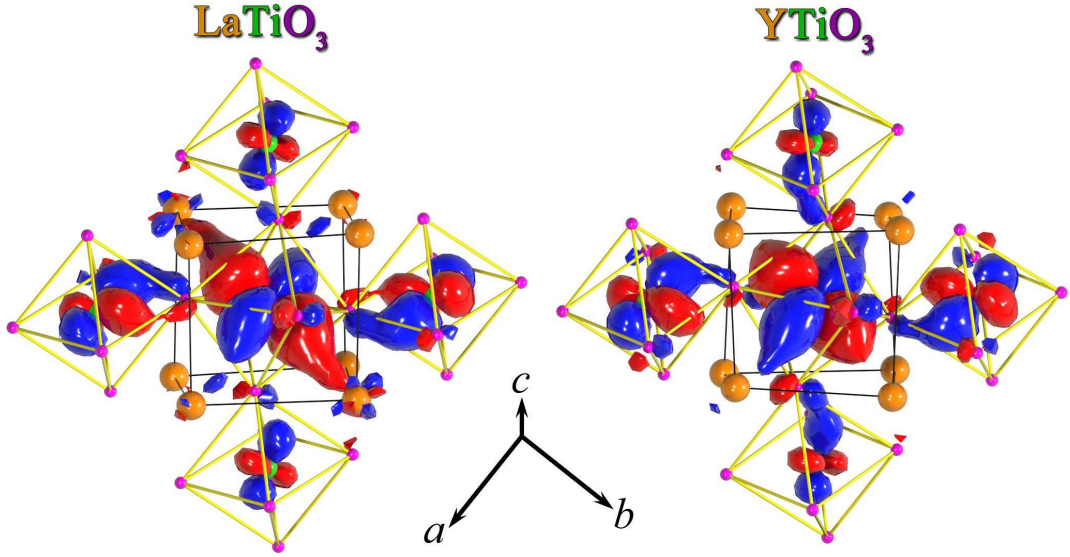}}
\end{center}
\caption{Extended crystal-field NMTO $\left\vert \mathrm{I}\right\rangle $
spanning the red LDA bands of figure \protect\ref{fig15} and with the same
local character as $\left\vert 1\right\rangle $ in figure \protect\ref{fig7}%
 Compared with that figure, the present viewpoint is more from the top,
looking down on a mirror-plane. The central part of the orbital is in
subcell 1 000 and the rest is in subcells 2: 010 to the left and
right, and 100 up and down. The amplitude in subcell 3 001
is smaller and has been truncated for the sake of clarity. The contour is
the same as in figure \protect\ref{fig7}. This orbital (before
orthogonalization) has 63\% of its charge density on the central A$_{8}$BO$%
_{6}$ unit, and this holds for both titanates. When moved to the
neighbouring Ti sites, this orbital follows the space-group symmetry, like
orbital $\left\vert 1\right\rangle $ in figure \protect\ref{fig8} (orbital
order).}
\label{fig16}
\end{figure}

The way we arrived at this result was, first, to see whether with the NMTO
method we could produce \emph{one} Wannier function which describes the four
lowest orthorhombic bands and has the character of the lowest crystal-field
orbital. The result of this attempt is demonstrated in figure \ref{fig15}.
The dashed black bands are the correct LDA bands, as obtained with a large
NMTO basis set, and the 4 red bands are obtained with a \emph{truly minimal}
NMTO \emph{basis set} containing only \emph{one} $t_{2g}$ orbital per site,
the one with the symmetry of the lowest crystal-field orbital, \textit{i.e.}%
\thinspace with the eigenvector $\left\vert 1\right\rangle $ defined in
equations (\ref{eq13}) and (\ref{eq15}) for LaTiO$_{3}$ and YTiO$_{3},$
respectively. Since this orbital is \emph{not} the partner of two other $%
t_{2g}$ orbitals, which together with it would span the entire $t_{2g}$
band, the $\left\vert 2\right\rangle $ and $\left\vert 3\right\rangle $
characters (figure \ref{fig7}) are downfolded into the tail of this new, 
\emph{extended} $\left\vert 1\right\rangle $ orbital, which we shall name $%
\left\vert \mathrm{I}\right\rangle $ and show in figure \ref{fig16}.
Moreover, the energy mesh of $\left\vert \mathrm{I}\right\rangle $ was
chosen to span only the occupied part of the $t_{2g}$ band. The blue bands
in figure \ref{fig15} are obtained from \emph{another} truly minimal NMTO 
\emph{basis set} with \emph{two} $t_{2g}$ orbitals per site, $\left\vert 
\mathrm{II}\right\rangle $ and $\left\vert \mathrm{III}\right\rangle $, with
respectively $\left\vert 2\right\rangle $ and $\left\vert 3\right\rangle $
character, and with $\left\vert 1\right\rangle $ downfolded in the tails.
Its energy mesh is mostly in the empty part of the $t_{2g}$ band (%
appendix A). 
The marvel is that orbital $\left\vert \mathrm{I}\right\rangle $ picks 4
of the 12 $t_{2g}$ bands, that orbitals $\left\vert \mathrm{II}\right\rangle 
$ and $\left\vert \mathrm{III}\right\rangle $ together pick the remaining 8 $%
t_{2g}$ bands, and that the 4th band near $\Gamma $ is the lowest of the
blue bands and has exclusively $\left\vert \mathrm{III}\right\rangle $
character. The fact that such a decomposition of the bandstructure is
possible is not trivial, at least not from the point of view of the
orthorhombic bands. Since all the LDA Bloch states are orthonormal, so are
the three extended crystal-field Wannier functions, to the extent that they
span the LDA bands, and --as figure \ref{fig15} demonstrates-- they do this
with good accuracy.

Comparison of the bands in figure \ref{fig15} with the ones decorated with
the local 1,2,3 characters in figure \ref{fig3b} confirms that band I picks
four low bands with dominant 1-character. The orbital which describes those
four bands has 1-character in its head and the minority characters 2 and 3
in its tail, on the neighbouring Ti sites. The way in which it acquires a
low energy is for a given neighbour to pick that linear combination of
orbitals 2 and 3 which provides the \emph{same} orientation of the oxygen $p$
function to be shared, and then to add this linear combination with the 
\emph{opposite} phase and an appropriate amplitude. In this way, the
resulting oxygen $p$ function will \emph{bond} with the Ti characters on the
neighbours.\medskip

The Hamiltonian for band I and the one for bands II and III are now
orthonormalized and Fourier transformed. The resulting on-site and hopping
matrix elements are given in tables \ref{tableLaLong} and \ref{tableYLong}. 
Remember that, by virtue of our NMTO construction, there is no LDA
interaction between orbital I and the two other orbitals. We note that the
splitting between level I and the nearly degenerate levels II and III is
huge, 0.7 eV. The bands have therefore really been separated and what we
have is extreme orbital order at the LDA level. For LaTiO$_{3}$ (YTiO$_{3}), 
$ the energy of orbital I has been lowered by 0.35 (0.30) eV from that of
orbital 1, and the energies of orbitals II and III have been raised from
those of orbitals 2 and 3 by, in total, the same amount. This has been
achieved by adding to orbital 1 \emph{bonding }2- and 3-characters at the
neighbouring sites, as was mentioned above, and by adding to orbitals 2 and
3 \emph{antibonding} 1-character at the neighbours. It is therefore also
obvious that we have got rid of inter-orbital hopping at the expense of
inter-site Coulomb repulsion and Hund's-rule coupling, without which the
ferromagnetism of YTiO$_{3}$ cannot be explained. For the set of extended
crystal-field orbitals, the Fermi level almost coincides with the lowest
level; it is slightly below in LaTiO$_{3}$ and a bit above in YTiO$_{3}$.
This is because the 4th band at $\Gamma $ (band III) dips deeper below the
Fermi level in LaTiO$_{3}$ than in YTiO$_{3}.$

One might argue that these Wannier orbitals are far too extended to be of
any use for describing correlated electrons. The way to think about these
orbitals is, however, as the initial, $\Sigma $=0 orbitals in a
self-consistent LDA+DMFT approach like the one recently suggested \cite%
{Anisimov05}. As self-consistency is approached, the self-energy will
separate the bands and the extended Wannier orbitals will localize so that $%
\left\vert \mathrm{I}\right\rangle \rightarrow \left\vert 1\right\rangle ,$ $%
\left\vert \mathrm{II}\right\rangle \rightarrow \left\vert 2\right\rangle ,$
and $\left\vert \mathrm{III}\right\rangle \rightarrow \left\vert
3\right\rangle $. So what the initial orbitals do is to tell us in which
way, if any, \textquotedblleft the chemistry\textquotedblright\ wants to
break the symmetry. In the present case of $t_{2g}^{1}$ titanates, it is
clear that the chemistry tells us that we have \emph{one} half-filled band,
rather than three degenerate $\frac{1}{6}$-full bands. For this one band the
DMFT will finally create a self-energy which will separate it into a lower
and an upper Hubbard band, provided that the width of this band is
sufficiently small.

For orbital I, the 1st-nearest neighbour hopping integral perpendicular to
the mirror plane is positive, while those parallel to the plane are
negative, and merely half the size. Moreover, the hopping integrals in YTiO$%
_{3}$ are half the size of those in LaTiO$_{3}.$ A way to analyze the
hoppings would be to express $\left\vert \mathrm{I}\right\rangle $ as a
linear combination of $\left\vert 1\right\rangle ,$ $\left\vert
2\right\rangle ,$ and $\left\vert 3\right\rangle $ on the same site and on
the 6 neighbour sites. This would, for instance, explain why the hopping is
twice as strong in the $z$ direction as in the parallel direction, despite
the fact --obvious from figure \ref{fig16}-- that the orbital is most
extended in the $y$ direction and least in the $z$ direction. As we have
experienced before, hopping integrals are sums of terms with alternating
signs.

For LaTiO$_{3}$ and YTiO$_{3}$ we have thus succeeded in decoupling \emph{one%
} orbital and in removing the coupling between the two other orbitals
approximately. The consequence of this is that we can \emph{fold} the band
structure \emph{out} in a \emph{pseudo-cubic} Brillouin zone (see figures \ref%
{fig1e} and \ref{fig5}). The reason is that the group of covering operations
for a \emph{single} orbital on the B sites is \emph{cyclic,} provided that
the orbital is defined as in the \emph{bottom} row of figure \ref{fig4}. As
a result, we can now let $\exp i\mathbf{k\cdot R}$ be the irreducible
representations of this cyclic group with $\mathbf{R}$ being primitive
monoclinic translations. The meaning of $\mathbf{k}$ in the large, primitive
monoclinic BZ is thus different from that in section \ref{DistBS}, where we
combined the B-centred orbitals, $\left\vert \mathbf{R},m\right\rangle ,$
into Bloch sums following the middle-row convention of figure \ref{fig4}. In
the present, so-called \emph{pseudo-cubic} scheme, we still use equations (%
\ref{4Q}) and (\ref{Blochcub}), but there is only a single orbital shape, $%
\left\vert \mathbf{R}\right\rangle ,$ and there will be \emph{no} coupling
between the four $\mathbf{Q}$ vectors.

\begin{figure}[t]
\par
\begin{center}
\rotatebox{270} {\includegraphics[height=\textwidth]{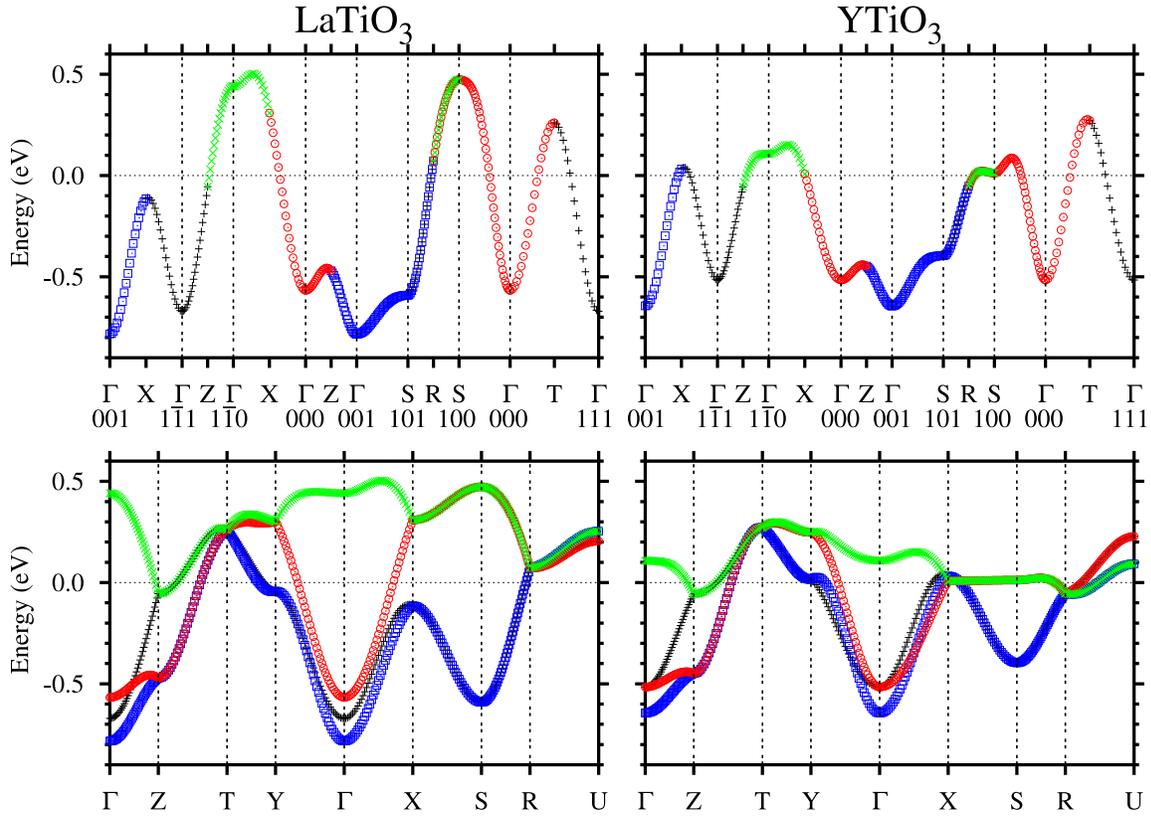}}
\end{center}
\caption{Folding band I out from the orthorhombic (bottom) to the
pseudo-cubic BZ (top). See figure 9. 
In the orthorhombic zone with 4 sites, $\mathbf{R}$,
the eigenvectors are $\frac{1}{2}\exp i\mathbf{Q\cdot R}$ with $\mathbf{Q}%
=000$ (red), $00\protect\pi $ (blue), $\protect\pi \protect\pi 0$ (green),
and $\protect\pi \protect\pi \protect\pi $ (black). The unit along the
abscissa is $\protect\pi .$ The bands were produced using tables \protect\ref%
{tableLaLong} and \protect\ref{tableYLong}. The pseudo-cubic bands are given
by equation (\protect\ref{eq8}). The letters denote high-symmetry points in
the orthorhombic BZ and the path taken along the simple monoclinic BZ is the
same as in figure \protect\ref{fig3a}, but the pseudo-cubic $\mathbf{k}$
includes the orbital order.}
\label{fig19}
\end{figure}

For band I of LaTiO$_{3}$ and YTiO$_{3}$ the unfolding is demonstrated in
figure \ref{fig19}. For a single orbital, the band dispersion is given by
equation (\ref{eq8}), an amazingly simple result considering the complicated
orthorhombic bandstructures at the bottom of figure \ref{fig19}, not to
speak of figure \ref{fig3}. Since the dominating nearest-neighbour hopping
integral, $t^{001},$ is positive and the perpendicular ones, $t^{010}\mathrm{%
=}t^{100},$ are negative, the bottom of the band is at $00\pi $ and the top
is in the $k_{z}$=0 plane near $\pi \pi 0$ and $\pi 00.$ This may seem
unfamiliar, but is related to the necessity of using the orbital-order
convention at the bottom row of figure \ref{fig4} instead of the physical
convention in the middle row. We shall return to this later when we consider
the cubic $t_{2g}$ bands

\begin{figure}[t]
\par
\begin{center}
\rotatebox{270} {\includegraphics[height=%
\textwidth]{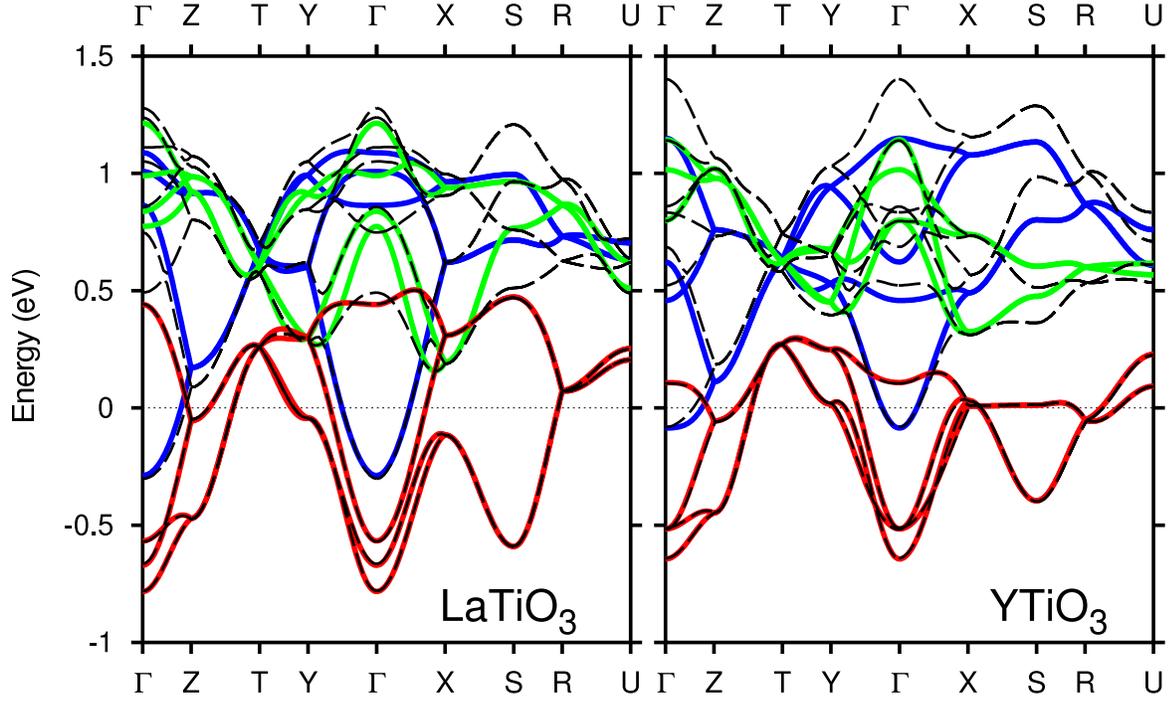}}
\end{center}
\caption{Hybridized (dashed black) and unhybridized I (red), II (green), and
III (blue) bands produced from the on-site and hopping integrals in tables 
\protect\ref{tableLaLong} and \protect\ref{tableYLong}.}
\label{fig21}
\end{figure}

In figure \ref{fig21} we include the hybridized bands II and III, this time
in the orthorhombic zone because the hybridization cannot be folded out. The
result is the dashed black band structure, which is the $t_{2g}$ band
structure, but with all hops longer than to the 120 neighbours truncated.
The red band structure is that of band I, which is identical to the four
lowest black bands, and to the bands in the bottom panel of figure \ref%
{fig19}. The green and blue bands are, respectively, the unhybridized bands
II and III, and they are seen to be reasonably accurate below, and up to 0.3
eV above, the Fermi level.

\begin{figure}[t]
\par
\begin{center}
\includegraphics[width=0.86\textwidth]{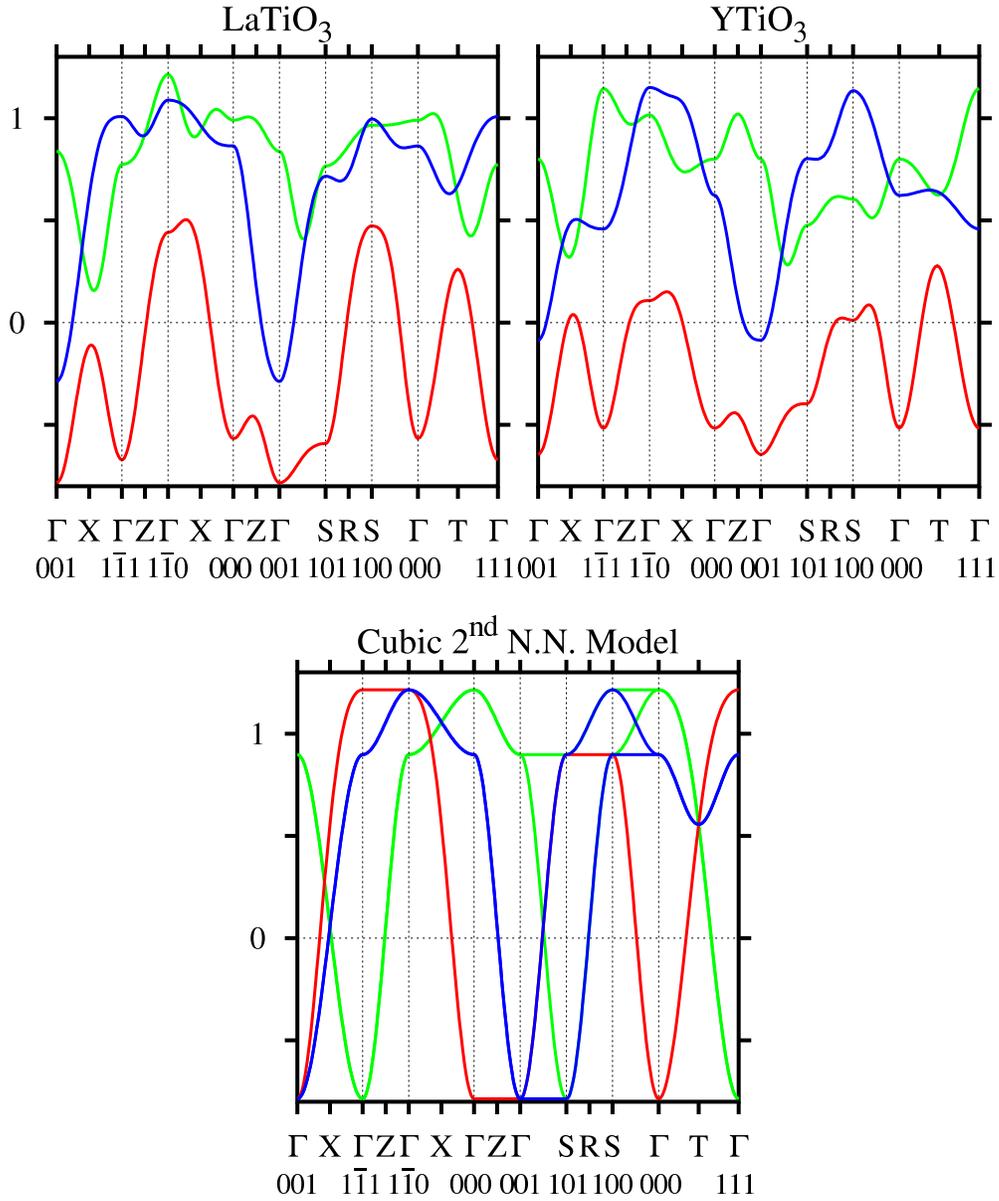}
\end{center}
\caption{Bandstructures in the pseudo-cubic BZ (figure \protect\ref{fig5}).
For the titanates, the red band (I) is the same as in figure \protect\ref%
{fig19}, and the green (II) and blue (III) bands are identical with those in
figure \protect\ref{fig21}, but folded out. The hybridization between II and
III is neglected. The bands are described by equation (\protect\ref{eq8})
with the parameters given in tables \protect\ref{tableLaLong} and  \protect
\ref{tableYLong}. For the cubic model, given by equation (\protect\ref{eq9})
with $t_{\protect\pi }=-250$ meV, $t_{\protect\sigma }^{\prime }/t_{\protect%
\pi }=r=0.34$ and $t_{\protect\delta }=0,$ the red band (I) is $xy,$ the
green bands are $\protect\varepsilon _{yz}\left( \mathbf{\bar{k}}\!-\! 
\protect\pi \protect\pi \protect\pi \right) $ and $\protect%
\varepsilon _{xz}\left( \mathbf{\bar{k}}\!-\! \protect\pi \protect\pi 
\protect\pi \right) ,$ and the blue bands are $\protect\varepsilon %
_{yz}\left( \mathbf{\bar{k}}\!-\! 00\protect\pi \right) $ and $%
\protect\varepsilon _{xz}\left( \mathbf{\bar{k}}\!-\! 00\protect\pi %
\right) $. These $yz$ and $xz$ bands are degenerate in the $\left(
\pm 110\right) $ planes containing $00\protect\pi $, $000$, $\pm \protect\pi 
\protect\pi 0$, and $\pm \protect\pi \protect\pi \protect\pi .$ Here, they
therefore equal their respective averages, namely $\protect\varepsilon %
\left( \mathbf{\bar{k}-} \protect\pi \protect\pi \protect\pi 
\right) ,$ the green $yz+xz$ band II, and $\protect\varepsilon \left( 
\mathbf{\bar{k}-} 00\protect\pi \right) ,$ the blue $yz-xz$
band III. The cubic bands are roughly those of SrVO$_{3}$ and are shown
along the same path in reciprocal space as in figure \protect\ref{fig3a}.
The unit along the abscissa is $\protect\pi .$}
\label{fig22}
\end{figure}

With the hybridization between bands II and III neglected, also these bands
may be folded out: all three bands are then given by equation (\ref{eq8}),
and they are shown in figure \ref{fig22}. We see that the lowest band is
separated from the two others by a direct gap. Hence, the \emph{pseudo-gap}
is a real gap in pseudo-cubic $\mathbf{k}$ space, but with the lowest and the
upper bands overlapping in energy.

Before continuing the discussion, let us first explain how the pseudo-gap
arises under increasing GdFeO$_{3}$-type distortion. 
This we did in section \ref%
{DistBS} by using conventional $\mathbf{Q}$ coupling and considering the
orthorhombic S$_{o}$ point. Now, we want to use the pseudo-cubic $\mathbf{k}$
representation and connect back to the cubic limit, with its three
degenerate, non-interacting, two-dimensional bands. That limit is given in
the bottom row of figure \ref{fig22}, where we used the
2nd-nearest-neighbour model (\ref{eq9}) with $t_{\delta }=0.$ However, the
pseudo-cubic representation is valid only if we neglect hybridization between
the three Bloch waves defined \emph{with orbital order} (bottom row of
figure \ref{fig4}). This means that we first have to break the cubic
symmetry and prepare the cubic bands for the GdFeO$_{3}$-type distortion. In
other words, we have to pick band I, and this will of course be the $xy$
band because this is the band for which the orbital order does not break the
physical coupling (for $xy$, the bottom and middle-row definitions in figure %
\ref{fig4} are the same). The $k_{x},k_{y}$ dispersion is thus given by
equation (\ref{eq9}), also in the pseudo-cubic zone; this is the red band in
figure \ref{fig22}. With $t_{\delta }=0,$ this band has no $k_{z}$
dispersion and its minimum is at $00k_{z} ,$ the
saddle-points at $\pi 0k_{z}$ and $0\pi k_{z} ,$ 
and the maximum at $\pi\pi k_{z}.$
The bandwidth is $8\left\vert t_{\pi }\right\vert ,$ and with increasing $%
t_{\sigma }^{\prime }/t_{\pi }\equiv r,$ the energy of the saddle-points
shifts away from the minimum towards the maximum, \textit{i.e.} the lower
part of the band is stretched. The value of $r$ chosen for the figure is
that of SrVO$_{3},$ whose cubic band structure was shown along the same path
in figure \ref{fig3a}. Moreover, $t_{\pi }=-250$ meV, which is numerically a
bit smaller than that of SrVO$_{3},$ but is like those of CaVO$_{3}.$

Next, we consider the $yz$ and $xz$ bands. Lets us be pedagogical and
perform their unfolding explicitly. The hopping Hamiltonian in the \emph{%
four-site orthorhombic} Bloch representation is given by:%
\[
\fl\fbox{$%
\begin{array}{rccccccccc}
H/\left( 2\left\vert t_{\pi }\right\vert \right) &  & yz\left( 1\right) & 
xz\left( 2\right) & -yz\left( 3\right) & -xz\left( 4\right) & xz\left(
1\right) & yz\left( 2\right) & -xz\left( 3\right) & -yz\left( 4\right) \\ 
&  &  &  &  &  &  &  &  &  \\
yz\left( 1\right) &  & 0 & 0 & c_{z} & 0 & 0 & -c_{y} & 0 & 2rc_{y}c_{z} \\ 
xz\left( 2\right) &  & 0 & 0 & 0 & c_{z} & -c_{x} & 0 & 2rc_{x}c_{z} & 0 \\ 
-yz\left( 3\right) &  & c_{z} & 0 & 0 & 0 & 0 & 2rc_{y}c_{z} & 0 & -c_{y} \\ 
-xz\left( 4\right) &  & 0 & c_{z} & 0 & 0 & 2rc_{x}c_{z} & 0 & -c_{x} & 0 \\ 
xz\left( 1\right) &  & 0 & -c_{x} & 0 & 2rc_{x}c_{z} & 0 & 0 & c_{z} & 0 \\ 
yz\left( 2\right) &  & -c_{y} & 0 & 2rc_{y}c_{z} & 0 & 0 & 0 & 0 & c_{z} \\ 
-xz\left( 3\right) &  & 0 & 2rc_{x}c_{z} & 0 & -c_{x} & c_{z} & 0 & 0 & 0 \\ 
-yz\left( 4\right) &  & 2rc_{y}c_{z} & 0 & -c_{y} & 0 & 0 & c_{z} & 0 & 0%
\end{array}%
$} 
\]%
where the rows and columns have been ordered so that the first 4 Bloch waves
belong to the middle, and the last 4 Bloch waves to the right column in the
bottom row of figure \ref{fig4}. After the unitary transformation (\ref%
{Blochcub}), $H/\left( 2\left\vert t_{\pi }\right\vert \right) $ becomes:%
\[
\fl\fbox{$\setlength{\arraycolsep}{.2pt}%
\begin{array}{cccccccc}
yz\!\left( 000\right) & xz\!\left( 00\pi \right) & -\!yz\!\left( \pi \pi
0\right) & -\!xz\!\left( \pi \pi \pi \right) & xz\!\left( 000\right) & 
yz\!\left( 00\pi \right) & -\!xz\!\left( \pi \pi 0\right) & -\!yz\!\left(
\pi \pi \pi \right) \\ 
&  &  &  &  &  &  &  \\ 
c_{z} & 0 & 0 & 0 & -\!\left( \!1\!\!-\!\!2rc_{z}\!\right) \!u & 0 & 
-\!\left( \!1\!\!-\!\!2rc_{z}\!\right) \!v & 0 \\ 
0 & -c_{z} & 0 & 0 & 0 & -\!\left( \!1\!\!+\!\!2rc_{z}\!\right) \!u & 0 & 
-\!\left( \!1\!\!+\!\!2rc_{z}\!\right) \!v \\ 
0 & 0 & c_{z} & 0 & \left( \!1\!\!-\!\!2rc_{z}\!\right) \!v & 0 & \left(
\!1\!\!-\!\!2rc_{z}\!\right) \!u & 0 \\ 
0 & 0 & 0 & -c_{z} & 0 & \left( \!1\!\!+\!\!2rc_{z}\!\right) \!v & 0 & 
\left( \!1\!\!+\!\!2rc_{z}\!\right) \!u \\ 
-\!\left( \!1\!\!-\!\!2rc_{z}\!\right) \!u & 0 & \left(
\!1\!\!-\!\!2rc_{z}\!\right) \!v & 0 & c_{z} & 0 & 0 & 0 \\ 
0 & -\!\left( \!1\!\!+\!\!2rc_{z}\!\right) \!u & 0 & \left(
\!1\!\!+\!\!2rc_{z}\!\right) \!v & 0 & -c_{z} & 0 & 0 \\ 
-\!\left( \!1\!\!-\!\!2rc_{z}\!\right) \!v & 0 & \left(
\!1\!\!-\!\!2rc_{z}\!\right) \!u & 0 & 0 & 0 & c_{z} & 0 \\ 
0 & -\!\left( \!1\!\!+\!\!2rc_{z}\!\right) \!v & 0 & \left(
\!1\!\!+\!\!2rc_{z}\!\right) \!u & 0 & 0 & 0 & -c_{z}%
\end{array}%
$}\;, 
\]%
where for simplicity of notation we have defined%
\[
u\equiv \frac{1}{2}\left( c_{x}+c_{y}\right) ,\quad \mathrm{and}\quad
v\equiv \frac{1}{2}\left( c_{x}-c_{y}\right) . 
\]%
Along the diagonal we only have the $k_{z}$ dispersion, as is also obvious
from the bottom row of figure \ref{fig4}. We may, however, get rid of the
off-diagonal elements proportional to $u$ by transformation to 
\[
\left\vert \mathrm{III}\right\rangle =\frac{1}{\sqrt{2}}\left( \left\vert
yz\right\rangle -\left\vert xz\right\rangle \right) \quad \mathrm{and}\quad
\left\vert \mathrm{II}\right\rangle =\frac{1}{\sqrt{2}}\left( \left\vert
yz\right\rangle +\left\vert xz\right\rangle \right) , 
\]%
where $\left\vert yz\right\rangle $ and $\left\vert xz\right\rangle $ now
refer to subcell 1, or, equivalently, follow the notation of the bottom row
in figure \ref{fig4}. This yields a block-diagonal Hamiltonian with the
following four blocks:%
\begin{eqnarray}
\fl\qquad && \fbox{$%
\begin{array}{cc}
\mathrm{II}\left( 000,\mathbf{k}\right) & \mathrm{III}\left( \pi \pi 0,%
\mathbf{k}\right) \\ 
&  \\ 
\varepsilon \left( \mathbf{k}\!-\! \pi \pi \pi  \right) & 
\delta \left( \mathbf{k}\!-\! \pi \pi \pi  \right) \\ 
\delta \left( \mathbf{k}\!-\! \pi \pi \pi  \right) & 
\varepsilon \left( \mathbf{k}\!-\! \pi \pi \pi \right)%
\end{array}%
$}\;,\quad \fbox{$%
\begin{array}{cc}
\mathrm{II}\left( \pi \pi 0,\mathbf{k}\right) & \mathrm{III}\left( 000,%
\mathbf{k}\right) \\ 
&  \\ 
\varepsilon \left( \mathbf{k}\!-\! 00\pi  \right) & \delta
\left( \mathbf{k}\!-\! 00\pi  \right) \\ 
\delta \left( \mathbf{k}\!-\! 00\pi  \right) & \varepsilon
\left( \mathbf{k}\!-\! 00\pi  \right)%
\end{array}%
$}\;,  \label{H0} \\*
\fl \qquad && \fbox{$%
\begin{array}{cc}
\mathrm{II}\left( 00\pi ,\mathbf{k}\right) & \mathrm{III}\left( \pi \pi \pi ,%
\mathbf{k}\right) \\ 
&  \\ 
\varepsilon \left( \mathbf{k}\!-\! \pi \pi 0 \right) & \delta
\left( \mathbf{k}\!-\! \pi \pi 0 \right) \\ 
\delta \left( \mathbf{k}\!-\! \pi \pi 0 \right) & \varepsilon
\left( \mathbf{k}\!-\! \pi \pi 0 \right)%
\end{array}%
$}\;,\quad \fbox{$%
\begin{array}{cc}
\mathrm{II}\left( \pi \pi \pi ,\mathbf{k}\right) & \mathrm{III}\left( 00\pi ,%
\mathbf{k}\right) \\ 
&  \\ 
\varepsilon \left( \mathbf{k}\right) & \delta \left( \mathbf{k}\right) \\ 
\delta \left( \mathbf{k}\right) & \varepsilon \left( \mathbf{k}\right)%
\end{array}%
$}\;,  \label{H1}
\end{eqnarray}%
where according to equation (\ref{eq9}),%
\begin{eqnarray*}
\varepsilon \left( \mathbf{k}\right) &\equiv &\frac{1}{2}\left[ \varepsilon
_{yz}\left( \mathbf{k}\right) +\varepsilon _{xz}\left( \mathbf{k}\right) %
\right] =2t_{\pi }\left( u+c_{z}\right) +4t_{\sigma }^{\prime }uc_{z}, \\*
\delta \left( \mathbf{k}\right) &\equiv &\frac{1}{2}\left[ \varepsilon
_{yz}\left( \mathbf{k}\right) -\varepsilon _{xz}\left( \mathbf{k}\right) %
\right] =-2t_{\pi }\left( 1+2rc_{z}\right) v,
\end{eqnarray*}%
are, respectively, the average and half the difference of the $yz$ and $xz$
band dispersions with $\mathbf{k}$ in the orthorhombic zone. The 
eight $yz$ and $xz$ bands are thus obtained by hybridization between pairs
of degenerate average bands, $\mathrm{II}\left( \mathbf{Q,k}\right) $ and $%
\mathrm{III}\left( \mathbf{Q}+ \pi \pi 0 \mathbf{,k}\right) $.
There exists such a pair of hybridized bands, $\varepsilon _{yz}\left( 
\mathbf{k-Q}\right) $ and $\varepsilon _{xz}\left( \mathbf{k-Q}\right) ,$
for each of the four $\mathbf{Q}$ vectors. These are 8 of the 12 bands seen
in figure \ref{fig3a} for SrVO$_{3}$, and the remaining 4 are the $%
\varepsilon _{xy}\left( \mathbf{k+Q}\right) $ bands. Note that the $%
\varepsilon _{x_{i}x_{j}}\left( \mathbf{k}\right) $ and $\varepsilon
_{x_{i}x_{j}}\left( \mathbf{k-Q}\right) $ bands are degenerate at the
corresponding boundary of the orthorhombic BZ (red in figure \ref{fig5}).

%
%

The orthorhombic zone may now be folded out in the $k_{z}$ direction because
the 1st and 2nd Hamiltonians (\ref{H1}) are identical with, respectively,
the 1st and 2nd Hamiltonians (\ref{H0}), if considered as functions of the
wave vector $\mathbf{\vec{k}\equiv k+Q}_{z}.$ If we therefore let the latter
run over the orthorhombic zone \emph{doubled} in the $k_{z}$ direction, the
Hamiltonians (\ref{H1}) should be dropped. This is \emph{exact} for \emph{any%
} band structure, and examination of any set of orthorhombic bands in this
paper will reveal that they \emph{can} be folded out along any vertical
path, \textit{i.e.}\thinspace a pair of bands along $\Gamma _{o}Z_{o}$ can
be folded out to one band along $000$ -- $00\pi ,$
and similarly for Y$_{o}$T$_{o}\rightarrow \frac{\pi }{2}\frac{\pi }{2%
}0 $ -- $ \frac{\pi }{2}\frac{\pi }{2}\pi ,$ and S$_{o}$R$%
_{o}\rightarrow  \pi 00$ -- $\pi 0\pi$. This is
also the reason why all orthorhombic bands are at least 2-fold degenerate on
the horizontal face $k_{z}=\frac{\pi }{2}.$ The cubic bands folded out to
the double orthorhombic zone are the $yz$ and $xz$ bands (\ref{eq9})
translated to $\pi \pi \pi $ and the same bands translated to $00\pi $, as
can be seen from the Hamiltonians (\ref{H0}).

Folding out to a pseudo-cubic BZ is \emph{not} exact for a Hamiltonian, but
only for a single band. Consider, \textit{e.g.}, the cubic band Hamiltonians
(\ref{H0}) as functions of the pseudo-cubic wave vector $\mathbf{\bar{k}%
\equiv \vec{k}+Q}_{xy}=\mathbf{k}-\mathbf{Q:}$%
\[
\fbox{$%
\begin{array}{cc}
\mathrm{II}\left( 000,\mathbf{k}\right) & \mathrm{III}\left( \pi \pi 0,%
\mathbf{k}\right) \\ 
&  \\ 
\varepsilon \left( \mathbf{\bar{k}}\!-\! \pi \pi \pi  \right) & 
\delta \left( \mathbf{k}\!-\! \pi \pi \pi  \right) \\ 
\delta \left( \mathbf{k}\!-\! \pi \pi \pi  \right) & 
\varepsilon \left( \mathbf{\bar{k}}\!-\! 00\pi  \right)%
\end{array}%
$}\;,\;\;\fbox{$%
\begin{array}{cc}
\mathrm{II}\left( \pi \pi 0,\mathbf{k}\right) & \mathrm{III}\left( 000,%
\mathbf{k}\right) \\ 
&  \\ 
\varepsilon \left( \mathbf{\bar{k}}\!-\! \pi \pi \pi  \right) & 
\delta \left( \mathbf{k}\!-\! 00\pi \right)  \\ 
\delta \left( \mathbf{k}\!-\! 00\pi \right)  & \varepsilon
\left( \mathbf{\bar{k}}\!-\!  00\pi \right) %
\end{array}%
$}\;. 
\]%
We have obviously succeeded in folding out the diagonal elements, but the 
\emph{hybridization} cannot be folded out. In the pseudo-cubic zone, we have
two bands, namely the average band, $\varepsilon \left( \mathbf{k}\right) ,$
translated to the sites $00\pi $ and $\pi \pi \pi .$ But there is no way in
which these two bands can hybridize in the pseudo-cubic zone to yield the
proper cubic bands. That can only occur after translation by $\pi \pi 0,$ 
\textit{i.e.}\thinspace by returning to the double orthorhombic zone.
Although important for nearly cubic bands, the neglected coupling between
bands II and III causes no problem for the lowest $\frac{1}{3}$ of the $%
t_{2g}$ band in the titanates. This must be so because we have succeeded in
downfolding band I correctly with the NMTO method.

The cubic bands are given in the bottom part of figure \ref{fig22}. In red $%
\varepsilon _{xy}\left( \mathbf{\bar{k}}\right) $, in green $\varepsilon
_{yz}\left( \mathbf{\bar{k}}- \pi \pi \pi  \right) $ and $%
\varepsilon _{xz}\left( \mathbf{\bar{k}}- \pi \pi \pi  \right)
, $ and in blue $\varepsilon _{yz}\left( \mathbf{\bar{k}}- 00\pi
 \right) $ and $\varepsilon _{xz}\left( \mathbf{\bar{k}}\!-\!
00\pi \right) $. These $yz$ and $xz$ bands are degenerate, and
therefore equal their respective averages, $\varepsilon \left( \mathbf{\bar{k%
}-} \pi \pi \pi  \right) $ with wave function $yz+xz$ and $%
\varepsilon \left( \mathbf{\bar{k}-} 00\pi  \right) $ with wave
function $yz-xz,$ in the $\left( \pm 110\right) $ planes containing $00\pi $%
, $000$, $\pm \pi \pi 0$, and $\pm \pi \pi \pi $.

In this pseudo-cubic representation we may now follow the development of the
bands as a function of the GdFeO$_{3}$-type distortion. For this purpose it is
helpful to use the orthorhombic labelling at the bottom of figure \ref{fig22}
to relate the simple pseudo-cubic bands to the cubically averaged bands in
figure \ref{fig3a} and to the projected orthorhombic bands in figure \ref%
{fig3b}. The pseudo-cubic bands of CaVO$_{3}$ are intermediate between those
of the cubic 2nd-nearest neighbour model and those of LaTiO$_{3}.$ Starting
from the cubic bands, the 45\thinspace meV coupling between the $yz$ and $xz$
orbitals in CaVO$_{3}$ (table \ref{tableCa}), produces a 90 meV gap between
the green and blue bands at X$_{o}$ $\frac{\pi }{2}\frac{-\pi }{2}\pi
$, and similarly at Y$_{o}$ $\frac{\pi }{2}\frac{\pi }{2}\pi
$, just above the Fermi level. The couplings between the other pairs
of orbitals are $\sim $30\thinspace meV and the 3-fold degenerate
red-green-blue level at T$_{o}$ $\frac{\pi }{2}\frac{\pi }{2}\frac{\pi }{%
2}$ splits well above $\varepsilon _{F}$. Along $\Gamma _{o}$
$00\pi$ -- S$_{o}$ $\pi 0\pi$ -- R$_{o}$ $\pi 0\frac{\pi 
}{2}$ -- S$_{o}$ $\pi 00$ there is a splitting between the
red $xy$ and the green band almost at the Fermi level, and at high energy
there is a splitting of all three bands. Moreover, the $xy$-band is lowered
by $\sim $70 meV due to the crystal-field splitting. Finally, going along $%
\Gamma _{o}$ $\pi \bar{\pi} \pi $ -- Z$_{o}$ $\pi \bar{\pi} \frac{\pi 
}{2}$ -- $\Gamma _{o}$ $\pi \bar{\pi} 0$ -- X$_{o}$ $\frac{%
\pi }{2}\frac{\bar{\pi} }{2}0 $ -- $\Gamma _{o}$ $ 000$ there
is a red-blue, a green-blue, and a red-green splitting. Of these, the latter
splitting between the $xy$ and the $yz+xz$ bands is at the lowest energy. In
conclusion, due to the coupling between the $xy,$ $yz,$ and $xz$ orbitals in
CaVO$_{3}$, there are small splittings ($<$100 meV) at the crossings of the
cubic bands, and this separates off a lowest band. The top of this band is a
red-green maximum between X$_{o}$ $\frac{\pi }{2}\frac{\bar{\pi} }{2}0 
$ and $\Gamma _{o}$ $000$, and the corresponding width of the
lowest band can be read from figure \ref{fig3} or \ref{fig3b} as being 1.80
eV. This is substantially smaller than the $t_{2g}$ bandwidth (2.45 eV)
listed in table \ref{tableW}. Due to the sharp avoided crossings, the
Wannier function for the lowest band in CaVO$_{3}$ has very long range and
is certainly not a suitable basis for a Hubbard model.


We now go to the titanates, for which the red, green, and blue bands in
figure \ref{fig22} are the I, II, and III bands corresponding to the
extended Wannier functions defined earlier in this section. The strongest
nearest-neighbour hopping integral coupling between the $xy,$ $yz$ and $xz$
orbitals (tables \ref{tableLaCwik} and \ref{tableY}) has increased to 75 meV
in LaTiO$_{3}$ and 83 meV in YTiO$_{3}$, and the avoided crossings at X$%
_{o}$ $\frac{\pi }{2}\frac{\bar{\pi} }{2}\pi $, Y$_{o}$ $\frac{%
\pi }{2}\frac{\pi }{2}\pi $, and T$_{o}$ $\frac{\pi }{2}\frac{%
\pi }{2}\frac{\pi }{2}$ have increased correspondingly to 150 and
166 meV. This can be seen in figures \ref{fig3} or \ref{fig3b}, but not
really in figure \ref{fig22} where the truncation of the separate Fourier
series (\ref{eq8}) for the red and green bands has rounded off the avoided
crossings. Moreover, as may be seen from figure \ref{fig15}, numerical
truncations in the NMTO calculation of the extended Wannier functions also
produce some error in the II+III Hamiltonian near sharp avoided crossings,
most noticeably at X$_{o}$. Along $\Gamma _{o}$ $000$ -- Z$%
_{o}$ $00\frac{\pi }{2}$ -- $\Gamma _{o}$ $00\pi$
there is now a red-blue avoided crossing which was not noticed in the
2nd-nearest-neighbour model where the $xy$ band neither disperses along $%
k_{z},$ nor couples to the $yz$ and $xz$ bands$.$ As seen from figure \ref%
{fig3b}, this avoided crossing is between bands I and III near $\Gamma
_{o}$ $00\pi .$ Here, band III is purely 3-like and much more
dispersive in LaTiO$_{3}$ than in YTiO$_{3},$ and it also lies higher
because the 1-3 crystal field splitting is 205 meV in LaTiO$_{3},$ but 330
meV in YTiO$_{3}.$ These differences in dispersion and position of band 3
might be the underlying cause for the differences in the onset of optical
absorption calculated in subsection \ref{Opt} and discussed most recently by
R\"{u}ckamp \textit{et al.} \cite{Grueninger}.

Avoided crossings give rise to small tongues in the joint density of states,%
\[
J_{f}\left( \omega \right) =\sum\nolimits_{\mathbf{\vec{k}}}\delta \left\{
\omega +\varepsilon _{\mathrm{I}}\left( \mathbf{\vec{k}}\right) -\varepsilon
_{f}\left( \mathbf{\vec{k}}\right) \right\} , 
\]%
extending down to $\sim $0.15 eV, but the strong onset of $J_{f}\left(
\omega \right) $ occurs at higher energies: at 0.24\thinspace eV 
in LaTiO$_{3},$ and at 0.33\thinspace eV 
in YTiO$_{3}.$ This relatively large gap is
due to the combined effects of orbital order, strong hopping from orbital 1
to orbitals 2 and 3, and comparatively weak hopping between orbitals 1
(table \ref{tableLaYXJ}). This gives band I a relatively small width, and
it causes orbital I to have a much lower energy (0.7\thinspace eV) than
orbitals II and II. This energy is \emph{the number of nearest neighbours}
times twice a typical hopping integral between orbital 1 and orbital 2 or 3,
times some reduction factor due to the localization of the extended Wannier
functions. This may be realized from figure~\ref{fig16} and the discussion
given at the beginning of this section. Even in the basis of $xy,yz,xz$
orbitals, the inter-orbital hopping increases strongly along the series, a
fact mentioned in connection with table \ref{tabletrms}. Going now the
crystal-field basis dramatically enhances inter-orbital hopping.
Specifically, the ratio $\tau $ defined in equation (\ref{tau}) is
larger for YTiO$_{3}$ than for LaTiO$_{3}$ (table \ref{tableLaYXJ}). This is
what makes YTiO$_{3}$ ferromagnetic \emph{and} what tends to make the
relative width of band I small.

\begin{table}[tbp]
\caption{Width of the lowest band, I, in eV and the ratio $W_{\mathrm{I}}/W$}
\label{tableWI}
\par
\begin{center}
\smallskip 
\begin{tabular}{cccccc}
\br & SrVO$_{3}$\cite{srstr} & CaVO$_{3}$\cite{castr} & LaTiO$_{3}$\cite%
{lastr} & LaTiO$_{3}$\cite{Cwik03} & YTiO$_{3}$\cite{ystr} \\ 
$W_{\mathrm{I}}$ & $2.85$ & $1.80$ & $1.40$ & $1.30$ & $0.95$ \\ 
$W_{\mathrm{I}}/W$ & 1.00 & 0.74 & 0.63 & 0.62 & 0.50 \\ 
\br &  &  &  &  & 
\end{tabular}%
\end{center}
\end{table}

The pseudo-gap penetrates the density of states in figure \ref{fig6} better
in YTiO$_{3}$ than in LaTiO$_{3},$ because there is less indirect band
overlap in YTiO$_{3}$. The chemical reason is that the maxima of band I
between $\Gamma _{o}$ $ \pi \bar{\pi} 0 $ and X$_{o}$ $\frac{\pi 
}{2}\frac{\bar{\pi} }{2}0 ,$ near S$_{o}$ $\pi 00 ,$ and at T$%
_{o}$ $\frac{\pi }{2}\frac{\pi }{2}\frac{\pi }{2}$ are pushed
down by the A\thinspace $d_{3z_{111}^{2}-1}$ and $d_{xy}$ characters, as was
demonstrated in figure \ref{fig11}. In terms of effective hopping integrals,
this is what makes $t_{yz,yz}^{001}$ in YTiO$_{3}$ anomalously small, as was
explained in connection with equation (\ref{eq7}) and referred to many
times. This then points to the common reason why YTiO$_{3}$ is
ferromagnetic, both according to super-exchange theory and in the Stoner
band picture. Finally we can read off the width, $W_{\mathrm{I}},$ of the
lowest band in figures \ref{fig3} and \ref{fig3b}, and list them in table %
\ref{tableWI} together with the relative subband widths $W_{\mathrm{I}}/W$%
.
\bigskip

\section{Summary and Outlook\label{Concl}}

By means of Wannier functions obtained from \textit{ab initio}
density-functional (LDA) calculations we have studied the series of
orthorhombic perovskites ABO$_{3}$ = SrVO$_{3},$ CaVO$_{3},$ LaTiO$_{3},$and
YTiO$_{3}$ in which, under the influence of an increasing GdFeO$_{3}$-type
distortion, the single B $t_{2g}$ electron becomes increasingly localized
and undergoes a Mott transition between CaVO$_{3}$ and LaTiO$_{3}.$ The
energy bands in figure \ref{fig1} and the Wannier functions for the oxygen 2$%
p$ band in figures \ref{fig1b} and \ref{fig1c} show that covalency between
the occupied O $p$ and the empty large-cation A $s$ and, in particular, $d$
states is an important mechanism of this distortion (figure \ref{fig1d}).
The A $d$ states $pd\sigma $-bond with the \emph{same} oxygen $p$ orbitals as
those with which the empty B $t_{2g}$ states $pd\pi $-bond and, as a result,
the B $t_{2g}$ Wannier functions have, not only oxygen $pd\pi $ but also
residual A $d_{3z_{111}^{2}-1}$ and $d_{xy}$ characters (figures \ref{fig4}, %
\ref{fig11} and \ref{fig12}). This situation is very different from the one
in $e_{g}^{1}$ perovskites such as LaMnO$_{3},$ where the $e_{g}$ orbitals $%
pd\sigma $-bond --and thus cause strong JT distortion-- to \emph{different}
oxygen orbitals than those bonding to A $d$.
Through the series of $t_{2g}^{1}$ perovskites, the
increasing misalignment of the $xy,yz,$and $zx$ orbitals and the theft of O $%
p$ character by the A ions lead to a decrease of the $t_{2g}$ bandwidth, $W,$
by about 50 \% (tables \ref{tabletrms} and \ref{tableW} 
and figures \ref{fig3} and \ref{fig3a}), and the increasing
A $d_{3z_{111}^{2}-1}$ and $d_{xy}$ character leads to increasing
inter-orbital coupling, splitting of the $t_{2g}$ degeneracy (figure \ref%
{fig7} and equations (\ref{eq10})--(\ref{eq15})), and to the development of a
lowest subband (figures \ref{fig15}, \ref{fig19} and \ref{fig22}) 
with a width, $W_{I},$ which decreases
from $W$ to $\frac{1}{2}W$ (table \ref{tableWI}). To the LDA low-energy
Hamiltonian defined by the $t_{2g}$ Wannier functions we have finally added
the on-site Coulomb interaction terms.

In order to calculate the correlated spectral densities and, hence, to study
the Mott transition, Pavarini \textit{et al.} \cite{Eva03} solved this
Hubbard Hamiltonian in the single-site dynamical mean-field approximation
(DMFT) at temperatures well above those where magnetic orderings occur in
the titanates. The critical values, $U_{c}^{\prime },$ of the average
Coulomb interaction, $U^{\prime },$ required to drive the materials into the
Mott insulating state decrease by a factor two through the series (figure %
\ref{fig8a} and table \ref{tableUc}), thus reflecting not only the
decreasing $W$, but also the increasing crystal-field splittings and the
decreasing width of the lowest subband. For the titanates, the Mott
transition occurs essentially in the lowest subband, that is, the orbital
degeneracy decreases from 3 in SrVO$_{3}$ to $\sim $1 in the titanates
(figure \ref{fig10}) and, for $U^{\prime }\geq U_{c}^{\prime },$ the
occupied orbital is the Wannier function for the lowest crystal-field level.
Using $U^{\prime }\approx 3.7$\thinspace eV for all four materials, the 
spectral densities in figure \ref{fig9} not only reproduce the trend that
SrVO$_{3}$ and CaVO$_{3}$ are metals with increasing mass enhancements and
LaTiO$_{3}$ and YTiO$_{3}$ Mott insulators with increasing gaps, but also
the quantitative features of photoemission and BIS spectra. In the
titanates, the orbital polarization is nearly complete (0.91 in LaTiO$_{3}$
and 0.96 in YTiO$_{3})$ and the crystal-field splittings and orbital orders
agree well with recent experiments. The difference between the orbital
orders in La and Y titanate (figure \ref{fig8}) is quantitative, rather than
qualitative, and is caused by the dominating A character being $%
d_{3z_{111}^{2}-1}$ in LaTiO$_{3}$ and 
being $d_{xy}$ in the more heavily GdFeO$%
_{3}$-type distorted YTiO$_{3}.$ The difference in the onsets of optical
conductivities in the titanates is reasonably well accounted for (figure \ref%
{fig18}). The volume reduction needed to make LaTiO$_{3}$ a metal is
reproduced and that for YTiO$_{3}$ is predicted (figure \ref{fig13a}).

In order to calculate the magnetic orderings in the insulators at low
temperature we have assumed complete orbital order and have applied
conventional super-exchange theory to our Hubbard Hamiltonians. In accord
with experiments, we find LaTiO$_{3}$ to be a G-type antiferromagnet with
fairly isotropic exchange coupling constants, but with values three times
smaller than those obtained by from spin-wave spectra (table \ref%
{tableLa8KXJ}). We correctly find YTiO$_{3}$ to be a ferromagnet with very
small coupling constants, but their anisotropy is too large (table \ref%
{tableLaYXJ} and figure \ref{fig14}). The reason for the trend from
antiferro- towards ferromagnetism with increasing GdFeO$_{3}$-type
distortion is that the hopping between the nearest-neighbour Wannier
functions for the lowest crystal-field level decreases compared with the
hoppings to the higher-level Wannier functions. This is also the mechanism
for creating a lowest subband of reduced width. At the point where the
super-exchange coupling becomes weak -- because it changes sign -- the JT
distortion becomes the controlling factor for the magnetic order, although
it hardly influences the orbital order. We find that YTiO$_{3}$ without JT
distortion should be a C-type antiferromagnet and that YTiO$_{3}$ at 16%
\textbf{\thinspace }GPa pressure, where the JT distortion is strongly
reduced, should become an A-type antiferromagnet below $\sim 100$\thinspace
K (table \ref{tablenoJTYXJ}).

Our results concerning the role of the GdFeO$_{3}$-type and JT distortions
for the crystal-field splittings, orbital orders, and magnetic couplings are
in accord with those obtained from studies of model Hamiltonians by
Mochizuki and Imada \cite{Mochizuki04}. Both theoretical works,
together with an increasing number of experimental works, point to a
crystal-field splitting in the titanates at the order of 200 meV, \textit{%
i.e.} much larger that $kT$ and the spin-orbit splitting. This puts doubts
on the applicability to the titanates of the recent orbital-liquid theory 
\cite{Khal1,Khal2}, which is based on the assumption the  $t_{2g}$
orbitals are nearly degenerate. Although the picture we have presented
appears consistent from the chemical point of view, the fine balance between
O-A, O-B, A-B, and B-B covalencies is hardly reproduced with sufficient
accuracy to describe the magnetic exchange couplings well in these $%
t_{2g}^{1}$ materials. This may be due to the LDA, to our use of the
atomic-spheres approximation to generate the LDA potentials, and to our NMTO 
$t_{2g}$ Wannier functions being too extended for use as basis functions for
the Hubbard model. Moreover, our assumption of complete orbital order for
LaTiO$_{3}$ may be too crude considering the fact that the measured magnetic
moment is only 0.57\thinspace $\mu _{B}$ in this material. Only other types
of calculations, for instance spin-polarized LDA+DMFT calculations for LaTiO$%
_{3},$ can throw light on this in the future.

In order to prepare the grounds for future experimental and theoretical
clarification, and also to ease the applicability of the present theoretical
work to the wealth of interesting, similar $t_{2g}$ systems, we have
presented a great deal of detail about why our results come out the way they
do, and we have tabulated the LDA on-site and hopping matrix elements
(tables \ref{tableSr}, \ref{tableCa}, \ref{tableLaCwik}, and \ref%
{tableY}). Moreover, in section \ref{DistBS} we have given an analytical
expression in the $\mathbf{k}+\mathbf{Q}$ representation for the
orthorhombic Hamiltonian, into which the self-energy may be included, once
it has been extracted from calculations or experiments. In the present
LDA+DMFT calculations the $\mathbf{k}$-dependence of the self-energy was
totally neglected, but our success in explaining the Mott transition in the
series of $t_{2g}^{1}$ perovskites indicates that, for these systems, the
single-site DMFT is indeed a good approximation. 
On the other hand, since the GdFeO$_3$-type distortion severely influences
not only the on-site matrix elements, but also the hopping integrals in 
$t_{2g}$ systems, this might induce a $\mathbf{k}$-dependence of the
self-energy not found in $e_{g}$ systems.
In the near future it would
be useful to get hold of the self-energy matrix for \emph{real} frequencies,
because this would give us the correlated bandstructure ($\mathbf{k}$%
-dependent spectral function), which could then be compared with for
instance angle-resolved photoemission and dHvA experiments. For metals,
theoretical work on the electron liquid \cite{Takada} indicates that the
self-energy does have a significant $\mathbf{k}$-dependence, and that could
be checked. In the future it may be possible to use cluster-DMFT
calculations to evaluate the $\mathbf{k}$-dependence of the self-energy
for real systems.

In this paper we have also demonstrated the use of the new NMTO downfolding
technique \cite{nmto} as a tool for generating truly minimal basis sets,
Wannier functions in particular. It is of course possible to generate sets
of Wannier functions which span also the oxygen $p$ and transition-ion $e_{g}
$ LDA bands as needed when describing high-energy excitations. As another
extreme, it is sometimes possible to downfold to even fewer than three $t_{2g}
$ functions and, in such cases, discover tendencies towards symmetry
breaking, which may be exploited by the Coulomb correlations (figure %
\ref{fig15}). We may, in general, follow the localization process more
closely and accurately by generating the Wannier functions self-consistently
during the cause of a DMFT calculation \cite{Anisimov05}.

We are convinced that the rich physics of the materials studied in this
paper, as well as the computational techniques used to do so, will remain 
active fields of research for years to come.

\section*{Acknowledgments}

We are indebted to A. I. Lichtenstein, A. Georges, S. Biermann, and A.
Poteryaev for having initiated and taken part in this research at its
earlier stages. Our interest in these materials was aroused by G.
Khaliullin, B. Keimer, and C. Ulrich. The use of software developed by T.
Saha-Dasgupta as well as discussions with her are gratefully acknowledged.
With Olle Gunnarsson and Erik Koch we have enjoyed many enlightening
discussions and from them we have received more fruitful suggestions than
from anyone else. I. Loa and K. Syassen informed us about their
high-pressure experiments and allowed us to make use of their results prior
to publication. Contacts with M. Gr\"{u}ninger, D. D. Sarma and K. Maiti are
also acknowledged. N. A. Spalding kindly referred us to the work of
Woodward. Finally, we would like to thank the KITP Santa Barbara for
hospitality (NSF Grant No. PHY99-07949) and the INFM-Iniziativa Calcolo
Parallelo for support.

\addcontentsline{toc}{section}{Acknowledgements}
\appendix

\section*{Appendix A. NMTO basis sets\label{NMTO}}
\setcounter{section}{1}
\addcontentsline{toc}{section}{Appendix A. NMTO basis sets}

The $N$th-order muffin-tin orbitals (NMTO) method \cite{nmto} is more
intelligible, flexible, and accurate than its predecessor, the linear
muffin-tin orbitals (LMTO) method \cite{LMTO84}, a so-called fast
band-structure method. In the present paper, we use the NMTO method for
generation of localized Wannier functions for the Kohn-Sham band-structure
problem.

The method constructs a set of local-orbital basis functions which span the
solutions of Schr\"{o}dinger's equation --actually, the (scalar) Dirac
equation-- for a local potential, written as a superposition, $%
\sum_{R}v_{R}\left( \left\vert \mathbf{r-R}\right\vert \right) ,$ of
spherically symmetric potential wells with ranges $s_R$, 
a so-called overlapping
muffin-tin potential. This is done by first solving the radial equations
(numerically) to find $\varphi _{Rl}\left( \epsilon _{n},\left\vert \mathbf{r%
}-\mathbf{R}\right\vert \right) $ for all angular momenta, $l,$ with
non-vanishing phase-shifts, for all potential wells, $R,$ and for a chosen
set of energies spanning the region of interest, $\epsilon _{n}=\epsilon
_{1},....,\epsilon _{N}$. These energies are the ones shown on the
right-hand side of figure \ref{fig1a} for the oxygen 2$p$ bands and on the
right-hand side of figure \ref{fig3} for the B\thinspace 3$d\left(
t_{2g}\right) $ bands.

The partial-wave channels, $Rlm,$ are now partitioned into \emph{active} and
passive. The active ones are those for which one wants to have orbitals in
the basis set; \textit{i.e.,} they are the chosen one-electron degrees of
freedom. For the red bands in figures \ref{fig2} and \ref{fig3}, the B $%
d\left( t_{2g}\right) $ channels are active, while for the red bands in
figure \ref{fig1a}, the oxygen $p$ channels are active. In all figures, the
black bands are calculated with a large NMTO set having the O\thinspace $p,$
B\thinspace $spd,$ and A\thinspace $spd$ channels active.

For each active channel, $\bar{R}\bar{l}\bar{m},$ a so-called \emph{kinked}
partial wave, $\phi _{\bar{R}\bar{l}\bar{m}}\left( \epsilon _{n},\mathbf{r}%
\right) ,$ is constructed from \emph{all} the partial waves, $\varphi
_{Rl}\left( \epsilon _{n},\left\vert \mathbf{r}-\mathbf{R}\right\vert
\right) Y_{lm}\left( \widehat{\mathbf{r-R}}\right) ,$ inside the
potential-spheres, and from \emph{one} solution, $\psi _{\bar{R}\bar{l}\bar{m%
}}\left( \epsilon _{n},\mathbf{r}\right) ,$ of the wave-equation in the
interstitial, a so-called screened spherical wave. The construction is such
that the kinked partial wave is a solution of Schr\"{o}dinger's equation at
energy $\epsilon _{n}$ in all space, except at so-called hard
screening-spheres --which are concentric with the potential-spheres, but
have no overlap-- where it is allowed to have radial kinks in the \emph{%
active} channels. It is now clear that if we can form a linear combination
of such kinked partial waves with the property that all kinks cancel, we
have found a solution of Schr\"{o}dinger's equation with energy $\epsilon
_{n}.$ In fact, this kink-cancellation condition leads to the classical
method of Korringa, Kohn and Rostoker \cite{KKR} (KKR), but in a general,
so-called screened representation and valid for overlapping MT
potentials to leading order in the potential overlap. The screened KKR
equations are a set of energy-dependent, homogeneous linear equations, with
a matrix, $K_{\vec{R}\vec{l}\vec{m},\bar{R}\bar{l}\bar{m}}\left( \varepsilon
\right) ,$ whose rows and columns are labelled by the active channels. We do
not solve this set of secular equations, but proceed a bit differently:

The major computational task in the screened KKR and NMTO methods is to
construct the set of \emph{envelope} functions for the kinked partial waves,
the set of screened spherical waves, as superpositions of spherical Hankel
functions. This is known as the real-space calculation of the screened
structure-matrix, which is the non-diagonal part of the KKR matrix, $K\left(
\epsilon _{n}\right) $. Now, $\psi _{\bar{R}\bar{l}\bar{m}}\left( \epsilon
_{n},\mathbf{r}\right) $ must join smoothly onto all the \emph{passive}
partial waves, \textit{i.e.}\thinspace it must have the proper phase shifts.
For all the \emph{active} channels, except the eigenchannel, $\bar{R}\bar{l}%
\bar{m},$ it can be forced to \emph{vanish} at the screening spheres, and it
is this confinement which makes it \emph{localized,} provided that this is
possible with the actual choices of energy, partition between active
and passive channels, and screening-radii, $a_R$. Since the screened spherical wave
is required to vanish merely for the \emph{other} active channels, but not
for the eigenchannel, it is an impurity solution for the hard-sphere solid.
In order to obtain maximal localization, the hard spheres are usually chosen
to be nearly touching. The passive channels are said to be \emph{downfolded.}

As an example, let us consider the set of O\thinspace 2$p$ kinked partial
waves at the energy $\epsilon _{1}$ in figure \ref{fig1a} and assume that
this set equals the set of NMTOs shown in figures \ref{fig1b} and \ref{fig1c}%
, which is approximately true. This set consists of the three $p_{x},$ $%
p_{y},$ and $p_{z}$ orbitals on \emph{all} oxygens in the solid. The
O1\thinspace $p_{x}$ orbital shown in the 3rd column and upper row of figure %
\ref{fig1b} therefore satisfies the following conditions: at its \emph{own}
site, the $p_{y}$- and $p_{z}$-projections vanish, at all \emph{other}
oxygen sites in the solid, \emph{all} three oxygen $p$-projections vanish,
and all \emph{remaining} projections, \textit{i.e.} O\thinspace $sd..,$ La $%
spdf..,$ and Ti $spd..,$ are smooth solutions of Schr\"{o}dinger's equation
at energy $\epsilon _{1}.$ Also the oxygen $p$-projections are solutions of
Schr\"{o}dinger's equation at energy $\epsilon _{1},$ but they have kinks at
the screening-spheres.

Another example are the B $d_{xy},$ $d_{yz},$ and $d_{xz}$ NMTOs shown
figure \ref{fig4}. In order to generate the orbitals with the convention
used in the second row, the active channels were specified simply as $%
d_{xy}, $ $d_{yz},$ and $d_{xz}$ on each of the four B-sites, with $x,$ $y,$
and $z$ referring to the \emph{global} axes. This is possible when the
structure is nearly cubic, because then the orbitals can adjust their
orientation due to the freedom of, say, the $d_{xy}$-NMTO to contain any
on-site character, except $d_{yz}$ and $d_{xz},$ such as for instance $%
d_{x^{2}-y^{2}}$ and $d_{3z^{2}-1}.$

The set of NMTOs is formed as a superposition of the kinked-partial-wave
sets for the energies, $\epsilon _{1},....,\epsilon _{N}$:%
\begin{equation}
\chi _{\vec{R}\vec{l}\vec{m}}^{\left( N\right) }\left( \mathbf{r}\right)
=\sum_{n=0}^{N}\sum_{\bar{R}\bar{l}\bar{m}}\phi _{\bar{R}\bar{l}\bar{m}%
}\left( \epsilon _{n},\mathbf{r}\right) L_{n;\bar{R}\bar{l}\bar{m},\vec{R}%
\vec{l}\vec{m}}^{\left( N\right) }.  \label{eq1}
\end{equation}%
Note that the size of this NMTO basis set is given by the number of active
channels and is independent of the number, $N+1,$ of energy points. The
coefficient matrices, $L_{n}^{\left( N\right) },$ in equation (A.1)
are determined by the condition that the set of NMTOs span the
solutions, $\Psi _{i}\left( \varepsilon _{i},\mathbf{r}\right) ,$ of Schr%
\"{o}dinger's equation\emph{\ }with an error%
\begin{eqnarray}
\Psi _{i}^{\left( N\right) }\left( \mathbf{r}\right) -\Psi _{i}\left(
\varepsilon _{i},\mathbf{r}\right) &=&c^{\left( N\right) }\left( \varepsilon
_{i}-\epsilon _{0}\right) \left( \varepsilon _{i}-\epsilon _{1}\right)
...\left( \varepsilon _{i}-\epsilon _{N}\right)  \label{eq2} \\* 
&&+o\left( \left( \varepsilon _{i}-\epsilon _{0}\right) \left( \varepsilon
_{i}-\epsilon _{1}\right) ...\left( \varepsilon _{i}-\epsilon _{N}\right)
\right) .  \nonumber
\end{eqnarray}%
This is polynomial approximation for the Hilbert space of Schr\"{o}dinger
solutions and $L_{n}^{\left( N\right) }$ are the coefficients in the
corresponding Lagrange interpolation formula. An NMTO with $N>0$, has no
kinks, but merely discontinuities in the $(2N+1)$st radial derivatives at
the screening-spheres for the active channels.

The Lagrange coefficients, $L_{n}^{\left( N\right) },$ as well as the
Hamiltonian and overlap matrices in the NMTO basis are expressed solely in
terms of the KKR resolvent, $K\left( \varepsilon \right) ^{-1},$ and its
first energy derivative, $\dot{K}\left( \varepsilon \right) ^{-1},$
evaluated at the energy mesh, $\varepsilon =\epsilon _{1},...,\epsilon _{N}.$
Variational estimates of the one-electron energies, $\varepsilon _{i},$ such
as the red bands in figures \ref{fig2}, \ref{fig1a}, and \ref{fig3} may be
obtained from the generalized eigenvalue problem,%
\[
\left( \left\langle \chi ^{\left( N\right) }\left\vert \mathcal{H}%
\right\vert \chi ^{\left( N\right) }\right\rangle -\varepsilon
_{i}\left\langle \chi ^{\left( N\right) }\mid \chi ^{\left( N\right)
}\right\rangle \right) \mathbf{v}_{i}=\mathbf{0,} 
\]%
with%
\[
\mathcal{H}\equiv -\Delta +\sum_{R}v_{R}\left( \left\vert \mathbf{r-R}%
\right\vert \right) , 
\]%
or as the eigenvalues of the one-electron Hamiltonian matrix,%
\begin{equation}
H^{LDA}=\left\langle \chi ^{\left( N\right) \perp }\left\vert \mathcal{H}%
\right\vert \chi ^{\left( N\right) \perp }\right\rangle  \label{eq3}
\end{equation}%
in the basis of \emph{symmetrically orthonormalized} NMTOs:%
\begin{equation}
\left\vert \chi ^{\left( N\right) \perp }\right\rangle ~\equiv ~\left\vert
\chi ^{\left( N\right) }\right\rangle ~\left\langle \chi ^{\left( N\right)
}\mid \chi ^{\left( N\right) }\right\rangle ^{-\frac{1}{2}}.  \label{eq4}
\end{equation}

The prefactor, $c^{\left( N\right) },$ of the leading error of an NMTO set (%
\ref{eq2}) depends on the size of the set, and the larger the set, the
smaller the prefactor. For that reason we have not bothered to indicate the
energy mesh used for the black bands in figures \ref{fig2}, \ref{fig1a}, and %
\ref{fig3}. Assuming that they are exact, the red bands will touch the black
bands quadratically at the energy points used to generate the red-band set.
That they touch rather than cross is by virtue of the variational principle.

For an isolated set of bands, like the oxygen $p$-bands or the B\thinspace $%
t_{2g}$-bands, and for energy meshes spanning the range of those bands, as
the number of energy points increases and the distances between them
decrease, the set of truly minimal NMTOs will converge and be exact. Upon
orthonormalization, they will therefore form a set of localized Wannier
functions.

The construction of a minimal NMTO basis set is different from standard L%
\"{o}wdin downfolding. The latter partitions a \emph{given}, large (say
orthonormal) basis into active $\left( A\right) $ and passive $\left(
P\right) $ subsets, then finds the downfolded Hamiltonian matrix as:%
\begin{equation}
\fl \left\langle A\left( \varepsilon \right) \left\vert \mathcal{H}%
\right\vert A\left( \varepsilon \right) \right\rangle =\left\langle
A\left\vert \mathcal{H}\right\vert A\right\rangle +\left\langle A\left\vert 
\mathcal{H}\right\vert P\right\rangle \left\langle P\left\vert \varepsilon -%
\mathcal{H}\right\vert P\right\rangle ^{-1}\left\langle P\left\vert \mathcal{%
H}\right\vert A\right\rangle ,  \label{eq6}
\end{equation}%
and finally removes the $\varepsilon $-dependence of the downfolded basis by 
\emph{linearizing} $\left\langle P\left\vert \mathcal{H}-\varepsilon
\right\vert P\right\rangle ^{-1}$ and treating the term linear in $%
\varepsilon $ as an overlap matrix. Obviously, since the NMTO set is exact
at $N+1$ energy points, it is more accurate. Nevertheless, since truly
minimal NMTOs can be fairly complicated functions, we often use
equation\thinspace (\ref{eq6}) to \emph{interpret} their (orthonormalized)
Hamiltonian in terms of the Hamiltonian represented in a larger basis set
whose orbitals are simpler and more localized.

For crystals, all calculations except the generation of the screened
structure matrix are performed in the Bloch $\mathbf{k}$-representation%
\begin{equation}
\chi _{\bar{R}\bar{l}\bar{m}}^{\left( N\right) }\left( \mathbf{k,r}\right) =%
\frac{1}{\sqrt{L}}\sum_{T}\chi _{\bar{R}\bar{l}\bar{m}}^{\left( N\right)
}\left( \mathbf{r-T}\right) e^{i\mathbf{k\cdot }\left( \mathbf{\bar{R}+T}%
\right) },  \label{Bloch}
\end{equation}%
where $T$ labels the $L\ \left( \rightarrow \infty \right) $ lattice
translations and $\bar{R}$ the active sites in the primitive cell. In order
to obtain the orbitals and the Hamiltonian in configuration space,
Fourier-transformation over the Brillouin zone is performed.

In the present paper, the orbitals shown are NMTOs \emph{before}
orthonormalization (\ref{eq4}) because they are (slightly) more localized
than the orthonormalized ones. The hopping integrals and on-site elements
given in the tables are of course matrix elements of the \emph{%
orthonormalized} Hamiltonian (\ref{eq3}).

\appendix
\section*{Appendix B. Technical details of the LMTO-ASA potential 
          calculations\label{technical}}
\setcounter{section}{2}
\addcontentsline{toc}{section}{Appendix B. Technical details of the LMTO-ASA
potential calculations}

Since our present NMTO code is not self-consistent, we used the current
Stuttgart TB-LMTO-ASA code \cite{StgLMTO} to generate the LDA potentials.
Such a potential in the atomic-spheres approximation (ASA) is an overlapping
MT-potential, like the one handled by the NMTO method, but with the relative
overlaps, 
\begin{equation}
\omega _{RR^{\prime }}\equiv \frac{s_{R}+s_{R^{\prime }}}{\left\vert \mathbf{%
R}-\mathbf{R}^{\prime }\right\vert }-1,  \label{ovl}
\end{equation}%
limited to about 20\%. This limitation comes from the LMTO-ASA+cc method\cite{LMTO84},
which solves Schr\"{o}dinger's equation by treating the overlap as a
perturbation (the socalled combined-correction term, cc) \emph{and} uses
screened spherical waves of \emph{zero} kinetic energy in the $s$%
-interstitial. Poisson's equation is solved for the output charge density,
spherically symmetrized inside the \emph{same} atomic $s$-spheres. 
For a given potential, the hopping integrals obtained with the NMTO method
are more accurate than those obtained with the LMTO method,
first of all because the
NMTOs do not use the zero-kinetic-energy approximation in the interstitial
region and, secondly, because we use $N>1$ with well-chosen energy meshes.
However, since the on-site matrix elements of the LDA Hamiltonian (\ref{eq3}%
) turned out to be crucial for the present study, it is possible that our
ASA treatment of the potential leads to an \emph{under}estimation of the
electrostatic contribution to the crystal field.

We now specify our computational set-up. 
The radii of the potential spheres, $s_{R},$
were dictated by our use of the LMTO-ASA method to generate the LDA
potentials. In order to limit the overlaps defined by equation (\ref{ovl}),
interstital --or empty-- spheres (E) were inserted in the non-cubic
structures. Table \ref{tables1} gives the radii of the
potential spheres. As a result, the overlap between atomic spheres was 
$<$16\%, between atomic and empty spheres $<$18\%, and
between empty spheres $<$20\%.
We used the guidance given by the current version of the code in choosing 
the potential spheres appropriately.  

\begin{table}[tbp]
\caption{Radii $s_R$ of potential spheres in Bohr atomic units.}
\label{tables1}
\begin{center}
{\setlength{\tabcolsep}{4pt} 
\begin{tabular}{cccccccccccc}
\br
 ABO$_3$ &  & A & B & O1 & O2 & E & E1 & E2 & E3 & E4 & E5\\ 
  &  &  &  &  &  &  &  &  &  &  &\\ 
 SrVO$_{3}$ && 3.97 & 2.29 & 1.77 &   &   &   &   &   &   &  \\ 
 CaVO$_{3}$ && 3.34 & 2.33 & 1.80 & 1.80 & 1.28 & 1.25 & 1.17 & 1.22 &1.24& \\ 
 LaTiO$_{3}$\cite{lastr} 
            && 3.37 & 2.51 & 1.90 & 1.90 & 1.50 & 1.41 & 1.31 & 1.29 & 1.19 & 1.13 \\ 
 LaTiO$_{3}$\cite{Cwik03} 
            && 3.31 & 2.52 & 1.93 & 1.93 & 1.61 & 1.49 & 1.42 & 1.28 & 1.08 & 1.07 \\ 
 YTiO$_{3}$ && 2.95 & 2.51 & 1.92 & 1.90 & 1.93 & 1.83 & 1.48 & 1.44 &   &   \\%
\br
\end{tabular}
}
\end{center}
\end{table}

With these reasonably large oxygen spheres, the oxygen 2$s$ electrons could
be treated as part of the core. 
The self-consistent valence-electron densities were
calculated with the LMTO bases listed in Table~\ref{tables2}.
In order to describe properly the A--O--B covalency 
we found it important to downfold the oxygen $d$
partial waves, rather than to neglect them (\textit{i.e.} to approximate
them by spherical Bessel functions when solving Schr\"{o}dingers equation,
and to neglect them in the charge density). 
Since the LMTO calculations were used
to produced the self-consistent charge densities, the energies, $\epsilon
_{Rl},$ for the linear $\phi _{Rl},\dot{\phi}_{Rl}$ expansions were chosen
at the centres of gravity of the \emph{occupied} parts of the respective DOS 
$Rl$-projections.

With the benefit of hindsight, we should have inserted the 12 E spheres above
the octahedron edges also in cubic SrVO$_3$. 
That would have reduced $s_{\mathrm{Sr}}$ to 3.46 a.u., 
a value closer to those for the
other systems.
As a result, the $t_{2g}$ bandwidth for cubic SrVO$_3$ would be reduced by
8\%.
This seems to be the largest computational ``error'' of the present
calculations.
This inaccuracy only concerns cubic SrVO$_3$ and it reduces the decrease 
of rms bandwidth (tables 6 and 8) when going from SrVO$_3$ to CaVO$_3$, 
from 19 to 11\%. 
Hence, the trend that CaVO$_3$ is a more correlated metal than SrVO$_3$ is 
somewhat diminished. 
All remaining results, such as those concerning the Mott transition and 
the properties of the titanates, of course remain valid.

After completion of the calculations, we also found that the optimal 
trade off between the errors caused by the confinement of the O 2$s$ 
electrons to the sphere and by the overlap of spheres, is obtained with 
a larger oxygen radius, 2.04 a.u.. 
This leads to a small downwards shift of the O 2$p$ band and a 4\%  
decrease of the $t_{2g}$ bandwidth, $W$, for all four materials. 
Since all materials are influenced the same way, and since our value of $U$ 
was fitted relatively to $W$, this error  has no effect, except that 
it might influence the sensitive values of the exchange-coupling 
constants calculated in section \ref{M}.

\begin{table}[tb]
\caption{LMTO basis sets
used in the self-consistent calculation of LDA potential.
$(l)$ means that the $l$-partial waves were downfolded within
in the LMTO-ASA+cc.}
\label{tables2}
\begin{center}
\begin{textit}
{\setlength{\tabcolsep}{4pt} 
\begin{tabular}{ccllllllllll}
\br
&  & {\rm A} & {\rm B} & {\rm O1} & {\rm O2} & 
     {\rm E} & {\rm E1} & {\rm E2} & {\rm E3} & {\rm E4} & {\rm E5}\\ 
  &  &  &  &  &  &  &  &  &  &  &\\ 
 {\rm SrVO$_{3}$} && s{\rm (}p{\rm )}d{\rm (}f{\rm )} & spd & 
 {\rm (}s{\rm )}p{\rm (}d{\rm )} &   &   &   &   &   &   &  \\ 
 {\rm CaVO$_{3}$} && s{\rm (}p{\rm )}d & spd & 
 {\rm (}s{\rm )}p{\rm (}d{\rm )} & {\rm (}s{\rm )}p{\rm (}d{\rm )} & 
 s{\rm (}p{\rm )} & s{\rm (}p{\rm )} & s{\rm (}p{\rm )} & s{\rm (}p{\rm )} & 
 s{\rm (}p{\rm )}&  \\ 
 {\rm LaTiO$_{3}$}&& s{\rm (}p{\rm )}df & spd & 
 {\rm (}s{\rm )}p{\rm (}d{\rm )} & {\rm (}s{\rm )}p{\rm (}d{\rm )} & 
 s{\rm (}p{\rm )} & s{\rm (}p{\rm )} & s{\rm (}p{\rm )} & s{\rm (}p{\rm )} & 
 s{\rm (}p{\rm )}&s{\rm (}p{\rm )} \\ 
 {\rm YTiO$_{3}$} && s{\rm (}p{\rm )}d{\rm (}f{\rm )} & spd & 
 {\rm (}s{\rm )}p{\rm (}d{\rm )} & {\rm (}s{\rm )}p{\rm (}d{\rm )} & 
 s{\rm (}pd{\rm )} & s{\rm (}pd{\rm )} & s{\rm (}p{\rm )} & 
 s{\rm (}p{\rm )} &  &   \\ 
\br
\end{tabular}
}
\end{textit}
\end{center}
\end{table}

Finally, in the NMTO calculations, the hard-sphere radii, $a_{R},$ for the
active channels were chosen as 0.7$s_{R}.$ 

\section*{References}
\addcontentsline{toc}{section}{References}


\end{document}